\documentclass{article}

\usepackage[T1]{fontenc}
\usepackage[english]{babel}

\usepackage{amsfonts}
\usepackage{amssymb}
\usepackage{amstext}
\usepackage{amsmath,bm,bbm}
\usepackage{enumerate}
\usepackage{algorithm}
\usepackage{algorithmic}
\usepackage[normalem]{ulem}
\usepackage{comment}

\usepackage{amsthm}
\usepackage{thmtools}
\usepackage{thm-restate}
\usepackage[dvipsnames,svgnames,table]{xcolor}
\definecolor{dartmouthgreen}{rgb}{0.05, 0.5, 0.06}
\definecolor{ceruleanblue}{rgb}{0.16, 0.32, 0.75}

\usepackage{graphicx}
\usepackage{subcaption}
\usepackage{cancel}
\usepackage{multirow, makecell}
\usepackage{tablefootnote}
\usepackage{framed}
\usepackage{enumitem}

\usepackage[colorlinks=true,pdfpagemode=UseNone,citecolor=dartmouthgreen,linkcolor=ceruleanblue,urlcolor=BrickRed,pdfstartview=FitH]{hyperref}
\usepackage{cleveref}

\addto\extrasenglish{%
}


\DeclareMathAlphabet{\mathsfit}{T1}{\sfdefault}{\mddefault}{\sldefault}
\SetMathAlphabet{\mathsfit}{bold}{T1}{\sfdefault}{\bfdefault}{\sldefault}

\newenvironment{proofof}[1]{{\medbreak\noindent \em Proof of #1.  }}{\hfill\qed\medbreak}

\newcommand{\C}{\ensuremath{\mathbb C}}
\newcommand{\N}{\ensuremath{\mathbb N}}
\newcommand{\R}{\ensuremath{\mathbb R}}
\newcommand{\Z}{\ensuremath{\mathbb Z}}
\newcommand{\A}{\ensuremath{\mathcal A}}

\newcommand{\cR}{\ensuremath{\mathcal R}}
\newcommand{\cL}{\ensuremath{\mathcal L}}
\newcommand{\F}{\ensuremath{\mathcal F}}
\newcommand{\M}{\ensuremath{\mathcal M}}

\newcommand{\bZ}{\boldsymbol{Z}}
\newcommand{\id}{\operatorname{Id}}

\def\eps{{\varepsilon}}

\newcommand{\inner}[2]{\langle #1, #2 \rangle} % inner product
\newcommand{\norm}[1]{\left\lVert#1\right\rVert} % norms

\renewcommand{\Pr}{\mathbb{P}}
\newcommand{\E}{\mathbb{E}}
\newcommand{\1}{\mathbbm{1}}
\newcommand{\Xf}{X_{\mathrm{free}}}
\newcommand{\XF}{\bar{X}_{\mathrm{free}}}
\newcommand{\tXf}{\Tilde{X}_{{\rm free}}}

\DeclareMathOperator{\tr}{tr} % trace
\DeclareMathOperator{\tot}{tr\otimes \tau}
\DeclareMathOperator{\Tr}{Tr} % trace
\DeclareMathOperator{\poly}{poly}
\DeclareMathOperator{\polylog}{polylog}

\DeclareMathOperator{\vecc}{vec} % vectorization
\DeclareMathOperator{\re}{Re}
\DeclareMathOperator{\im}{Im}

\DeclareMathOperator{\supp}{supp} % support
\DeclareMathOperator{\spec}{spec} % spectrum

\DeclareMathOperator{\cov}{Cov}

\renewcommand{\preceq}{\preccurlyeq}
\renewcommand{\succeq}{\succcurlyeq}

\newtheorem{theorem}{Theorem}[section]
\newtheorem*{theorem*}{Theorem}
\newtheorem{fact}[theorem]{Fact}
\newtheorem*{fact*}{Fact}
\newtheorem{lemma}[theorem]{Lemma}
\newtheorem{definition}[theorem]{Definition}
\newtheorem*{definition*}{Definition}
\newtheorem{corollary}[theorem]{Corollary}
\newtheorem{proposition}[theorem]{Proposition}
\newtheorem*{proposition*}{Proposition}
\newtheorem{claim}[theorem]{Claim}
\newtheorem*{claim*}{Claim}

\newtheorem*{problem*}{Problem}

\newtheorem*{remark*}{Remark}

\newtheorem*{observation*}{Observation}
\newtheorem{example}[theorem]{Example}



\newcommand{\bigip}[2]{\bigl\langle #1, #2 \bigr\rangle}

\newcommand{\bignorm}[1]{\bigl\lVert #1 \bigr\rVert}
\newcommand{\Bignorm}[1]{\Bigl\lVert #1 \Bigr\rVert}
\newcommand{\biggnorm}[1]{\biggl\lVert #1 \biggr\rVert}

\title{Derandomizing Matrix Concentration Inequalities from\\ Free Probability}

\author{
Robert Wang\footnote{Cheriton School of Computer Science, University of Waterloo, Canada. Email: \href{mailto:robert.wang2@uwaterloo.ca}{robert.wang2@uwaterloo.ca}}
\and Lap Chi Lau\footnote{Cheriton School of Computer Science, University of Waterloo, Canada. Research supported by an NSERC Discovery Grant.  Email: \href{mailto:lapchi@uwaterloo.ca}{lapchi@uwaterloo.ca}.}
\and Hong Zhou\footnote{School of Mathematics and Statistics, Fuzhou University, China. Research supported in part by National Key R\&D Program of China, Natural
Science Foundation of China, and Fuzhou University startup funding. Email: \href{mailto:hong.zhou@fzu.edu.cn}{hong.zhou@fzu.edu.cn}.}}

\date{}

\begin{document}

% Give front-matter pages distinct PDF page labels.
\pagenumbering{roman}
\pagestyle{empty}

\maketitle

\begin{abstract}
Recently, sharp matrix concentration inequalities~\cite{BBvH23,BvH24} were developed using the theory of free probability.
In this work, we design polynomial time deterministic algorithms to construct outcomes that satisfy the guarantees of these inequalities.
As direct consequences, we obtain polynomial time deterministic algorithms for the matrix Spencer problem~\cite{BJM23} and for constructing near-Ramanujan graphs. 
Our proofs show that the concepts and techniques in free probability are useful not only for mathematical analyses but also for efficient computations.
\end{abstract}

\thispagestyle{empty}

\clearpage

\newpage

\setcounter{tocdepth}{2}
\tableofcontents

\newpage

\clearpage

% This command resets the page counter to 1 automatically.
\pagenumbering{arabic}
\pagestyle{plain}

\section{Introduction}

Matrix concentration inequalities~\cite{T12,T15,vH17} consider the sum of independent matrix-valued random variables $X = \sum_i X_i$ and show that its spectral statistics such as its operator norm $\norm{X}$ or the moments of its eigenvalues $\tr(X^p)^{\frac{1}{p}}$ are close to their expected values.
These inequalities have found various important applications in theoretical computer science and mathematics, such as spectral sparsification of graphs~\cite{SS11}, 
randomized numerical linear algebra~\cite{Woodruff}, 
analysis of semidefinite programs~\cite{AK16}, 
probabilistic constructions of expander graphs~\cite{Fri08}, 
and many more~\cite{T15}.
Derandomization of these concentration inequalities leads to deterministic algorithms for these problems and deterministic constructions of pseudorandom objects~\cite{WX08,AK16,MOP20}.
From the technical perspective, matrix concentration inequalities are more challenging to prove than the classical concentration inequalities for real-valued random variables because of the non-commutative nature of matrices.

Consider a general Gaussian model $X=\sum_{i=1}^n g_i A_i$, where $A_i \in \C^{d \times d}$ are arbitrary self-adjoint matrices and $g_i$ are i.i.d.~standard real Gaussian variables. 
Define the standard deviation $\sigma(X)$ as  
\begin{equation} \label{e:sigma}
\sigma(X)^2 := \norm{\E[X^2]} = \Bignorm{\sum_{i=1}^n A_i^2}.
\end{equation}
The non-commutative Khintchine inequality of Lust-Piquard and Pisier implies that
\begin{equation} \label{e:Khintchine}
\sigma(X) \lesssim \E[\norm{X}] \lesssim \sqrt{\log d} \cdot \sigma(X).
\end{equation}
It is known that both the lower and upper bounds are asymptotically tight:
The lower bound is of the correct order when $X$ has independent entries,
such that each $A_i$ is supported on a distinct entry.
The upper bound is of the correct order when $X$ is a diagonal matrix,
or more generally when all $A_i$ commute so that they can be simultaneously diagonalized.
The dimension-dependent logarithmic factor between the lower and upper bounds is suboptimal and undesirable in some applications.
This motivates the question of whether there is a more refined quantity, based on the structure of $A_i$, that provides a tighter approximation to $\E[\norm{X}]$.

The progress to this question was initiated by Tropp~\cite{T18} and significantly advanced by Bandeira, Boedihardjo, and van Handel~\cite{BBvH23}.
Informally, they demonstrated a new phenomenon that when the matrices $A_i$ are sufficiently non-commutative, then the logarithmic factor can be removed.
To formalize this, \cite{BBvH23} defined the $d^2 \times d^2$ covariance matrix $\cov(X)$ where 
\begin{equation} \label{e:nu}
\cov(X)_{ij,kl} = \E[X_{ij} \overline{X_{kl}}]
\quad \textrm{and} \quad
\nu(X)^2 := \norm{\cov(X)} = \Bignorm{\sum_{i=1}^n \vecc(A_i) \vecc(A_i)^*},
\end{equation} 
and proved a strong matrix concentration inequality that implies
\begin{equation} \label{e:BBvH-norm}
\E[\norm{X}] \leq 2\sigma(X) + \sigma(X)^{\frac12} \cdot \nu(X)^{\frac12} \cdot (\log d)^\frac34.
\end{equation}
Note that the parameter $\nu(X)$ is much smaller than $\sigma(X)$ in the case where $X$ has i.i.d.~entries,
and more generally $\nu(X)$ is small when the randomness of $X$ is more well-spread over its entries. 
In~\cite[Section~3]{BBvH23}, it was shown that $\nu(X) \cdot (\log d)^{\frac32} \lesssim \sigma(X)$ in many interesting settings, for which the new inequality provides an asymptotic sharp upper bound that $\E[\norm{X}] \lesssim \sigma(X)$. 

Notably, this new inequality was at the heart of a recent major progress towards proving the matrix Spencer conjecture~\cite{BJM23}, with applications in improved spectral sparsification for directed graphs~\cite{JSSTZ25}.

More recently, Brailovskaya and van Handel~\cite{BvH24} extended these matrix concentration inequalities to a much larger class of random matrices.
The random matrix model is $Z = A_0 + \sum_{i=1}^n Z_i$,
where $Z_1, \ldots, Z_n$ are independent Hermitian random matrices with $\E[Z_i] = 0$ and each $\norm{Z_i}$ bounded. 
This captures many commonly studied random graph models,
with arbitrary dependency pattern in their entries. 
They proved a general non-commutative universality principle, that the spectral statistics of $Z$ are closely approximated by those of a Gaussian matrix $X$. 

\subsection{Technical Review}

{\bf Free Probability}: The inequality in \eqref{e:BBvH-norm} was obtained through a novel connection to the concepts and results in free probability -- 
a theory for non-commutative random variables developed by Voiculescu~\cite{NS06,MS17}.
The key idea in~\cite{BBvH23} is to compare the Gaussian model $X = \sum_i g_i A_i$ with the free model 
\begin{equation} \label{e:Xf}
\Xf := \sum_i A_i \otimes s_i,
\end{equation}
where $s_i$ are freely independent semicircular elements (self-adjoint infinite-dimensional operators with ``semicircle distributions''; see \autoref{s:prelim} for definitions).
A concrete way to understand the free model is through the Gaussian models 
\begin{equation} \label{e:XN}
X_N := \sum_i A_i \otimes G_i^N,
\end{equation}
where each $G_i^N$ is an independent $N \times N$ standard Wigner matrix
(self-adjoint matrix with independent Gaussian variables). 
The free model $\Xf$ can be understood as the limiting object of the Gaussian models $X_N$:
The weak asymptotic freeness property proven by Voiculescu~\cite{Voi91} states that for any $p \in \N$,
\begin{equation} \label{e:weak}
\lim_{N \to \infty} \E[\tr(X_N^p)]
= \tr \otimes \tau(\Xf^p),
\end{equation}
where $\tau$ is an abstract trace acting on the algebra generated by $s_1,\ldots,s_n$.
The strong asymptotic freeness property of Haagerup and Thorbjornsen~\cite{HT05} states that
\begin{equation*} %\label{e:strong}
\lim_{N \to \infty} \E\big[\bignorm{X_N}\big]
= \bignorm{\Xf}.
\end{equation*}
Despite being infinite-dimensional and more abstract, an important advantage of the free model is that its spectral statistics are easier to analyze using tools from operator theory:
Lehner derived an exact formula for $\lambda_{\max}(\Xf)$ 
and Pisier showed that
\begin{equation} \label{e:Pisier}
\sigma(X) \leq \bignorm{\Xf} \leq 2 \sigma(X).
\end{equation}
This provides an avenue to establish that $\E[\norm{X}]$ is close to $\sigma(X)$ if one can compare $\E[\norm{X}]$ to $\norm{\Xf}$.
See \autoref{s:prelim} for more background on free probability.

{\bf Interpolation:}
The comparison approach in~\cite{BBvH23} is to interpolate between $X_N$ and $X$ by considering 
\begin{equation} \label{e:interpolation}
X_t^N := \sum_i A_i \otimes \Big( \sqrt{t} \cdot D_i^N + \sqrt{1-t} \cdot G_i^N \Big),
\end{equation}
where $D_i^N$ are independent $N \times N$ diagonal matrices with i.i.d.~standard Gaussians on the diagonal.
Note that $X_0^N = X^N$ whose moments converge to those of $\Xf$ as $N \to \infty$ by \eqref{e:weak},
and $X_1^N$ has the same moments as $X$ such that $\E[\tr(X^p)] = \E[\tr((X_1^N)^p)]$ for every $p, N \in \N$.
Therefore, the difference of the $2p$-moments of the eigenvalues of $X$ and $\Xf$ can be written as
\[
\Big| \E\big[\tr(X^{2p})\big] - \tr \otimes \tau(\Xf^{2p}) \Big|
= \bigg| \lim_{N \to \infty} \int_0^1 \frac{d}{dt} \E\big[ \tr\big((X_t^N)^{2p}\big) \big] dt \bigg|.
\]
By a direct calculation of the derivative and some subtle observations,
the integrand $\frac{d}{dt} \E[ \tr((X_t^N)^{2p}) ]$ can be written as a weighted sum of differences of the form
\begin{equation} \label{e:pushed-inside}
\E\Big[ \tr\Big( Y \big(X_t^N\big)^k Y \big(X_t^N\big)^{2p-2-k} \Big) \Big] 
-  \tr \Big( Y \cdot \E\Big[\big(X_t^N\big)^k\Big] \cdot Y \cdot \E\Big[\big(X_t^N\big)^{2p-2-k}\Big] \Big),
\end{equation}
for some matrix $Y$, where the second term is similar to the first term but with the expectation ``pushed inside''.
This allows the use of the Gaussian covariance identity (see \autoref{s:prelim}) and some complex analysis to relate the sum of these differences to Tropp's matrix alignment parameter in \cite{T18}, defined as
\begin{equation*} %\label{e:Tropp}
w(X_0^N,X_1^N) := \sup_{U,V,W \textrm{ unitary}} \bignorm{ \E[X_0^N U X_1^N V X_0^N W X_1^N]}^\frac14.
\end{equation*}
Finally, some linear algebraic arguments are used to bound the matrix alignment parameter by
\begin{equation*} %\label{e:friendly}
w(X_0^N,X_1^N)^4 \leq \nu(X_0^N) \cdot \sigma(X_0^N) \cdot \nu(X_1^N) \cdot \sigma(X_1^N).
\end{equation*}
Putting together these key steps, the conclusion in~\cite[Theorem 2.7]{BBvH23} is that
\begin{equation} \label{e:BBvH-moments}
\Big| \E\bignorm{X}_{2p} - \bignorm{\Xf}_{2p} \Big|
= \Big| \E\big[\tr(X^{2p})\big]^{\frac{1}{2p}} - \tr \otimes \tau(\Xf^{2p})^\frac{1}{2p} \Big| 
\leq 2p^\frac34 \cdot \sigma(X)^\frac12 \cdot \nu(X)^\frac12.
\end{equation}
{\bf Spectrum:} To extend the arguments to bound the spectrum of $X$ (not just the moments of the eigenvalues of $X$), the resolvent $R_X(z) := (zI - X)^{-1}$ of $X$ at $z \in \C$ is considered.
It is relatively straightforward to show that if $\norm{|zI-X_0^N|^{-1}} \approx \norm{|zI-X_1^N|^{-1}}$ for every point $z$ in a sufficiently fine net, then $\spec(X_0^N) \approx \spec(X_1^N)$.
To establish that $\norm{|zI-X_0^N|^{-1}} \approx \norm{|zI-X_1^N|^{-1}}$ for a fixed $z \in \C$, the large moments of the resolvents $\E[\tr(|zI-X_0^N|^{-2p})]^\frac{1}{2p}$ and $\E[\tr (|zI-X_1^N|^{-2p})]^\frac{1}{2p}$ are compared using the same key steps as above.
In a slightly simplified form, the main result in~\cite[Theorem 2.1]{BBvH23} is that
\begin{equation} \label{e:BBvH-spectrum}
\Pr\Big[ \spec(X) \subseteq \spec(\Xf) + C \cdot \sigma(X)^\frac12 \cdot \nu(X)^\frac12 \cdot \big((\log d)^\frac34 + \alpha \big) [-1,1] \Big] \geq 1 - e^{-\alpha^2},
\end{equation}
for all $\alpha \geq 0$, where $C$ is a universal constant. 
The inequality in \eqref{e:BBvH-norm} follows as a corollary of \eqref{e:BBvH-spectrum}.

{\bf Universality:}
Brailovskaya and van Handel~\cite{BvH24} consider the general model $X := A_0 + \sum_{i=1}^n Z_i$, where $A_0$ is a deterministic Hermitian $d \times d$ matrix and $Z_1, \ldots, Z_n$ are independent $d \times d$ random self-adjoint matrices with $\E[Z_i]=0$ for $1 \leq i \leq n$.
They defined the corresponding Gaussian matrix model as $G := A_0 + \sum_{i=1}^m g_i A_i$ for some $d \times d$ (deterministic) self-adjoint matrices $A_1,\ldots,A_m$ such that $\E[G] = \E[X]$ and $\cov(G) = \cov(X)$.
They proved the universality principle that, informally,
if each $\norm{Z_i}$ is small, 
then the spectrum of $X$ is close to that of $G$ such that
\begin{equation} \label{e:universality-informal}
\E[\norm{X}_{2p}] \approx \E[\norm{G}_{2p}], 
\quad
\E[(zI - X)^{-1}] \approx \E[(zI - G)^{-1}],
\quad
\textrm{and } \spec(X) \approx \spec(G). 
\end{equation}
This allows them to extend the results in \cite{BBvH23} to the general model, with many applications beyond the Gaussian model as the general model captures discrete random variables.

An interpolation as in \eqref{e:interpolation} is used to compare the spectral statistic of the Gaussian and the general model,
but the calculations are considerably more involved with several new technical ingredients such as the moment-cumulant formula, a new trace inequality~\cite[Proposition 5.1]{BvH24}, and Poincar\'{e}'s lemma.

\subsection{Our Results}

We derandomize the general results in \cite{BBvH23,BvH24} by designing deterministic polynomial time algorithms to find outcomes that satisfy the guarantees in the probabilistic statements.
We show applications in designing deterministic algorithms for matrix discrepancy, spectral sparsification, and constructions of expander graphs.

\subsubsection{Norms of Eigenvalues}

Our first result is a one-sided derandomization of \eqref{e:BBvH-moments}.
By standard Gaussian concentration inequalities, 
$\norm{g}^2 \asymp n$ with high probability. 
In the context of randomized algorithms, 
\eqref{e:BBvH-moments} guarantees that with high probability, 
we can sample a vector $x\in \R^n$ with $\norm{x}^2 \asymp n$ such that
\[
\Bignorm{\sum_{i=1}^n x(i) \cdot A_i}_{2p}
\leq \norm{\Xf}_{2p} + 2p^{\frac34} \cdot \sigma(X)^\frac12 \cdot \nu(X)^\frac12
\leq 2\sigma(X) + 2p^{\frac34} \cdot \sigma(X)^\frac12 \cdot \nu(X)^\frac12,
\] 
where the last inequality follows from \eqref{e:Pisier}. 
Such a result can be useful in designing randomized algorithms,
but often there are additional requirements such as $x \in \{\pm 1\}^n$ or $x$ satisfies some linear constraints.
The following theorem provides a derandomization of the upper bound of \eqref{e:BBvH-moments} incorporating additional constraints.

\begin{theorem}[Deterministic Partial Coloring, Simplified Version of {\autoref{t:partial-coloring-full}}] \label{t:partial-coloring}
Let $A_1, \ldots, A_n$ be $d \times d$ Hermitian matrices. Let $\eps,\delta\in (0,1)$ such that $\eps+\delta+3/n<1$.
Let $\mathcal{H} \subseteq \R^n$ be a linear subspace of dimension greater than $(1-\eps)n$.
For any integer $p \geq 2$, there is a deterministic polynomial time algorithm to find a vector $x \in [-1,1]^n \cap \mathcal{H}$ with $|\{i \mid x(i) \in \{\pm1\}\}| \geq \delta n$ such that
\[
\Bignorm{\sum_{i=1}^n x(i) \cdot A_i}_{2p} 
\leq \sqrt{K_{\delta,\eps}} \cdot \norm{\Xf}_{2p} + O\big(p^\frac34 \cdot \sigma(X)^\frac12 \cdot \nu(X)^\frac12 \big),
\quad \textrm{where} \quad
K_{\delta,\eps}: = \frac{1}{1-(\eps+\delta) - \frac3n} 
\]
and $\sigma(X), \nu(X)$ are defined as in \eqref{e:sigma}, \eqref{e:nu} respectively.
\end{theorem}
In the full version of \autoref{t:partial-coloring} in \autoref{t:partial-coloring-full}, when there are no box constraints $x \in [-1,1]^n$ and linear subspace constraints $x \in \mathcal{H}$, the algorithm returns a vector $x \in \R^n$ with $\norm{x}^2 \asymp n$ and no multiplicative constant $K_{\delta,\eps}$, 
matching the upper bound of~\cite[Theorem 2.7]{BBvH23} stated in \eqref{e:BBvH-moments}.

\subsubsection{Deterministic Matrix Discrepancy}

\autoref{t:partial-coloring} can be interpreted as a general ``partial coloring'' result in matrix discrepancy theory.
We demonstrate that it can be directly applied to obtain a deterministic algorithm for the matrix Spencer problem, matching the randomized algorithm in~\cite{BJM23}.

\begin{theorem}[Deterministic Matrix Spencer] \label{t:matrix-Spencer}
Given $n\times n$ symmetric matrices $A_1, \ldots, A_n$ with $\norm{A_i}\leq 1$ and $\norm{A_i}_F^2\leq r$ for $1 \leq i \leq n$, and an arbitrary integer $p \geq 2$,
there is a deterministic polynomial time algorithm to compute a coloring $x: [n]\rightarrow \{-1,1\}$ such that
\[
\Bignorm{\sum_{i=1}^n x(i) \cdot A_i}_{2p} \lesssim \sqrt{n} + p^{\frac34} \cdot n^{\frac14} \cdot r^{\frac14}.
\]
In particular, when $r\lesssim n/\log^{3}n$ and $p \asymp \log{n}$, 
then $\norm{\sum_{i=1}^n x(i) \cdot A_i} \lesssim \norm{\sum_{i=1}^n x(i) \cdot A_i}_{2p} \lesssim \sqrt{n}$.
\end{theorem}

In \cite{BJM23}, the inequality in \eqref{e:BBvH-norm} was used to lower bound the Gaussian measure of the norm ball $\mathcal{K} := \{ x \in \R^n \mid \norm{\sum_{i=1}^n x(i) \cdot A_i} \leq 1\}$, so that Rothvoss' result~\cite{Rot17} can be applied to obtain a partial coloring.
In \autoref{t:matrix-Spencer}, we apply \autoref{t:partial-coloring} to obtain a partial coloring without using Rothvoss' result, providing a simpler and more direct approach to matrix discrepancy.

The matrix discrepancy result in \cite{BJM23} has found an interesting application in spectral sparsification of directed graphs~\cite{JSSTZ25}, and \autoref{t:matrix-Spencer} implies a deterministic algorithm for this application.
See \autoref{ss:application-discrepancy} for details.

\subsubsection{Full Spectrum}

We also derandomize the result in \eqref{e:BBvH-spectrum}, deterministically finding an outcome with the full spectrum close to that of $\Xf$.

\begin{theorem}[Deterministic Full Spectrum, Simplified Version of \autoref{t:spectrum-full}] \label{t:spectrum}
Let $A_1,\ldots,A_n$ be $d\times d$ Hermitian matrices.
There is a deterministic polynomial time algorithm to compute a vector $x \in \R^n$ with $\norm{x}^2 \leq n$ such that
\begin{align*}
\spec\Big(\sum_{i=1}^n x(i) \cdot A_i\Big) \subseteq \spec(\Xf) + [-\eta, \eta]
\quad \textrm{and} \quad
\spec(\Xf) \subseteq \spec\Big(\sum_{i=1}^n x(i) \cdot A_i\Big) + [-\eta, \eta],
\end{align*}
where $\eta \lesssim \sigma(X)^\frac12 \cdot \nu(X)^\frac12 \cdot (\log d)^{\frac34}$ and 
$\sigma(X), \nu(X)$ are defined as in \eqref{e:sigma}, \eqref{e:nu} respectively.
\end{theorem}

As an application of this result, we give a deterministic analog of the spiked detection model. 
Given a set of unit vectors $v_1,\ldots,v_n$ with signal values $\theta_1,\ldots,\theta_n >0$, we deterministically construct a pseudorandom matrix $W$ such that for all $j$, the maximum eigenvector of $\theta_j v_jv_j^\top +W$ has nontrivial correlation with $v_j$ if and only if $\theta_j > 1$. 
This is a derandomization of \cite[Theorem 3.1]{BCS+24}. 
When $\theta_j \leq 1$, the vectors $v_j$ are ``efficiently hidden'' by $W$ with respect to the spectral detection algorithm (analogous to how a planted clique of size $\sqrt{d}$ is ``hidden'' by a random graph of $d$ vertices). 
This application shows that \autoref{t:spectrum} can be used to construct pseudorandom matrices that capture the behavior of not only the eigenvalues of random matrices but also their eigenvectors.
See \autoref{s:application-spectrum} for details.

\subsubsection{General Model}

Let $X := A_0 + \sum_{i=1}^n Z_i$, where $A_0$ is a deterministic Hermitian $d \times d$ matrix and $Z_1, \ldots, Z_n$ are independent $d \times d$ random self-adjoint matrices with $\E[Z_i]=0$ for $1 \leq i \leq n$.
For derandomization, we assume that each $Z_i$ has discrete support so that it can be enumerated efficiently, which is satisfied in random graph models.

The following are two related models.
Let $G := A_0 + \sum_{i=1}^m g_i A_i$ be the corresponding Gaussian model of $X$ with $\E[G]=\E[X]$ and $\cov(G) = \cov(X)$.
Note that the distribution of $G$ is uniquely defined but the representation as $A_0 + \sum_{i=1}^m g_i A_i$ is not unique.
Let $\Xf := A_0 \otimes 1 + \sum_{i=1}^m A_i \otimes s_i$
be the corresponding free model of $G$ as defined in \eqref{e:Xf}, where $s_1, \ldots, s_m$ is a free semicircular family and $1$ is the identity element.

We provide one-sided derandomization of the norm bounds in \cite{BvH24}.

\begin{theorem}[Deterministic Moment Universality, Simplified Version of \autoref{t:moment-uni}] \label{t:moment-uni-intro}
Let $X := A_0 + \sum_{i=1}^n Z_i$, where $A_0$ is a deterministic Hermitian $d \times d$ matrix, and $Z_1, \ldots, Z_n$ are independent $d \times d$ random self-adjoint matrices with $\E[Z_i]=0$ for $1 \leq i \leq n$.
Assume that $\norm{Z_i} \leq r$ with probability one for $1 \leq i \leq n$,
and the support size of each $Z_i$ is polynomially bounded.
Then, for any $p \in \N$, there is a deterministic polynomial time algorithm to find a matrix $Z_i' \in \supp(Z_i)$ for $1 \leq i \leq n$ such that\footnote{The notation $\Tilde{O}(\cdot)$ hides some logarithmic term.} 
\[
\tr\Big(\Big(A_0 + \sum_{i=1}^n Z_i'\Big)^{2p}\Big)^{\frac{1}{2p}} 
\leq \tr\otimes\tau\big(\Xf^{2p}\big)^{\frac{1}{2p}} 
+ \Tilde{O}\big(p^{\frac34} \cdot \sigma(X)^\frac12 \cdot \nu(X)^\frac12 
+ p^{\frac23} \cdot \sigma(X)^\frac23 \cdot r^\frac13 \big).
\]
\end{theorem}

The following is a stronger bound for the operator norm, with the leading constant being one, which is important for applications such as constructing near-Ramanujan graphs.

\begin{theorem}[Deterministic Norm Universality, Simplified Version of \autoref{t:norm-uni}] \label{t:resolvent-uni}
Under the same setting in \autoref{t:moment-uni-intro}, there is a deterministic polynomial time algorithm to find a matrix $Z_i'\in \supp(Z_i)$ for $1 \leq i \leq n$ such that 
\[
\biggnorm{ A_0 + \sum_{i=1}^n Z_i'} \leq \norm{\Xf} + \Tilde{O}\big(\sigma(X)^\frac12 \cdot \nu(X)^\frac12 + \sigma(X)^\frac23 \cdot r^\frac13 \big).
\]
\end{theorem}

\subsubsection{Deterministic Expander Constructions}

Brailovskaya and van Handel~\cite[Section 3.2]{BvH24} showed that their results for the general model can be applied to give new probabilistic constructions of expander graphs.
Our deterministic algorithms in \autoref{t:moment-uni-intro} and \autoref{t:resolvent-uni} imply the following consequences:
\begin{enumerate} 
    \item (Edge-Signing Model, \autoref{theorem:2-lift}): For any $k$-regular graph $G = (V,E)$, with $k\geq \polylog{|V|}$, there is a deterministic algorithm to find an edge signing of $G$ whose signed adjacency matrix has eigenvalues bounded by $2\sqrt{k}\cdot (1+\Tilde{O}(k^{-\frac16}))$.
    
    \item (Permutation Model, \autoref{theorem:perm}): For any $k\geq \polylog{d}$, there is a deterministic algorithm to compute $2k$ perfect matchings over $d$ vertices such that their union has spectral radius $2\sqrt{2k} \cdot (1+\Tilde{O}(k^{-\frac16}))$.
    
    \item (Lift Model, \autoref{theorem:lift}): Let $G = (V,E)$ be a simple (possibly nonregular) graph with maximum degree $k_{\max} \geq \polylog(|V|)$. 
    There is a deterministic algorithm to compute an $m$-lift of $G$ whose new eigenvalues are bounded by $\lambda \cdot (1+\Tilde{O}(k_{\max}^{-\frac16}))$ where $\lambda$ is the spectral radius of the universal cover of $G$.
    
    \item (Group-Labeled Lift Models, \autoref{theorem:group}): Let $G=(V,E)$ be a $k$-regular simple graph and $\Gamma$ be a finite group with $k \geq \polylog(|\Gamma|,|V|)$. There is a deterministic algorithm to compute a $\Gamma$-lift of $G$ whose new eigenvalues are bounded by $2\sqrt{k} \cdot (1+\Tilde{O}(k^{-\frac16}))$
\end{enumerate}

To our knowledge, these are the first polynomial time deterministic algorithms for all these settings when $k \geq \polylog(|V|)$. 
See \autoref{ss:uni-application} for background and details.

\subsection{Proof Overview} \label{s:overview}

Our proofs show that the concepts and techniques in the theory of free probability are not only useful for mathematical analyses, but also useful for efficient computations.

A classical technique for derandomization is the method of conditional expectation. 
Given a set of variables $X_1,\ldots, X_n$, and a function $f(x_1,\ldots, x_n)$, we wish to find an outcome $x_1,\ldots, x_n \in \supp(X_1,\ldots, X_n)$ such that $f(x_1,\ldots, x_n)\leq \E[f(X_1,\ldots, X_n)]$. 
The method of conditional expectation allows us to find such an outcome as long as we can compute the expectation of $f$ conditioned on the outcomes of any subset of variables. 
However, for moments or norms of general random matrices, this is typically difficult to compute in polynomial time. 
For example, if $X$ is a $d\times d$ random matrix with independent entries, computing $\E[\tr(X^{\log{d}})]$ would take $d^{\log{d}}$ time. 
Thus, previous derandomizations of matrix concentration inequalities (e.g., \cite{WX08}) use ``pessimistic estimators'' to estimate the conditional expectation of various spectral statistics of random matrices. 
By nature, these cannot recover the sharp matrix concentration results in \cite{BBvH23} and \cite{BvH24}, as these involve inequalities such as Golden-Thompson, which, just like \eqref{e:Khintchine}, cannot distinguish between the commutative and non-commutative settings. 

In this work, we observe that using the ``non-crossing'' structure in the theory of free probability, the moments of the free model $\Xf$ can be computed efficiently in polynomial time via a natural recursive formula; see \autoref{s:computation} for computational aspects. 
This suggests the following framework for de-randomizing the new matrix concentration inequalities: 
start with the operator $\Xf$, and at each iteration, replace a ``little bit'' of it with a finite-dimensional random matrix. 
If this random matrix is simple enough (e.g., with pairwise independent entries), then we can deterministically find an outcome to replace the random matrix. 
This method can be viewed as a free version of the method of conditional expectation, where instead of computing the expectation of a spectral statistic on $X$, we compute the spectral statistic on the corresponding operator $\Xf$,
which is a main theme in this work. Our ideas for derandomization are inspired by algorithms for matrix discrepancy minimization used for spectral sparsification \cite{LRR17,LWZ25}.

\subsubsection{Derandomizing Random Gaussian Matrix Models by Brownian Walks}

We elaborate more technical ideas in the derandomization of the random Gaussian matrix model.
A natural derandomization strategy is to interpolate from $X_0^N$ to $X_1^N$ in \eqref{e:interpolation} as was done in \cite{BBvH23}, where in each iteration $t$ we replace a little bit of each $G_i^N$ by a deterministic diagonal matrix $(D_i^N)_t$.
The choice of lifting each Gaussian variable $g_i$ to a random matrix $D_i^N$ with independent diagonal entries was used in \cite{BBvH23} to apply the multivariate Gaussian integration-by-parts formula in a nice way to compute the derivative $\frac{d}{dt}\E[\tr(f(X_{N,t}))]$. 
However, for the purposes of derandomization, this lifting technique introduces some issues. 
The major one is that the expectation of the Gaussian matrix moments are not easy to compute, as previously noted. 
It is also not clear what should be the scalar value of $x(i)$ given a deterministic sampling of a diagonal matrix $D_i^N$.\footnote{ 
If the potential function $f$ is convex, then Jensen's inequality gives $f(\tr(D_i^N)) \leq \tr(f(D_i^N))$, so one could potentially take $x(i)$ to be the average of the entries in $D_i^N$. 
However, if $f$ is not convex, then it becomes unclear how $x(i)$ should be sampled. 
We note that the proof of \autoref{t:spectrum} requires interpolating non-convex functions to control the full spectrum.
}

{\bf Interpolation with $\mathbf{X_\mathbf{free}}$}: 
In our analysis, we use an alternative interpolation method that avoids lifting the random $d \times d$ matrix $X$ to a $dN \times dN$ random matrix. 
Instead, we directly interpolate between $X$ and $\XF = \sum_{i=1}^n A_i \otimes s_i$.
For any vector $x\in \R^n$, we define $A(x) := \sum_{i=1}^n x(i) \cdot A_i$,
and the mixed operator 
\[
A_t(x) := A_0 \otimes 1  + A(x) \otimes 1 + \sqrt{1-t} \cdot \XF.
\]
By definition, $A_0(\vec{0}) = \Xf$ and $A_1(g) = X$ when $g\sim \mathcal{N}(0,I)$. 
Given a potential function, say $\Phi(t,x) = \tr\otimes\tau(A_t(x)^{2p})^{\frac{1}{2p}}$, we can evaluate it at any point using dynamic programming, because the ``non-crossing structure'' in free probability gives recursive formulas for computing moments (while such formulas do not exist for random matrices).
See \autoref{s:computation} for these computational aspects after the required background is introduced in \autoref{s:prelim}.

{\bf Brownian Walks and Pairwise Independent Updates}:
To interpolate between $\vec{0}$ and $g$, we consider the stochastic process $x_t \sim \mathcal{N}(0, tI)$, where $x_0 = \vec{0}$ and $x_1 = g$. 
Rather than integrating only the derivative with respect to $t$ as in \cite{BBvH23}, we integrate with respect to the stochastic process $x_t \sim \mathcal{N}(0, tI)$. 
Given the potential function $\Phi(t,x)$, we evaluate
\[
\E[\Phi(1,x_1)] - \Phi(0, x_0) = \int_{0}^1 \E[d \Phi(t,x_t) ].
\]
The main task is to bound $\E[d\Phi(t,x_t)]$. By Ito's lemma,
\[
\E[d\Phi(t,x_t)] 
= \E[ \partial_t \Phi(t,x_t)dt] 
+ \E\big[\nabla\Phi(t,x_t)^\top dx_t\big] 
+ \frac12 \E \big[dx_t^\top \nabla^2\Phi(t,x_t) dx_t \big],
\]
where $\nabla\Phi (t,x)$ and $\nabla^2\Phi(t,x)$ are the gradient and Hessian of $\Phi(t,x)$ as a function of $x \in \R^n$. 
The key to derandomizing this process is to observe that $dx_t$ need not be a Gaussian vector. 
Since the update formula only depends on the first and second order statistics of the entries of $dx_t$, 
it suffices to take $dx_t$ to be a vector with \textit{pairwise independent} entries.

We derandomize this interpolation process by discretizing the stochastic integral. 
In particular, we break the interval $[0,1]$ into steps $\eta,2\eta,\ldots,T\eta$ where $\eta = 1/T$. 
At each step $t$, we deterministically find an update vector $y_t$ such that $\Phi(t+\eta,x_t + \sqrt{\eta} \cdot y_t)\leq \Phi(t,x_t)$. 
Such an update can be computed in polynomial time because 
1) the potential function $\Phi$ is in terms of $\Xf$, and so can be evaluated efficiently, 
and 2) the \textit{expected} potential change is small over a pairwise independent distribution of update vectors, so we only need to search over polynomially many possibilities for $y_t$. 
In addition to being easily derandomizable, this method of ``Brownian interpolation''\footnote{
We note that this method can be interpreted as a geometric version of the method of conditional expectation, since at each time $t$, the operator $\sqrt{1-t} \cdot \Xf$ is an infinite dimensional approximation of the random matrix $\sqrt{1-t}\sum_i A_ig_i$, which captures the randomness that has yet to be derandomized. The $\tr\otimes\tau$ operator then approximates the expectation over this randomness. 
The selection of an update to minimize the potential function is analogous to selecting a deterministic outcome that minimizes the expectation of the random part conditioned on the outcome we selected.
} can also be used to handle linear constraints on $x_t$ using the sticky Brownian walk method of Lovett and Meka \cite{LM15}. 
These linear constraints are required for many applications, especially for matrix discrepancy problems. 
The sticky Brownian walk method can be derandomized using the eigenspace methods in \cite{LRR17}. This is similar to pairwise independent updates but with stronger guarantees, which will be elaborated in \autoref{section:discrepancy}. 

{\bf Bounding the Expected Change}:
The key to bounding the expected change in potential lies in the property that the sum of freely independent semicircular random variables is also semicircular. This means that
\[
\sqrt{1-t} \cdot \XF \overset{d}{=} \sqrt{1-(t+\eta)} \cdot \XF + \sqrt{\eta} \cdot \XF',
\]
where $\XF'$ is a freely independent copy of $\XF$. 
Under this discretization,
\begin{align*}
\partial_t \Phi(t,x_t)dt 
&~\approx~ \Phi(t+\eta,x_t) - \Phi(t,x_t)
\\
&~\approx~ -\tr\otimes\tau \big( \big(A_t(x_t) + \sqrt{\eta} \cdot \Xf'\big)^{2p}\big)^{\frac{1}{2p}} + \tr\otimes\tau \big(A_t(x_t)^{2p}\big)^{\frac{1}{2p}}. 
\end{align*}
Using second-order Taylor expansion, the derivatives in terms of $dx_t$ can be expressed as
\begin{align*}
\nabla\Phi(t,x_t)^\top dx_t + \frac{1}{2}dx_t^\top \nabla^2\phi(t,x_t) dx_t 
&~\approx~ \Phi(t, x_t + \sqrt{\eta}y_t) - \Phi(t, x_t)
\\
&~=~ \tr\otimes\tau \big(\big(A_t(x_t) + \sqrt{\eta} \cdot A(y_t)\big)^{2p})^{\frac{1}{2p}}-\tr\otimes\tau \big(A_t(x_t)^{2p}\big)^{\frac{1}{2p}}
\end{align*}
Thus, we can interpret this step as transferring one unit of mass from the free part of the mixed operator $\sqrt{1-t} \cdot \Xf$ 
to the finite-dimensional part $A(x_t)$. 
The term $\nabla\Phi(t,x_t)^\top dx_t + \frac{1}{2}dx_t^\top \nabla^2\phi(t,x_t) dx_t$ is the loss incurred by adding more mass to the finite dimensional part, 
while the term $\partial_t \Phi(t,x_t)dt$ is the gain incurred by taking away mass from the free part. 

{\bf Free Probability and Gaussian Analysis}:
To show that the loss and gain terms approximately cancel each other out, there are two steps. 
First, we apply Taylor expansion on the potential function to show that both expressions are dominated by the second order term in the Taylor expansion. 
From this second order expansion, using basic properties of free independence, we show that the difference of the loss term and the gain term is exactly a weighted sum of differences of the form in~\eqref{e:pushed-inside}.
This provides a better insight on the subtle step where the expectation is ``pushed inside'' in~\cite{BBvH23}.
Once we reached this form, we can reuse the Gaussian analysis and the linear algebraic arguments proven in~\cite{BBvH23} to bound the difference.

To summarize, the Brownian walk method uses techniques developed in discrepancy theory to ensure that the outcome satisfies box constraints and linear constraints, while the Gaussian and complex analysis developed in~\cite{BBvH23} is used in bounding the potential increase in each infinitesimal step where we replace a bit of $\Xf$ by a bit of $A(y_t) \otimes 1$. 

{\bf Full Spectrum and Multiplicative Weight Update}: Given this Brownian interpolation method, the proof of \autoref{t:spectrum} has a very similar structure, where the potential function is changed to the moments of the resolvent such that
\[
\Phi^z_t(x_t) = \tr\otimes\tau\big(|z I \otimes 1 -A_0 \otimes 1 -A(x_t) \otimes 1 - \sqrt{1-t} \cdot \XF|^{-2p}\big)^\frac{1}{2p}.
\]
To control the full spectrum, the probabilistic analysis in \cite{BBvH23} bounds the moments of the resolvent for many different points $z \in \C$ and applies a union bound.
For derandomization, an extra ingredient in this proof is to use the multiplicative weight update method to combine many potential functions into one.

\subsubsection{Derandomizing General Random Matrix Models by Random Swap}

In \cite{BvH24}, Brailovskaya and van Handel showed, via interpolation arguments, that the spectral statistics of the general random matrix model $X = A_0 + \sum_{i=1}^n Z_i$ is close to those of the corresponding Gaussian matrix model. 
Such a method does not directly lead to a polynomial time derandomization scheme for two reasons. 
First, as in the previous setting, the expected Gaussian moments are not easy to compute in general. 
To address this issue, we will bypass the Gaussian approximation of $Z$, and instead directly compare the spectral statistics of $Z$ with those of the free model
\[
\Xf : = A_0 + X_1+X_2+ \cdots + X_n,
\]
where $X_1, \ldots, X_n$ are freely independent and satisfy $\id \otimes \tau[X_i] = \E[Z_i]$ and $\cov(X_i) = \cov(Z_i)$ for all $i\in [n]$. 
This operator $\Xf$ is exactly the free approximation of the Gaussian matrix corresponding to the general model $Z$.
See \autoref{ss:statements-uni} for the construction of this free model.

The second issue is that unlike in the Gaussian model, where we can approximately decompose a Gaussian vector into the sum of pairwise independent random vectors, the general random matrix model $Z$ admits no such simple decomposition. 
Thus, we make the assumption that each $Z_i$ is a discrete random matrix with polynomial support size. 
This still encapsulates many important applications such as expander graph constructions. 
For example, if $Z$ is the signing of the adjacency matrix of a graph, then each $Z_i$ corresponds to the signing of an edge, with only two outcomes.

{\bf Random Swap}:
The non-Gaussian structure of $Z$ makes it difficult to apply the Brownian walk method to the general matrix model. 
Our assumption of polynomial support size of each $Z_i$ suggests a more direct application of the method of conditional expectation. Consider the following ``random swap'' procedure, in which we randomly select an index $i\in [n]$ and perform the swap $X_i\gets Z_i \otimes 1$. 
Then, we find an outcome of $Z_i$ that minimizes the desired potential function. For example, if our potential is the $2p$-th moment, then we want to bound the quantity
\begin{align}
    \frac{1}{n}\sum_{i=1}^n\E_{Z_i}\big[\tr\otimes\tau\big((\Xf + Z_i \otimes 1-X_i)^{2p}\big)\big]
-\tr\otimes\tau(\Xf^{2p}) \label{eqn:swap-formula}
\end{align}
There are two new technical ingredients in our proofs.
An important step in our analysis is to derive a ``{\em semicircular integration by parts formula}'' (see \autoref{proposition:free-ibp}). 
This is crucially used to compute the potential update for our interpolation with the free model, where the formula allows us to replace ``dependent'' random variables by freely independent random variables in the Taylor expansion of \eqref{eqn:swap-formula}.
One can view this as a replacement of the moment-cumulant formula used in \cite{BvH24} to carry out the interpolation with the Gaussian model.
A technical remark is that we exploited the property that $\max_i \norm{X_i}$ is bounded, while it does not hold in the Gaussian interpolation setting where $\norm{G_i}$ is unbounded.

Another technical ingredient is the use of the barrier method, developed in \cite{BSS12}, to derandomize the norm universality result.
We analyze the potential function $\tot((\lambda I - X)^{-2p})$ where $\lambda > \lambda_{\max}(X)$. 
In each iteration of the random swap algorithm, we update the barrier $\lambda\gets\lambda +\delta$ and show that the potential function does not increase. 
This allows us to bypass the step of having to simultaneously control the resolvent norm at many points in \cite{BvH24}, which was much more challenging than in the Gaussian setting of \cite{BBvH23} and required proving new concentration inequalities for the general model.
Again, we exploited the property that $\max_i \norm{X_i}$ is bounded for the barrier method to be well-defined.
These considerations further highlight the advantages of directly interpolating between $Z$ and the free model in our analysis.

\section{Free Probability and Random Matrices} \label{s:prelim}

Free probability is a theory for non-commutative random variables, which are often characterized by the spectral distributions of matrices and operators. 
In \autoref{s:intro-free}, we introduce the basic notations and definitions and some fundamental results in free probability, then we present the connection to random matrix theory in \autoref{s:RMT-free}.

This section is relatively long, as we aim to provide a friendly introduction of free probability with more background and intuition.
Readers who are familiar with free probability could skip ahead and only come back when necessary.

\subsection{An Introduction to Free Probability} \label{s:intro-free}

Everything in this subsection can be found in \cite{NS06}.

We start with the basic definitions of a non-commutative probability space and the distribution of a non-commutative random variable.
Then, we present the concept of free independence,
and the free central limit theorem, 
which states that the sum of freely independent non-commutative random variables converges to a semicircular element.
Finally, we show the formulas for computing joint distributions using non-crossing partitions, which is the underlying combinatorial structure that allows for efficient derandomization.

\subsubsection{Algebra and Probability Space}

\begin{definition}[Non-Commutative Probability Space]
A non-commutative probability space is a pair $(\mathcal{A},\tau)$ where $\mathcal{A}$ is an associative unital algebra over the field $\mathbb{C}$ (i.e. containing a multiplicative identity element $1$) and $\tau : \mathcal{A} \to \mathbb{C}$ is a unital linear functional with $\tau(1) = 1$.
The elements $a \in \mathcal{A}$ are called non-commutative random variables in $(\mathcal{A},\tau)$.
A non-commutative probability space $(\mathcal{A},\tau)$ is called tracial if $\tau$ satisfies the trace property that $\tau(ab) = \tau(ba)$ for all $a,b \in \mathcal{A}$. 
\end{definition}

For the purpose of this paper, the most important example is the space of random matrices.

\begin{example}[Random Matrices]
Let $\A$ be the algebra of $d\times d$ random matrices where each entry has finite moments of all orders. 
Let $\tau$ be the expectation functional such that 
\[
\tau(X) := \E[\tr(X)] := \frac{1}{d} \E[\Tr(X)] := \frac1d \sum_{i=1}^d \E[X_{i,i}].
\]
Then $(\A,\E \tr)$ is a tracial non-commutative probability space.
\end{example}

The space of random matrices, equipped with the conjugate operation and the operator norm, has some additional structure and is a $C^*$-probability space.

\begin{definition}[$C^*$-Probability Space] 
Let $(\A,\tau)$ be a non-commutative probability space.
$\A$ is called a $*$-algebra if $\A$ is equipped with an anti-linear map $*: \A \rightarrow \A$ such that $(a^*)^* = a$ and $(ab)^* = b^*a^*$.  $(\A,\tau)$ is called a $*$-probability space if $\A$ is a $*$-algebra, and $\tau$ satisfies $\tau(a^*) = \overline{\tau(a)}$ and $\tau(a^*a) \geq 0$ for all $a\in \A$.

If $\A$ is equipped with a norm $\norm{\cdot}$ such that the topology over $\A$ induced by the distance $d(x,y) = \norm{x-y}$ is complete, $\A$ is called a Banach algebra if $\norm{xy} \leq \norm{x} \norm{y}$ for $x,y \in \A$.

$\A$ is called a $C^*$-algebra if it is a $*$-algebra and a Banach algebra, with norm identity satisfying $\norm{a^*a} = \norm{a}^2$.
$(\A, \tau)$ is called a $C^*$-probability space if $\A$ is a unital $C^*$-algebra and $(\A, \tau)$ forms a non-commutative probability space.

An element $a$ in a $*$-algebra $\A$ is called normal if $a^* a = aa^*$, self-adjoint if $a=a^*$, unitary if $aa^* = a^*a = 1$, and positive if $a = b^*b$ for some $b \in \A$.

If $\A$ is a set of bounded linear operators, we will use $C^*(\A)$ to denote the $C^*$ algebra generated by $\A$ under the closure of the operator norm.
\end{definition}

Let the spectrum of an element $a$ be 
\[
\spec(a) = \{\lambda \in \C: \lambda 1-a \text{ is not invertible}\}.
\]
A classical and useful result in matrix analysis is the spectral theorem, which says that normal matrices are unitarily diagonalizable.
Then, for any normal matrix $M$ and any function $f:\spec(M) \rightarrow\C$, one can uniquely define the matrix $f(M)$ having the same eigenvectors as $M$ with eigenvalues $\{f(\lambda): \lambda\in \spec(M)\}$.
Some common examples include $e^{M}$, $|M|$, and $\sqrt{M}$ for a PSD $M$.
This functional calculus can be generalized to normal elements in an arbitrary $C^*$-algebra.

% \begin{theorem}[Functional Calculus and Spectral Mapping Theorem] \label{t:functional-calculus}
% Let $\A$ be a unital $C^*$-algebra. 
% For every $a\in \A$, its spectrum $\spec(a)$ is a compact set contained in the unit disc of radius $\norm{a}$ in $\C$. 

% If $a$ is normal and $f: \spec(a) \to \C$ is a continuous function, 
% then there is a unique element $f(a) \in \A$ with $\spec(f(a)) = f(\spec(a))$.
% In words, the spectrum of $f(a)$ is equal to the image of the spectrum of $a$ under $f$.
% In particular, $\norm{f(a)} = \sup\{|f(z)| \mid z \in \spec(a)\}$.
% \end{theorem}

% Note that if $a$ is self-adjoint, then the spectrum is supported in $\R$, so in that case, we can view $f$ as a real-valued function. 

\begin{theorem}[Continuous Functional Calculus and Spectral Mapping]
\label{t:functional-calculus}
Let $\A$ be a unital $C^*$-algebra. For every $a\in\A$, the spectrum
$\spec(a)$ is a nonempty compact subset of
\[
\{z\in\C: |z|\leq \norm{a}\}.
\]

Now suppose that $a$ is normal, and let
$\iota:\spec(a)\rightarrow\C$ be the coordinate function
$\iota(z)=z$. There exists a unique unital $*$-isomorphism $\Phi_a:C(\spec(a))\longrightarrow C^*(1,a)$, where $C^*(1,a)$ denotes the unital $C^*$-subalgebra of $\A$ generated by $a$, such that $\Phi_a(\iota)=a$.
For every $f\in C(\spec(a))$, define $f(a):=\Phi_a(f)$.
Then $\spec(f(a))=f(\spec(a))$ and $\norm{f(a)}=\max_{z\in\spec(a)}|f(z)|$.
Moreover, if $f$ is real-valued, then $f(a)$ is self-adjoint.
\end{theorem}

If $a$ is self-adjoint, then $\spec(a)\subseteq\R$. 
Thus the continuous functional calculus may be applied to continuous functions on the compact subset $\spec(a)\subseteq\R$.
In particular, if $f:\spec(a)\rightarrow\R$ is continuous, then $f(a)$ is self-adjoint.

\subsubsection{Distributions and Convergence}

In classical probability theory, the distribution of compactly supported random variables can be obtained from its moments via techniques from Fourier analysis.
As such, the distributions of such random variables are often characterized by their moments. 
In a similar way, the distribution of a non-commutative random variable in a $C^*$-probability space is defined by its moments.

\begin{definition}[Analytical Distribution] \label{d:analytical-distribution}
Let $a$ be a normal random variable in a $C^*$-probability space $(\A,\tau)$. 
The distribution of $a$ is the unique Borel probability measure $\mu$ on $\C$ (supported on $\spec(a)$) such that, for all $p,q \in \N$,
\[
\int_{\C} z^p\bar{z}^{q}d(\mu(z)) = \tau(a^p(a^*)^q).
\]
\end{definition}

We can think of the distribution of $a$ as a distribution of a classical random variable $Z$ satisfying $\E[Z^p\bar{Z}^q] = \tau(a^p(a^*)^q)$. 
If $a$ is normal and $\tau$ is a faithful trace, then it can be proved that the support of $\mu$ is equal to the spectrum $\spec(a)$. 
Thus, the distribution of $Z$ is exactly the eigenvalue distribution (or empirical spectral measure) of $a$, 
so $Z$ can be thought of as a ``uniformly sampled eigenvalue'' of $a$.

This distribution provides a way to compute the norm using the moments, which we will state for self-adjoint elements.

\begin{proposition}[Norm from Moments]
Let $(\A,\tau)$ be a $C^*$-probability space. 
Let $a\in \A$ be a self-adjoint element and $\tau$ be a faithful trace. 
Then 
\[
\norm{a} = \lim_{p\rightarrow\infty}\tau(a^{2p})^{\frac{1}{2p}}.
\]
\end{proposition}

Unlike in the classical setting, the joint distribution over a set of random variables $a_1, \ldots, a_k$ cannot be represented by a probability distribution over $\C^k$, as such a distribution would define commutative random variables by definition. Instead, the joint distribution of $a_1, \ldots,a_k$ is characterized by the evaluation of $\tau$ on all possible non-commutative polynomials of $a_1, \ldots,a_k$, denoted by $\C\langle a_1, \ldots, a_k \rangle$.

\begin{definition}[Joint Distribution of Random Variables] % \label{definition:distribution}
The distribution of $a_1, \ldots, a_n$ is the linear function $\Phi: \mathbb{C}\langle x_1, \ldots, x_n, x_1^*, \ldots, x_n^*\rangle \rightarrow \C$ given by 
\[
\Phi(p) = \tau(p(a_1, \ldots, a_n)).
\]
\end{definition}

This distribution gives a way to study the convergence of random matrices to operators in an arbitrary $C^*$-probability space.

\begin{definition}[Convergence in Distribution] \label{d:convergence-distribution}
Let $(\A,\tau), (\A_N, \tau_N)_{N=1}^\infty$ be $C^*$-probability spaces. 
Let $a^N_1, \ldots, a^N_n\in \A_N$ and $a_1, \ldots, a_n\in \A$. We say that $a_1^N,\ldots, a_n^N$ converges in distribution to $a_1, \ldots, a_n$ if for every non-commutative polynomial $p\in \C\langle x_1,\ldots, x_n, x_1^*,\ldots, x_n^*\rangle$, it holds that
\[
\lim_{N\rightarrow \infty}\tau_N(p(a_1^N,\ldots, a^N_n)) = \tau(p(a_1,\ldots, a_n)).
\]
We denote this convergence by $a^N_1,\ldots, a^N_n \rightarrow _D a_1,\ldots, a_n$.
\end{definition}

Moreover, the functional calculus (\autoref{t:functional-calculus}) allows us to pass from convergence in moments to convergence of expectation of broader classes of continuous functions.

\begin{lemma}\label{lem:weak-converge-cont}
    Let $X_N$ be a sequence of self adjoint random matrices such that for all $p$, $\limsup_N \E\tr(|X_N|^{p}) < \infty$. If $X_N \rightarrow_D X$ for some self-adjoint random variable $X\in (\mathcal{A},\tau)$ in a $C^*$ probability space, then for all continuous functions $f:\R\rightarrow \C$ with polynomial growth rate $|f(x)|\leq O(|x|^p)$ for some finite $p$, we have $\E\tr(f(X_N))\rightarrow \tau(f(X))$.
\end{lemma}

\subsubsection{Free Independence} \label{s:free-independence}

A central concept in free probability is free independence, a notion of independence for non-commutative random variables. In the classical setting, if we are given independent random variables $X_1, \ldots, X_n$, we can compute their joint distribution of the random variables from the individual distributions. 
Free independence, similarly, characterizes the joint distribution of a set of freely independent non-commutative random variables.

\begin{definition}[Free Independence] \label{def:free-ind}
Let $\A_1,\A_2,\ldots, \A_n$ be a set of unital subalgebras in a probability space $(\A,\tau)$. 
We say $\A_1,\ldots, \A_n$ are freely independent from each other if
\[
\tau(a_1) = \cdots = \tau(a_m) = 0 \quad \implies \quad \tau(a_1 \cdots a_m) = 0,
\]
whenever $a_j \in \A_{i_j}$ for $j \in [m]$ and neighboring elements are from different subalgebras such that $i_j \neq i_{j+1}$ for $1 \leq j < m$.

Given a set of random variables $a_1,a_2, \ldots, a_n$, we say they are freely independent (or just ``free'') if the algebras that they generate are freely independent.
\end{definition}

This definition provides an elegant way to compute the joint distribution. 
For a particular monomial expression $\tau(a_1a_2\cdots a_n)$, 
we can assume that each $a_i$ and $a_{i+1}$ are from different subalgebras by combining them into the same variable if they are not. 
Then, we can write each $a_i$ as $\bar{a}_i + \tau(a_i) \cdot 1$, where $\bar{a}_i:= a_i -\tau(a_i)1$ is centered such that $\tau(\bar{a}_i)=0$. 
One can then verify that by expanding out everything, we get a polynomial where every term either evaluates to $0$ under $\tau$ by \autoref{def:free-ind} or is a scalar multiple of $1$. 
The following are some simple examples. 

\begin{example}\label{example:moment-computations}
Let $\A$ and $\A'$ be two subalgebras in a probability space that are freely independent. Let $a,a'\in \A$ and $b,b'\in \A'$. Then,
\begin{enumerate}
    \item $\tau(ab) = \tau(a)\tau(b)$.
    \item $\tau(aba') = \tau(aa')\tau(b)$.
    \item $\tau(aba'b') = \tau(aa')\tau(b)\tau(b') + \tau(a)\tau(a')\tau(bb') - \tau(a)\tau(a')\tau(b)\tau(b')$.
\end{enumerate}
\end{example}

We note that only the last equation is different from the classical commutative setting.
In the classical setting, if $a,a'$ are independent from $b,b'$, then $\E[aba'b'] = \E[aa'] \cdot \E[bb']$.
We will use the last equation in our proofs.

\subsubsection{Free Central Limit Theorem and Semicircular Element}

In classical probability, the central limit theorem states that the normalized sum of $n$ identically distributed independent random variables converges to the Gaussian distribution as $n \to \infty$.
In free probability, the free central limit theorem states that the normalized sum of $n$ identically distributed freely independent random variables converges to the semicircle distribution as $n \to \infty$.

\begin{theorem}[Free Central Limit Theorem]
Let $a_1, a_2, \ldots, \in \A$ be a sequence of freely independent, identically distributed, self-adjoint random variables in a non-commutative probability space $(\A, \tau)$ such that $\tau(a_i)=0$ and $\tau(a_i^2) = 1$ for all $i$.
Then the normalized sum $s_n = (a_1 + \cdots + a_n)/\sqrt{n}$ converges in distribution to the semicircle distribution such that
\[
\lim_{n \to \infty} \tau(s_n^{p}) = \int_{-2}^2 x^{p} \cdot \frac{\sqrt{4-x^2}}{2\pi} dx.
\]
\end{theorem}

A random variable with its analytical distribution being semicircle is called a semicircular element.

\begin{definition}[Semicircular Element] \label{d:semicircular-element}
Let $(\A,\tau)$ be a non-commutative probability space.
An element $s \in \A$ is called a standard semicircular element if it is self-adjoint and its analytical distribution $\mu$ with respect to $\tau$ (see \autoref{d:analytical-distribution}) has the density 
\[
\mu(dx) = \begin{cases}
          \frac{1}{2\pi} \sqrt{4-x^2} & |x| \leq 2\\
          0 & \text{otherwise}
          \end{cases}
\]
In particular, a standard semicircular element has mean $\tau(s) = 0$ and variance $\tau(s^2)=1$.
\end{definition}

We note that a standard semicircular element with mean $0$ and variance $1$ can be interpreted as the free analog of the standard Gaussian $\mathcal{N}(0,1)$, but an important difference with $\mathcal{N}(0,1)$ is that the semicircular distribution is compactly supported.

An interesting property of the semicircle distribution is that its odd moments are zero while its even moments are given by the Catalan numbers.

\begin{fact}[Catalan Moments]
Let $(\A,\tau)$ be a non-commutative probability space
and $s \in \A$ be a standard semicircular element.
Then 
\[
    \tau(s^p)=\begin{cases}
         C_{p/2} &\text{$p$ is even}\\
         0 &\text{$p$ is odd},
	 \end{cases}
\]
where $C_k = \frac{1}{k+1} \binom{2k}{k}$ is the $k$-th Catalan number.
\end{fact}

These Catalan moments underlie a combinatorial approach to free probability (see \cite{NS06}). 

\subsubsection{Joint Distributions and Non-Crossing Partitions}

An important aspect of probability theory is the characterization of moments of random variables using the combinatorics of partitions. 

\begin{definition}[Partitions]
Let $P[n]$ be the set of partitions of $n$ elements, and $P_k[n]$ be the set of partitions of $[n]$ elements where each part has size $k$. 
If $\pi = (V_1,\ldots,V_q)$ is a partition of $[n]$, we write $u\sim_\pi v $ if there exists $l\in [q]$ such that $u,v\in V_l$.
\end{definition}

 Wick's formula states that if $g_1, \ldots, g_n$ are jointly Gaussian random variables with $\E[g_i]=0$ for all $i$, then their moments are characterized as:
\begin{equation} \label{e:Wick}
\E[g_1g_2 \cdots g_n] = \sum_{\pi\in P_2[n]}\prod_{(i,j)\in \pi}\E[g_{i}g_{j}].
\end{equation}

In a non-commutative probability space, one can define ``jointly-semicircular'' random variables in an analogous manner, with free independence replacing classical independence. This follows from the key fact that the sum of free semicircular random variables is still semicircular, just as the sums of independent Gaussians remain Gaussian.

\begin{definition}[Jointly Semicircular Distribution] \label{d:jointly-semi}
A set of semicircular random variables $s_1, \ldots, s_n$ is jointly semicircular if
there exist freely independent semicircular variables $s_1', \ldots, s'_m$, and $M\in \R^{n\times m}$, $b\in \R^n$ such that $s_1,\ldots, s_n$ have the same distribution as $(\sum_{j=1}^m M_{i,j}s_j' + b_i)_{i=1}^n$. Note that the matrix $M$ in this representation does not need to be unique.
\end{definition}

An important property of jointly semicircular random variables is that their moments can be characterized by an analog of \eqref{e:Wick}, but with the restriction that the summation is just over pairings that contain no crossings.

\begin{definition}[Non-Crossing Partitions] \label{d:varphi-pi} 
Given a partition $\pi \in P[n]$, we say $\pi$ has a crossing, 
if there exists $i<j<k<l$ such that $i\sim_\pi k$ and $j\sim_\pi l$. 
We say $\pi$ is non-crossing if no such crossing exists. 
We let $NC[n]$ denote the set of all non-crossing partitions of $[n]$, 
and $NC_m[n]$ denote the set of all non-crossing partitions where each part has size $m$.
\end{definition}

\begin{theorem}[Semicircular Wick Formula] \label{t:free-Wick}
Let $(\A,\tau)$ be a non-commutative probability space, and $s_1, \ldots, s_n \in \A$ be a set of self-adjoint random variables. Then $s_1,\ldots, s_n$ are jointly semicircular if and only if for all $p\in \N$, $i_1,\ldots i_p\in [n]$
\[
\tau\big((s_{i_1}-\tau(s_{i_1})1) \cdot (s_{i_2}-\tau(s_{i_2})1) \cdots (s_{i_p}-\tau(s_{i_p})1)\big) 
= \sum_{\pi\in NC_2[p]}\prod_{(u,v)\in \pi}\tau\big((s_{i_u}-\tau(s_{i_u})1) \cdot (s_{i_v}-\tau(s_{i_v})1)\big).
\]
\end{theorem}

We can see from \autoref{t:free-Wick} that the distribution of jointly semicircular $s_1,\ldots s_n$ is fully determined by the quantities $\{\tau(s_i)\}_{i=1}^n$ and $\{\tau((s_i - \tau(s_i)1)(s_j - \tau(s_j)1))\}_{i,j\in [n]}$ for $i,j\in [n]$. These are exactly the entries of $b$ and $MM^*$ respectively in \autoref{d:jointly-semi}.

The non-crossing partitions have a simple recursive structure similar to that for the Catalan numbers.
This is the key reason that the moments of the free model $\Xf$ can be computed efficiently in polynomial time using dynamic programming.
See \autoref{s:computation}.

So far the semicircular elements are abstract objects,
but we note that they can be concretely realized as infinite-dimensional operators in a Hilbert space.

\begin{theorem}[Fock Space] \label{theorem:concrete-realization}
There exists a Hilbert space $\mathcal{H}$, a linear functional $\tau: B(\mathcal{H})\rightarrow \C$ satisfying $\tau(1)=1$, and self-adjoint operators $\{s_i\}_{i\in \N} \in B(\mathcal{H})$ such that each $s_i$ is a semicircular element and
 $s_1,s_2, \ldots$ are freely independent.
\end{theorem}

The explicit construction of the Fock space is simple and elegant, but the details are not relevant to this paper. 
We mention the explicit construction because it is important in the proofs of formulas and bounds for norms of free objects that we will see in the next subsection.
Because of this construction, sometimes we also call a semicircular element a semicircular operator.

\subsection{Random Matrices and Semicircular Matrices} \label{s:RMT-free}

A fundamental result in random matrix theory is Wigner's semicircle law, 
which states that the limiting distribution of the eigenvalues of many random symmetric matrices is the semicircle distribution in \autoref{d:semicircular-element}.
This motivated Voiculescu to build a connection between the theory of free probability and random matrix theory.

\subsubsection{Gaussian Random Matrices and Semicircular Family}

We begin by defining the Gaussian Orthogonal Ensemble and outlining its relationship to the family of freely independent semicircular elements. 
This will allow us to formally define the free model $\Xf$ and study its relationship to the Gaussian model. 

\begin{definition}[Gaussian Orthogonal Ensemble (GOE)]
A GOE matrix of dimension $d$ can be expressed as
\[
G= \sum_{1 \leq i\leq j \leq d} \frac{g_{i,j}}{\sqrt{d}} E_{i,j},
\quad 
\]
where $g_{i,j} \sim \mathcal{N}(0,1)$ and
\[
E_{i,j} =\begin{cases}
    \chi_i\chi_j^\top + \chi_j\chi_i^\top &i\neq j\\
    \sqrt{2}\chi_i\chi_i^\top &i=j
\end{cases}. 
\]
\end{definition}
In other words, a GOE matrix is a random symmetric matrix with independent Gaussian entries on the upper-diagonal. Each diagonal entry has variance $2/d$ and each off-diagonal entry has variance $1/d$. Under this normalization, we have $\E[\tr(G)] = 0$ and $\E[\tr(G^2)] = 1+1/d$. 

A classical result in random matrix theory is that if $G^{(N)}$ is a Gaussian GOE matrix of dimension $N$, then for all $p\in \N$,
\[
\lim_{N\rightarrow\infty}\E\tr((G^{(N)})^p)  = \begin{cases}
         C_{p/2} &\text{$p$ is even}\\
         0 &\text{$p$ is odd},
\end{cases}
\]
where $C_k$ is the $k^{th}$ Catalan number. 
This coincides with the moments of a standard semicircular element.
It follows that in the limit as $N\rightarrow\infty$, the empirical eigenvalue distribution of $G^{(N)}$ tends towards that of a semicircle distribution in \autoref{d:semicircular-element}. 

An important result by Voiculescu is that independent GOE matrices are asymptotically freely independent.
More formally, let $G_1, \ldots, G_m$ be independent $N \times N$ GOE matrices, $p_1, \ldots, p_k \in \N$, and $i_1, \ldots, i_k \in [m]$ be such that $i_1 \neq i_2, i_2 \neq i_3, \ldots, i_{k-1} \neq i_k$.
Then,
\[
\lim_{N \to \infty} \E \tr\Big[ \big( G_{i_1}^{p_1} - \E[\tr G_{i_1}^{p_1}] I \big) \big( G_{i_2}^{p_2} - \E[\tr G_{i_2}^{p_2}] I \big) \cdots \big( G_{i_k}^{p_k} - \E[\tr G_{i_k}^{p_k}] I \big) \Big] = 0.
\]  
As free independence determines the joint distribution of a set of random variables (see \autoref{s:free-independence}), it follows that independent GOE matrices converge in distribution to freely independent semicircular elements.

\begin{theorem}[Weak Convergence of GOE Matrices] \label{theorem:weak-convergence}
Let $G_1^{(N)}, \ldots, G_n^{(N)}$ be independent $N \times N$ GOE matrices. 
Let $(\A,\tau)$ be a probability space where $\A$ is generated by a family of freely independent semicircular elements $s_1, \ldots, s_n$. 
Then, for any non-commutative polynomial $p\in \C\langle x_1, \ldots, x_n\rangle$,
\[
        \lim_{N\rightarrow \infty}\E\tr(p(G_1^{(N)}, \ldots, G_n^{(N)})) = \tau(p(s_1, \ldots, s_n)).
\]    
\end{theorem}

For concreteness, we can think of the infinite-dimensional operators $s_1,s_2, \ldots$ in \autoref{theorem:concrete-realization} as a representation of the limit objects $\{ \lim_{N\rightarrow\infty}(G_i^{(N)}) \}_{i \in \N}$.
We refer to the $C^*$-algebra generated by $s_1,s_2, \ldots$ as the semicircular algebra.

Using the weak convergence in \autoref{theorem:weak-convergence} and properties of Gaussian matrices, we can establish analogous properties of semicircular elements.
For example, the following property follows as the sum of independent Gaussian matrices is also a Gaussian matrix. 

\begin{fact}[Sum of Freely Independent Semicircular Elements] \label{fact:semi-conv}
Suppose $s_1$ and $s_2$ are freely independent semicircular elements. 
Then $s_1+s_2$ is a semicircular element with variance $\tau(s_1^2) + \tau(s_2^2)$.
\end{fact}

\subsubsection{Semicircular Matrices}

We have seen that a semicircular operator is a good approximation for a large GOE matrix. 
What should be the free operator approximation of general Gaussian matrices? 
The main result of \cite{BBvH23} is that under certain ``intrinsic freeness'' conditions, a random matrix with Gaussian entries can be approximated by a matrix with semicircular elements.
 
% Formally, let $(\A,\tau)$ be the semicircular algebra, 
% and $\M_d(\C)$ be the set of $d\times d$ complex matrices. 
% We consider the algebra $\M_d(\C)\otimes \A$, which can be represented as the space of $d\times d$ matrices whose entries are elements in $\A$. 
% We say that an operator $X\in \M_d(\C)\otimes \A$ is a ``semicircular matrix'' if each of its matrix entries has the semicircular distribution. 
% This is analogous to a Gaussian matrix in the classical setting. 
% By \autoref{d:jointly-semi}, these operators can be characterized as sums of freely independent semicircular elements with matrix coefficients.

Let $s_1,\ldots,s_n$ be a free standard semicircular family.
The free analogue of a Gaussian matrix $A_0+\sum_{i=1}^n g_iA_i$ is the operator-valued linear pencil $A_0\otimes1+\sum_{i=1}^n A_i\otimes s_i$.
Note that this should not in general be interpreted as a matrix whose entries are individually semicircular. 
In particular, an off-diagonal entry need not be self-adjoint. 
% The relevant structure is instead determined by the coefficient matrices and the associated operator-valued covariance map.

\begin{definition}[Semicircular Matrix] \label{d:semicircular-matrix}
Let $X\in \M_d(\C)\otimes \A$. 
$X$ is called a semicircular matrix if there exist matrices $A_0,A_1, \ldots, A_n$ such that $X = A_0 \otimes 1 + \sum_{i=1}^nA_i\otimes s_i$ where $s_1, \ldots, s_n$ are freely independent semicircular elements. 
We say $X$ is centered if $A_0 = 0$. 
\end{definition}

The main result in \cite{BBvH23}, as stated in \eqref{e:BBvH-spectrum}, is that the Gaussian model $X = A_0 + \sum_{i=1}^n g_i A_i$ where $g_1, \ldots, g_n$ are independent Gaussians can be approximated by the free model $\Xf = A_0 \otimes 1 + \sum_{i=1}^n A_i \otimes s_i$ where $s_1, \ldots, s_n$ are freely independent semicircular elements.
In their approach, $\Xf$ is approximated by $Nd\times Nd$ random matrix $X_N$ for large enough $N$, defined as
\[
    X_N = A_0 \otimes I_N + \sum_{i=1}^n A_i\otimes G_i^N
\]
where $G_1^N, \ldots, G_n^N$ are i.i.d.~GOE matrices. 
As $N\rightarrow\infty$, the spectral distribution of $X^N$ tends towards that of $\Xf$. 
This follows from \autoref{theorem:weak-convergence}, as for each $p\in \N$, $(X^N)^{p}$ and $(\Xf)^p$ can be expressed as matrices whose entries are polynomials in $G_1^N, \ldots, G_n^N$ and $s_1, \ldots, s_n$ respectively. 
For appropriate classes of continuous function $f$, there is a sequence of polynomials converging to $f$, so that \autoref{theorem:weak-convergence} implies
\begin{equation}
\tr\otimes\tau (f(\Xf) ) = \lim_{N\rightarrow\infty} \E \tr[f(X_N)].
\end{equation}

A key advantage of working with the free model is that there are formulas for bounding and calculating the norms of semicircular matrices. 
These formulas are derived using the concrete realizations of the free semicircular random variables as operators in a Hilbert space as described in \autoref{theorem:concrete-realization}.

\begin{theorem}[Pisier's Inequality] \label{lemma:pisier}
Let $\Xf = A_0 \otimes 1 + \sum_{i=1}^n A_i\otimes s_i$ be a semicircular matrix. Then
\[
\norm{\Xf} \leq \norm{A_0} + \Bignorm{\sum_{i=1}^n A_i^*A_i}^{\frac12}  + \Bignorm{\sum_{i=1}^n A_iA_i^*}^{\frac12}.
\]
\end{theorem}

\begin{theorem}[Lehner's Formula \cite{Leh99}] \label{lemma:lehner}
    Let $\Xf = A_0 + \sum_{i=1}^n A_i\otimes s_i$ be a semicircular matrix. Then
    \begin{align*}
        \norm{\Xf} = \inf_{\substack{Y \succ 0\\}} \max_{\eps\in \{\pm 1\}}\lambda_{\max}\Big(\eps A_0 + Y^{-1} + \sum_{i=1}^n A_i^*YA_i\Big).
    \end{align*}
\end{theorem}

A corollary~\cite[Eq 1.6]{BCS+24} is that if $X$ is a random matrix with the same covariance profile as $\Xf$, then  
\begin{equation} \label{e:Lehner-corollary}
        \norm{\Xf} = \inf_{\substack{Y \succ 0\\}} \max_{\eps\in \{\pm 1\}} \lambda_{\max} \big(\eps \E[X] + Y^{-1} + \E\big[ (X-\E[X])\cdot Y \cdot (X-\E[X])\big]\big).
\end{equation}

\subsubsection{Partial Trace of Matrices and Operators}

For random variables with matrix coefficients, we extend $\tau$ to the partial trace operator
\[
\varphi := \id \otimes \tau.
\]
This can be viewed abstractly as a special case of an operator-valued conditional expectation, but we just need some simple properties such as $\varphi[(A \otimes 1)(X)(B \otimes 1)] = A \cdot \varphi(X) \cdot B$ for $X \in \M_d(\C)\otimes\A$.

The following identity is from~\cite[Lemma 5.2]{BBvH23}, which will be used to approximate the partial trace of a semicircular matrix. 

\begin{lemma}[Partial Trace of Random Gaussian Matrices] \label{lemma:cond-exp}
Let $\A_N$ be the algebra generated by independent $N \times N$ GOE matrices $G_1^N, G_2^N,\ldots$. 
For any $Y_N\in \M_d(\C)\otimes\A_N$, 
\[
\E[Y_N] = \E[ (\id_d\otimes \tr) (Y_N)] \otimes I_N.
\] 
Moreover, for any self adjoint $A_1,\ldots A_n$, $X_N = \sum_i A_i\otimes G_i^N$, and continuous $h:\C\rightarrow \R$, we have
\begin{align*}
    \E[h(X_N)] =   \E[ (\id_d\otimes \tr) h(X_N)] \otimes I_N
\end{align*}
\end{lemma}

Applying \autoref{lemma:cond-exp} on monomials gives the following extension of the weak convergence in \autoref{theorem:weak-convergence}.

\begin{corollary}[Weak Convergence of GOE Matrices] \label{corollary:cond-convergence}
Let $G_1^N, \ldots, G_n^N$ be i.i.d.~$N \times N$ GOE matrices and $s_1, \ldots, s_n$ be freely independent semicircular elements. 
Let $P\in \M_d(\C)\langle x_1,\ldots, x_n\rangle$ be a non-commutative polynomial in $n$ variables with matrix coefficients in $\M_d(\C)$. Then, 
\[
\E[(\id_d \otimes \tr) P(G_1^N, \ldots, G_n^N)] \rightarrow  \varphi[P(s_1, \ldots, s_n)].
\]

\end{corollary}

Moreover, since GOE matrices have uniformly bounded moments, a similar analog to \autoref{lem:weak-converge-cont} holds.

\begin{corollary}\label{cor:cond-convergence-cont}
     Let $A_1,\ldots A_n\in M_d(\C)$ be self adjoint, $X_N = \sum_i A_i\otimes G_i^N$, and $h:\C\rightarrow \R$ be a continuous function with polynomial growth rate. Then we have
\begin{align*}
    \E[(\id_d \otimes \tr) h(X_N)] \rightarrow  \varphi[h(X)],
\end{align*}
where $X = \sum_iA_i\otimes s_i$. 
\end{corollary}

The reason why we work with the partial trace operator is that it provides a convenient way to extend the free Wick formula for semicircular elements in \autoref{t:free-Wick} to semicircular matrices.
For this, we need the following notation for a non-crossing partition. 

\begin{definition}[Definition 2.1.1 in~\cite{Spe98}]
Let $X_1,X_2, \ldots, X_n$ be semicircular matrices.
Let $\pi \in NC[n]$ for some $n\in \N$. 
Let $\mathcal{I} = \{i,i+1, \ldots, i+\ell\}$ be an interval contained in $\pi$.
Define
\begin{align*}
\varphi_\pi(X_1,\ldots,X_n) = \begin{cases}
    \varphi[X_1\cdots X_n] &\text{if $\mathcal{I} = [n]$}, \\
    \varphi_{\pi\backslash \mathcal{I}}(X_1,\ldots,X_{i-1},(\varphi[X_i\cdots X_{i+\ell}] \otimes 1) \cdot X_{i+\ell+1},\ldots,X_n) &\text{otherwise}.
    \end{cases}
\end{align*}
\end{definition}
For example, if $\pi=(\{1,4\},\{2,3\})$, then
\[
\varphi_\pi(X_1,X_2,X_3,X_4) 
= \varphi_{(\{1,4\})}(X_1, (\varphi[X_2X_3] \otimes 1) \cdot X_4) 
= \varphi[X_1 \cdot (\varphi[X_2X_3] \otimes 1) \cdot X_4].
\]
With this notation at hand, the free Wick formula extends nicely to the matrix-coefficient setting.

\begin{theorem}[Free Wick Formula for Semicircular Matrices] \label{lemma:op-wick}
Let $X_1,X_2, \ldots, X_n$ be centered semicircular matrices. Then
    \begin{align*}
        \varphi[X_1\cdots X_n] = \sum_{\pi \in NC_2[n]}\varphi_\pi(X_1,X_2,\ldots,X_n).
    \end{align*}
\end{theorem}

The following is a slightly more general formula for computing the joint moments of semicircular matrices.

\begin{lemma}[Joint Moments of Freely Independent Semicircular Matrices] \label{lemma:sc-joint-moments}
Let $X_1,\ldots, X_n \in \M_d(\C)\otimes \A$ be freely independent centered semicircular matrices. 
Let $m\in \N$, $B_1,\ldots, B_{m-1}\in \M_d(\C)$ and $i_1,\ldots, i_m\in [n]$. 
Given a non-crossing pair partition $\pi \in NC_2[m]$, 
we say that $\pi$ is consistent with the indices $i_1, \ldots, i_m \in [n]$ if $i_u = i_v$ whenever $u\sim_\pi v$ and denote it by $i\sim \pi$. 
Then, 
\[
\varphi[X_{i_1}(B_1 \otimes 1) \cdots X_{i_{m-1}}(B_{m-1} \otimes 1)X_{i_m}] = \sum_{\substack{\pi\in NC_2[m]\\i\sim\pi}} \varphi_\pi (X_{i_1} (B_1 \otimes 1),\ldots, X_{i_{m-1}}(B_{m-1} \otimes 1), X_{i_m}).
\]
\end{lemma}
\begin{proof}
By \autoref{lemma:op-wick},
\[
\varphi[X_{i_1}(B_1 \otimes 1)\cdots X_{i_{m-1}}(B_{m-1} \otimes 1)X_{i_m}] 
= \sum_{\substack{\pi\in NC_2[m]}} \varphi_\pi (X_{i_1} (B_1 \otimes 1),\ldots, X_{i_{m-1}}(B_{m-1} \otimes 1), X_{i_m}).
\]
For any matrix $B\in \M_d(\C)$, note that $\varphi[X_i (B \otimes 1) X_j] = 0$ for $i\neq j$ as $X_i$ and $X_j$ are freely independent and centered. 
Indeed, one can verify this by decomposing $X_i = \sum_{k}A_{i,k}\otimes s_k$ and $X_j = \sum_\ell A_{j,\ell}\otimes s'_\ell$,
and noting that $\tau(s_ks'_\ell) = 0$ for all $k,\ell$. 
Thus, for any $\pi$ with $u\sim_\pi v$ but $i_u \neq i_v$, 
the expression $\varphi_\pi (X_{i_1}(B_1 \otimes 1),\ldots, X_{i_{m-1}}(B_{m-1}\otimes 1), X_{i_m})$ must evaluate to $0$. 
Therefore, we can restrict the summation to those $\pi$ that are consistent with $i_1, \ldots, i_m$.
\end{proof}

The following is a simple formula that is important in our proofs, for analyzing the second order terms involving semicircular matrices.

\begin{lemma}[Partial Trace of Second-Order Terms] 
\label{lemma:free-moments}
Let $X = \sum_{i=1}^n A_i\otimes s_i$ be a centered semicircular matrix. 
Let $\A'$ be a unital algebra freely independent from the algebra generated by $\{s_1,\ldots,s_n\}$. 
For any $Y,Z\in \M_d(\C)\otimes \A'$,
\[
\varphi[X YXZ] = \varphi[X \cdot (\varphi[Y] \otimes 1) \cdot X] \cdot \varphi[Z].
\]
\end{lemma}
\begin{proof}
Since partial trace is multi-linear, it suffices to prove the equality for $Y = B\otimes b$, $Z = C\otimes c$ for $B,C\in \M_d(\C)$ and $b,c\in \A'$. 
The left hand side is
\[
\varphi[XYXZ] = \sum_{i=1}^n \tau(s_ibs_ic) \cdot A_iBA_iC  = \sum_{i=1}^n \tau(s_i^2)\tau(b)\tau(c) \cdot A_iBA_iC,
\]
where the second equality follows from item 3 in \autoref{example:moment-computations}. 
The right hand side is
\[
\varphi[X(\varphi[Y] \otimes 1)X] \varphi[Z] 
= \tau(b)\tau(c) \cdot \varphi[X(B \otimes 1)X]C 
= \tau(b)\tau(c) \cdot \sum_{i,j=1}^n \tau(s_i s_j) \cdot A_i B A_j C 
= \sum_{i=1}^n \tau(s_i^2) \tau(b)\tau(c) \cdot A_iBA_iC,
\]
where the last equality is by free independence.
\end{proof}

Using partial trace, we can conveniently define the notion of a covariance matrix of a semicircular matrix $X \in \M_d(\C) \otimes \A$.
In the derandomization of general random matrix model in \autoref{section:uni}, we will use this notion to define a semicircular matrix with the same covariance matrix as a general random matrix $Z \in {\mathcal M}_d(\C)$.

% \begin{definition}[Covariance of Semicircular Matrix] \label{d:cov-profile} 
%     Let $X = A_0 \otimes 1 + \sum_{k=1}^n A_k \otimes s_k$ be a semicircular matrix, where $s_1, \ldots, s_n$ are freely independent semicircular elements. 
%     The covariance matrix, $\cov(X)$, of $X$ is a $d^2 \times d^2$ matrix such that the $((i_1,j_1), (i_2, j_2))$-th entry is
%     \[
%         \tau\big(\bar{X}(i_1,j_1) \cdot \bar{X}(i_2,j_2)^* \big),
%     \]
%     where $i_1, j_1, i_2, j_2 \in [d]$ and $\bar{X} = X-A_0\otimes 1$, i.e. $\bar{X}(i,j) = \sum_{k=1}^n A_k(i,j) \cdot s_k$.
    
%     Let $Z \in \M_d(\C)$ be a self-adjoint random matrix, we say self-adjoint $X$ and $Z$ have the same covariance if
%     \[
%         %\tau\big(X(i_1, j_1) \cdot X(i_2,j_2)\big) = \E\big[Z(i_1,j_1) \cdot Z(i_2,j_2)\big], \quad \text{for all pairs $(i_1, j_1), (i_2, j_2) \in [d] \times [d]$}.
%         \cov(X) = \cov(Z)
%     \]
%     Using partial trace, this can be compactly written as
%     \begin{align} \label{eq:cov-profile}
%         \varphi[\bar{X} (M \otimes 1) \bar{X}] = \E[(Z-\E Z) M (Z-\E Z)] \quad \text{for all $M \in \M_d(\C)$}.
%     \end{align}
% \end{definition}

\begin{definition}[Covariance of Semicircular Matrix] \label{d:cov-profile} 
Let $(\A,\tau)$ be a tracial $*$-probability space, let
$\varphi:=\id_d\otimes\tau$, and let
$X\in\M_d(\C)\otimes\A$ have finite second moments. 
The covariance matrix $\cov(X)\in\M_{d^2}(\C)$ is defined by
\[
\cov(X)_{(i,j),(k,\ell)}
:=
\tau\big((X - \varphi[X] \otimes 1)_{i,j} (X - \varphi[X] \otimes 1)_{k,\ell}^*\big).
\]
Equivalently, the same second-order information is encoded by the
covariance map
\begin{equation}
\label{eq:operator-valued-covariance-map}
\varphi\big[ (X - \varphi[X] \otimes 1) (M\otimes1) (X - \varphi[X] \otimes 1)^*\big],
\qquad M\in\M_d(\C).
\end{equation}
since
\[
\cov(X)_{(i,j),(k,\ell)} = \varphi\big[ (X - \varphi[X] \otimes 1) (E_{j,\ell} \otimes 1) (X - \varphi[X] \otimes 1)^*\big]_{i,k}.
\]

% We define
% \begin{align*}
% \sigma(X)^2
% &:=
% \max\left\{
% \norm{\varphi[(X^\circ)^*X^\circ]},
% \norm{\varphi[X^\circ(X^\circ)^*]}
% \right\},
% \\
% \nu(X)^2
% &:=
% \norm{\cov(X)},
% \\
% \sigma_*(X)^2
% &:=
% \sup_{\norm{u}=\norm{v}=1}
% \tau\left[
% \big(u^*X^\circ v\big)^*
% \big(u^*X^\circ v\big)
% \right].
% \end{align*}
% Here $u^*X^\circ v$ denotes
% \[
% (u^*\otimes1)X^\circ(v\otimes1)\in\A.
% \]
% By traciality of $\tau$ and positivity of the covariance map,
% \begin{equation}
% \label{eq:sigma-star-covariance-map}
% \sigma_*(X)^2
% =
% \sup_{\norm{v}=1}\norm{\eta_X(vv^*)}
% =
% \sup_{\substack{M\succeq0\\ \Tr(M)=1}}
% \norm{\eta_X(M)}.
% \end{equation}
% Thus $\sigma_*(X)$ depends only on the covariance profile of $X$
% and is unchanged by adding a deterministic matrix in
% $\M_d(\C)\otimes1$.

% If $X$ is self-adjoint, then $(X^\circ)^*=X^\circ$ and
% \[
% \sigma(X)^2=\norm{\eta_X(I_d)}.
% \]

If $Z$ is a classical random matrix, the same definitions are used
with $\tau=\E$. We say that $X$ and $Z$ have the same covariance
profile if $\cov(X)=\cov(Z)$, or equivalently if
\begin{equation}
\label{eq:cov-profile}
\varphi\big[ (X - \varphi[X] \otimes 1) (M \otimes 1) (X - \varphi[X] \otimes 1)^*\big]
=
\E\big[(Z-\E Z)M(Z-\E Z)^*\big]
\qquad\text{for every }M\in\M_d(\C).
\end{equation}
For self-adjoint $X$ and $Z$, the adjoints in
\eqref{eq:operator-valued-covariance-map} and
\eqref{eq:cov-profile} may be omitted.
\end{definition}

A simple but very useful fact is that if the operator norm of a random matrix $Z$ is almost surely bounded, then the semicircular matrix with the same covariance as $Z$ would also have bounded operator norm.
We note that this is not true for the finite dimensional Gaussian matrices that have the same covariance as $Z$. 

\begin{lemma}[Bounded Semicircular Operator Norm] \label{lemma:semicircular-norm}
    Let $Z$ be a centered $d$-dimensional Hermitian random matrix such that $\|Z\| \leq \rho$ with probability one.
    Let $X \in \M_d(\C)\otimes \A$ be a centered self-adjoint semicircular matrix with the same covariance as $Z$.
    Then
    \[ \|X\| \leq 2\rho. \]
\end{lemma}
\begin{proof}
    By \autoref{d:semicircular-matrix}, any centered semicircular matrix $X \in \M_d(\C)\otimes \A$ can be written in the form of $X = \sum_{j=1}^{m} A_j \otimes s_j$ where $A_1, \ldots, A_{m}$ are $d\times d$ Hermitian matrices and $s_1, \ldots, s_m$ are freely independent semicircular elements. It follows by Pisier's inequality in \autoref{lemma:pisier} that
    \begin{align*}
        \norm{X} \leq  2 \bigg\|\sum_{j=1}^m A_j^2 \bigg\|^{\frac12} = 2\sqrt{\norm{\varphi[X^2]}}.
    \end{align*}
    Since $X$ has the same covariance as $Z$, it holds that
    \begin{align*}
        \norm{\varphi[X^2]} = \norm{\E[Z^2]} \leq \E[\|Z^2\|] \leq \rho^2,
    \end{align*}
    where the second last inequality follows from Jensen's inequality.
\end{proof}

Finally, we define some intrinsic matrix parameters, which are analogs of the $\sigma, \nu, \sigma_*$ parameters for Gaussian matrices.

\begin{definition}[Intrinsic Matrix Parameters]
    For $X\in\M_d(\C)\otimes\A$, we define
\begin{align*}
\sigma(X)^2
&:=
\max\left\{
\norm{\varphi[(X - \varphi[X] \otimes 1)^* (X - \varphi[X] \otimes 1)]},
\norm{\varphi[(X - \varphi[X] \otimes 1) (X - \varphi[X] \otimes 1)^*]}
\right\},
\\
\nu(X)^2
&:=
\norm{\cov(X)},
\\
\sigma_*(X)^2
&:=
\sup_{\norm{u}=\norm{v}=1}
\tau\left[
\big(u^* (X - \varphi[X] \otimes 1) v\big)^*
\big(u^* (X - \varphi[X] \otimes 1) v\big)
\right].
\end{align*}
Here $u^* (X - \varphi[X] \otimes 1) v$ denotes
\[
(u^*\otimes1) (X - \varphi[X] \otimes 1) (v\otimes1)\in\A.
\]
\end{definition}

% By traciality of $\tau$ and positivity of the covariance map,
% \begin{eqnarray*}
% % \label{eq:sigma-star-covariance-map}
% \sigma_*(X)^2
% & = &
% \sup_{\norm{v}=1}\norm{ \varphi\big[ (X - \varphi[X] \otimes 1) (v v^* \otimes 1) (X - \varphi[X] \otimes 1)^*\big] } \\
% & = &
% \sup_{\substack{M\succeq0\\ \Tr(M)=1}}
% \norm{\varphi\big[ (X - \varphi[X] \otimes 1) (M \otimes 1) (X - \varphi[X] \otimes 1)^*\big]}.
% \end{eqnarray*}
% Thus, $\sigma_*(X)$ depends only on the covariance profile of $X$ and is unchanged by adding a deterministic matrix in $\M_d(\C)\otimes1$.

Note that, for $X = A_0 \otimes 1 + \sum_{i=1}^n A_i \otimes s_i \in \M_d(\C)\otimes \A$, we have
\[
\sigma(X)^2 = \max\bigg\{\biggnorm{\sum_{i=1}^n A_i A_i^*}, \biggnorm{\sum_{i=1}^n A_i^* A_i} \bigg\}
\]
and
\[
\sigma_*(X)^2 = \sup_{\norm{u}=\norm{v}=1} \tau\bigg( \bigg( \sum_{i=1}^n u^* A_i v \otimes s_i\bigg)^* \bigg( \sum_{i=1}^n u^* A_i v \otimes s_i\bigg) \bigg) = \sup_{\norm{u}=\norm{v}=1} \sum_{i=1}^n (u^* A_i v)^2.
\]
Both are consistent with those for the corresponding Gaussian matrix $A_0 + \sum_{i=1}^n g_i A_i$.

We also note that $\sigma_*(X) \le \min\{\sigma(X), \nu(X)\}$ (see, e.g., Section 2.1 in~\cite{BBvH23}).
Finally, we make an observe that $\sigma_*$ is a seminorm, which is helpful in our later analysis.
\begin{lemma}[Triangle Inequality of $\sigma_*$, cf. Section 8.1.2 in~\cite{BBvH23}] \label{l:sigma*}
    For any $X, Y \in \M_d(\C) \otimes \A$, we have $\sigma_*(X+Y) \leq \sigma_*(X) + \sigma_*(Y)$.
\end{lemma}

\section{Technical Tools} 

In this section, we collect some formulas and inequalities that will be used in the proofs.
Many of these were used in \cite{BBvH23,BvH24} for finite-dimensional random matrices.
Since we work directly with general non-commutative random variables in $\M_d(\C) \otimes \A$ for derandomization purposes, we need to generalize some statements in \cite{BBvH23,BvH24} to the infinite-dimensional setting.
The proofs are relatively straightforward by using the finite-dimensional approximation method, but we include them for completeness.

This section is a reference section, which may be skipped or quickly glossed over at the first reading and is only referred to when some results are needed.  

\subsection{Calculus for Banach Space}

In this subsection, we review some basic facts about differentiating and bounding functions on arbitrary normed (not necessarily finite-dimensional) vector spaces. 
While these definitions and theorems are stated abstractly, we will see concrete applications of them in the main results. 
For more detailed review of this subject, we refer the reader to \cite{Car71}. 

\begin{definition} [Banach Space and Algebra] 
A Banach space $Y$ is a vector space over $\C$ or $\R$ equipped with a norm $\norm{\cdot}$ such that the topology on $Y$ induced by the distance $d(x,y) = \norm{x-y}$ is complete. 
If $Y$ is also an associative algebra over the same field (i.e., closed under multiplication) satisfying $\norm{xy}\leq \norm{x}\norm{y}$, then $Y$ is a Banach algebra. 
\end{definition}

A canonical example of a Banach space is the vector space $\C^n$, equipped with the standard $L_2$ norm. 
A canonical example of a Banach algebra is the set of linear operators $\cL(\C^n;\C^n)$, equipped with the operator norm. 

\begin{definition}[Linear Maps]
Let $Y,Z$ be Banach spaces. 
Let $\cL(Y;Z)$ be the set of linear maps from $Y$ to $Z$. 
Moreover, given $Y_1, \ldots,Y_k,Z$, let $\cL(Y_1, \ldots, Y_k;Z)$ be the set of $k$-linear maps from $Y_1\times Y_2\times \cdots \times Y_k$ to $Z$. 
Finally, denote $\cL(\underbrace{Y,Y, \ldots, Y}_{\text{$k$ times}};Z)$ as $\cL^k(Y;Z)$.
\end{definition}

\subsubsection{Derivatives}

Next, we see the definitions of derivatives of functions over Banach spaces and algebras.

\begin{definition}[Derivative]
Let $f:U\rightarrow Z$ be a continuous function, and $U\subseteq Y$ be an open set. 
Given a point $a\in U$, $f$ is differentiable at $a$ if there exists a linear map $L \in \cL(Y;Z)$ so that
\[
\lim_{\norm{h}\rightarrow0} \frac{f(a+h) - f(a) - Lh}{\norm{h}} = 0.
\]
The linear map $L$ is the derivative (or Frechet derivative) of $f$ at $a$, which is denoted by $Df(a)$. 
The function $f$ is differentiable everywhere if it is differentiable on an open set around every $a\in U$.
\end{definition}

\begin{definition}[Directional Derivative]
Suppose $f$ is differentiable at $a$. Given a vector $h\in Y$, the
directional derivative of $f$ at $a$ in the direction $h$ is the
element $Df(a)h\in Z$ given by
\[
Df(a)h
=
\lim_{t\to 0}
\frac{f(a+th)-f(a)}{t}.
\]
Here, $Df(a)\in\cL(Y;Z)$ is the derivative of $f$ at $a$.
\end{definition}

\begin{definition} [Partial Derivative]
Given an $m$-variable function $f:Y_1\times \cdots \times Y_m \rightarrow Z$, 
let $\partial_{i}f$ be the partial derivative of $f$ with respect to the $i$-th variable such that
\[
\partial_{i}f(a_1,\ldots,a_m)h := \lim_{t\rightarrow 0}\frac{f(a_1, \ldots, a_i+th, \ldots, a_m)-f(a_1, \ldots, a_i, \ldots, a_m)}{t}.
\]
\end{definition}

To compute derivatives of functions on Banach spaces, 
we can use the basic rules of calculus, translated to this general setting.

\begin{fact}[Chain Rule] \label{fact:chain-rule}
Let $Y_1,Y_2,Y_3$ be Banach spaces and $f:Y_1\rightarrow Y_2$, $g: Y_2\rightarrow Y_3$ be continuously differentiable functions. 
Let $a\in Y_1$ and $U,V$ be open sets such that $a\in U\subseteq Y_1$ and $f(a)\in V\subseteq Y_2$. 
Then, for any $h\in Y_1$,
\[
D(g\circ f)(a)h = Dg(f(a)) \cdot Df(a)h.
\]
\end{fact}

% \begin{fact}[Product Rule] \label{fact:prod-rule}
% Let $Y,Z$ be Banach spaces and 
% $f: Y\times \cdots \times Y\rightarrow Z$ be a $m$-linear map. 
% Then, for any $h\in E$,
% \[
% Df(a_1,a_2, \ldots, a_m)h = \sum_{k=1}^{m}f(a_1,\ldots,a_{k-1},h,\ldots,a_m).
% \]
% \end{fact}

\begin{fact}[Multilinear Product Rule]
\label{fact:prod-rule}
Let $Y_1,\ldots,Y_m,Z$ be Banach spaces, and let $f \in\cL(Y_1,\ldots,Y_m;Z)$ be a $m$-linear map.
Then for any $a=(a_1,\ldots,a_m) \in Y_1 \times \cdots \times Y_m$ and $h=(h_1,\ldots,h_m) \in Y_1 \times \cdots \times Y_m$, we have
\[
D f(a) h
=
\sum_{k=1}^m
f(a_1,\ldots,a_{k-1},h_k,a_{k+1},\ldots,a_m).
\]

% More generally, let $E$ be a Banach space, let $U\subseteq E$ be open, and let $F_k:U\rightarrow Y_k$ be differentiable at $x\in U$ for $1\leq k\leq m$. Define $F(u):=f(F_1(u),\ldots,F_m(u))$.
% If $F$ is differentiable at $x$, then
% \[
% D F(x) h
% =
% \sum_{k=1}^m
% f\big(
% F_1(x),\ldots,F_{k-1}(x),
% DF_k(x) h,
% F_{k+1}(x),\ldots,F_m(x)
% \big).
% \]

In particular, let $E$ be a Banach algebra, $U \subseteq E$ be open, and $F_1,\ldots,F_m: U \rightarrow E$ be differentiable. Let $a_1, \ldots, a_m \in U$ and $f(a_1, \ldots, a_m) = F_1(a_1) \cdots F_m(a_m)$. Then, for any $x \in U$ and $h \in E$,
\[
D(F_1\cdots F_m)(x) h
=
\sum_{k=1}^m
F_1(x)\cdots F_{k-1}(x)
(DF_k(x) h)
F_{k+1}(x)\cdots F_m(x).
\]
\end{fact}

\begin{fact}[Derivative of Inverses] \label{fact:inv} 
Let $A$ be an invertible element in a Banach algebra, and $f(A) = A^{-1}$. Then 
\[
Df(A)H = -A^{-1}HA^{-1}.
\]
\end{fact}

\subsubsection{Higher Order Derivatives}

The map $Df:U\rightarrow \cL(Y;Z)$ can be differentiated further. 
For each $a\in U$, the second derivative $D^2f(a)$ is in $\cL(Y; \cL(Y;Z))$.  
We can identify each linear map $L\in \cL(Y; \cL(Y;Z))$ with a bi-linear map in $\cL(Y,Y;Z)$ by writing $L(g,h) = (Lg)h$. 
Thus, we can interpret $D^2f(a)$ as an element in $\cL(Y,Y;Z)$. 
It can be shown that this map is symmetric, meaning that $D^2f(a)(h,g) = D^2f(a)(g,h)$ for all $a$ (see \cite[Theorem 5.1.1]{Car71}). 
This is analogous to $\partial_x\partial_yf(x,y) = \partial_y\partial_xf(x,y)$ for functions on $\R^2$.
 
Extending further, we can define the general $k$-th order derivative of $f$ as a $k$-linear map as follows.

\begin{definition}[Higher Order Derivatives]
A function $f: U\rightarrow Z$ for $U\subseteq Y$ is $k$-times differentiable if $f,Df, \ldots, D^{k-1}f$ are differentiable everywhere in $U$. 
For each $a\in U$, the $k$-th order derivative $D^kf(a)$ at $a$ can be identified with a multi-linear map $\cL^k(Y,Z)$. 
Furthermore, this map is symmetric for all $a$.
\end{definition}

Given the definition of higher-order derivatives, we state some formulas that will be useful in the proofs. 
The first is the generalized product rule on the product of $n$ functions, which follows from applying the product rule in \autoref{fact:prod-rule} $n$ times.

\begin{fact}[Generalized Product Rule] \label{fact:gen-prod-rule}
Let $\mathcal{A}$ be a Banach Algebra and $F_1,\ldots, F_n:\mathcal{A}\rightarrow \mathcal{A}$ be $\ell$-times differentiable functions. 
Let $\mathcal{P}_\ell$ be the set of permutations of $\ell$ elements. Given integers $k_1,\ldots k_n\geq 0$, define the partial sum $K_0 = 0$ and $K_i = k_1+k_2+\ldots k_i$.
For any $X\in \mathcal{A}$, 
\[
D^\ell(F_1\cdots F_n)(X)(H_1,\ldots, H_\ell) 
= \sum_{\substack{k_1+k_2+\ldots k_n= \ell}}\frac{1}{k_1!\cdots k_n!}\cdot \sum_{\sigma\in \mathcal{P}_\ell} 
\prod_{j=1}^n D^{k_j}F_j(X)(H_{\sigma(K_{j-1}+1)},\ldots, H_{\sigma(K_{j})}).
\]
\end{fact}

The formula for derivatives of moments is a special case of \autoref{fact:gen-prod-rule}. 

\begin{fact}[Higher Derivatives of Moments] \label{lemma:moment-higher-der}
Let $X$ be an element of a Banach Algebra and let $F(X) = X^p$. 
For any $\ell \leq p$,
\[
D^\ell F(X)(H_1, \ldots, H_\ell) = \sum_{\sigma{\in \mathcal{P}_{\ell}}}\sum_{\substack{k_1+\cdots+k_{\ell+1} = p-\ell\\k_1,k_2,\ldots,k_{\ell+1}\geq 0}} X^{k_1}H_{\sigma(1)}\cdots X^{k_\ell}H_{\sigma(\ell)}X^{k_{\ell+1}}.
\]
\end{fact}

The formula for derivatives of resolvents is obtained by applying the product rule and \autoref{fact:inv} repeatedly.

\begin{lemma}[Higher Derivatives of Resolvents] \label{lemma:res-higher-der}
Let $X$ be an element in a Banach algebra over $\C$ and let $M_z(X) = (z1 - X)^{-1}$. 
For any $\ell \geq 1$,
\[
D^{\ell}M_z(X)^p(H_1,\ldots, H_\ell) = 
\sum_{\sigma\in \mathcal{P}_\ell}
\sum_{\substack{k_1+\ldots+k_{\ell+1}=p+\ell\\k_1,k_2\ldots,k_{\ell+1}\geq 1}} 
M_z(X)^{k_1}H_{\sigma(1)}\cdots M_z(X)^{k_\ell}H_{\sigma(\ell)}M_z(X)^{k_{\ell+1}}
\]
    For each permutation $\sigma$, the summation inside has exactly $\binom{p+\ell-1}{\ell}$ terms.
\end{lemma}

\subsubsection{Taylor Approximation Theorem}

The Taylor approximation theorem is crucial to our analysis.

\begin{theorem}[Taylor Approximation Theorem~{\cite[Theorem 5.6.1]{Car71}}] \label{theorem:taylor-general}
Let $Y,Z$ be Banach spaces and $U\subseteq Y$ be an open set. 
Given a function $f:U\rightarrow Z$, that is $(n+1)$-times differentiable, define the $n$-th order approximation of $f$ around $a$ to be the function
\[
p^{(n)}_a(h) = f(a) + Df(a)h + \frac{1}{2}D^{2}f(a)(h,h) + \cdots + \frac{1}{n!}D^nf(a)(\underbrace{h,h,\ldots,h}_{\text{$n$ times}}).
\]
If the interval $[a,a+h]$ is contained in $U$, then we have
\[
f(a+h) - p^{(n)}_a(h) 
= \int_{0}^1\frac{(1-t)^n}{n!}D^{n+1}f(a+th)(\underbrace{h,h,\ldots,h}_{\text{$n+1$ times}}) dt.
\]
In particular, if for all $t\in [0,1]$, $\norm{D^{n+1}f(a+th)(h,\ldots,h)} \leq R$, then 
\[
\norm{f(a+h) - p^{(n)}_a(h)} \leq \frac{R}{(n+1)!}.
\]
\end{theorem}

For convenience, we write down the expressions for the first three terms of the Taylor expansion of a function that we will use.

\begin{lemma}[Third Derivatives] \label{lemma:derivatives}
Let $Y$ be a Banach space, 
$g:Y\rightarrow Y$ be a smooth operator-valued function,
and $f:\C\rightarrow\C$ be a smooth complex-valued function. 
Let $\tau:Y\rightarrow \C$ be a bounded linear functional. 
For any $a,h\in Y$,
\begin{align*}
\frac{d}{dt}f(\tau(g(a+th)))\Big|_{t=0} &= f'(\tau(g(a))) \cdot \tau(Dg(a)(h)),
\\
\frac{d^2}{dt^2}f(\tau(g(a+th)))\Big|_{t=0}&= f''(\tau(g(a))) \cdot \tau(Dg(a)(h))^2 + f'(\tau(g(a))) \cdot \tau(D^2g(a)(h,h)),
\\
\frac{d^3}{dt^3}f(\tau(g(a+th)))\Big|_{t=0}&= f'''(\tau(g(a))) \cdot \tau(Dg(a)(h))^3
\\
         &\quad+ 3f''(\tau(g(a))) \cdot \tau(Dg(a)(h)) \cdot \tau(D^2g(a)(h,h))\\
         &\quad+f'(\tau(g(a))) \cdot \tau(D^3g(a)(h,h,h)).
\end{align*}
\end{lemma}

\subsection{Gaussian Analysis}

In this subsection, we review some key properties of Gaussian random variables, which are crucial in the analysis of the Gaussian matrix model in~\cite{BBvH23}. 

We start with the basic Gaussian integration-by-parts formula.

\begin{lemma}[Gaussian Integration-by-Parts] \label{lemma:gaussian-int}
Let $g\sim \mathcal{N}(0,1)$ and $f$ be a smooth function. Then
\[
\E[gf(g)] = \E[f'(g)].
\]
\end{lemma}

A key lemma in \cite[Corollary 4.12]{BBvH23} is the Gaussian covariance identity. 
Given a Gaussian vector $x$ and smooth functions $f,h$, the identity bounds the quantity
\[
\big| \E[f(x) \cdot h(x)] - \E[f(x)] \cdot \E[h(x)] \big|.
\]
In other words, it measures the extent to which the functions $f$ and $g$ are uncorrelated over the Gaussian measure. 
Since this lemma was only stated in the vector setting, we provide a more compact and self-contained proof of it in the matrix setting for our applications.

\begin{lemma}[Gaussian Covariance Identity] \label{lemma:gaussian-covariance}
Let $A_0,A_1,\ldots,A_n$ be arbitrary $d\times d$ matrices. 
Let $X=A_0+\sum_{i=1}^n x_i A_i$ be a Gaussian random matrix where $x \sim \mathcal{N}(0,I_n)$. 
Let $F,G:\M_d(\C)\rightarrow \M_d(\C)$ be matrix-valued smooth functions. 
Let $y,z$ be i.i.d.~copies of $x$. 
Define the interpolation matrix
\[
X_t = A_0 + t\sum_{i=1}^n x_iA_i + \sqrt{1-t^2}\sum_{i=1}^n y_iA_i.
\]
Let $Z = \sum_{i=1}^n z_iA_i$ be an independent and centered copy of $X - A_0$. Then,
\[
\E\big[ \tr\big(F(X) \cdot G(X)\big) \big] - \tr \big(\E[F(X)] \cdot \E[G(X)]\big) 
= \int_{0}^1\E\big[\tr(D(F(X))(Z) \cdot D(G(X_t))(Z))\big] dt.
\]
\end{lemma}

\begin{proof}
Note that $X_t$ has the same distribution as $X$ for any $t \in [0,1]$. 
At the endpoints, $X_1 = X$ and $X_0$ are independent copies of $X$. 
Thus,
\[
\E\tr[F(X)G(X)] - \tr(\E[F(X)]\E[G(X)]) 
= \E\tr[F(X)G(X_1)] - \E\tr[F(X)G(X_0)] 
=\int_{0}^1 \;\frac{d}{dt}\E\tr[F(X)G(X_t)]dt.
\]
To prove the lemma, it suffices to show that the derivative terms matches the integrand in the lemma statement. 
To compute this derivative, we evaluate
\[
\frac{d}{dt}\E\tr[F(X)G(X_t)] 
= \E\tr\bigg[\frac{d}{dt}F(X)G(X_t)\bigg] 
= \sum_{i=1}^n\E\tr\bigg[F(X)D(G(X_t))\bigg(x_i A_i - \frac{t}{\sqrt{1-t^2}} y_iA_i \bigg) \bigg].
\]
For each $i$, we apply the Gaussian integration-by-parts formula in \autoref{lemma:gaussian-int} to obtain
\begin{align*}
\E[x_i \tr(F(X)D(G(X_t))(A_i))] 
&= \E\big[\partial_{x_i}\tr\big[F(X) \cdot D(G(X_t))(A_i)\big]\big]\\
&= \E\big[\tr\big[D(F(X))(A_i) \cdot D(G(X_t))(A_i)\big]\big] 
+ t \cdot \E\big[\tr\big[F(X) \cdot D^2(G(X_t))(A_i,A_i)\big]\big].
\end{align*}
Using the Gaussian integration-by-parts formula and noting that only $G(X_t)$ depends on $y$,
\begin{align*}
\frac{t}{\sqrt{1-t^2}} \cdot \E\big[y_i\tr\big[F(X)D(G(X_t))(A_i)\big]\big] 
&= \frac{t}{\sqrt{1-t^2}} \cdot \E\big[\partial_{y_i}\tr\big[F(X) \cdot D(G(X_t))(A_i)\big]\big] \\
&= \frac{t}{\sqrt{1-t^2}}\cdot \sqrt{1-t^2} \cdot \E\big[\tr\big[F(X) \cdot D^2(G(X_t))(A_i,A_i)\big]\big].
\end{align*}
Therefore, by linearity, the terms involving $D^2(G(X_t))$ canceled out, 
and we conclude that
\[
\frac{d}{dt}\E\tr\big[F(X) \cdot G(X_t)\big] 
= \sum_{i=1}^n\E\tr\big[D(F(X))(A_i) \cdot D(G(X_t))(A_i)\big] 
= \E\tr\big[ D(F(X))(Z) \cdot D(G(X_t))(Z) \big],
\]
where the last equality follows by expanding the right hand side and using $z \sim \mathcal{N}(0,I_n)$.
\end{proof}

Another useful property of Gaussians is that the expectation of convex functions of Gaussians do not increase under projection.
The following lemma will be used in the deterministic algorithm for matrix discrepancy.
 
\begin{lemma}[Convex Functions of Gaussians] \label{lemma:sum-of-ind}
Let $y,z\in \mathbb{R}^n$ be independent and centered Gaussian random vectors.
Let $f: \R^n\rightarrow \R$ be a convex function. Then
\[
\E[f(y)] \leq \E[f(y+z)]
\]
Let $g\sim \mathcal{N}(0,I_n)$ be a standard Gaussian vector and $P$ be an $n\times n$ orthogonal projection matrix such that $P^2 = P$, and $P = P^\top$. Then
\[
\E[f(Pg)] \leq \E[f(g)]
\]
\end{lemma}
\begin{proof}
Since $y$ and $z$ are independent and centered Gaussian vectors, it follows that $y = \E[y+z | y]$. 
Thus, by applying Jensen's inequality for conditional expectations,
\[
\E_y[f(y)] = \E_y[f(\E[y+z|y])] \leq \E_{y}\E[f(y+z)|y] = \E[f(y+z)].
\]
For the second inequality, let $g$ and $g'$ be independent standard Gaussians. 
Let $y = Pg$ and $z = (I-P)g'$. 
Using $P^2=P$ and $g,g' \sim \mathcal{N}(0,I_n)$, 
we see that $\E[(y+z)(y+z)^\top]= \E[yy^\top + zz^\top] = I$, 
meaning $y+z$ has the same distribution as $g$. 
Thus, by the first inequality,
\[
\E[f(Pg)] = \E[f(y)] \leq \E[f(y+z)] = \E[f(g)].
\]
\end{proof}

\subsection{Trace Inequalities} \label{s:trace}

Given a $C^*$-probability space $(\A, \tau)$, the expectation functional $\tau$ naturally defines an $L_p$ norm on $\A$:
\[
\norm{a}_p := \tau(|a|^p)^{\frac{1}{p}}.
\]
H\"older's inequality is generalized in this setting: For any $p,q > 0$ such that $\frac{1}{p} + \frac{1}{q} = 1$,
\begin{equation} \label{e:Holder}
|\tau(ab)| \leq \norm{a}_p\norm{b}_q.
\end{equation}
In this paper, we will require several generalized versions of these H\"older-type trace inequalities for the product of more than two variables. 
All of the trace inequalities that we use are standard, 
or have appeared in \cite{BBvH23} and \cite{BvH24} for finite dimensional random matrices. 
We will show that they naturally extend to the algebra $\M_d(\C)\otimes \A$ by approximating operators in $\M_d(\C)\otimes \A$ with finite dimensional random matrices, then apply the weak convergence result in \autoref{theorem:weak-convergence} by taking limits.
We remark that these trace inequalities can also be proved directly in the infinite dimensional setting by extending $\A$ to a Von-Neumann algebra of operators, but we will use the finite dimensional approximation method instead to avoid introducing heavier operator-algebraic machinery.

\begin{lemma}[Trace Inequality] \label{lemma:matrix-square}
Let $X\in \M_d(\C)\otimes \A$ be a self-adjoint semicircular matrix and $f:\R\rightarrow \C$ be a function bounded and continuous on $\spec(X)$. Let $A_1,\ldots, A_n\in M_d(\C)$ be Hermitian matrices. 
Then 
\[
\sum_{i=1}^n|\tr\otimes\tau[f(X) (A_i \otimes 1)]|^2
\leq \biggnorm{\sum_{i=1}^nA_i^2} \cdot \tr\otimes\tau[|f(X)|]^2,
\]
where $|f(X)|$ is defined by functional calculus in \autoref{t:functional-calculus}.
\end{lemma}

\begin{proof}
First, we prove the case where $f$ is a polynomial. $X = \bar{A}_0 \otimes 1+ \sum_{i=1}^n \bar{A}_i \otimes s_i$ by \autoref{d:semicircular-matrix}.
Let $(X_N)_{N=1}^\infty = \bar{A}_0 \otimes I_N + \sum_{i=1}^n \bar{A}_i \otimes G_i^N$ be a sequence of $dN\times dN$ self-adjoint random matrices in $\M_d(\C) \otimes \A_N$ where $G_i^N$ are independent GOE matrices. 
For each random matrix $f(X_N)$, let $U$ be a unitary matrix such that $f(X_N) = U \cdot |f(X_N)|$ is its polar decomposition. 
By Holder's inequality in \eqref{e:Holder} with $p=q=2$,
\begin{align*}
|\E[\tr(f(X_N) \cdot (A_i\otimes I_N))]|^2 &= |\E[\tr(U \cdot |f(X_N)| \cdot (A_i\otimes I_N))]|^2
\\
&\leq |\E[\tr(U \cdot |f(X_N)| \cdot U^*)]| \cdot \E[\tr((A_i\otimes I_N) \cdot |f(X_N)| \cdot (A_i\otimes I_N))]
\\ 
&= \E[\tr(|f(X_N)|)] \cdot \E[\tr(|f(X_N)| \cdot (A_i^2\otimes I_N))].
\end{align*} 
Now, we can apply the weak convergence result in \autoref{corollary:cond-convergence} to obtain
\begin{align*}
\sum_{i=1}^n|\tr\otimes\tau[f(X) (A_i \otimes 1)]|^2 
&= \lim_{N\rightarrow\infty}\sum_{i=1}^n|\E[\tr(f(X_N) \cdot (A_i\otimes I_N))]|^2
\\
&\leq \lim_{N\rightarrow\infty}  \E[\tr(|f(X_N)|)] \cdot \sum_{i=1}^n \E\tr(|f(X_N)| \cdot(A_i^2\otimes I_N))
\\
&\leq \lim_{N\rightarrow\infty}~\biggnorm{\sum_{i=1}^n A_i^2} \cdot \E[\tr(|f(X_N)|)]^2
\\
&= \biggnorm{\sum_{i=1}^nA_i^2}[\tr\otimes\tau(|f(X)|)]^2. 
\end{align*} 
Since any continuous function on $\spec(X)$ is uniformly approximated by polynomials, the result immediately extends to bounded continuous $f$.
\end{proof}

The next trace inequality bounds the trace of products of many operators and is crucial for bounding the Taylor approximation error in our analysis. 
As in the proof of \autoref{lemma:matrix-square}, we start with the finite-dimensional version of this inequality.

\begin{lemma}[{\cite[Lemma 5.3]{BvH24}}]\label{lemma:matrix-holder}
Let $Y_1,Y_2, \ldots, Y_k$ be $d\times d$ random matrices and $p_1,p_2, \ldots, p_k\geq 1$ be such that $\sum_{i=1}^k\frac{1}{p_i} = 1$. 
Then
\[
\big|\E[\tr (Y_1Y_2 \cdots Y_k)]\big| 
\leq \prod_{i=1}^{k}\E\big[\tr\big(|Y_i|^{p_i}\big)\big]^{\frac{1}{p_i}}.
\]
\end{lemma}

We extend \autoref{lemma:matrix-holder} to the infinite dimensional setting.
One caveat to note is that the Schatten $\infty$-norm of finite dimensional Gaussian matrices are always unbounded, and thus do not converge to the $\infty$-norm of semicircular operators. 
This issue can be handled by first approximating the $\infty$-norm with a large finite $q$-norm and then taking $q$ to $\infty$. 

\begin{lemma}[Generalized H\"older's Inequality] \label{corollary:general-holder}
Let $X, Y, Y_1, \ldots, Y_m$ be elements in $\M_d(\C)\otimes \A$ where $\A$ is the $C^*$-algebra generated by semicircular elements $s_1, s_2, \ldots$. 
Let $k_1,k_2, \ldots, k_m\in \N$ satisfy $k_1+k_2+ \cdots +k_m=p$. 
Suppose that $|Y_i|\preceq |Y|^{k_i}$ for each $i\in [m]$. Then
\[
|\tr\otimes\tau(XY_1XY_2 \cdots XY_m)|\leq \norm{X}^m \cdot \tr\otimes\tau(|Y|^p).
\]
\end{lemma}
\begin{proof}
Each element in $\M_d(\C)\otimes\A$ can be written as the limit of polynomials of semicircular elements under the operator norm. Thus, it suffices to prove the lemma in the case where each $Y_i$ is a non-commutative polynomial: $Y_i\in \M_d(\C)\langle s_1,\ldots, s_n\rangle$.  
By the weak convergence result in \autoref{theorem:weak-convergence},
there are finite-dimensional random matrices $X^N, Y_1^N, \ldots, Y_m^N$ in the algebra generated by independent GOE matrices such that
\[
X^N, Y_1^N, \ldots, Y_m^N \rightarrow_D X, Y_1,\ldots, Y_m\;\; \text{as } N \to \infty,
\]
where $\rightarrow_D$ denotes convergence in distribution as defined in \autoref{d:convergence-distribution}.

Let $\eps \in (0,1)$ be arbitrary. 
For all $N\in \N$, we will apply \autoref{lemma:matrix-holder} with $p_{2i-1} = \frac{m}{\eps}$ and $p_{2i} = \frac{p}{k_i(1-\eps)}$ for all $i \in [m]$ (where we take $p/0$ to mean $\infty$). 
Verify that $\sum_{i=1}^{2m} \frac{1}{p_i} = m\cdot \frac{\eps}{m} + \sum_{i=1}^m \frac{k_i(1-\eps)}{p} = 1$. 
Thus, by \autoref{lemma:matrix-holder},
\[
\big|\E\big[\tr\big(X^NY_1^N \cdots X^NY_m^N\big)\big]\big| 
\leq \E\Big[ \tr\big( \big|X^N\big|^{\frac{m}{\eps}}\big) \Big]^\eps \cdot 
\prod_{i=1}^m\E\Big[ \tr\big( |Y_i^N|^{\frac{p}{k_i(1-\eps)}} \big) \Big]^\frac{k_i(1-\eps)}{p}.
\]
Using the weak convergence result in \autoref{theorem:weak-convergence}, and \autoref{lem:weak-converge-cont},
\begin{align*}
\big|\tr\otimes\tau(XY_1 \cdots XY_m)\big| 
&= \lim_{N\rightarrow\infty} \big|\E\tr\big(X^NY_1^N \cdots X^NY_m^N\big) \big|
\\
&\leq
\lim_{N\rightarrow\infty} 
\E\Big[ \tr\big( \big|X^N\big|^{\frac{m}{\eps}}\big) \Big]^\eps \cdot 
\prod_{i=1}^m\E\Big[ \tr\big( |Y_i^N|^{\frac{p}{k_i(1-\eps)}} \big) \Big]^\frac{k_i(1-\eps)}{p}.
\\
&= \tr\otimes\tau\big(|X|^{\frac{m}{\eps}}\big)^\eps \cdot \prod_{i=1}^m \tr\otimes\tau\Big(|Y_i|^{\frac{p}{k_i(1-\eps)}}\Big)^{\frac{k_i(1-\eps)}{p}}
\\
&\leq \norm{|X|^m} \cdot \tr\otimes\tau\big(|Y|^{\frac{p}{(1-\eps)}}\big)^{\sum_{i=1}^m\frac{k_i(1-\eps)}{p}}
\\
&=\norm{X}^m \cdot \tr\otimes\tau\big(|Y|^{\frac{p}{(1-\eps)}}\big)^{1-\eps}
\end{align*}
where the assumption $|Y_i| \preceq |Y|^{k_i}$ for all $i$ is used in the second-to-last line. Finally, since the above inequality holds for all $\eps \in (0,1)$, we can take the limit as $\eps \rightarrow 0$ to obtain that
\[
\big|\tr\otimes\tau(XY_1 \cdots XY_m)\big| 
\leq \norm{X}^m \cdot \tr\otimes\tau(|Y|^{p}). \qedhere
\]
\end{proof}

Then, we show a semicircular analog of \autoref{lemma:sum-of-ind} which will be useful for matrix discrepancy.

\begin{lemma}[Trace of Projection]
\label{lemma:free-proj}
Let $A_0,A_1, \ldots, A_n$ be symmetric matrices and $s_1, \ldots, s_n$ be freely independent semicircular elements. 
Let $f:\R\rightarrow\R$ be a continuous convex function with polynomial growth rate. 
Let $P$ be an $n\times n$ orthogonal projection matrix, and $\Tilde{A}_1, \ldots, \Tilde{A}_n$ be projected coefficient matrices such that $\Tilde{A}_i = \sum_{j=1}^n P(i,j)A_j$.
Then
\[
\tr\otimes \tau \bigg( f\bigg(A_0 \otimes 1 + \sum_{i=1}^n \Tilde{A}_i\otimes s_i \bigg) \bigg)
\leq \tr\otimes \tau\bigg( f\bigg(A_0 \otimes 1+ \sum_{i=1}^n A_i\otimes s_i \bigg) \bigg). 
\]
\end{lemma}
\begin{proof}
Again, we use finite-approximation of $s_i$ by independent GOE matrices $G_i^{N}$. By \autoref{theorem:weak-convergence},     
\[
\tr\otimes \tau \bigg( f\bigg(A_0 \otimes 1 + \sum_{i=1}^n \Tilde{A}_i\otimes s_i \bigg) \bigg) 
=  \lim_{N\rightarrow\infty}\E\tr \bigg( f\bigg(A_0 \otimes I_N + \sum_{i=1}^n \Tilde{A}_i\otimes G_i^{N}\bigg)\bigg).
\]
Fix $N$ and denote $G^N_i =\frac{1}{\sqrt{N}}\sum_{1\leq r\leq s\leq N}E_{r,s}g_{i,r,s}$, where $g\in  \R^n\otimes \R^{\binom{N+1}{2}}$ is the vector of i.i.d. Gaussian entries. Define the function $h(g) := \tr(f(A_0\otimes I_N+\frac{1}{\sqrt{N}}\sum_{i,r,s}A_i\otimes E_{r,s}g_{i,r,s}))$. Observe that
\begin{align*}
    \sum_{i=1}^n \Tilde{A}_i\otimes G_i^{N} 
&= \frac{1}{\sqrt{N}}\sum_{i=1}^n \sum_{j=1}^n\sum_{r\leq s} P(i,j) \cdot A_j\otimes  E_{r,s}g_{i,r,s} \\
&=\frac{1}{\sqrt{N}}\sum_{j=1}^n\sum_{r\leq s}A_j\otimes E_{r,s}\sum_{i=1}^nP(i,j)g_{i,r,s}\\
&=\frac{1}{\sqrt{N}}\sum_{j=1}^n \sum_{r\leq s}A_j \otimes E_{r,s}\cdot ((P\otimes I_{q_N})g)_{j,r,s}
\end{align*}
where $q_N = \binom{N+1}{2}$ is the number of independent Gaussians in a single GOE matrix. By the second inequality in \autoref{lemma:sum-of-ind}, and the fact that $h$ is convex, we have
\begin{align*}
 \E\tr\bigg(f\bigg(A_0 \otimes I_N + \sum_{i=1}^n \Tilde{A}_i\otimes G_i^{N}\bigg)\bigg)=\E h((P\otimes I_{q_N})g)\leq \E [h(g)] =  \E\tr\bigg(f\bigg(A_0 \otimes I_N + \sum_{i=1}^n A_i\otimes G_i^{N}\bigg)\bigg)
\end{align*}
Taking limits in $N$, and applying weak convergence (\autoref{theorem:weak-convergence}, \autoref{lem:weak-converge-cont}) then gives the desired bound on the free semicirculars.
\end{proof}

\subsection{Intrinsic Freeness} \label{s:intrinsic}

A key part of the analysis in \cite{BBvH23} is to prove a version of trace H\"older's inequality that bounds the crossing terms in the product of random matrices. 
This underlies a key phenomenon called ``intrinsic freeness'', in which the non-commutative structure of random matrices suppresses the crossing terms between themselves. 

\begin{lemma}[Intrinsic Freeness~\cite{BBvH23}] \label{lemma:crossing-bound}
Let $H $ and $H'$ be independent centered random matrices. 
For any normal random matrices $Y_1,Y_2,Y_3,Y_4$ independent of $H,H'$ and any $p_1,p_2,p_3,p_4 > 0$ satisfying $\sum_i \frac{1}{p_i} = 1$, 
\[
\big| \E[\tr(Y_1 HY_2H'Y_3HY_4H') ] \big| 
\leq \sigma(H) \cdot \nu(H) \cdot \sigma(H') \cdot \nu(H') \cdot \prod_{i=1}^4 \E\big[\tr(|Y_i|^{p_i})\big]^{\frac{1}{p_i}},
\]
where $\sigma(H)$ and $\nu(H)$ are as defined in \eqref{e:sigma} and \eqref{e:nu} respectively.
\end{lemma}

We restate it in the following form for ease of our applications.

% \begin{corollary}[Intrinsic Freeness] \label{corollary:crossing-bound}
% Let $H,H',Y_1,Y_2,Y_3,Y_4$ be as stated in \autoref{lemma:crossing-bound}, and let $k_1,k_2,k_3,k_4 \geq 0$ be integers such that $k_1+k_2+k_3+k_4=p$. 
% Suppose there exists a PSD random matrix, $Y$, such that for all $1 \leq i \leq 4$, we have $\E[|Y_i|]\preceq \E[Y^{k_i}]$. 
% Then 
% \[
% \big|\E[\tr(Y_1 H Y_2 H' Y_3 H Y_4 H') ]\big| 
% \leq \sigma(H) \cdot \nu(H) \cdot \sigma(H') \cdot \nu(H') \cdot \E[\tr(|Y|^{p})].
% \]
% \end{corollary}

\begin{corollary}[Intrinsic Freeness]
\label{corollary:crossing-bound}
Let $H,H',Y_1,Y_2,Y_3,Y_4$ be as stated in
\autoref{lemma:crossing-bound}. Let
$k_1,k_2,k_3,k_4\in\mathbb Z_{\geq 0}$, and set $p:=k_1+k_2+k_3+k_4$.
Suppose there exists a PSD random matrix $W$ of the same dimension
such that, for every $1\leq i\leq 4$ and every real $q\geq 1$, $\E[|Y_i|^q]\preceq \E[W^{k_iq}]$, where all the moments are finite and $W^0:=I$.
Then
\[
\big|\E[\tr(Y_1 H Y_2 H' Y_3 H Y_4 H')]\big|
\leq
\sigma(H)\nu(H)\sigma(H')\nu(H')
\cdot \E[\tr(W^p)].
\]
\end{corollary}

Then, we extend \autoref{lemma:crossing-bound} to the infinite dimensional setting, which will allow us to bound the crossing terms that arise when interpolating between free operators and finite-dimensional random matrices. 
We first prove the moment bound which will be crucially used in \autoref{section:discrepancy}.

\begin{proposition}[Intrinsic Freeness of Moments] \label{proposition:moment-crossing-bound2} 
Let $X \in \M_d(\C) \otimes \A$ be a self-adjoint semicircular matrix
and $Y$ be a $d\times d$ centered, self-adjoint random matrix, and $p$ be an integer $\geq 2$. 
Then  
\[
\bigg| \sum_{k=0}^{2p-2} \E_Y\Big[\tr\otimes\tau\big((Y \otimes 1)X^k (Y \otimes 1) X^{2p-2-k}\big) 
- \tr\big(Y\varphi[X^{k}]Y\varphi[X^{2p-2-k}]\big)\Big] \bigg| 
\lesssim p^3 \cdot \Tilde{\nu}(X)^2 \cdot \Tilde{\nu}(Y)^2 \cdot \tr\otimes\tau(X^{2p-4}),
\]
where $\Tilde{\nu}(X)$ is defined such that $\Tilde{\nu}(X)^2 = \sigma(X) \cdot \nu(X)$.
\end{proposition}

\begin{proof}
We approximate $X$ by finite dimensional Gaussian matrices as usual.
Let 
\begin{equation} \label{e:XXN}
X = A_0 \otimes 1  + \sum_{i=1}^n A_i\otimes s_i
\quad \textrm{and} \quad
X_N = A_0\otimes I_N + \sum_{i=1}^n A_i\otimes G^N_i,
\end{equation}
where $s_1, \ldots, s_n$ are freely independent semicircular elements and
$G_1^N, \ldots, G_n^N$ are independent $N\times N$ GOE matrices. 
By \autoref{theorem:weak-convergence}, for any fixed $Y\in \M_d(\C)$,
\[
\sum_{k=0}^{2p-2}\tr\otimes\tau\big((Y \otimes 1) X^{k} (Y \otimes 1) X^{2p-2-k})\big) 
= \lim_{N\rightarrow\infty} \sum_{k=0}^{2p-2} \E_{X_N} \big[\tr (Y\otimes I_N) X_{N}^k(Y\otimes I_N)X_N^{2p-2-k}\big]. 
\]
Similarly, by \autoref{corollary:cond-convergence}, 
\[
\sum_{k=0}^{2p-2} \tr \big(Y \cdot \varphi[X^{k}] \cdot Y \cdot \varphi[X^{2p-2-k}]\big) 
= \lim_{N\rightarrow\infty}\sum_{k=0}^{2p-2}\tr((Y\otimes I_N) \cdot \E_{X_N}[X_N^{k}] \cdot (Y\otimes I_N) \cdot \E_{X_N}[X_N^{2p-2-k}]).
\]
To bound the difference, we apply the Gaussian covariance identity in \autoref{lemma:gaussian-covariance}. 
In particular, we define the interpolation matrix, 
$X_{N,t}$ and centered copy $\bar{X}_N$ as
\begin{equation} \label{e:XNtXN}
X_{N,t} := A_0\otimes I_N + t \cdot \sum_{i=1}^nA_i\otimes G_i^N + \sqrt{1-t^2} \cdot \sum_{i=1}^nA_i\otimes G_i^{'N}
\quad \textrm{and} \quad
\bar{X}_N := \sum_{i=1}^n A_i\otimes G_i^{''N}
\end{equation}
where $G_i^{'N}$ and $G_i^{''N}$ are independent copies of $G_i^N$. 
By substituting $F(X_N) = (Y\otimes I_N)(X_N)^k$ and $G(X_N) = (Y\otimes I_N)(X_N)^{2p-2-k}$, it follows from \autoref{lemma:gaussian-covariance} that
\begin{eqnarray*}
&& \sum_{k=0}^{2p-2} \E_Y \Big[ \E_{X_N} \big[\tr \big((Y\otimes I_N) X_{N}^k(Y\otimes I_N)X_N^{2p-2-k} \big) \big] 
- \tr\big((Y\otimes I_N) \cdot \E_{X_N}[X_N^{k}] \cdot (Y\otimes I_N) \cdot \E_{X_N}[X_N^{2p-2-k}]\big)\Big] 
\\ 
&=& \int_{0}^1 \sum_{0 \leq a+b+c \leq 2p-4}\E_{Y, X_N, X_{N,t}, \bar{X}_N} \big[\tr\big((Y\otimes I_N)X_N^{a} \bar{X}_N X_N^{b}(Y\otimes I_N)X_{N,t}^{c} \bar{X}_N X_{N,t}^{2p-4-a-b-c}\big) \big] dt
\\
&\lesssim& p^3 \cdot \Tilde{\nu}(Y\otimes I_N)^2 \cdot \Tilde{\nu}(X_N)^2 \cdot \E\tr\big(X_N^{2p-4}\big),
\end{eqnarray*}
where the last line follows by applying \autoref{corollary:crossing-bound} to each summand with $H := Y\otimes I_N$, $H' := \bar{X}_N$, and $W = |X_N|$,
and noting that $X_{N}$ and $X_{N,t}$ have the same distribution and so $\E[|X_{N,t}|^q] = \E[|X_{N}|^q]$ for all real $q \geq 1$ and there are $O(p^3)$ summands.

Finally, \cite[Lemma 5.5]{BBvH23} showed that $\sigma(Y \otimes I_N)=\sigma(Y)$, $\nu(Y \otimes I_N)= \sqrt{N} \cdot \nu(Y)$, $\sigma(X_N) = \sqrt{1 + 1/N} \cdot \sigma(X)$, and $\nu(X_N) = \sqrt{2/N} \cdot \nu(X)$.
These imply that
\[
\widetilde{\nu}(Y\otimes I_N)^2
\widetilde{\nu}(X_N)^2
=
\sqrt{2(1+1/N)} \cdot \,
\widetilde{\nu}(Y)^2
\widetilde{\nu}(X)^2,
\]
and the proposition follows.
\end{proof}

We then prove an analogous bounds for resolvents $(z1-X)^{-1}$ that will be used in both \autoref{section:resolvent} and \autoref{section:uni}.
 
\begin{proposition}[Intrinsic Freeness of Resolvents] \label{proposition:res-crossing-bound2}
Let $X \in \M_d(\C) \otimes \A$ be a self-adjoint semicircular matrix
and $Y$ be a $d\times d$ centered, self-adjoint, random matrix.
Let $z\in \C$ with $\im(z) > 0$. 
Let $M_z(X) = (z1-X)^{-1}$ be the resolvent of $X$. 
For any $p_1,p_2,q_1,q_2 \in \N$ such that $p_1+p_2+q_1+q_2=p$, 
\begin{eqnarray*}
& & \Big|\E_Y\tot \big(M_z(X)^{p_1} \cdot M_{\bar{z}}(X)^{p_2} \cdot (Y \otimes 1) \cdot M_z(X)^{q_1} \cdot M_{\bar{z}}(X)^{q_2} \cdot (Y \otimes 1) \big)
\\
& & \quad \quad - \E_Y\tr\big(\varphi[M_z(X)^{p_1}M_{\bar{z}}(X)^{p_2}] \cdot Y \cdot \varphi[M_z(X)^{q_1}M_{\bar{z}}(X)^{q_2}] \cdot Y \big)\Big| 
\\
& & \quad \quad \quad \quad \quad \quad \lesssim (p_1+p_2) \cdot (q_1+q_2) \cdot \Tilde{\nu}(X)^2 \cdot \Tilde{\nu}(Y)^2 \cdot \tot(|M_z(X)|^{p+2}).
\end{eqnarray*}
\end{proposition}

\begin{proof}
As in the proof of \autoref{proposition:moment-crossing-bound2}, we define the $dN \times dN$ Gaussian approximation $X_N$ for $X$ as in \eqref{e:XXN}, 
with matrix parameters $\sigma(X_N)^2 = (1+1/N)\sigma(X)^2$ and $\nu(X_N)^2 = 2\nu(X)^2/N$. 
By \autoref{theorem:weak-convergence} and \autoref{corollary:cond-convergence}, the expression on the left hand side of the statement can be written as $\lim_{N\rightarrow \infty} T_N$ where
\begin{align*}
T_N = \Big|\E_Y\tr &\big(M_z(X_N)^{p_1} \cdot M_{\bar{z}}(X_N)^{p_2} \cdot (Y\otimes I_N) \cdot M_z(X_N)^{q_1} \cdot M_{\bar{z}}(X_N)^{q_2} \cdot (Y\otimes I_N)\big)
\\
&- \E_Y\tr\big(\E[M_z(X_N)^{p_1}M_{\bar{z}}(X_N)^{p_2}] \cdot (Y\otimes I_N) \cdot \E[M_z(X_N)^{q_1}M_{\bar{z}}(X_N)^{q_2}] \cdot (Y\otimes I_N)\big)\Big|. 
\end{align*}
To compute the expression on the right hand side of the statement, we apply \autoref{lemma:gaussian-covariance} with $F(X_N) := M_z(X_N)^{p_1}M_{\bar{z}}(X_{N})^{p_2}(Y\otimes I_N)$ and $G(X_N) := M_z(X_N)^{q_1}M_{\bar{z}}(X_{N})^{q_2}(Y\otimes I_N)$. 
Notice that
\begin{align*}
D(F(X_N))(H) = &\sum_{k=1}^{p_1}M_{z}(X_N)^{k} \cdot H \cdot M_{z}(X_N)^{p_1-k+1} \cdot M_{\bar{z}}(X_N)^{p_2} \cdot (Y\otimes I_N)\\ 
        &+ \sum_{k=1}^{p_2} M_{z}(X_N)^{p_1} \cdot M_{\bar{z}}(X_N)^{k} \cdot H \cdot M_{\bar{z}}(X_N)^{p_2-k+1} \cdot (Y\otimes I_N),
\end{align*}
and $D(G(X_N))(H)$ can be computed similarly.
We also define our interpolation matrix $X_{N,t}$ and independent centered copy $\bar{X}_N$ as in \eqref{e:XNtXN} in the proof of \autoref{proposition:moment-crossing-bound2}. 
To simplify the notation, we write 
\[
M := M_{z}(X_N), \quad 
M_t := M_z(X_{N,t}), \quad 
M^* = M_{\bar{z}}(X_N), \quad
M_t^* = M_{\bar{z}}(X_{N,t}).
\]
Then, the Gaussian covariance identity in \autoref{lemma:gaussian-covariance} implies that
\begin{align}
T_N & = \Bigg|\int_{0}^1\E\tr\bigg[\sum_{k=1}^{p_1}\sum_{\ell=1}^{q_1}M^k \bar{X}_N M^{p_1-k+1}M^{*p_2}(Y\otimes I_N)M_t^{\ell}\bar{X}_N M_t^{q_1-\ell+1}M_t^{*q_2}(Y\otimes I_N) \nonumber 
\\
& \qquad \qquad \qquad + \sum_{k=1}^{p_1}\sum_{\ell=1}^{q_2}M^k \bar{X}_N M^{p_1-k+1}M^{*p_2}(Y\otimes I_N)M_t^{q_1}M_t^{*\ell} \bar{X}_N M_t^{*q_2-\ell+1}(Y\otimes I_N) \nonumber 
\\
& \qquad \qquad \qquad + \sum_{k=1}^{p_2} \sum_{\ell=1}^{q_1} M^{p_1}M^{*k} \bar{X}_N M^{*p_2-k+1}(Y\otimes I_N)M_t^{\ell} \bar{X}_N M_t^{q_1-\ell+1}M_t^{*q_2}(Y\otimes I_N) \nonumber 
\\
& \qquad \qquad \qquad + \sum_{k=1}^{p_2} \sum_{\ell=1}^{q_2} M^{p_1}M^{*k} \bar{X}_N M^{*p_2-k+1}(Y\otimes I_N)M_t^{q_1}M_t^{*\ell} \bar{X}_N M_t^{*q_2-\ell+1}(Y\otimes I_N) \bigg] dt\Bigg| \label{eq:TN}
\end{align}
where the $\E$ denotes expectation over all random matrices appearing in the expression. Now, we are going to apply \autoref{corollary:crossing-bound} to each of the terms in the summation. We take
\[
M^k \bar{X}_N M^{p_1-k+1}M^{*p_2}(Y\otimes I_N)M_t^{\ell} \bar{X}_N M_t^{q_1-\ell+1}M_t^{*q_2}(Y\otimes I_N)
\]
as an example, where we are going to set $k_1 := k,\; k_2 := p_1-k+1 + p_2,\; k_3 := \ell$ and $k_4 := q_1 - \ell + 1 + q_2$ with $k_1 + k_2 + k_3 + k_4 = p+2$. We take $H := \bar{X}_N$, $H' := (Y \otimes I_N)$, and
\begin{align*}
    &Y_1 := M^k,\; |Y_1| \preccurlyeq |M|^{k_1},~~
Y_2 := M^{p_1-k+1} M^{*p_2}, \;|Y_2|\preccurlyeq |M|^{k_2}\\  
&Y_3 = M_t^{\ell},\; |Y_3| \preccurlyeq |M_t|^{k_3},~~
\text{and~} Y_4 = M_t^{q_1-\ell+1} M_t^{*q_2}\; |Y_4| \preccurlyeq |M_t|^{k_4}.
\end{align*}

Then, noting that $M$ and $M_t$ have the same distribution, we have $\E[|M_t|^q] = \E[|M|^q]$ for all real $q \geq 1$, so in the notation of \autoref{corollary:crossing-bound}, the dominating
matrix can be chosen as $W = |M|$. Thus, by applying \autoref{corollary:crossing-bound}, we have
\[
\big|\E \tr \big(M^k\bar{X}_NM^{p_1-k+1}M^{*p_2}(Y\otimes I_N)M_t^{\ell}\bar{X}_NM_t^{q_1-\ell+1}M_t^{*q_2}(Y\otimes I_N) \big) \big|
\leq \Tilde{\nu}(X_N)^2 \cdot \Tilde{\nu}(Y \otimes I_N)^2 \cdot \E\tr(|M|^{p+2}).
\]
All the terms in the summation of \eqref{eq:TN} can be handled in a similar way.
Therefore, 
\begin{eqnarray*}
T_N 
& \leq & 
\int_{0}^1(p_1q_1+p_1q_2 +p_2q_1+p_2q_2) \cdot \Tilde{\nu}(X_N)^2 \cdot \Tilde{\nu}(Y\otimes I_N)^2 \cdot \E\tr(|M|^{p+2})dt
\\
&\lesssim&(p_1+p_2)(q_1+q_2) \cdot \Tilde{\nu}(X)^2 \cdot \Tilde{\nu}(Y)^2 \cdot \E\tr(|M_z(X_N)|^{p+2}).
\end{eqnarray*}
Finally, letting $N \to \infty$, it holds that
\begin{eqnarray*}
\lim_{N\rightarrow \infty}T_N 
&\lesssim& 
\lim_{N\rightarrow\infty} (p_1+p_2)(q_1+q_2)\cdot \Tilde{\nu}(X)^2 \cdot \Tilde{\nu}(Y)^2 \cdot \E\tr(|M_z(X_N)|^{p+2})
\\
& = & (p_1+p_2)(q_1+q_2) \cdot \Tilde{\nu}(X)^2 \cdot \Tilde{\nu}(Y)^2 \cdot \tot(|M_z(X)|^{p+2}).
\end{eqnarray*}
\end{proof}

\autoref{proposition:res-crossing-bound2} can be extended to $z$ on the real line that is not in the spectrum of $X$, which will be useful in our barrier method argument in \autoref{section:uni}.

\begin{corollary}[Intrinsic Freeness of Resolvents on Reals] \label{cor:res-crossing-bound}
Let $X \in \M_d(\C) \otimes \A$ be a self-adjoint semicircular matrix and $Y\in \M_d(\C)$ be a centered, self-adjoint, random matrix. Suppose $\lambda \in \R$ is such that $\lambda > \lambda_{\max}(X)$. Then for any $p,q\in \N$, $p+q\geq 4$, we have
\begin{align*}
        | \E\tot(M_{\lambda}(X)^{p} \cdot (Y\otimes 1) \cdot M_\lambda(X)^{q} \cdot (Y \otimes 1)) - & \E\tr(\varphi[M_\lambda(X)^p] \cdot Y \cdot \varphi[M_\lambda(X)^q] \cdot Y) | \\
        & \lesssim p \cdot q \cdot \Tilde{\nu}(X)^2 \cdot \Tilde{\nu}(Y)^2 \cdot \tot(M_{\lambda}(X)^{p+q+2}).
\end{align*}
\end{corollary}
\begin{proof}
Since $\lambda$ is bounded away from $\spec(X)$, we have $M_\lambda(X) = \lim_{\delta\rightarrow 0}M_{\lambda + \delta i}(X)$. Thus, we can apply \autoref{proposition:res-crossing-bound2} with $z = \lambda + \delta i$ for arbitrarily small $\delta > 0$. Taking the limit as $\delta \rightarrow 0$ then gives the desired result.
\end{proof}

\subsection{Ultracontractivity Bounds}

For a $d\times d$ matrix $M$, we can relate the operator norm of $M$ with its Schatten $p$-norm by the inequality 
\[
d^{-\frac{1}{2p}} \cdot \norm{M} \leq \tr[|M|^{2p}]^{\frac{1}{2p}} \leq \norm{M}.
\] 
In this section, we will review infinite-dimensional analogs of this bound in the algebra $\M_d(\C)\otimes \A$, which were proven in \cite{BCS+24}. These bounds extend the concentration of spectrum result in \cite{BBvH23} to a two-sided bound. 
In the following results, let $\Xf = A_0 + \sum_{i=1}^nA_i\otimes s_i$ be an arbitrary semicircular matrix. 

\begin{theorem}[Ultracontractivity Bound for Polynomials {\cite[Theorem 4.1]{BCS+24}}]\label{lemma:poly-ultra}
Let $P\in \M_d(\C)\langle x\rangle$ be a polynomial of degree $k$. 
For any $q\geq 4$,
\[
\norm{P(\Xf)} \leq (d(4qk+1))^{\frac{3}{4q}} \cdot \norm{P(\Xf)}_{4q}.
\]
\end{theorem}

This result is extended to the norm of the resolvent by using polynomial approximations of the function $(z1 - x)^{-1}$.

\begin{theorem}[Ultracontractivity Bound for Resolvent {\cite[Corollary 4.4]{BCS+24}}] \label{lemma:res-ultra}
Let $\Xf \in \M_d(\C) \otimes \A$.
For $p \gtrsim \log{d}$ and $\im(z) > 0$, 
\[
\norm{(z1-\Xf)^{-1}} \lesssim \tr\otimes\tau \big[|z1 - \Xf|^{-2p}\big]^{\frac{1}{2p}} +\frac{\sigma_*(\Xf)}{\im(z)^2},
\]
where $\sigma_*(X)$ is defined as $\sigma_*(X)^2 := \sup_{\norm{u} = \norm{v} = 1}\E|\inner{u}{Xv}|^2$.
\end{theorem}

Using the same polynomial approximation technique, we can also derive ultra-contractivity bounds for resolvent when $z\in \R$ and bounded away from the spectrum of $\Xf$. 
We restate it here as this was not explicitly stated in \cite{BCS+24}. 
The proof is the same as in \cite{BCS+24} so we omit it. 

\begin{lemma}[Ultracontractivity Bound for Resolvent] \label{lemma:bar-ultra}
Suppose $\lambda \geq \lambda_{\max}(\Xf) + \eps$. 
For any $q \geq 4$ and $r\geq 1$, 
\[
        \norm{(\lambda 1 - \Xf)^{-1}} \leq (d(4qr+1))^{\frac{3}{4q}}\Big(\norm{(\lambda 1-\Xf)^{-1}}_{4q}+ \frac{24\sigma_*(\Xf)}{r\eps^2}\Big).
\]
\end{lemma}

\section{Moment Concentration Inequalities with Linear Constraints}\label{section:discrepancy}

The goal in this section is to prove \autoref{t:partial-coloring} and its application to matrix discrepancy in \autoref{t:matrix-Spencer}.

\subsection{Technical Statements}

Let $X = \sum_{i=1}^n g_iA_i$ be the Gaussian model, where each $A_i \in \C^{d \times d}$ is an arbitrary self-adjoint matrix.
The non-commutative Khintchine inequality of Lust-Piquard and Pisier states that
\[
\E\big[\Tr(X^{2p})\big]^{\frac{1}{2p}} \leq \sqrt{2p-1} \cdot \Tr \bigg( \bigg( \sum_{i=1}^n A_i^2 \bigg)^{p} \bigg)^{\frac{1}{2p}}.
\]
Since $\E[{\norm{X}}] \leq \E[\norm{X}^{2p}]^{\frac{1}{2p}} \asymp \E[\Tr[X^{2p}]]^{\frac{1}{2p}}$ for $p \asymp \log d$, this implies that
\[
\E[\norm{X}] \lesssim \sqrt{\log{d}} \cdot \sigma(X)
\quad \textrm{where} \quad \sigma(X)^2 := \norm{E[X^2]} = \Bignorm{\sum_{i=1}^n A_i^2}.
\]
This upper bound can be achieved for diagonal matrices, where $A_1, \ldots, A_n$ are commutative with each other.
However, it is far from tight for non-commutative matrices, e.g., $\E[\norm{X}] \asymp \sigma(X)$ when $X$ is a GOE matrix. 
A main result in~\cite{BBvH23} quantifies the non-commutativeness of $X$ by a new parameter $\nu(X)$ and demonstrates that the $\sqrt{\log d}$ factor can be removed when $\nu(X)$ is sufficiently small.
Their approach is by comparing the Gaussian model to the free matrix model in \eqref{e:Xf}.
The following is the formal statement of \eqref{e:BBvH-moments} in the introduction.

\begin{theorem}[{\cite[Theorem 2.7]{BBvH23}}] \label{theorem:bbvh-p-norm}
Let $A_0,A_1,\ldots,A_n$ be $d \times d$ self-adjoint matrices.
Let $X = A_0 + \sum_{i=1}^n g_iA_i$ be the Gaussian model and $\Xf = A_0 + \sum_{i=1}^nA_i\otimes s_i$ be the corresponding free model. 
Then, for any $p \in \N$,
\[
\Big| \E\bignorm{X}_{2p} - \bignorm{\Xf}_{2p} \Big|
= \Big| \E\big[\tr(X^{2p})\big]^{\frac{1}{2p}} - \tr \otimes \tau(\Xf^{2p})^\frac{1}{2p} \Big| 
\leq 2p^\frac34 \cdot \sigma(X)^\frac12 \cdot \nu(X)^\frac12,
\]
where $\cov(X)$ is the $d^2 \times d^2$ matrix with 
\[
\cov(X)_{ij,kl} = \E[X_{ij} \overline{X_{kl}}]
\quad \textrm{and} \quad
\nu(X)^2 := \norm{\cov(X)} = \Bignorm{\sum_{i=1}^n \vecc(A_i) \vecc(A_i)^*}.
\]
\end{theorem}

This result implies that $\E[\norm{X}] \lesssim \sigma(X)$ when $\nu(X) \cdot (\log d)^{\frac32} \lesssim \sigma(X)$.
This consequence is the key of the recent major progress in the matrix Spencer problem~\cite{BJM23}, which is used to lower bound the Gaussian measure of the norm ball $\mathcal{K} := \{ x \in \R^n \mid \norm{\sum_{i=1}^n x_i \cdot A_i} \leq 1\}$ so that Rothvoss' result~\cite{Rot17} can be applied to obtain a partial coloring.

Using techniques from both free probability and algorithmic discrepancy theory, we provide a direct and simpler approach to obtain a partial coloring that satisfies additional box constraints and linear constraints.
The following is the full version of \autoref{t:partial-coloring}.

\begin{theorem}[Deterministic Partial Coloring] \label{t:partial-coloring-full}
Let $A_0,A_1, \ldots, A_n$ be $d\times d$ Hermitian matrices with matrix parameters\footnote{We assume that $\max_{i=0}^n \norm{A_i} \leq \poly(d)$, $\max_{i=1}^n \norm{A_i}\geq 1/\poly(d)$.  This assumption ensures that the matrix parameters $\sigma,\nu,\sigma_*$ are all upper bounded by $\poly(n,d)$ and lower bounded by $1/\poly(n,d)$.}:
\[
\sigma^2 = \norm{\sum_{i=1}^nA_i^2}, 
\quad \nu^2 = \norm{\sum_{i=1}^n\vecc(A_i)\vecc(A_i)^*},
\quad \sigma_*^2 = \sup_{\norm{y},\norm{z}=1}\sum_{i=1}^n |\inner{y}{A_iz}|^2.
\]
Let $\XF := \sum_{i=1}^nA_i\otimes s_i$ be the centered free matrix model for our input matrices. 
Let $b,c\in \R^n$ be box constraint vectors such that $b\leq 0\leq c$. 
Let $\eps,\delta \geq 0$ be constants such that $\eps+\delta + 3/n < 1$. 
Let $\mathcal{H} \subseteq \R^n$ be a linear subspace of dimension greater than $(1-\eps)n$. 
Then, for any integer $p \geq 2$, there is a deterministic algorithm with running time $\poly(n,d,p)$ %$\poly(n, \frac{\sigma\nu}{\sigma_*},p) \leq \poly(n,d,p)$
that finds a vector $x\in \R^n$ satisfying
\begin{enumerate}
\item {\em (Box constraints and linear constraints:)} $b \leq x \leq c$ and $x\in \mathcal{H}$.
\item {\em (Partial coloring:)} Either $|\{i \mid x_i = b_i \textrm{ or } x_i = c_i\}| \geq \delta n$ or $\norm{x}^2 \ge n$.
\item {\em ($2p$-norm bound:)} Let $A(x) := \sum_{i=1}^n x_i A_i$.  Then
\[
\norm{A_0+A(x)}_{2p} \leq \norm{A_0 \otimes 1 +\sqrt{K_{\delta,\eps}} \cdot \XF}_{2p} + O\big(p^{\frac34}\sqrt{\sigma\nu}\big)
\quad \textrm{where} \quad K_{\delta,\eps} :=\frac{1}{1- \eps - \delta - \frac{3}{n}}.
\]
\end{enumerate}
\end{theorem}

We remark that, when there are no box constraints and no linear constraints (i.e., $b = -\infty \cdot 1$, $c = + \infty \cdot 1$, and $\eps = \delta = 0$), \autoref{t:partial-coloring-full} returns a vector $x$ with $\norm{x}^2 \ge n$ in polynomial time such that
\begin{align*}
\norm{A_0 + A(x)}_{2p} \leq \norm{A_0 \otimes 1 + \XF}_{2p} + O(p^{\frac34}\sqrt{\sigma\nu} + \sigma/\sqrt{n}). 
\end{align*}
This provides a one-sided derandomization of \autoref{theorem:bbvh-p-norm},
with an essentially negligible additional error term of $\sigma/\sqrt{n}$. 

In the general case when $\eps,\delta > 0$, 
there is a multiplicative factor loss of $K^{1/2}_{\eps,\delta}$ on top of the additive loss of $O(p^{3/4}\sqrt{\sigma\nu})$ in \autoref{theorem:bbvh-p-norm}. 
In many applications such as for matrix discrepancy, a constant multiplicative loss is tolerable.
In exchange, we can handle additional linear and box constraints (e.g., $x \in [-1, 1]^n$), which broadens the scope of where such a bound can be applied.

\subsection{Brownian Walk Algorithm}

We provided an overview of the algorithm in \autoref{s:overview},
using a modified interpolation from $X$ to $\XF$ and the Brownian walk approach in algorithmic discrepancy theory to implement the interpolation.

In the algorithm, we use the following potential function to govern the moments of our mixed operators. 
Given a vector $x\in \R^n$, define the potential function\footnote{
The $2/p$ in the exponent normalizes the potential to be the $4$-th power of the $2p$-norm. 
This choice of taking the $4$-th power is only to make calculations more convenient. 
The analysis would still work if we use $1/2p$ in the exponent, as we will do later in \autoref{ss:moment-uni}.
}
\begin{equation*} %\label{e:potential-trace}
\Phi(t,x) := \tot\Big( \Big(A_0 \otimes 1 + A(x) \otimes 1 + \sqrt{K_{\delta,\eps} \cdot (1-t)} \cdot \XF \Big)^{2p} \Big)^{\frac{2}{p}}
=  \Bignorm{ A_0 \otimes 1 + A(x) \otimes 1 + \sqrt{K_{\delta,\eps} \cdot (1-t)} \cdot \XF }_{2p}^4.
\end{equation*}
The idea of the Brownian walk algorithm is to divide the interval $[0,1]$ into small discrete steps of size $\eta$.
We start with $t = 0$, $m = 0$, and $x = 0$. 
At the $m$-th step, we update $t \gets t+\eta/M$ and $x_{m+1} \gets x_m + y_m$, where the update $y_m$ is chosen so that $\norm{x_m}^2$ increases by a small but non-trivial amount, while subject to the box and linear-subspace constraints required to satisfy conditions (1) and (2) in \autoref{t:partial-coloring-full}.
The multiplicative factor $K_{\delta,\eps}$ ensures that the decrease of the potential function due to the shrinking of the free part is sufficiently large enough to offset the increase of the potential function caused by the update $y_m$. 
The key in the analysis is to control the increase of the potential function. 

\begin{proposition}[Potential Increase] \label{proposition:moment-interpolation}
Let $\eta > 0 $ be a sufficiently small step size,
and the potential function $\Phi(t,x)$ be defined with respect to an integer $p \geq 2$.
Given the setting in \autoref{t:partial-coloring-full},
for any $x\in \R^n$ with $\norm{x}^2\leq n$ and any $t\in [0,1-\eta/n)$,
there exists a subspace $\mathcal{H}' \subseteq \R^n$ of dimension at least $(\eps +\delta)n + 2$, such that for any $y\in \mathcal{H}'$ with $\norm{y}^2=1$, 
\[
\Phi\Big(t+\frac{\eta}{n},~x+\sqrt{\eta} \cdot y \Big) \leq \Phi(t,x) + O\Big(\frac{\eta}{n} \cdot p^3 \cdot \sigma^2 \cdot \nu^2\Big).
\]
In particular, the inequality holds as long as
\begin{equation} \label{e:eta}
\eta \leq  \min \Big\{\frac{\sigma^4 \cdot \nu^4}{n^3 \cdot \rho^6 \cdot \rho_0^2},~\frac{\sigma^2 \cdot \nu^2}{n^3 \cdot \rho^4},~\frac{1}{n^2}\Big\} \leq \frac{1}{\poly(n,d)}
\quad \textrm{where} \quad
\rho := \max \Big\{ \frac{2 \cdot\sigma}{\sqrt{n}}, \sigma_* \Big\} 
\quad \textrm{and} \quad 
\rho_0 = \norm{A_0} + \rho\sqrt{n}.
\end{equation}
\end{proposition}

\begin{framed}{\bf Deterministic Sticky Brownian Walk Algorithm}
    \begin{itemize}
        \item Let $\eta$ be a small enough step length as defined in \eqref{e:eta}.
    
        \item Initialize $m \gets 0$, $x_0 \gets 0$, and set $\F_0 \gets [n]$ as the set of alive/active coordinates.
        
        \item While $|\F_{m}| > (1-\delta)n$ and $\norm{x_m}^2 < n$ Do
        \begin{enumerate}
            \item Let $\mathcal{H}_m$ be the subspace defined by the linear constraints
            \[
                \mathcal{H}_m := \mathcal{H}\cap\{\inner{y}{x_m}=0 \textrm{~and~} y(i) = 0 \textrm{ for all } i \notin \mathcal{F}_m\}.
            \]
            
            \item Apply \autoref{proposition:moment-interpolation} with $t = \sum_{k=0}^{m-1} \eta_k/n$ and $x=x_m$ to find a large subspace $\mathcal{H}'$ such that, for any unit vector $y \in \mathcal{H}' \cap \mathcal{H}_m$
            \begin{equation} \label{e:potential-increase}
                \Phi\Big(t + \frac{\eta}{n},~x_{m} + \sqrt{\eta} \cdot y\Big) \leq \Phi\Big(t,x_m\Big) + O\Big(\frac{\eta}{n} \cdot p^3 \cdot \sigma^2 \cdot \nu^2 \Big).
            \end{equation}

            \item Let $\eta_m > 0$ be the largest step length such that 
            \[ \eta_m \leq \eta, \quad \norm{x_m + \sqrt{\eta_m} \cdot y}^2 \leq n, \quad \text{and} \quad x_m + \sqrt{\eta_m} \cdot y \in [b,c]. \]
            
            \item 
            Update $x_{m+1}\gets x_m+\sqrt{\eta_m} \cdot y$,  $\mathcal{F}_{m+1} \gets \{i \in \F_m: b_i < x_{m+1}(i) < c_i\}$, and $m \gets m+1$.
            
            % \item If $|\F_{m+1}| < (1-\delta)n$, then round each $i\notin \mathcal{F}_m$ to $b(i)$ or $c(i)$ depending on which one is closer to $x(i)$. Then return $x_{m+1}$ and terminate the algorithm.
        \end{enumerate}
        \item Return $x_{m}$. 
    \end{itemize}
\end{framed}

We first prove \autoref{t:partial-coloring-full} by analyzing this algorithm assuming \autoref{proposition:moment-interpolation}, 
which will then be proved in the next subsection.

\begin{proofof}{\autoref{t:partial-coloring-full}}
First, we argue that Step (2) in the while loop of the algorithm always succeeds.
Since the algorithm has not terminated, the number of alive variables is at least $|\mathcal{F}_m| \geq (1-\delta)n$.
Thus, the dimension of the (bad) subspace $\mathcal{H}_m^{\perp}$ is at most $(\delta + \eps)n+1$, 
with at most $\eps n$ constraints from $\mathcal{H}$, at most $\delta n$ constraints from $\{y(i) = 0 \textrm{ for all } i \notin \mathcal{F}_m\}$, and one constraint from $\inner{y}{x_m}=0$.
Note that the algorithm always maintains that $\norm{x_m}^2 < n$. Thus, we can apply \autoref{proposition:moment-interpolation} to show that there exists a (good) subspace $\mathcal{H}'$ of dimension at least $(\eps+\delta)n+2$ such that any $y \in \mathcal{H}'$ with $\norm{y}_2=1$ satisfies \eqref{e:potential-increase}.
Therefore, the subspace $\mathcal{H}'\cap \mathcal{H}_m$ is nontrivial,
and hence Step (2) always succeeds. 

Then, we check that each of the three conditions in \autoref{t:partial-coloring-full} are satisfied.

\begin{itemize}
    \item The box constraints $x_m \in [b,c]$ are satisfied due to Step (3) in the while loop. The linear subspace constraint is satisfied as each update step $y \in \mathcal{H}_m \subseteq \mathcal{H}$ by Step (1) in the while loop of the algorithm.

    \item The partial coloring condition is satisfied due to the termination criteria of the while loop.

    \item To verify the $2p$-norm condition, we first show that $\sum_{m=0}^{T-1} \eta_m \leq n$, where $T$ is the last iteration in which the algorithm terminated. Since in each iteration the incremental $\sqrt{\eta_m} y \perp x_m$, it holds that $\sum_{m=0}^{T-1} \eta_m = \norm{x_T}^2$. It follows from $\norm{x_T}^2 \leq n$ that $\sum_{m=1}^{T-1} \eta_m \leq n$, as desired.

    Then, we apply \autoref{lemma:free-proj} to control the $2p$-norm of the final solution by
    \[
    \norm{A_0+A(x_{T})}_{2p}^4
    \leq \biggnorm{A_0 \otimes 1 +A(x_{T}) \otimes 1 + \sqrt{K_{\eps,\delta} \cdot \Big(1-\sum_{m=0}^{T-1} \frac{\eta_m}{n}\Big)} \cdot \XF}^4_{2p}.
    \]
    Repeatedly applying \autoref{proposition:moment-interpolation}, we obtain that
    \begin{eqnarray*}
    \norm{A_0+A(x_{T})}_{2p}^4 & \leq & \norm{A_0 \otimes 1 + \sqrt{K_{\eps,\delta}} \cdot \XF}_{2p}^4 + O\bigg( \sum_{m=0}^{T-1} \frac{\eta_m}{n} p^3\sigma^2\nu^2 \bigg) \\
    & \leq & \norm{A_0 \otimes 1 + \sqrt{K_{\eps,\delta}} \cdot \XF}_{2p}^4 + O(p^3\sigma^2\nu^2 ),
    \end{eqnarray*}
    where the last inequality follows as $\sum_{m=0}^{T-1} \eta_m \leq n$.
    
    Using the inequality that $y-x \leq (y^4 - x^4)^{\frac14}$ for real numbers $y > x > 0$, we conclude that
    \[
    \norm{A_0+A(x_{T})}_{2p} \leq \norm{A_0 \otimes 1 + \sqrt{K_{\eps,\delta}} \cdot \XF}_{2p}+ O(p^{\frac34}\sqrt{\sigma \nu}).
    \]
\end{itemize}

Finally, we show that the while loop of the algorithm always terminates by providing an upper bound on $T$.
First note that only the final iteration would have $\norm{x_m}^2 = n$. In each iteration of the while loop, if $\eta_m < \eta$ and $\norm{x_m}^2 < n$, then we can reduce the size of $\F_m$ by at least one. Thus, there are at most $\delta n$ iterations such that $\eta_m < \eta$ and $\norm{x_m}^2 < n$.
Therefore, combining with the facts that each incremental $\sqrt{\eta_m} y \perp x_m$ and $x_T \in [b,c]$, it holds that
\[ (T-\delta n - 1)\eta \leq \sum_{m=0}^{T-1} \eta_m = \norm{x_T}^2 \leq n. \]
Therefore, the while loop terminates within $T \leq n/\eta + \delta n + 1$ steps.
By choosing the step length $\eta$ appropriately to satisfy \eqref{e:eta}, the solution can be returned in $\poly(n,d,p)$ time.
% For the partial coloring condition, 
% since $y\perp x_m$, it follows that $\norm{x_{m+1}}^2 = \norm{x_m}^2 + \eta$.
% If the algorithm terminates after the for-loop finished, then $\norm{x_{\lfloor n/\eta \rfloor}}^2 \geq n-\eta$, and we can round it to a vector $x$ such that $\norm{x}^2 = n$. 
% Note that the increase of the potential function due to the rounding is negligible, as $\norm{A(x) - A(x_{n/\eta})}_{2p} \leq \sigma_* \sqrt{\eta} \leq \sqrt{\sigma \cdot \nu}$, using the facts that $\norm{A(y)}_{2p} \leq \sigma_* \norm{y}$ and $\sigma_* \leq \min\{\sigma,\nu\}$. 
% If the algorithm terminates before the for-loop finished, then there are at least $\delta n$ frozen coordinates, so the rounding in Step (5) of the algorithm ensures that the values of these coordinates are either $b(i)$ or $c(i)$, satisfying the partial coloring condition. 
% The increase of the potential function is again negligible, as $\norm{A(x) - A(x_T)}_{2p}\leq \sigma_* \norm{x-x_T} \leq \sigma_* \leq \sqrt{\sigma \nu}$ since $|x(i) - x_T(i)|\leq 1/n$ for all $i$.
\end{proofof}

\subsection{Interpolation Analysis for Moments} \label{s:interpolation-moment}

We prove \autoref{proposition:moment-interpolation} in this subsection. 
As outlined in the technical overview in \autoref{s:overview}, 
we bound the change in our potential function $\Phi(t,x)$ by comparing the moments of our mixed operators before and after the update using the intermediate operator
\[
X_{t,\eta} := A_0 \otimes 1 + A(x) \otimes 1 + \sqrt{K_{\eps,\delta} \cdot \Big(1-t-\frac{\eta}{n}\Big)} \cdot \XF.
\]
{\bf Interpolation:} 
The first step is to interpret the update as replacing a small increment of $\XF$ by a small increment of the finite deterministic part $A(y)$. 
Using \autoref{fact:semi-conv}, observe that
\[
\sqrt{K_{\eps,\delta} \cdot (1-t)} \cdot \XF
\quad \textrm{and} \quad
\sqrt{K_{\eps,\delta} \cdot \Big(1-t-\frac{\eta}{n}\Big)} \cdot \XF + \sqrt{K_{\eps,\delta} \cdot \frac{\eta}{n}} \cdot \XF'
\]
have the same distribution, where $\XF'$ is a freely independent copy of $\XF$. 
Thus, we can rewrite the potential change as
\begin{equation} \label{e:potential-change-trace}
\Phi\Big( t+\frac{\eta}{n},~x + \sqrt{\eta} \cdot y \Big) - \Phi(t,x)
= \underbrace{\tr\otimes\tau \big(\big(X_{t,\eta}+ \sqrt{\eta} \cdot A(y)\big)^{2p}\big)^{\frac{2}{p}}}_{(1)} 
- \underbrace{\tr\otimes\tau \Big(\Big(X_{t,\eta}+ \sqrt{K_{\eps,\delta} \cdot \frac{\eta}{n}} \cdot \XF'\Big)^{2p}\Big)^{\frac{2}{p}}}_{(2)}.
\end{equation}
{\bf Taylor Expansion:}
We use Taylor's Approximation \autoref{theorem:taylor-general} to approximate both (1) and (2) up to second-order terms.
Consider $\tr\otimes\tau ((X_{t,\eta}+  H)^{2p})^{\frac{2}{p}}$ for $H\in \{\sqrt{\eta}A(y), \sqrt{K_{\eps,\delta} \cdot \frac{\eta }{n}} \cdot \XF'\}$. 
Applying \autoref{lemma:derivatives} (with $f(x) := x^\frac2p$, $\tau := \tot$, and $g(x) = x^{2p}$) and \autoref{lemma:moment-higher-der} to compute the first and second derivatives, the Taylor approximation is
\begin{align*}
\tr\otimes\tau ((X_{t,\eta}+H)^{2p})^{\frac{2}{p}} 
= \tr\otimes&\tau \big(X_{t,\eta}^{2p}\big)^{\frac{2}{p}} 
+ 4\tr\otimes\tau\big(X_{t,\eta}^{2p}\big)^{\frac{2}{p}-1} \cdot \tot\big(X_{t,\eta}^{2p-1} \cdot H\big)
\\
&+4 p \cdot \Big(\frac{2}{p}-1\Big) \cdot \tot\big(X_{t,\eta}^{2p}\big)^{\frac{2}{p}-2} \cdot \tr\otimes\tau\big(X_{t,\eta}^{2p-1} \cdot H\big)^2
\\
&+ 2\tr\otimes\tau\big(X_{t,\eta}^{2p}\big)^{\frac{2}{p}-1} \cdot \sum_{k=0}^{2p-2}\tr\otimes\tau\big(H \cdot X_{t,\eta}^k \cdot H \cdot X_{t,\eta}^{2p-2-k}\big)
\\
&\pm O(\cR_m(H)),
\end{align*}
where $\cR_m(H):= \sup_{\substack{0\leq r \leq 1}}|D^3\big(\tr\otimes\tau\big((X_{t,\eta} +  rH)^{2p}\big)^{\frac{2}{p}}\big)(H,H,H)|$. 

{\bf The Free Term:} We expand the free term (2) in \eqref{e:potential-change-trace}.
Substitute $H := \sqrt{K_{\eps,\delta} \cdot \frac{\eta}{n}} \cdot \XF'$ in the Taylor expansion above.
Since $\XF'$ and $\XF$ are freely independent and centered, 
$\tot(X_{t,\eta}^{2p-1} \cdot \XF')=0$ and thus the first-order term vanishes.
For the second-order term, we apply \autoref{lemma:free-moments} to compute\footnote{We remark that this calculation provides a simple and natural explanation why the expectation is ``pushed inside'' in~\eqref{e:pushed-inside}, which is a subtle calculation in \cite[Lemma~5.2 and Corollary~5.3]{BBvH23} using Gaussian matrices. 
}
\begin{eqnarray*}
\tot (\XF' \cdot X_{t,\eta}^k \cdot \XF' \cdot X_{t,\eta}^{2p-2-k}) 
&=& \tr \varphi \big(\XF' \cdot X_{t,\eta}^k \cdot \XF' \cdot X_{t,\eta}^{2p-2-k} \big)
\\
&=&\tr\big( \varphi[\XF' \cdot (\varphi[X_{t,\eta}^k] \otimes 1) \cdot \XF'] \cdot \varphi[X_{t,\eta}^{2p-2-k}]\big)
\\
&=& \sum_{i=1}^n \tr \big(A_i \cdot \varphi[X_{t,\eta}^k] \cdot A_i \cdot \varphi[X_{t,\eta}^{2p-2-k}] \big),
\end{eqnarray*}
where the last equality follows from $\varphi[\XF' \cdot (M \otimes 1) \cdot \XF'] = \sum_{i=1}^n A_iMA_i$ as $s_1,\ldots,s_n$ are freely independent with $\tau(s_i)=0$ and $\tau(s_i^2)=1$ for $i \in [n]$.
Therefore, the free term can be expanded as 
\begin{align*}
\tot& \Big( \Big(X_{t,\eta}+ \sqrt{K_{\eps,\delta} \cdot \frac{\eta}{n} } \cdot \XF' \Big)^{2p}\Big)^{\frac{2}{p}} = \tr\otimes\tau \big(X_{t,\eta}^{2p} \big)^{\frac{2}{p}} 
\\
& 
+ \frac{2 \eta}{n} \cdot K_{\eps,\delta} \cdot \tr\otimes\tau \big(X_{t,\eta}^{2p} \big)^{\frac{2}{p}-1} \sum_{k=0}^{2p-2} \sum_{i=1}^n \tr\big(A_i \cdot \varphi[X_{t,\eta}^{k}] \cdot A_i \cdot \varphi[X_{t,\eta}^{2p-2-k}]\big)
+  O\Big(\cR_m \Big( \sqrt{K_{\eps,\delta} \cdot \frac{\eta}{n}} \cdot \XF' \Big) \Big).
\end{align*}

{\bf The Deterministic Term and The Subspace $\mathcal{H}'$:}
Next, we expand the deterministic term (1) in \eqref{e:potential-change-trace} and define the subspace $\mathcal{H}'$ in \autoref{proposition:moment-interpolation}.
Define the auxiliary function $\phi(y) := \tr\otimes\tau\big((X_{t,\eta} + A(y)\otimes 1)^{2p}\big)$.
We compute
\[
\partial_{y_i}\phi(0) = 2p \cdot \tr\otimes\tau\big(X^{2p-1}_{t,\eta} \cdot (A_i \otimes 1)\big)
\quad \textrm{and} \quad
\partial_{y_i}\partial_{y_j}\phi(0) 
= 2p \cdot \sum_{k=0}^{2p-2}\tr\otimes\tau\big(X^{k}_{t,\eta} \cdot (A_i \otimes 1) \cdot X_{t,\eta}^{2p-k-2} \cdot (A_j \otimes 1)\big).
\]
Note that the factors in the first and second order terms in the Taylor expansion can be written as
\[
2p \cdot \tr\otimes\tau \big(X_{t,\eta}^{2p-1} \cdot (A(y) \otimes 1) \big) =\nabla\phi(0)^\top y 
\quad \textrm{and} \quad  
\sum_{k=0}^{2p-2}\tr\otimes\tau\big( (A(y) \otimes 1) \cdot X_{t,\eta}^k \cdot (A(y) \otimes 1) \cdot X_{t,\eta}^{2p-2-k}\big) =\frac{1}{2p}y^\top\nabla^2\phi(0)y. 
\]  
Since $\phi$ is convex, the Hessian matrix $\nabla^2\phi(0)$ is positive semidefinite, meaning its eigenvalues are non-negative. 
Let $\mathcal{H}_{K}$ be the top $n/K$ eigenspace (with largest eigenvalues) of $\nabla^2 \phi(0)$.
Define
\[
\mathcal{H}' := \{ x \in \R^n \mid x \perp \nabla \phi(0) \textrm{ and } x \perp \mathcal{H}_{K_{\eps,\delta}} \}
\]
be the subspace orthogonal to $\nabla \phi(0)$ and the top $n/K_{\eps,\delta}$ eigenspace of $\nabla^2 \phi(0)$.
The dimension of $\mathcal{H}'$ is
\[
\dim(\mathcal{H}') = n - 1 - \frac{n}{K_{\eps,\delta}} 
= n - 1 - n\Big(1-\eps-\delta-\frac3n\Big)
= (\eps + \delta) n + 2.
\]
Since all eigenvalues of the Hessian $\nabla^2 \phi(0)$ are non-negative,
by Markov's inequality, for any $K > 0$,
there are at most $n/K$ eigenvalues which are at least $\frac{K}{n}\Tr(\nabla^2 \phi(0))$. 
This implies that, for any $y \in \mathcal{H}'$ with $\norm{y}=1$, it holds that $\nabla\phi(0)^\top y = 0$ and
\[
y^\top \nabla^2\phi(0)y 
\leq \frac{1}{n} \cdot K_{\eps,\delta} \cdot \Tr(\nabla^2\phi(0))
= \frac{2p}{n} \cdot K_{\eps,\delta} \cdot \sum_{i=1}^n \sum_{k=0}^{2p-2} \tr\otimes\tau( (A_i \otimes 1) \cdot X_{t,\eta}^k \cdot (A_i \otimes 1) \cdot X_{t,\eta}^{2p-2-k}).
\]
Using these bounds in the Taylor expansion,
we obtain that for any $y \in \mathcal{H}'$ with $\norm{y}=1$,
\begin{align*}
\tr\otimes\tau& \big(\big(X_{t,\eta}+ \sqrt{\eta} \cdot (A(y) \otimes 1)\big)^{2p}\big)^{\frac{2}{p}} 
\leq \tr\otimes\tau\big(X_{t,\eta}^{2p}\big)^{\frac{2}{p}} 
\\
& + \frac{2\eta}{n} \cdot K_{\eps,\delta} \cdot \tr\otimes\tau\big(X_{t,\eta}^{2p}\big)^{\frac{2}{p}-1}\sum_{k=0}^{2p-2}\sum_{i=1}^n\tr\otimes\tau\big( (A_i \otimes 1) \cdot X_{t,\eta}^k \cdot (A_i \otimes 1) \cdot X_{t,\eta}^{2p-2-k}\big)
+  O(\cR_m(\sqrt{\eta} \cdot (A(y) \otimes 1))).
\end{align*}

{\bf The Error Terms:} To bound the error terms, we first show that
\begin{equation} \label{e:Taylor-error}
\cR_m(H) \lesssim p^2 \cdot (\norm{X_{t,\eta}}_{2p}\norm{H}^3 + \norm{H}^4).
\end{equation}
Let $X$ be a self-adjoint element. By computing the derivatives of $X^{2p}$ using \autoref{lemma:moment-higher-der} and then applying the generalized H\"older's inequality in \autoref{corollary:general-holder} term-by-term (with $\ell! \cdot \binom{2p}{l}$ terms), it follows that 
\[
\big|\tot \big( D^\ell((X)^{2p})(H,\ldots, H) \big)\big| \leq \frac{(2p)!}{(2p-\ell)!} \cdot \norm{H}^\ell \cdot \tot(|X|^{2p-\ell}).
\]
By Jensen's inequality, $\tot(|X|^{2p-\ell})\leq \tot(X^{2p})^{1-\frac{\ell}{2p}}$ for any $1\leq \ell\leq 2p$. 
Together with \autoref{lemma:derivatives} and \autoref{corollary:general-holder}, 
\begin{align*}
|D^3\big(\tr\otimes\tau[X^{2p}]^{\frac{2}{p}}\big)(H,H,H)| 
& \lesssim \norm{H}^3 \cdot p^2 \cdot \Big(\tot(X^{2p})^{\frac{2}{p}-3} \cdot \tot(|X|^{2p-1})^3
\\
& \hspace{2.5cm} + \tot(X^{2p})^{\frac{2}{p}-2} \cdot \tot(|X|^{2p-2}) \cdot \tot(|X|^{2p-1})\\
& \hspace{2.5cm} + \tot(X^{2p})^{\frac{2}{p}-1} \cdot \tot(|X|^{2p-3})\Big)\\
&\lesssim \norm{H}^3 \cdot p^2 \cdot \tot(X^{2p})^{\frac{1}{2p}}.
\end{align*}
Substituting $X = X_{t,\eta} + rH$ and bounding $\tot(X^{2p})^{\frac{1}{2p}} \leq \norm{X}\leq \norm{X_{t,\eta}} + \norm{H}$ as $r\leq 1$
establishes \eqref{e:Taylor-error}.

It remains to bound $\norm{H}$ and $\norm{X_{t,\eta}}$.
Recall that $\norm{A(y)}\leq \sigma_*$ for $\norm{y}=1$.
By Pisier's \autoref{lemma:pisier}, $\norm{\XF'}\leq 2\sigma$. 
Since we defined $\rho := \max\{\sigma_*, 2\sigma/\sqrt{n}\}$, it follows that
\[
\Bignorm{\sqrt{\frac{\eta}{n} \cdot K_{\eps,\delta}} \cdot \XF'} 
\lesssim \frac{\sqrt{\eta} \cdot \sigma}{\sqrt{n}} \leq \sqrt{\eta} \cdot \rho 
\quad \textrm{and} \quad
\norm{\sqrt{\eta} \cdot A(y)} \leq \sqrt{\eta} \cdot \rho.
\]  
By triangle inequality, 
$\norm{X_{t,\eta} } = \norm{A_0 +\sqrt{1-t} \cdot \XF + A(x_t)} 
\leq \norm{A_0} + 2\sigma + \sigma_*\sqrt{n} \leq  \norm{A_0} + \rho\sqrt{n} = \rho_0$.
Therefore, by \eqref{e:Taylor-error},
\[
\cR_m(\sqrt{\eta} \cdot (A(y) \otimes 1))
\lesssim p^2(\norm{X_{t,\eta}} \cdot \norm{\sqrt{\eta} \cdot A(y)}^3 + \norm{ \sqrt{\eta} \cdot A(y)}^4) 
\leq p^2(\eta^{\frac32}\rho^3\rho_0 + \eta^2\rho^4) 
\lesssim \frac{\eta}{n} p^2\sigma^2\nu^2
\]
where the last inequality is by our assumption on $\eta$ in \eqref{e:eta}.
The same bound holds for $\cR_m ( \sqrt{K_{\eps,\delta} \cdot \frac{\eta}{n}} \cdot \XF')$.

{\bf Putting Together and Bounding Crossing Terms}:
Putting together the free term, the deterministic term, and the error terms into \eqref{e:potential-change-trace},
for any $y \in \mathcal{H}'$,
the potential change is
\begin{eqnarray*}
& & \tr\otimes\tau \big(\big(X_{t,\eta}+ \sqrt{\eta} \cdot A(y)\big)^{2p}\big)^{\frac{2}{p}} - \tr\otimes\tau \Big(\Big(X_{t,\eta}+ \sqrt{\frac{\eta}{n} \cdot K_{\eps,\delta}} \cdot \XF'\Big)^{2p}\Big)^{\frac{2}{p}}
\\
&\leq & 
\frac{2\eta}{n} \cdot K_{\eps,\delta} \cdot \tr\otimes\tau\big(X_{t,\eta}^{2p}\big)^{\frac{2}{p}-1}
\sum_{k=0}^{2p-2} \sum_{i=1}^n \Big( \tr\otimes\tau\big((A_i \otimes 1) \cdot X_{t,\eta}^{k} \cdot (A_i \otimes 1) \cdot X^{2p-2-k}_{t,\eta}\big)-\tr(A_i \cdot \varphi[X_{t,\eta}^{k}] \cdot A_i \cdot \varphi[X_{t,\eta}^{2p-2-k}])\Big)
\\
& & \hspace{1cm} +~O\Big(\frac{\eta}{n} p^2\sigma^2\nu^2\Big)
\end{eqnarray*}
Finally, we bound the second-order crossing terms. 
Define the random matrix $Y = \sum_{i=1}^n y'(i) \cdot A_i$, where $y'$ is a random vector with i.i.d.~$\pm 1$ entries. 
Then, 
\begin{eqnarray*}
& & \sum_{k=0}^{2p-2}\sum_{i=1}^n\Big(\tr\otimes\tau\big( (A_i \otimes 1) \cdot X_{t,\eta}^{k} \cdot (A_i \otimes 1) \cdot X^{2p-2-k}_{t,\eta}\big)-\tr(A_i \cdot \varphi[X_{t,\eta}^{k}] \cdot A_i \cdot \varphi[X_{t,\eta}^{2p-2-k}])\Big)
\\
&=& \sum_{k=0}^{2p-2} \E\big[\tr\otimes\tau\big( (Y \otimes 1) \cdot X_{t,\eta}^{k} \cdot (Y \otimes 1) \cdot X^{2p-2-k}_{t,\eta}\big)
-\tr(Y \cdot \varphi[X_{t,\eta}^{k}] \cdot Y \cdot \varphi[X_{t,\eta}^{2p-2-k}])\big]
\\
&\lesssim & p^3 K_{\eps,\delta}(1-t-\eta/n) \cdot \sigma^2 \cdot \nu^2 \cdot \tr\otimes\tau(X_{t,\eta}^{2p-4})
\\
& \leq & p^3 K_{\eps,\delta}(1-t-\eta/n) \cdot \sigma^2 \cdot \nu^2 \cdot \tr\otimes\tau(X_{t,\eta}^{2p})^{1-\frac{2}{p}}, 
\end{eqnarray*}
where the second-to-last inequality is by applying the intrinsic freeness \autoref{proposition:moment-crossing-bound2} with $X = X_{t,\eta}$ and noting that $\Tilde{\nu}(X_{t,\eta})^2 = K_{\eps,\delta}(1-t-\eta/n)\sigma\nu$ and $\Tilde{\nu}(Y)^2 = \sigma\nu$,
and the last inequality follows from Jensen's inequality. 
This completes the proof of \autoref{proposition:moment-interpolation}.

\subsection{Applications to Matrix Discrepancy} \label{ss:application-discrepancy}

An important tool in discrepancy theory is Rothvoss's algorithmic partial coloring theorem for convex bodies. 
The partial coloring theorem states that if a symmetric convex body has large enough Gaussian volume, then it contains a point with a constant fraction of entries in $\pm 1$. 
Rothvoss proved a constructive version of this result using a randomized Gaussian projection algorithm, with wide-ranging applications in algorithmic discrepancy theory, spectral sparsification, and numerical linear algebra.

\begin{theorem}[Randomized Partial Coloring for Convex Body~\cite{Rot17}]
Let $\eps < \frac{1}{60000}$, $\delta \asymp \log{\frac{1}{\eps}}$ be constants and $U\subseteq \R^n$ be a subspace of dimension at least $(1-\delta )n$. Let $\mathcal{K}\subseteq \R^n$ be a symmetric, convex set, and $x_0\in [-1,1]^n$. 
Suppose the Gaussian measure of $\mathcal{K}$ is at least $e^{-\eps n}$.
There is a randomized polynomial time algorithm to find a vector $x$ satisfying
\begin{enumerate}
\item $x \in \mathcal{K}$, 
\item $x+x_0 \in [-1,1]^n$, 
\item at least $\eps n$ coordinates of $x+x_0$ are in $\{-1,1\}$.
\end{enumerate}
\end{theorem}

Using \autoref{t:partial-coloring-full}, we provide a derandomization of Rothvoss's result when $\mathcal{K}$ is the $p$-norm ball of matrices (with constant Gaussian measure).
The proof is based on the idea in~\cite{BJM23} to apply partial coloring on projected matrices.

\begin{proposition}[Deterministic Partial Coloring for Matrix Norms] \label{lemma:matrix-partial-coloring}
Let $A_1,\ldots, A_n$ be symmetric matrices with matrix parameters $\norm{\sum_{i=1}^nA_i^2} = \sigma^2$ and $\sum_{i=1}^n\norm{A_i}_F^2 = nf^2$. Let $x_0 \in [-1,1]^n$ be an initial vector. For any $p \geq 2$, there is a deterministic polynomial time algorithm to find a partial coloring $x$ satisfying
\begin{enumerate}
\item $\norm{A(x)}_{2p} \lesssim \sigma + p^{\frac34}\sqrt{\sigma f}$,
\item $x\perp x_0$ and $x_0 + x \in [-1,1]^n$,
\item at least $\frac{n}{4}$ coordinates of $x_0 + x$ are in $\{-1,1\}$.
\end{enumerate}
\end{proposition}
\begin{proof}
Let $M$ be the $n \times n$ matrix where $M(i,j) = \inner{A_i}{A_j}$. 
Let $\mathcal{H}$ be the subspace orthogonal to the top $\frac{n}{3}$ eigenspace of $M$.
Let $P$ be the projection matrix into $\mathcal{H}$.
Define the projected matrices as
\[
\Tilde{A}_{i} = \sum_{j=1}^n P(i,j) \cdot A_j.
\]
We apply \autoref{t:partial-coloring-full} with input matrices $A_0=0,\Tilde{A}_1,\ldots, \Tilde{A}_n$, linear subspace constraint $\mathcal{H}\cap \{x_0\}^{\perp}$, box constraints $b(i) = -1-x_0(i)$, $c(i)= 1-x_0(i)$, and parameter $\delta=\frac14$. 
Specify the remaining input parameters as follows.
Since $\dim\big(\mathcal{H} \cap \{x_0\}^\perp\big) \geq \frac23 n-1$, we set $\eps = \frac{1}{3}+\frac{1}{n}$ so that $\eps+\delta < 1$. 
Let $\Tilde{M}$ be the $n \times n$ matrix where $\Tilde{M}(i,j) = \inner{\Tilde{A}_i}{\Tilde{A_j}}$. 
Note that $\Tilde{M} = PMP \succeq 0$, with maximum eigenvalue at most $\frac{3}{n}\Tr(M) = 3f^2$. 
The input matrices satisfy the matrix parameters
\[
\sigma^2 = \Bignorm{\sum_{i=1}^n \Tilde{A}_i^2} \leq \Bignorm{\sum_{i=1}^n A_i^2} =\sigma^2\quad \textrm{and} \quad 
\nu^2 = \Bignorm{\sum_{i=1}^n \vecc(\Tilde{A}_i)\vecc(\Tilde{A}_i)^{\top}} = \norm{\Tilde{M}} \leq 3f^2.
\]
Let $x$ be the output vector guaranteed by \autoref{t:partial-coloring-full}.  
Since $x\in \mathcal{H}$, we have
\[
A(x) 
= A(Px)
= \sum_{i=1}^n \Big(\sum_{j=1}^n P(i,j) \cdot x(j) \Big) \cdot A_i 
= \sum_{j=1}^n x(j) \cdot \Big(\sum_{i=1}^n P(i,j) \cdot A_i \Big)  
= \sum_{j=1}^n x(j) \cdot \Tilde{A_j} = \Tilde{A}(x),
\]
where the second-to-last equality uses that $P$ is a symmetric matrix.
Let $\XF = \sum_i \Tilde{A}_i\otimes s_i$ be the free model in \autoref{t:partial-coloring-full}. 
By \autoref{lemma:pisier}, $\norm{\XF}_{2p}\leq \norm{\XF}\leq 2\sigma$. 
Thus, the $2p$-norm bound guarantee in \autoref{t:partial-coloring-full} implies that
\[
\bignorm{\Tilde{A}(x)}_{2p} \lesssim \sigma + p^{\frac34}\sqrt{\sigma f}.
\]
The box constraints that $b\leq x\leq c$ ensures that $x+x_0 \in [-1,1]^n$.
The linear subspace constraint ensures that $x \perp x_0$.
The partial coloring condition in \autoref{t:partial-coloring-full} ensures that there are at least $\frac{n}{4}$ coordinates of $x+x_0$ are in $\{-1,1\}$ (as the possibility that $\norm{x}^2=n$, together with
$x\perp x_0$ and the box constraint, implies that
$\norm{x+x_0}^2=n$ and hence that all coordinates are in
$\{-1,1\}$).
\end{proof}

\subsubsection{Matrix Spencer Problem}

In the matrix Spencer problem, we are given $n\times n$ matrices $A_1,\ldots,A_n$ with $\norm{A_i}\leq 1$ for $i \in [n]$. 
The goal is to find a coloring $x: [n]\rightarrow \{\pm 1\}$ so that $\norm{A(x)}\lesssim O(\sqrt{n})$. 
Bansal, Jiang, and Meka~\cite{BJM23} proved that such a coloring exists when each $A_i$ has $\norm{A_i}_F^2\leq O(n/\log^3{n})$, by using \eqref{e:BBvH-norm} from~\cite{BBvH23} to show that the Gaussian measure of the matrix $p$-norm ball is large and then applying Rothvoss' partial coloring theorem. 
We use \autoref{lemma:matrix-partial-coloring} to provide a deterministic polynomial time algorithm with the same guarantee.

\begin{proofof}{\autoref{t:matrix-Spencer}}
Consider the following procedure for finding a full coloring with low discrepancy.
\begin{framed}
\vspace{-3mm}
       \begin{itemize}
            \item Initialize $\mathcal{F}_0 = [n]$ and $x_0 = 0$.
            \item For $m = 0$ to $O(\log{n})$
            \begin{enumerate}
                \item Let $x^{|\mathcal{F}_m|}_m$ be the restriction of $x_m$ to the coordinates in $\mathcal{F}_m$.
                \item Apply \autoref{lemma:matrix-partial-coloring} to find a vector $y^{|\mathcal{F}_m|} \in \R^{|\mathcal{F}_m|}$ such that $x_0^{|\mathcal{F}_m|} + y^{|\mathcal{F}_m|}\in [-1,1]^{|\mathcal{F}_m|}$, with at least $\frac14 |\mathcal{F}_m|$ coordinates in $\{1,-1\}$.
                \item Update $x_{m+1} \gets x_m+ y^{|\mathcal{F}_m|}$.
Let $\mathcal{F}_{m+1}$ be the coordinates in $x_{m+1}$ with value in $(-1,1)$.
            \end{enumerate}
        \end{itemize}
\end{framed}
    
We apply the partial coloring algorithm iteratively until all coordinates become $\pm1$. 
In each iteration $m$, $\mathcal{F}_m$ is the set of coordinates of $x_m$ with value in $(-1,1)$. 
Let $\sigma^2_m = \norm{\sum_{i\in \mathcal{F}_m}A_i^2}$ and $f^2_m = \frac{1}{|\mathcal{F}_m|}\sum_{i\in \mathcal{F}_m}\norm{A_i}_F^2 $.  
\autoref{lemma:matrix-partial-coloring} guarantees that $|\mathcal{F}_{m+1}|\leq \frac34|\mathcal{F}_m|$ and
\[
\norm{A(x_{m+1}) - A(x_m)}_{2p} 
= \bignorm{ A(y^{|\mathcal{F}_m|})}_{2p}
\lesssim \sigma_m + p^{\frac34}\sqrt{\sigma_mf_m}.
\]
Since $f_m^2 \leq r$ and $\sigma_m^2\leq |\F_m| \leq (\frac34)^m \cdot n$, 
this gives a final discrepancy bound of
\[
\sum_{m} \big( \sigma_m+p^{\frac34}\sqrt{\sigma_mf_m} \big)
\lesssim        \sqrt{n} + p^{\frac34} \cdot n^{\frac14} \cdot r^{\frac14},
\]
as this is a geometric sum which is dominated by the first term.
\end{proofof}

\subsubsection{Spectral Sparsification of Eulerian Graphs}

Given a graph $G=(V,E)$, a spectral sparsifier is a sparse reweighted subgraph whose Laplacian approximates that of $G$. A classical result, due to Batson, Spielman, and Srivastava \cite{BSS12}, is that any undirected graph can be $\eps-$approximated by a sparsifier of size $O(n/\eps^2)$. Recently, Reiss and Rothvoss \cite{RR20} introduced a new perspective on spectral sparsification by framing it as a matrix discrepancy problem. This powerful framework allows the construction of linear-sized spectral sparsifiers satisfying additional constraints, and was further developed and derandomized in \cite{LWZ25}.

One advantage of the discrepancy framework for spectral sparsification is that it can be applied to directed graphs. Given a weighted Eulerian directed graph, $\vec{G}=(V,E, w)$, a spectral sparsifier of $\vec{G}$ is a sparse reweighted subgraph, $\vec{H}$, satisfying
\[
\norm{L_G^{\dagger/2} (\vec{L}_{\vec{G}} - \vec{L}_{\vec{H}}) L_G^{\dagger/2}} \leq \eps,
\]
where $\vec{L}_{\vec{G}}$ is the directed Laplacian matrix of $\vec{G}$. 

Jambulapati, Sachdeva, Sidford, Tian, and Zhao~\cite{JSSTZ25} combine the matrix discrepancy result from~\cite{BJM23} with an effective resistance decomposition technique to construct a Eulerian sparsifier of size $O(\eps^{-2} n \log n + \eps^{-4/3} n \log^{5/3} n)$ in randomized polynomial time. 
Using the deterministic partial coloring result in \autoref{t:partial-coloring-full} as a black box, we obtain a deterministic polynomial time algorithm with the same sparsity guarantee (but with a much slower runtime).
We refer the reader to \cite{JSSTZ25} for details.

\section{Concentration of Full Spectrum}\label{section:resolvent}

In this section, we build on the ideas from the previous analysis to derandomize the concentration-of-spectrum theorem in \cite{BBvH23} and \cite{BCS+24}.
The goal is to prove the following full version of \autoref{t:spectrum}.

\begin{theorem}[Deterministic Full Spectrum] \label{t:spectrum-full}
Let $A_0,A_1,\ldots,A_n$ be $d\times d$ Hermitian matrices satisfying $\frac{1}{\poly{d}}\leq \max_{i=1}^n \norm{A_i}\leq \poly{d}$, $\norm{A_0}\leq \poly{d}$. 
Define the matrix parameters
\[
\sigma^2 = \Bignorm{\sum_{i=1}^nA_i^2},
\quad 
\nu^2 = \Bignorm{\sum_{i=1}^n\vecc(A_i)\vecc(A_i)^*},
\quad
\sigma_*^2 = \sup_{\norm{y},\norm{z}=1}\sum_{i=1}^n |\inner{y}{A_iz}|^2.
\]
Let $\mathcal{U}\subseteq \mathbb{R}^n$ be an arbitrary linear subspace of dimension $n-r+1$, and $\eps = \Theta(\sqrt{\sigma\nu}\log^{\frac34}d + \sigma_*\sqrt{r\log d})$. 
There is a deterministic polynomial time algorithm to compute a vector $x$ with $\norm{x}^2 \leq  n$, satisfying $x\in \mathcal{U}$ and
\[
\spec(A_0+A(x)) \subseteq \spec(\Xf) + [-\eps, \eps]
\quad \textrm{and} \quad
\spec(\Xf) \subseteq \spec(A_0+A(x)) + [-\eps, \eps],
\]
where $\Xf = A_0 \otimes 1+ \XF$, $\XF = \sum_{i=1}^n A_i\otimes s_i$, and $s_1, \ldots, s_n$ are freely independent semicircular elements.
\end{theorem}

\subsection{Overview and Outline}

In \cite{BBvH23}, the concentration of the full spectrum was established by interpolating between the resolvent of a Gaussian matrix $X$ and the resolvent of the corresponding free model $\Xf$. 
In particular, they proved that for any $z\in \C$ with $\im(z) > 0$, 
$\norm{(zI-X)^{-1}} \approx \norm{(z1-\Xf)^{-1}}$ with high probability. 
Since the operator norm is difficult to analyze directly due to its lack of smoothness, 
this comparison was done indirectly through the $p$-norm of the resolvent for $p\approx \log{d}$.
Then, by taking a union bound over $z=\lambda + i\eps$ for a range of $\lambda$'s covering the spectrum $\spec(X)$, they established that $\spec(X) \subseteq \spec(\Xf)+[-\eps,\eps]$ with high probability. 

\cite{BBvH23} did not establish the other direction that $\spec(\Xf)\subseteq \spec(X) + [-\eps,\eps]$. 
This is because while it is known that $\norm{(zI-X)^{-1}} \asymp \norm{(zI-X)^{-1}}_{2p}$ for $p\asymp \log{d}$, the corresponding inequality was not known for the free model $\Xf$. 
Later, in \cite{BCS+24}, the ultracontractivity bounds in \autoref{lemma:poly-ultra} and \autoref{lemma:res-ultra} were established, which implied that $\norm{(zI-\Xf)^{-1}} \asymp \norm{(zI-\Xf)^{-1}}_{2p}$ when $p \asymp \log{d}$. 
This shows that the two-sided concentration of spectrum result follows directly from the proof in \cite{BBvH23}.

In our work, we will also use the resolvent method. 
As in \cite{BBvH23}, we will pick a set of points $z = \lambda +i\eps$, where $\lambda$ is a fine cover of $\spec(X)\cup \spec(\Xf)$. 
For each such $z$, we define a potential function $\Phi_z(t,x)$ to control the quantity $\norm{(zI-A(x))^{-1}}_{2p}$ in terms of $\norm{(z1-\Xf)^{-1}}_{2p}$ for $p\approx\log{d}$. 
To bound the expected change in each potential function, we will use similar methods as in \autoref{s:interpolation-moment}, where we use second-order Taylor approximation to express the potential change in terms of the crossing terms, which will then be bounded using the intrinsic freeness of resolvents in \autoref{proposition:res-crossing-bound2}. 
This can be seen as a derandomization of \cite[Theorem 6.1]{BBvH23}. 

To control all potential functions simultaneously and deterministically, 
we define an aggregate potential function that combines all of the resolvent potential functions. 
By controlling the increase in the aggregate potential function, we control the maximum of all potential functions with only a logarithmic factor loss in the number of potential functions. 
This procedure can be interpreted as a multiplicative weight update method that derandomizes the Gaussian Lipschitz concentration inequality. 
We note this procedure can be used to control an arbitrary set of potential functions, not just restricted to one set of inputs. 
We will elaborate more on this in the applications in \autoref{s:application-spectrum}. 

Finally, we remark that we can still handle some linear subspace constraints on $x$ in \autoref{t:spectrum-full}. 
In particular, we can force $x$ to be orthogonal to a subspace of dimension $r$, but incur a loss of $\sqrt{r \log d}$. 
This means that, in many settings, we will not be able to handle $\Omega(n)$ linear constraints as in \autoref{t:partial-coloring-full}. 
This is a necessary trade-off for the two-sided approximation guarantee and arises from the non-convex nature of the resolvent norms.

\subsubsection{Outline}

We now formally define our potential functions and present our bounds on the potential updates. 
Given $z\in \C$ with $\im(z) > 0$, we define the potential function
\begin{equation} \label{e:potential2}
\Phi_z(t,x_t) := \tot\big(|z\cdot (I \otimes 1) - A_0 \otimes 1 - A(x_t) \otimes 1- \sqrt{1-t} \cdot \XF|^{-2p}\big)^\frac{1}{2p}.
\end{equation}
We will analyze a \textit{random} update $x_{t+1} = x_t + \sqrt{\eta} \cdot y$, where $y$ is a random vector with pairwise independent coordinates. 
As long as the expected change in the potential function is small, we can efficiently find an update that does not increase the potential function by more than the expected change, as the sample space of random vectors with pairwise independent coordinates is polynomially bounded. 

The following are the precise bounds on the expected change of the potential function.

\begin{proposition}[Expected Potential Increase] \label{proposition:res-interpolation}
Let $A_0,A_1,\ldots,A_n\in \M_d(\C)$ be given as in the setting of \autoref{t:spectrum-full}. 
Let $z\in \C$ be such that $\im(z) > 0$. 
Let $\eta > 0$ be a sufficiently small step-size. 
Given any $t \in [0,1-\eta]$ and $x_t\in \R^{n}$, the potential change with respect to update $y \in \R^n$ is defined as
\[
\Delta_z(t,y) := \Phi_z(t+\eta,x_t+\sqrt{\eta} \cdot y) - \Phi_z(t,x_t).
\]
Let $y$ be a random vector with pairwise independent $\pm1$ Rademacher entries
and $P$ be an $n\times n$ projection matrix with rank $n-r$. 
Then, the expected change is bounded as
\[
|\E_y[\Delta_z(t,Py)]| \lesssim  \eta \cdot \Big( \frac{p^{3} \cdot \sigma^2 \cdot \nu^2}{\im(z)^{5}} + \frac{p \cdot (\sigma_* \cdot \sigma +\sigma_*^2 \cdot r)}{\im(z)^{3}}\Big).
\]
Furthermore, the second moment and the maximum change are bounded as
\[
\E_y[\Delta_z(t,Py)^2] \lesssim \eta\cdot\frac{\sigma_*^2}{\im(z)^4}
\quad \textrm{and} \quad
\max_{y\in \{\pm 1\}^n} |\Delta_z(t,Py)| \leq \sqrt{\eta}\cdot \frac{(\sigma_*\sqrt{n}+2\sigma)}{\im(z)^2}.
\]
In particular, the inequality holds as long as 
\[
\eta \leq \min\Big\{ \frac{\im(z)^2 \cdot \sigma_*^2}{32p^2 \cdot \rho^4},\frac{\sigma^4 \cdot \nu^4}{\im(z)^2 \cdot \rho^6} \Big\}
\quad \textrm{where} \quad 
\rho =\max\{\sigma_*\sqrt{n}, 2\sigma\}.
\]
\end{proposition}

The organization of the remainder of this section is as follows.
We first assume \autoref{proposition:res-interpolation} and prove \autoref{t:spectrum-full} using the multiplicative weight update algorithm.
Then we will prove \autoref{proposition:res-interpolation} in \autoref{section:inter-res}.

\subsection{Full Spectrum and Multiplicative Weight Update Algorithm} \label{section:spec}

In this subsection, we prove \autoref{t:spectrum-full} assuming \autoref{proposition:res-interpolation}. 
First, we restate the results from \cite{BBvH23} (with some modifications), which show how to control the whole spectrum by controlling the resolvent at a finite net of points. 
Then, we present the multiplicative weight update algorithm to aggregate the potential functions into one potential function.
Finally, we combine these to prove \autoref{t:spectrum-full}.

\subsubsection{Full Spectrum via Resolvent}

The proofs in this part are the same as in \cite{BBvH23}.
We present them for completeness as there are minor modifications.
In the following lemma, we need the additional $K_2$ term for our application.

\begin{lemma}[{\cite[Lemma 6.4]{BBvH23}}] \label{lemma:res-to-spec}
Let $X,Y$ be self-adjoint operators and $K_1,K_2, K_3 \geq 0$. 
Suppose 
\[
\norm{(z1-X)^{-1}} \leq C\norm{(z1-Y)^{-1}} + \frac{K_1}{\eps^2} + \frac{K_2}{\eps^3} + \frac{K_3}{\eps^5}
\]
for all $z = \lambda +\eps i$ with $\lambda \in \spec(X)$ and a fixed $\eps \geq 3 \cdot \max\{K_1,K_2^{\frac12},K_3^{\frac14}\}$. Then 
\[
\spec(X)\subseteq \spec(Y) + [-2C\eps, 2C\eps].
\]
\end{lemma}

\begin{proof}
For any $S \subseteq \C$, let $d(z,S) := \inf_{z' \in S} |z-z'|$. 
Note that 
\[
\norm{(z1 - X)^{-1}} = \frac{1}{d(z,\spec(X))}.
\]
It follows from our assumptions that 
\[
\frac{1}{\eps}
\leq\norm{(z1-X)^{-1}} 
\leq C\norm{(z1-Y)^{-1}} + \frac{1}{3\eps} + \frac{1}{3^2\eps} + \frac{1}{3^4\eps}  
\leq \frac{C}{\sqrt{\eps^2+d(\lambda,\spec(Y))^2}} + \frac{1}{2\eps}.
\]
Rearranging this inequality gives
$d(\spec(Y),\lambda) \leq\sqrt{(2C\eps)^2-\eps^2} \leq 2C\eps$.
Thus, any eigenvalue of $X$ must be within a distance of at most $2C\eps$ from some eigenvalue of $Y$.
\end{proof}

The following lemma shows that it is enough to control the resolvent at a finite net of points.
The arguments are in the proofs of Lemma 6.5 and Lemma 6.7 in \cite{BBvH23} but the lemma was not stated explicitly.

\begin{lemma}[Finite Net] \label{lemma:spec-cover}
Let $X,Y$ be self-adjoint operators with $\norm{X}\leq b$.
Let $\delta >0$ be a small interval size and 
$\mathcal{N} = \{-b + t\delta \mid t\in \{0,1,\ldots,\lceil \frac{2b}{\delta} \rceil\}\}$. 
Suppose for all $z' = \lambda+\eps i$ with $\lambda \in \mathcal{N}$, it holds that
\[
\norm{(z'1-X)^{-1}} \leq C\norm{(z'1-Y)^{-1}} + R.
\]
Then, for all $z = \lambda + \eps i$ with $\lambda \in \spec(X)$, 
\[
\norm{(z1-X)^{-1}} \leq C\norm{(z1-Y)^{-1}} + R + \frac{\delta \cdot (C+1)}{\eps^2}.
\]
\end{lemma}
\begin{proof}
Let $\lambda \in \spec(X)$ and $\lambda'\in \mathcal{N}$ with $|\lambda'-\lambda| \leq \delta$. 
Let $z = \lambda + \eps i$ and $z' = \lambda'+\eps i$. Then 
\begin{eqnarray*}
\big|\norm{(z1-X)^{-1} } -\norm{(z'1-X)^{-1}}\big| 
& \leq & \norm{(z1-X)^{-1}-(z'1-X)^{-1}}
\\
& = & \sup_{\mu\in \spec(X)} \Big|\frac{1}{z-\mu} - \frac{1}{z'-\mu}\Big|
~\leq~ \sup_{\mu\in \spec(X)} \Big|\frac{\delta}{(z-\mu)(z'-\mu)}\Big|
~\leq~ \frac{\delta}{\eps^2}.
\end{eqnarray*}
Similarly, $|\norm{(z1-Y)^{-1}} - \norm{(z'1-Y)^{-1}}| \leq \frac{\delta}{\eps^2}$. 
Therefore, by triangle inequality,
\begin{eqnarray*}
& & \norm{(z1-X)^{-1}} -C\norm{(z1-Y)^{-1}} 
\\
&\leq&  \norm{(z'1-X)^{-1}}-C\norm{(z'1-Y)^{-1}} + C\big|\norm{(z'1-Y)^{-1}}-\norm{(z1-Y)^{-1}}\big| +  \big|\norm{(z1-X)^{-1}} - \norm{(z'1-X)^{-1}}\big|
\\
&\leq& R + \frac{\delta \cdot (C+1)}{\eps^2}. 
\end{eqnarray*}
\end{proof}

\subsubsection{Multiplicative Weight Update Algorithm} \label{section:mw}

In this part, we use a multiplicative weight update algorithm to simultaneously control many potential functions at once. 
This can be viewed as a derandomization of the Gaussian Lipschitz concentration inequality, which is used in the probabilistic proof in \cite{BBvH23}. 

We state our result in a general form, not just restricted to the potential function in \eqref{e:potential2}.

\begin{proposition}[Multiple Potential Functions] \label{proposition:mw}
Let $\Phi_1,\ldots, \Phi_N: [0,1]\times \R^n\rightarrow\R$ be a set of arbitrary potential functions satisfying $\Phi_i(0,0) = 0$ for $i \in [N]$. 
Let $r\in [n]$ and $\delta \in (0,1)$. 
Let $\mathcal{U}\subseteq \R^n$ be a subspace of dimension $n-r+1$. 
Suppose there exist bounds $B(r),L(r),Q(r)$ such that for any $i\in [N]$, $\eta\leq \delta$,  $t\in [0,1-\eta)$, $x\in \R^n$, and projection matrix $P\in \M_n(\R)$ of rank at least $n-r$, we have
\begin{enumerate}
\item $\E_{y}\big[\Phi_i(t+\eta,x+\sqrt{\eta} \cdot Py)\big] - \Phi_i(t,x) \leq \eta \cdot B(r)$,
\item $\E_{y}\big[\big(\Phi_i(t+\eta,x+\sqrt{\eta} \cdot Py) - \Phi_i(t,x)\big)^2\big] \leq \eta \cdot L(r)$,
\item $\big|\Phi_i(t+\eta,x+\sqrt{\eta} \cdot Py) - \Phi_i(t,x)\big|\leq \sqrt{\eta} \cdot Q(r)$,
\end{enumerate}
where the expectation is taken over a random vector $y$ with pairwise independent $\pm 1$ entries. 
Then, there is a deterministic algorithm which computes a vector $x$ such that 
\[
x\in \mathcal{U}
\quad \textrm{and} \quad 
\norm{x}^2\leq n 
\quad \textrm{and} \quad 
\Phi_i(1,x) \leq B(r) + 2\sqrt{L(r) \cdot \log{N}} \textrm{~~for~all~} i \in [N].
\]
The runtime of the deterministic algorithms are at most $K\cdot \poly\big(\frac{1}{\delta}, n,N,\frac{Q^2(r)}{L(r)}\big)$, where $K$ is an upper bound on the time to compute a potential function $\Phi_i$ on an input. 
\end{proposition}

To prove~\autoref{proposition:mw}, we define the following softmax potential which aggregates all the potential functions $\Phi_i$ for $i \in [N]$. 
Suppose the interval $[0,1]$ is divided into $T = 1/\eta$ evenly spaced intervals. 
For $m=1,2,\ldots, T$ and a parameter $\alpha$, define
\[
\Phi_{A}(m,x) = \log \sum_{i=1}^N \exp(\alpha \cdot \Phi_i(m\eta,x)).
\]
The following is the deterministic algorithm for computing a vector $x$ such that $\Phi_{A}(T,x)$ is not too large.

\begin{framed}{\bf{Multiplicative Weight Update Algorithm}}
    \begin{itemize}
        \item Initialize $x_0 = 0$, $\alpha = \sqrt{\frac{\log{N}}{L}}$, and $\eta = \min(\delta,\frac{L}{Q^2\log{N}})$.
        \item For $m=0,1,2,\ldots,\lfloor\frac{1}{\eta}\rfloor-1$
        \begin{enumerate}
            \item Let $P$ be the $n\times n$ projection matrix onto the $n-r$ dimensional space $\mathcal{U}\cap \{ y \mid y \perp x_m\}$.
            \item Choose $y$ from a pairwise independent distribution that minimizes $\Phi_{A}(m+1,~x_m + \sqrt{\eta} \cdot Py)$. 
            \item Update $x_{m+1}\gets x_m+\sqrt{\eta}Py$.
        \end{enumerate}
    \end{itemize}
\end{framed}

\begin{proofof}{\autoref{proposition:mw}}
By the assumption that $\Phi_i(0,0) = 0$ for all $i$,
the initial value of the aggregate function is $\Phi_{A}(0,0) = \log{N}$. 
At each iteration $m$, we will bound the expected increase to the aggregate potential function under a random update $y \in \{\pm1\}^n$ with pairwise independent coordinates. 
To simplify notation, we define the update and normalized weights as
\[
\Delta_{i}(m) = \Phi_{i}\big(\eta \cdot (m+1),~x_m+\sqrt{\eta} \cdot Py \big) - \Phi_i(\eta \cdot m ,~x_m)
\quad \textrm{and} \quad
w_i(m) = \frac{\exp(\alpha \cdot \Phi_i(m\eta,x_m))}{\sum_{j=1}^N \exp\big(\alpha \cdot \Phi_j(m\eta,x_m)\big)}.
\]  
We first note that $\alpha \cdot |\Delta_i(m)|\leq 1$ by our choice of $\eta$,
where the assumption on $Q$ is used here only. 
Now, using $e^{x}\leq 1+x+x^2$ for $|x|\leq 1$,
the potential update can be bounded as follows:
\begin{eqnarray*}
\Phi_{A}(m+1,x_m+\sqrt{\eta} \cdot Py) 
&=& \log\sum_{i=1}^N \exp\big(\alpha \cdot \Phi_i(m\eta,x_m) +\alpha \cdot \Delta_i(m)\big)
\\
&\leq&  \log\Big[\sum_{i=1}^N \Big( \exp\big(\alpha \cdot \Phi_i(m\eta,x_m)\big) \Big) \cdot \big(1+\alpha \cdot \Delta_i(m) + \alpha^2 \cdot \Delta_i(m)^2 \big)\Big]
\\
&=&  \log\Big[\Big(\sum_{j=1}^N \exp\big(\alpha \cdot \Phi_j(m\eta,x_m)\big) \Big) \Big(1+\sum_{i=1}^N w_i(m) \cdot \alpha \cdot \Delta_i(m) + \sum_{i=1}^N w_i(m) \cdot \alpha^2 \cdot \Delta_i(m)^2 \Big)\Big]
\\
&\leq& \Phi_{A}(m,x_m) +\sum_{i=1}^N w_i(m) \cdot \alpha \cdot \Delta_i(m) + \sum_{i=1}^N w_i(m) \cdot \alpha^2 \cdot \Delta_i(m)^2,
\end{eqnarray*}
where the last inequality is by $\log(1+x) \leq x$.
By our assumptions, $\E_y[\Delta_i(m)] \leq \eta \cdot B(r)$ and $\E_y[\Delta_i(m)^2] \leq \eta \cdot L(r)$ for $i \in [N]$. 
Therefore,
\[
\E_y\big[\Phi_{A}(m+1,~x_m+\sqrt{\eta} \cdot Py)\big] 
\leq \Phi_{A}(m,x_m) + \alpha \cdot \eta \cdot B(r) + \alpha^2 \cdot \eta \cdot L(r).
\]
Since we always choose an update with at most the expected value,
after $T=1/\eta$ iterations,
\[
\max_{i \in [N]} \Phi_i(1,x_T)
\leq \frac{1}{\alpha} \log \sum_{i=1}^N \exp\big(\alpha \cdot \Phi_i(1,x_T)\big) 
= \frac{1}{\alpha} \cdot \Phi_{A}(T,x_T) 
\leq \frac{\log{N}}{\alpha} + B(r) + \alpha \cdot L(r)\leq B(r)+2\sqrt{L(r)\log{N}}
\]
where we use the initial value $\Phi_A(0,0) = \log N$, and our choice of $\alpha = \sqrt{\log{N}/L}$. 
The projection constraint ensures that $Py \perp x_m$ and $Py\in \mathcal{U}$ at each iteration, and thus $\norm{x_T}^2 \leq n$ and $x_T\in \mathcal{U}$.

For the time complexity, there are $\frac{1}{\eta} = \max\{\frac{1}{\delta},\frac{Q(r)^2\log{N}}{L(r)}\}$ iterations.
In each iteration, we enumerate every vector $y$ in a sample space of pairwise independent $\pm 1$ bits of size $n^2$,
and evaluate the aggregate potential function $\Phi_A$ which consists of $N$ potential functions $\Phi_i$.
\end{proofof}

\subsubsection{Proof of \autoref{t:spectrum-full}}

We are ready to prove the main theorem in this subsection.
Define the net as
\[
\mathcal{N} = \Big\{-b + t\delta ~\Big|~ t\in \Big\{1,2,\ldots,2 \Big\lceil \frac{b}{\delta} \Big\rceil\Big\}\Big\}
\quad \textrm{where} \quad
b = \norm{A_0} + \max\big\{2\sigma, \sigma_*\sqrt{n}\big\} \textrm{~~and~~} \delta = \sigma_*.
\]
Let $\Phi_z(t,x)$ be the potential function in \eqref{e:potential2} with $p=\Theta(\log{d})$. 
Apply \autoref{proposition:mw} with the set of potential functions
\[
\big\{\Phi^+_z(t,x) := \Phi_z(t, x) - \Phi_z(0,0)
\textrm{~~and~~}
\Phi^-_z(t,x) := -\Phi_z(t, x) + \Phi_z(0,0) \mid z=\lambda + \eps i, \lambda \in \mathcal{N}\big\},
\]
where we fix $ \eps \asymp \sqrt{\log{|\mathcal{N}|}} \cdot \sigma_* + \sigma_* \cdot \sqrt{r\log{d}} + (\log{d})^{\frac34} \cdot \sqrt{\sigma\nu}$. 
By \autoref{proposition:res-interpolation}, 
these potential functions satisfy the properties in the hypotheses of \autoref{proposition:mw} with 
\[
B(r) \lesssim \frac{(\log^3d) \cdot \sigma^2 \cdot \nu^2}{\eps^5} + \frac{(\log{d}) \cdot (\sigma_* \cdot \sigma + \sigma_*^2 \cdot r)}{\eps^3},
\quad L(r) = \frac{\sigma_*^2}{\eps^4},
\quad Q(r) = \frac{\sigma_*\sqrt{n}+2\sigma}{\im(z)^2}.
\]
Let $x$ be the vector output by the algorithm in \autoref{proposition:mw}. 
Then, for all $z = \eps i  +\lambda$ with $\lambda \in \mathcal{N}$, 
\begin{eqnarray*}
|\Phi_z(1,x) - \Phi_z(0,0)| 
&=&\Big|\norm{(z1-A_0 -A(x))^{-1}}_{2p}-\norm{(z1-A_0-X_F)^{-1}}_{2p}\Big|
\\
&\lesssim& \frac{(\log^3d ) \cdot \sigma^2 \cdot \nu^2}{\eps^5} + \frac{(\log{d}) \cdot (\sigma_* \cdot \sigma + \sigma_*^2 \cdot r)}{\eps^3} + \frac{\sqrt{\log{|\mathcal{N}|}} \cdot \sigma_*}{\eps^2}.
\end{eqnarray*}    
Since $p=\Theta(\log{d})$, we have $\Phi_z(0,0) \asymp \bignorm{\big(z \cdot (I \otimes 1) - A_0 \otimes 1 - \XF\big)^{-1}}$ and $\Phi_z(1,x) \asymp \bignorm{\big(zI - A_0 - A(x)\big)^{-1}}$, where the first inequality follows from the ultracontractivity bound in \autoref{lemma:res-ultra}. 
Therefore, for all $z = \lambda + \eps i$ with $\lambda \in \mathcal{N}$,     
\[
\norm{(zI-A_0-A(x))^{-1}} \lesssim \norm{(z(I \otimes 1)-A_0 \otimes 1 - \XF)^{-1}} + \frac{\sigma^2\nu^2 \log^3d}{\eps^5}+ \frac{\log{d}(\sigma_*\sigma + \sigma_*^2r)}{\eps^3} + \frac{\sqrt{\log{|\mathcal{N}|}}\sigma_*}{\eps^2}
\]
\[
\norm{(z(I \otimes 1)-A_0 \otimes 1 - \XF)^{-1}} \lesssim \norm{(zI-A_0-A(x))^{-1}} + \frac{\sigma^2\nu^2\log^3d}{\eps^5}+ \frac{\log{d}(\sigma_*\sigma + \sigma_*^2r)}{\eps^3} + \frac{\sqrt{\log{|\mathcal{N}|}}\sigma_*}{\eps^2}.
\]
Since we are guaranteed that $\norm{x}^2\leq n$, 
the interval $[-b,b]$ contains $\spec(A_0 + \XF)$ and $\spec(A_0 + A(x))$. 
Applying \autoref{lemma:spec-cover} with $\delta = \sigma_*$ then guarantees that the above bounds hold for all $z\in \spec(A_0+A(x))\cup \spec(A_0 + \XF)$, while only losing an extra $\sigma_*/\eps^2$ term. 
Finally, applying \autoref{lemma:res-to-spec} with $K_3 = \sigma^2\nu^2 \log^3d$, $K_2 = \log{d}(\sigma_*\sigma + \sigma_*^2r)$ and $K_1 = \sqrt{\log |\mathcal{N}|}\sigma_*$ (and noting that $\sigma_*\leq \nu$) confirms the choice of $\eps$ and gives the spectrum guarantees in the statement in \autoref{t:spectrum-full}.  

Finally, note that our assumption $\frac{1}{\poly{d}}\leq\max_{i=1}^n\norm{A_i}\leq \poly(d), \norm{A_0}\leq \poly(d)$ imply that all matrix parameters are bounded above and below by a polynomial in $d$. 
It follows that $\log|\mathcal{N}| = \Theta(\log{d})$ and $B(r),L(r),Q(r) \leq \poly(d)$.
Assuming the potential function in \eqref{e:potential2} can be computed in $\poly(n,d)$ time,
which will be proved in \autoref{s:computation},
the total time complexity of the deterministic algorithm in \autoref{t:spectrum-full} is polynomial in $n$ and $d$.

\subsection{Interpolation Analysis for Resolvents} \label{section:inter-res}

We prove \autoref{proposition:res-interpolation} in this subsection. 
The proof structure is similar to that in \autoref{s:interpolation-moment}, 
where we do the interpolation analysis for moments.
The proof is a bit longer as the calculations are more involved.

\subsubsection{Interpolation}

We begin our analysis by proving the first and more difficult bound
\[
\big|\E_y[\Delta_z(t,Py)]\big| 
\lesssim \eta \cdot \Big( \frac{p^{3} \cdot \sigma^2 \cdot \nu^2}{\im(z)^{5}} + \frac{p \cdot (\sigma_* \cdot \sigma +\sigma_*^2 \cdot r)}{\im(z)^{3}}\Big).
\]
To shorten notation, we let
\[
M_z(X):=\big(z(I \otimes 1)-X\big)^{-1}
\]
be the resolvent of $X$ at the point $z$.

As in \autoref{s:interpolation-moment}, 
we define the intermediate mixed operator
\[
X_{t,\eta} = A_0 \otimes 1 + A(x_t) \otimes 1 + \sqrt{1-t-\eta} \cdot \XF.
\]
Using \autoref{fact:semi-conv} that
\[
\sqrt{1-t} \cdot \XF
\quad \textrm{and} \quad
\sqrt{1-t-\eta} \cdot \XF + \sqrt{\eta} \cdot \XF'
\]
have the same distribution where $\XF'$ is a freely independent copy of $\XF$,  
the expected change is
\begin{equation} \label{e:expected-change}
\E_y[\Delta_z(t,Py)] 
= \underbrace{\E\big[\tr\otimes\tau \big(|M_z(X_{t,\eta}+\sqrt{\eta} \cdot A(Py) \otimes 1|^{2p}\big)^{\frac{1}{2p}}\big]}_{(1)} 
-\underbrace{\tr\otimes\tau\big(|M_z(X_{t,\eta}+\sqrt{\eta} \cdot \XF')|^{2p}\big)^{\frac{1}{2p}}}_{(2)}.
\end{equation}

\subsubsection{Taylor Expansion} 

To bound the difference between terms $(1)$ and $(2)$, we use Taylor's Approximation \autoref{theorem:taylor-general} by computing the first and second order derivatives of the function $|M_z(X)|^{2p}$ as in~\cite{BBvH23}. 

To this end, we introduce the following notation to denote the derivatives of the moments of the resolvent:
\[
R_m(X,H_1,\ldots,H_m) = D^m (|M_z(X)|^{2p})(H_1,\ldots,H_m).
\]
Then, we write the Taylor expansion in \autoref{theorem:taylor-general} as
\begin{eqnarray}
\tr\otimes\tau\big(|M_z(X_{t,\eta}+H)|^{2p}\big)^{\frac{1}{2p}} 
&=& \tr\otimes\tau\big(|M_z(X_{t,\eta})|^{2p}\big)^{\frac{1}{2p}}
+\frac{1}{2p}\tr\otimes\tau\big(|M_z(X_{t,\eta})|^{2p}\big)^{\frac{1}{2p}-1} \cdot \tot(R_1(X_{t,\eta},H)) \nonumber
\\
&&+~\frac{1}{4p} \tr\otimes\tau\big(|M_z(X_{t,\eta})|^{2p}\big)^{\frac{1}{2p}-1} \cdot \tot ( R_2(X_{t,\eta},H,H) ) \nonumber
\\
&&-~\frac{1}{4p}\Big(1-\frac{1}{2p}\Big) \cdot \tr\otimes\tau\big(|M_z(X_{t,\eta})|^{2p}\big)^{\frac{1}{2p}-2} \cdot \tot\big(R_1(X_{t,\eta}, H)\big)^2 \nonumber
\\
&&\pm~O(\cR^{(3)}_{z}(H)) \label{e:Taylor2}
\end{eqnarray}
where the Taylor remainder term is
\begin{equation} \label{e:remainder}
\cR_z^{(m)}(H):= \sup_{\substack{r\in [0,1]}} \big|D^m\big(\tr\otimes\tau\big(|M_z(X_{t,\eta}+rH)|^{2p}\big)^{\frac{1}{2p}}\big)(H,\ldots, H)\big|.
\end{equation}

Now, we expand $\tot(R_1(X,H))$ and $\tot(R_2(X,H_1,H_2))$ for future calculations.
Since $|M_z(X)|^{2p} = M_z(X)^p \cdot M_{\bar{z}}(X)^{p}$, we can apply the product rule in \autoref{fact:prod-rule} and \autoref{lemma:res-higher-der} to compute
\begin{eqnarray}
\tot(R_1(X,H)) 
&=& p \cdot \tot\big(M_{z}(X)^{p+1} \cdot M_{\bar{z}}(X)^{p} \cdot H+M_z(X)^{p} \cdot M_{\bar{z}}(X)^{p+1} \cdot H\big) \nonumber
\\
&=& 2p \cdot \re \tot\big(M_z(X)^{p+1} \cdot M_{\bar{z}}(X)^p \cdot H\big), \label{e:R1}
\end{eqnarray}
where we used the fact that $M_{\bar{z}} = M_z^*$ commutes with $M_z$ and that $\tot$ is a trace in the first line, and the fact that $\tot(Y+Y^*) = 2\re\tot(Y)$ for all $Y$ in the second line. 
Taking the derivative of the above expression, we obtain
\begin{eqnarray} 
\tot(R_2(X,H_1,H_2)) &=& 2p \cdot \re \tot\bigg(\sum_{k=1}^{p+1} M_z(X)^{k} \cdot H_1 \cdot M_z(X)^{p-k+2} \cdot M_{\bar{z}}(X)^{p} \cdot H_2 \nonumber
\\
& & \hspace{2cm}+\sum_{k=1}^{p} M_z(X)^{p+1} \cdot M_{\bar{z}}(X)^{k} \cdot H_1 \cdot M_{\bar{z}}(X)^{p-k+1} \cdot H_2 \bigg).  \label{e:R2}
\end{eqnarray}

\subsubsection{Free Term}

We first compute term $(2)$ in \eqref{e:expected-change} by applying the Taylor expansion in the case where $H = \sqrt{\eta} \cdot \XF'$ in \eqref{e:Taylor2}. 
Since $\XF'$ and $\Xf$ are freely independent and centered, 
$\tot(R_1(X_{t,\eta}, \XF')) = 0$ and thus the first-order term in the Taylor expansion vanishes. 
Using \eqref{e:R2}, applying \autoref{lemma:free-moments}, and using $\varphi[\XF' \cdot (M \otimes 1) \cdot \XF'] = \sum_{i=1}^n A_iMA_i$ as in \autoref{s:interpolation-moment}, we have
\begin{align*}
\tot(R_2(X_{t,\eta},\XF',\XF')) 
= 2p \cdot \sum_{i=1}^n\re \tr\bigg(&\sum_{k=1}^{p+1} \varphi\big[M_z(X_{t,\eta})^{k}\big] \cdot A_i \cdot \varphi\big[M_z(X_{t,\eta})^{p-k+2} \cdot M_{\bar{z}}(X_{t,\eta})^{p}\big] \cdot A_i \\
    +&\sum_{k=1}^{p} \varphi\big[M_z(X_{t,\eta})^{p+1} \cdot M_{\bar{z}}(X_{t,\eta})^{k}\big] \cdot A_i \cdot \varphi\big[ M_{\bar{z}}(X_{t,\eta})^{p-k+1}\big] \cdot A_i\bigg)
\\
= 2p \cdot \re \E \bigg[ \tr\bigg(&\sum_{k=1}^{p+1} \varphi\big[M_z(X_{t,\eta})^{k}\big] \cdot A(y) \cdot \varphi\big[M_z(X_{t,\eta})^{p-k+2} \cdot M_{\bar{z}}(X_{t,\eta})^{p}\big] \cdot A(y)
\\
   +&\sum_{k=1}^{p} \varphi\big[M_z(X_{t,\eta})^{p+1} \cdot M_{\bar{z}}(X_{t,\eta})^{k}\big] \cdot A(y) \cdot \varphi\big[ M_{\bar{z}}(X_{t,\eta})^{p-k+1}\big] \cdot A(y)\bigg) \bigg]
\end{align*}
where the second equality can be seen by expanding out from the right hand side. 
Therefore,
\begin{align} \label{e:free2}
\tr\otimes&\tau\big( |M_z(X_{t,\eta}+\sqrt{\eta} \cdot \XF')|^{2p} \big)^{\frac{1}{2p}} 
= \tr\otimes\tau\big(|M_z(X_{t,\eta})|^{2p}\big)^{\frac{1}{2p}} \nonumber
\\
+&\frac{\eta}{2} \cdot \tr\otimes\tau\big(|M_z(X_{t,\eta})|^{2p}\big)^{\frac{1}{2p}-1} \cdot \re \E \bigg[\tr\Big(\sum_{k=1}^{p+1} \varphi[M_z(X_{t,\eta})^{k}] \cdot A(y) \cdot \varphi\big[M_z(X_{t,\eta})^{p-k+2} \cdot M_{\bar{z}}(X_{t,\eta})^{p}\big] \cdot A(y) \nonumber
\\
&\hspace{5.5cm}+\sum_{k=1}^{p} \varphi\big[M_z(X_{t,\eta})^{p+1} \cdot M_{\bar{z}}(X_{t,\eta})^{k} \big] \cdot A(y) \cdot \varphi\big[ M_{\bar{z}}(X_{t,\eta})^{p-k+1}\big] \cdot A(y)\Big) \bigg]   \nonumber
\\
\pm &O\big(\cR^{(3)}_z(\sqrt{\eta} \cdot \XF')\big).
\end{align}

\subsubsection{Finite-Dimensional Term}

Next, we bound the update from the finite-dimensional term $(1)$ in \eqref{e:expected-change} by applying Taylor expansion with $H = \sqrt{\eta} \cdot A(y) \otimes 1$ in \eqref{e:Taylor2}. 
As in \autoref{s:interpolation-moment}, we define an auxiliary function $\phi:\R^n\rightarrow\R$ given by
\[
\phi(y) = \tr\otimes \tau\big(|M_z(X_{t,\eta} + A(y) \otimes 1)|^{2p}\big),
\]
where the partial derivatives are
\[
\partial_i\phi(0) = \tot(R_1(X_{t,\eta},A_i \otimes 1))
\quad \textrm{and} \quad 
\partial_i\partial_j\phi(0) = \tot(R_2(X_{t,\eta}, A_i \otimes 1,A_j \otimes 1)).
\]
Note that $\E_y[\tot(R_1(X_{t,\eta}, A(Py) \otimes 1))] = 0$ as it is a linear function on $y$, so the first-order term in \eqref{e:Taylor2} vanishes.
Using the auxiliary function $\phi$ and \eqref{e:R1} and \eqref{e:R2}, observe that the factors in the second-order term in \eqref{e:Taylor2} can be written as 
\begin{eqnarray*}
\E_y\big[\tot\big(R_1(X_{t,\eta},A(Py) \otimes 1)\big)^2\big] & = & \norm{P \cdot \nabla\phi(0)}^2, \quad \textrm{and}\\
\E_y\big[\tot\big(R_2(X_{t,\eta},A(Py) \otimes 1,A(Py) \otimes 1)\big)\big] & = & \Tr\big(P \cdot \nabla^2\phi(0)\big).
\end{eqnarray*}
Therefore, the Taylor expansion can be expressed in terms of $\phi$ as
\begin{eqnarray*}
\E_y\big[\tr\otimes\tau \big(\big|M_z(X_{t,\eta}+\sqrt{\eta} \cdot A(Py) \otimes 1)\big|^{2p}\big)^{\frac{1}{2p}} \big]
&=& \tr\otimes\tau\big[|M_z(X_{t,\eta})|^{2p} \big]^{\frac{1}{2p}}
\\
&& +~\frac{\eta}{4p} \cdot \tr\otimes\tau\big[|M_z(X_{t,\eta})|^{2p}\big]^{\frac{1}{2p}-1} \cdot \Tr\big( P \cdot \nabla^2\phi(0)\big)
\\
&&-~\frac{\eta}{4p} \cdot \Big(1-\frac{1}{2p}\Big) \cdot \tr\otimes\tau\big[|M_z(X_{t,\eta})|^{2p}\big]^{\frac{1}{2p}-2} \cdot \norm{P \cdot\nabla\phi(0)}_2^2
\\
&& \pm~ O\big(\cR^{(3)}_{z}(\sqrt{\eta} \cdot A(Py) \otimes 1)\big).
\end{eqnarray*}
In order to achieve two-sided bounds on the spectrum, 
we need to bound the gradient norm and Hessian norm of $\phi$ using the following lemma.
\begin{lemma}[Gradient and Hessian Norm] \label{lemma:gradient-bound}
For all $t\in [0,1]$ and $z$ with $\im(z) >0$, the gradient norm of $\phi$ is bounded by\footnote{The reason that we need two different bounds on the norm of the gradient is that they can be combined to give a bound for the squared norm in terms of $\im(z)^{-3}$.
}
\[
\norm{\nabla\phi(0)} \leq \frac{2p \cdot \sigma_*}{\im(z)^2} \cdot \tr\otimes\tau\big[|M_z(X_{t,\eta})|^{2p}\big]^{1-\frac{1}{2p}}
\quad \textrm{and} \quad
\norm{\nabla\phi(0)} \leq \frac{2p \cdot \sigma}{\im(z)} \cdot \tr\otimes\tau\big(|M_z(X_{t,\eta})|^{2p}\big).
\]
For all $t\in [0,1]$ and $z$ with $\im(z) >0$, the Hessian norm of $\phi$ is bounded by
\[
\norm{\nabla^2\phi(0)} 
\leq \frac{p^2 \cdot \sigma_*^2}{\im(z)^3} \cdot \tot\big(|M_z(X_{t,\eta})|^{2p}\big)^{1-\frac{1}{2p}}.
\]
\end{lemma}
We first use \autoref{lemma:gradient-bound} to continue bounding the Taylor expansion, and then we will prove \autoref{lemma:gradient-bound}.
By combining the two gradient bounds in \autoref{lemma:gradient-bound}, and the fact that the 2-norm does not increase under projection, it follows that
\[
\Big|\frac{\eta}{4p} \cdot \Big(1-\frac{1}{2p}\Big) \cdot \tr\otimes\tau\big[|M_z(X_{t,\eta})|^{2p}\big]^{\frac{1}{2p}-2} \cdot \norm{P \cdot \nabla\phi(0)}_2^2 \Big| 
\lesssim \frac{\eta \cdot p \cdot \sigma_* \cdot \sigma}{\im(z)^3}.
\]
Next, we use the Hessian bound in \autoref{lemma:gradient-bound} to control $\Tr(P \cdot \nabla^2\phi(0))$.
Since $P$ is a projection onto a subspace of rank at least $n-r$, it follows that
$\big|\Tr\big(P \cdot \nabla^2\phi(0)\big) - \Tr\big(\nabla^2 \phi(0)\big) \big| \leq r\norm{\nabla^2\phi(0)}$.
So, by \autoref{lemma:gradient-bound},
\[
\Tr(P \cdot \nabla^2\phi(0))
\leq
\Tr( \nabla^2 \phi(0) )
+  \frac{r \cdot p^2 \cdot \sigma_*^2}{\im(z)^3} \cdot \tot\big(|M_z(X_{t,\eta})|^{2p}\big)^{1-\frac{1}{2p}}.
\]
By expanding $\Tr(\nabla^2\phi(0)) = \sum_{i=1}^n \partial_i^2 \phi(0) 
= \sum_{i=1}^n \tot(R_2(X_{t,\eta},A_i\otimes 1, A_i \otimes 1))$ and then expanding each summand using \eqref{e:R2} and putting these bounds in the Taylor expansion, we have 
\begin{align} \label{e:finite2}
\E_y\big[&\tr\otimes\tau \big(\big|M_z(X_{t,\eta}+\sqrt{\eta} \cdot A(Py) \otimes 1)\big|^{2p}\big)^{\frac{1}{2p}} \big]
= \tr\otimes\tau\big[|M_z(X_{t,\eta})|^{2p} \big]^{\frac{1}{2p}} \nonumber
\\
& +\frac{\eta}{2} \cdot \tot[|M_z(X_{t,\eta})|^{2p}]^{\frac{1}{2p}-1} 
\cdot \re \E_y\bigg[ \tot\bigg(\sum_{k=1}^{p+1} M_z(X_{t,\eta})^{k} \cdot A(y) \otimes 1 \cdot M_z(X_{t,\eta})^{p-k+2} \cdot M_{\bar{z}}(X_{t,\eta})^{p} \cdot A(y) \otimes 1
\nonumber \\
&\hspace{6.8cm}+\sum_{k=1}^{p} M_z(X_{t,\eta})^{p+1} \cdot M_{\bar{z}}(X_{t,\eta})^{k} \cdot A(y) \otimes 1 \cdot M_{\bar{z}}(X_{t,\eta})^{p-k+1} \cdot A(y) \otimes 1 \bigg) \bigg]
\nonumber \\
& + O\Big(\frac{\eta \cdot  p \cdot (\sigma_* \cdot \sigma + \sigma_*^2 \cdot r)}{\im(z)^3} + \cR^{(3)}_{z}\big(\sqrt{\eta} \cdot A(Py) \otimes 1\big)\Big).
\end{align}

This is the bound that we will use.  Now we prove \autoref{lemma:gradient-bound}.

\begin{proofof}{\autoref{lemma:gradient-bound}}
To prove the first inequality of the gradient, 
it suffices to show that the function $\phi(y)^{\frac{1}{2p}}$ is $L$-Lipschitz for $L = \frac{\sigma_*}{\im(z)^2}$. 
This can be seen in the proof of \cite[Lemma 6.5]{BBvH23}, but we provide a proof for completeness.  
Using the definition that $|M_z(X_{t,\eta})|^{2p}
= |z1-X_{t,\eta}|^{-2p}$,
\begin{eqnarray*}
\big|\phi(y)^{\frac{1}{2p}} - \phi(y')^{\frac{1}{2p}}\big| 
&=& \big|\norm{M_z(X_{t,\eta} + A(y) \otimes 1)}_{2p} - \norm{M_z(X_{t,\eta} + A(y') \otimes 1)}_{2p}\big|
\\
&\leq& \norm{M_z(X_{t,\eta} + A(y) \otimes 1)-M_z(X_{t,\eta} + A(y') \otimes 1)}_{2p}
\\
&\leq& \norm{M_z(X_{t,\eta} + A(y) \otimes 1)-M_z(X_{t,\eta} + A(y') \otimes 1)}
\\
&=& \norm{\big(z1-X_{t,\eta}-A(y) \otimes 1\big)^{-1} \cdot \big(A(y) \otimes 1-A(y') \otimes 1\big) \cdot \big(z1-X_{t,\eta}-A(y') \otimes 1\big)^{-1}}
\\
&\leq& \frac{1}{\im(z)^2} \cdot \norm{A(y) \otimes 1 - A(y') \otimes 1}
\\
&\leq& \frac{\sigma_* \cdot \norm{y-y'}}{\im(z)^2},
\end{eqnarray*}
where the fourth line uses the standard identity $(z1-X)^{-1} - (z1 - Y)^{-1} = (z1-X)^{-1}(X-Y)(z1-Y)^{-1}$. 
Now, by computing the gradient of the function $\phi^{\frac{1}{2p}}$, it follows that
\[
\norm{\nabla(\phi(0)^{\frac{1}{2p}})} 
= \frac{1}{2p} \cdot \phi(0)^{\frac{1}{2p}-1} \cdot \norm{\nabla\phi(0)} 
\leq \frac{\sigma_*}{\im(z)^2}.
\]
To obtain the second bound of the gradient, we use the fact that $\partial_{y_i}\phi(0) = \tot(R_1(X_{t,\eta},A_i \otimes 1))$. Then, 
\begin{eqnarray*}
\norm{\nabla\phi(0)}^2 &=&
4p^2 \cdot \sum_{i=1}^n \re\tot \big(M_{z}(X_{t,\eta})^{p+1} \cdot M_{\bar{z}}(X_{t,\eta})^p \cdot A_i \big)^2 
\\
&\leq& 4p^2 \cdot \tr\otimes\tau\big(|M_z(X_{t,\eta})|^{2p+1}\big)^{2} \cdot \sigma^2 
\\
&\leq& \frac{4p^2 \cdot \sigma^2}{\im(z)^2} \cdot \tr\otimes\tau\big(|M_z(X_{t,\eta})|^{2p}\big)^2,
\end{eqnarray*} 
where the second line follows from \autoref{lemma:matrix-square},
and the last line follows from $\norm{M_z(X)} \leq 1/\im(z)$. 
Taking square-root of both sides then gives the second bound.

To bound the operator norm of the Hessian, 
\begin{eqnarray*}
\norm{\nabla^2\phi(0)} 
&= & \sup_{\norm{x}=1} \Big|\sum_{i,j=1}^n\tot(R_2(X_{t,\eta}, A_i \otimes 1,A_j \otimes 1)) \cdot x(i) \cdot x(j)\Big|
\\
&=& \sup_{\norm{x}=1}\Big|\tot(R_2(X_{t,\eta}, A(x) \otimes 1,A(x) \otimes 1))\Big|\\
&\lesssim& \sup_{\norm{x}=1} \norm{A(x)}^2 \cdot p^2 \cdot \tot\big(|M_z(X_{t,\eta})|^{2p+2}\big)
\\
&\leq& \frac{p^2 \cdot \sigma_*^2}{\im(z)^3} \cdot \tot\big(|M_z(X_{t,\eta})|^{2p}\big)^{1-\frac{1}{2p}}, 
\end{eqnarray*}
where the third line follows from \autoref{corollary:general-holder}, and the last line follows from Jensen's inequality and the bound $\sup_{\norm{x}=1}\norm{A(x)}\leq \sigma_*$.
\end{proofof}

\subsubsection{Remainder Term}

We use the following claim to bound the Taylor-approximation error.

\begin{claim}[Taylor Remainder Term] \label{claim:res-remainder}
Suppose $H$ is self adjoint. Then, for each $m\in \{1,2,3\}$, 
\[\cR_z^{(m)}(H) \lesssim p^{m-1} \cdot \norm{H}^m \cdot \im(z)^{-m-1}.\]
\end{claim}

By using \autoref{claim:res-remainder} and the facts that $\norm{\XF} \leq 2\sigma$ and $\norm{A(Py)} \leq \sigma_*\sqrt{n}$, the Taylor remainder terms are
\begin{equation} \label{e:R3}
\cR^{(3)}_z\big(\sqrt{\eta} \cdot \XF' \big) + \cR^{(3)}_z\big(\sqrt{\eta} \cdot A(y)\otimes 1\big)
\lesssim \eta^{\frac32} \cdot p^2 \cdot (\sigma^3 + \sigma_*^3 \cdot n^{\frac32}) \cdot \im(z)^{-4} 
\leq \eta \cdot p^2 \cdot \sigma^2 \cdot \nu^2 \cdot \im(z)^{-5},
\end{equation}
where the last inequality uses our assumption that $\eta \leq \frac{\sigma^4 \cdot \nu^4}{\im(z)^2 \cdot \max\{\sigma,\sigma_* \cdot \sqrt{n}\}^6}$.

\begin{proofof}{\autoref{claim:res-remainder}}
We begin with a general formula for the higher order derivatives of the resolvent. 
By applying the derivative formula for powers of the resolvent in \autoref{lemma:res-higher-der}, 
\[
R_m(X,H,\ldots,H) = m!\sum_{k=0}^m\sum_{\substack{\ell_1+\cdots +\ell_{k+1} \\= p+k\\ \ell_1,\ldots, \ell_{k+1}\geq 1}} \sum_{\substack{r_1+\cdots+r_{m-k+1}\\=p+m-k\\r_1,\ldots, r_{m-k+1}\ge 1}} \Big(\prod_{j=1}^{k} M_{z}(X)^{\ell_j}H\Big)M_z(X)^{\ell_{k+1}}\Big(\prod_{j=1}^{m-k}M_{\bar{z}}(X)^{r_{j}}H\Big)M_{\bar{z}}(X)^{r_{m-k+1}}.
\]
While above expression is complicated, we just need to apply it for $m = 1,2,3$. 
In these cases, the number of terms are bounded by $O(p^m)$, and the power of $M_z$ is $2p+m$. 
Applying the generalized H\"older's inequality in \autoref{corollary:general-holder} to each summand, we obtain 
\[
|\tot(R_m(X,H,\ldots, H))| \leq \frac{(2p+m-1)!}{(2p-1)!} \cdot \norm{H}^m \cdot \tot(|M_z(X)|^{2p+m}).
\]
Note that $\norm{M_z(X)}\leq \im(z)^{-1}$ for any self-adjoint $X$. 
Using this and applying \autoref{lemma:derivatives},
\begin{eqnarray*}
\big|D\big(\tr\otimes\tau\big(|M_z(X)|^{2p}\big)^{\frac{1}{2p}}\big)(H)\big| 
&\leq&  \norm{H} \cdot \tot \big(|M_z(X)|^{2p}\big)^{\frac{1}{2p}-1} \cdot \tot\big(|M_z(X)|^{2p+1}\big)
\\
&\leq& \frac{1}{\im(z)^2} \cdot \norm{H} \cdot \tot \big(|M_z(X)|^{2p}\big)^{\frac{1}{2p}-1} \cdot \tot\big(|M_z(X)|^{2p-1}\big)
\\
&\leq& \frac{1}{\im(z)^2} \cdot \norm{H},
\end{eqnarray*}
where we used Jensen's inequality in the last line. By similar computations, we can also upper bound
\begin{eqnarray*}
\big|D^2\big(\tr\otimes\tau\big(|M_z(X)|^{2p}\big)^{\frac{1}{2p}}\big)(H,H)\big|
&\lesssim& p \cdot \norm{H}^2 \cdot \Big(\tot\big(|M_z(X)|^{2p}\big)^{\frac{1}{2p}-1} \cdot \tot\big(|M_z(X)|^{2p+2}\big)
\\ 
&&\hspace{1.5cm}+ \tot\big(|M_z(X)|^{2p}\big)^{\frac{1}{2p}-2} \cdot \tot\big(|M_z(X)|^{2p+1}\big)^2\Big)
\\
&&\lesssim \frac{p}{\im(z)^3} \cdot \norm{H}^2.
\\
\big|D^3\big(\tr\otimes\tau\big(|M_z(X)|^{2p}\big)^{\frac{1}{2p}}\big)(H,H,H)\big|
&\lesssim& 
p^2 \cdot \norm{H}^3 \cdot \Big(\tot\big(|M_z(X)|^{2p}\big)^{\frac{1}{2p}-1} \cdot \tot\big(|M_z(X)|^{2p+3}\big)
\\ 
&&\hspace{1.5cm}+ \tot\big(|M_z(X)|^{2p}\big)^{\frac{1}{2p}-2} \cdot \tot\big(|M_z(X)|^{2p+2}\big) \cdot \tot\big(|M_z(X)|^{2p+1}\big)
\\
&&\hspace{1.5cm}+ \tot\big(|M_z(X)|^{2p}\big)^{\frac{1}{2p}-3} \cdot \tot\big(|M_z(X)|^{2p+1}\big)^3\Big)
\\
&&\lesssim \frac{p^2}{\im(z)^4} \cdot \norm{H}^3. \qedhere
\end{eqnarray*}
\end{proofof}

\subsubsection{Putting Together and Bounding the Crossing Terms} 

Putting together the bounds on the free term in \eqref{e:free2}, the finite-dimensional term in \eqref{e:finite2}, and the Taylor remainder terms in \eqref{e:R3}, the expected potential change in \eqref{e:expected-change} is
\begin{eqnarray*}
& & \E_y[\Delta_z(t,Py)]
\\
& = & \frac{\eta}{2} \cdot \tot\big(|M_z(X_{t,\eta})|^{2p}\big)^{1-\frac{1}{2p}} \cdot 
\sum_{k=1}^{p+1}\re \bigg(\E_y\Big[ \tot\Big(M_z(X_{t,\eta})^{k} \cdot A(y)\otimes 1 \cdot M_z(X_{t,\eta})^{p-k+2} \cdot M_{\bar{z}}(X_{t,\eta})^{p} \cdot A(y)\otimes 1 \Big) \Big]
\\
& & \hspace{5.8cm}    -\E_y \Big[ \tr\Big(\varphi\big[M_z(X_{t,\eta})^{k}\big] \cdot A(y) \cdot \varphi\big[M_z(X_{t,\eta})^{p-k+2} \cdot M_{\bar{z}}(X_{t,\eta})^{p}\big] \cdot A(y) \Big) \Big] \bigg)
\\
& &+~\frac{\eta}{2} \tot\big(|M_z(X_{t,\eta})|^{2p}\big)^{1-\frac{1}{2p}} \cdot
\sum_{k=1}^{p}\re\bigg( \E_y \Big[ \tot\Big(M_z(X_{t,\eta})^{p+1} \cdot M_{\bar{z}}(X_{t,\eta})^{k} \cdot A(y) \otimes 1\cdot M_{\bar{z}}(X_{t,\eta})^{p-k+1} \cdot A(y) \otimes 1\Big) \Big]
\\
& & \hspace{6cm}   -\E_y \Big[ \tr\Big( \varphi\big[M_z(X_{t,\eta})^{p+1} \cdot M_{\bar{z}}(X_{t,\eta})^{k}\big] \cdot A(y) \cdot \varphi\big[ M_{\bar{z}}(X_{t,\eta})^{p-k+1}\big] \cdot A(y)\Big) \Big] \bigg)
\\
& & \pm~O\Big(\frac{\eta \cdot p \cdot (\sigma_* \cdot \sigma + \sigma_*^2 \cdot r)}{\im(z)^3} + \frac{\eta \cdot p^2 \cdot \sigma^2 \cdot \nu^2}{\im(z)^{5}}\Big).
\end{eqnarray*}
Now, we apply intrinsic freeness of resolvent in \autoref{proposition:res-crossing-bound2} with $Y = A(y)$ to each of the $2p+1$ crossing terms. 
For example, for each $k$, we apply the proposition to obtain that
\begin{eqnarray*}
& & \Big|\E_y \Big[ \tot\Big(M_z(X_{t,\eta})^{k} \cdot A(y)\otimes 1 \cdot M_z(X_{t,\eta})^{p-k+2} \cdot M_{\bar{z}}(X_{t,\eta})^{p} \cdot A(y)\otimes 1 \Big) \Big]
\\
& & \hspace{2cm} -~\E_y \Big[ \tr\Big(\varphi\big[M_z(X_{t,\eta})^{k}\big] \cdot A(y) \cdot \varphi\big[M_z(X_{t,\eta})^{p-k+2} \cdot M_{\bar{z}}(X_{t,\eta})^{p}\big] \cdot A(y) \Big) \Big] \Big|
\\
& \lesssim & (1-t-\eta) \cdot p^2 \cdot \sigma^2 \cdot \nu^2 \cdot \tot(|M_z(X_{t,\eta})|^{2p+4})
\\
& \leq & \frac{p^2 \cdot \sigma^2 \cdot \nu^2}{\im(z)^{5}} \cdot \tot(|M_z(X_{t,\eta})|^{2p-1})
\\
& \leq & \frac{p^2 \cdot \sigma^2 \cdot \nu^2}{\im(z)^{5}} \cdot \tot(|M_z(X_{t,\eta})|^{2p})^{1-\frac{1}{2p}},
\end{eqnarray*}
where we used $\sigma(Y) = \sigma$, $\nu(Y) = \nu$, $\sigma(X_{t,\eta }) = \sqrt{1-t-\eta} \cdot \sigma $, $\nu(X_{t,\eta}) = \sqrt{1-t-\eta} \cdot \nu$, $\norm{M_z(X)} \leq 1/\im(z)$, and Jensen's inequality. 
Applying these bounds, we arrive at the first conclusion in \autoref{proposition:res-interpolation} that
\[
\big|\E_y[\Delta_z(t,Py)]\big| \lesssim \frac{\eta \cdot p \cdot (\sigma_* \cdot \sigma + \sigma_*^2 \cdot r)}{\im(z)^3} + \frac{\eta \cdot p^3 \cdot \sigma^2 \cdot \nu^2}{\im(z)^{5}}.
\]

\subsubsection{Proof of \autoref{proposition:res-interpolation}}

It remains to verify the second and the third conclusions. 
We first bound the quantity $\E[\Delta_z(t,Py)^2]$. 
To do so, we simply use the first-order Taylor approximation of the potential such that
\begin{eqnarray*}
\Phi_z(t+\eta,x_t+\sqrt{\eta}Py) 
&=&
\tr\otimes\tau \big(|M_z(X_{t,\eta})|^{2p} \big)^{\frac{1}{2p}} 
\\
&&
+ \frac{\sqrt{\eta}}{2p} \cdot \tr\otimes\tau \big( |M_z(X_{t,\eta})|^{2p} \big)^{\frac{1}{2p}-1} \cdot \tot\big(R_1(X_{t,\eta},A(Py) \otimes 1)\big) 
\pm O\big(\cR_z^{(2)}(\sqrt{\eta} \cdot A(Py) \otimes 1)\big),
\\
\Phi_z(t,x_t)
&=& 
\tr\otimes\tau\big(|M_z(X_{t,\eta})|^{2p}\big)^{\frac{1}{2p}} \pm O\big( \cR_z^{(2)}(\sqrt{\eta} \cdot \XF')\big),
\end{eqnarray*}
where we used the fact that $\tot(R_1(X_{t,\eta},\XF'))$ vanishes in the second expression.  
By \autoref{claim:res-remainder}, $\cR_z^{(2)}(H) \leq 2p \cdot \norm{H}^2 \cdot \im(z)^{-3}$. 
For $H = \sqrt{\eta} \cdot A(Py)$ or $H= \sqrt{\eta} \cdot \XF'$, 
we have $\norm{H}\lesssim \sqrt{\eta} \cdot \rho$, where $\rho : = \max\{\sigma,\sigma_*\sqrt{n}\}$. 
Thus,
\begin{eqnarray*}
\E_y[\Delta_z(t,Py)^2] 
&=& \E_y\big[(\Phi_z\big(t+\eta,x_t+\sqrt{\eta}Py) -\Phi_z(t,x_t)\big)^2\big]
\\
&\leq& \E_y\Big[\Big(\frac{\sqrt{\eta}}{2p} \cdot \tr\otimes\tau \big( |M_z(X_{t,\eta})|^{2p} \big)^{\frac{1}{2p}-1} \cdot \tot\big(R_1(X_{t,\eta},A(Py) \otimes 1)\big) \pm 4\eta \cdot p \cdot \rho^2 \cdot \im(z)^{-3} \Big)^2\Big]
\\
&\leq& \frac{\eta}{2p^2} \cdot \tr\otimes\tau \big( |M_z(X_{t,\eta})|^{2p} \big)^{\frac{1}{p}-2} \cdot \E_y\big[\tot\big(R_1(X_{t,\eta},A(Py) \otimes 1\big)^2\big] + 32\eta^2 \cdot p^2 \cdot \rho^4 \cdot \im(z)^{-6}
\\
&\leq& \frac{\eta}{2p^2} \cdot \tr\otimes\tau\big(|M_z(X_{t,\eta})|^{2p}\big)^{\frac{1}{p}-2} \cdot \norm{\nabla\phi(0)}^2 + 32\eta^2 \cdot p^2 \cdot \rho^4 \cdot \im(z)^{-6}
\\
&\leq& 2\eta \cdot \sigma_*^2 \cdot \im(z)^{-4} + 32\eta^2 \cdot p^2 \cdot \rho^4 \cdot \im(z)^{-6}
\\
&\leq& 3\eta \cdot \sigma_*^2 \cdot \im(z)^{-4},
\end{eqnarray*}
where we used $(x+y)^2\leq 2x^2 + 2y^2$ in the third line, 
$\E_y\tot(R_1(X_{t,\eta},A(Py))^2) = \E_y\inner{\nabla \phi(0)}{P(y)}^2 \leq \norm{\nabla \phi(0)}^2$ in the fourth line,
the first bound in \autoref{lemma:gradient-bound} in the fifth line, 
and in the last line the assumption that $\eta \leq \frac{\im(z)^2\sigma_*^2}{32p^2\rho^4}$.  
    
Finally, to give an absolute bound on $\Delta_z$, we use a Lipschitz argument similar as in the proof of the first bound in \autoref{lemma:gradient-bound} such that
\begin{eqnarray*}
|\Delta_z(t,Py)| 
&= & \big|\norm{M_z(X_{t,\eta} + \sqrt{\eta} \cdot A(Py) \otimes 1)}_{2p} - \norm{M_z(X_{t,\eta}+\sqrt{\eta} \cdot \XF')}_{2p} \big|
\\
&\leq & \norm{M_z(X_{t,\eta} + \sqrt{\eta} \cdot A(Py) \otimes 1)-M_z(X_{t,\eta}+\sqrt{\eta} \cdot \XF')}
\\
& =& \norm{M_z(X_{t,\eta} + \sqrt{\eta} \cdot A(Py) \otimes 1) \cdot (\sqrt{\eta} \cdot \XF' - \sqrt{\eta} \cdot A(Py) \otimes 1) \cdot (M_z(X_{t,\eta}+\sqrt{\eta} \cdot \XF'))}
\\
&\leq& \sqrt{\eta} \cdot \im(z)^{-2} \cdot \big(\norm{\XF'} + \norm{A(Py)}\big)
\\
&\leq& \frac{\sqrt{\eta} \cdot (2\sigma+ \sigma_* \cdot \sqrt{n})}{\im(z)^2}.
\end{eqnarray*}
This completes the proof of \autoref{proposition:res-interpolation}.

\subsection{Application: Deterministic Planted Recovery Models} \label{s:application-spectrum}

In \cite{BCS+24}, the 2-sided spectrum concentration result was used to analyze spectral algorithms for the spiked Wigner model,
which is a fundamental framework for studying low rank signal detection in high dimensional matrices. 
In this model, we are given a unit vector $v$, and must recover it from the noisy observation $X_\theta = \theta vv^\top + G$, where $G$ is a Gaussian random matrix. 
This model captures many well-studied planted recovery problems, including the stochastic block model, planted clique, and Tensor PCA. 
A natural recovery algorithm for the spiked Wigner model is the spectral algorithm, which estimates $v$ from the top eigenvector of $X_\theta$. 
It is well known~\cite{BBP05} that the performance of the spectral algorithm depends on the sharp phase transition behavior of $\lambda_{\max}(X_\theta)$. 
In particular, with high probability,
\begin{align*}
    \lambda_{\max}(X_\theta)\approx B(\theta) : =\begin{cases}
        2 & \theta\leq 1\\
        \theta  + \frac{1}{\theta} &\theta\geq 1.
    \end{cases}
\end{align*}
The critical threshold is at $\theta=1$, 
below which the maximum eigenvalue of $X_\theta$ is indistinguishable from that of $G$. 
In \cite{BCS+24}, Bandeira et al. showed that the threshold for the spectral algorithm can be elegantly derived by comparing the eigenvalue of $X_\theta$ with the corresponding free model and using Lehner's \autoref{lemma:lehner} to explicitly compute the maximum eigenvalue in the free model. 
Their method extends to a broad range of models where the noise matrices can have dependent entries.

\begin{theorem}[{\cite[Theorems 2.7,\;2.9,\;3.1]{BCS+24}}] \label{theorem:bbp}
Let $P$ be an $d\times d$ projection matrix of rank $r$. Let $G$ be a symmetric Gaussian random matrix satisfying $\E[G] = 0$ and $\E[G^2] = I_d$. 
Let $X_\theta = \theta \cdot P + G$. 
Let $\eps (t) \asymp \nu(G)^{\frac12}\log^{\frac34}d +  \sigma_*(G) \cdot t^{1/2} + \sigma_*(G) \cdot \sqrt{r}$. 
If $\eps(t)\leq \theta^2$, then with probability at least $1-e^{-\Omega(t)}$, 
\[
|\lambda_{\max}(X_\theta) - B(\theta)| \leq \eps(t)
\quad\textrm{and}\quad 
\Big|\bigip{v_{\max}}{Pv_{\max}} - \Big(1-\frac{1}{\theta^2}\Big)_+\Big| \leq \sqrt{\eps(t)},
\]
where $v_{\max}$ denotes the top eigenvector of $X_{\theta}$.
\end{theorem}
In the case when $G$ is a $d \times d$ Wigner matrix (with i.i.d entries), 
then $\eps(t) = \Tilde{O}(d^{-1/4} + \sqrt{r}d^{-1/2})$, 
which gives a tight control on the eigenvalue- and eigenvector-correlation of $X_\theta$. 
A consequence of results like \autoref{theorem:bbp} is that if we have a matrix $P$ with some combinatorial structure, then $\theta\leq 1$ is the regime where adding the noise $G$ will obscure this structure from detection via the spectral method. 
For example, we may choose $P = \frac{1}{|S|}\chi_S\chi_S^\top $ for some $S\subseteq [d]$, and pick $\theta = \frac{|S|}{2\sqrt{d}}$ which gives a critical threshold of $2\sqrt{d}$. 
This is a close variant of the planted clique problem, with noise arising from $\mathcal{G}(d,1/2)$ being replaced by Gaussian noise of the same variance. 
It is conjectured that recovering planted cliques of size $o(\sqrt{d})$ is computationally hard. 
Such average case complexity assumptions are widely used in public key cryptography, where random graphs/matrices are used to hide a given signal, from polynomial time detection. 

Given this background, we ask the question whether we can construct \textit{pseudorandom} matrices that can hide given signals from detection by the spectral algorithm, which in many randomized settings, is conjectured to be optimal. 
Our derandomization results can produce a deterministic variant of \autoref{theorem:bbp} to construct matrices whose spectrum are close to those of random Gaussian matrices, so that they are indistinguishable by the spectral algorithm. 

\begin{theorem}[Deterministic Planted Models] \label{theorem:det-bbp}
Let $\theta_1,\ldots,\theta_k >0$ and $P_1, \ldots,P_k$ be $d\times d$ projection matrices of rank at most $r$. 
Let $A_1,\ldots, A_n$ be real symmetric matrices satisfying $\sum_iA_i^2 = I_d$, 
with parameters $\nu^2 = \norm{\sum_i\vecc(A_i)\vecc(A_i)^*}$ and $\sigma_*^2 = \sup_{\norm{y}=1}\sum_i\inner{y}{A_iy}^2$. 
Let $\eps \asymp \nu^{\frac12} \cdot \log^{\frac34}d+ (\sqrt{r} + \sqrt{\log{(kd)}}) \cdot \sigma_*$. 
If $\eps \leq \min_i \theta_i^2$,
then there is a polynomial time deterministic algorithm that computes a vector $x\in \R^{n}$ such that for all $j \in [k]$,
\[
|\lambda_{\max}(\theta_j P_j + A(x)) - B(\theta_j)|\leq \eps
\quad \textrm{and} \quad 
\Big|\bigip{v^{(j)}_{\max}}{P_j v^{(j)}_{\max}} - \Big(1-\frac{1}{\theta_j^2}\Big)_+\Big| 
\leq O(\sqrt{\eps}),\\
\]
where $v^{(j)}_{\max}$ is any vector in the top eigenspace of $\theta_jP_j+A(x)$. 
\end{theorem}
We can view the matrix $A(x)$ as a pseudorandom object that behaves like a typical Gaussian noise matrix $\sum_i g_i A_i$, when added to any $\theta_jP_j$. 
In particular, the two-sided spectral concentration controls both the eigenvalue of $\theta_jP_j + A(x)$ and how much correlation its maximum eigenvector has with $P_j$. 
As an example, consider the setting where we select subsets $S_1,S_2, \ldots, S_k\in \binom{[d]}{m}$ and set each $\theta_j = \frac{m}{2\sqrt{d}}$ and $P_j = \frac{1}{m}\chi_{S_j}\chi_{S_j}^\top$. 
If $m\leq 2\sqrt{d}$, then the maximum eigenvalue of the matrix $\frac{1}{2\sqrt{d}}\chi_{S_j}\chi_{S_j}^\top +A(x)$ is $2$ (up to $\eps$ error) for every $j$, while its corresponding eigenvector(s) have at most $\sqrt{\eps}$ squared correlation with $S_j$. 
Thus, $A(x)$ obscures all the cliques induced by the index sets $S_1,\ldots, S_k$. On the other hand, if $m > 2\sqrt{d}$, then for all $j$, the top eigenvector of $\frac{1}{2\sqrt{d}}\chi_{S_j}\chi_{S_j}^\top +A(x)$ has about $1-\frac{4d}{m^2} - \sqrt{\eps}$ squared correlation with $S_j$. 

We note that the matrices $A_1,\ldots, A_n$ can have arbitrary structure as long as $\sum_i A_i^2 = I$, which is known as the ``isotropic'' condition. 
This is analogous to how \autoref{theorem:bbp} can handle an error matrix $G$ with arbitrary covariance as long as $\E[G^2] = I$. 
In the unstructured setting, the matrices $A_1,\ldots, A_n$ would be taken as $\frac{1}{\sqrt{d+1}}(E_{i,j} + E_{j,i}): (i < j)$ and $\sqrt{\frac{2}{d+1}} E_{i,i}$, which would yield $\nu^{1/2} = \Theta(d^{-\frac14})$. 
A basic structured setting that is still isotropic would be when $A_1,\ldots, A_n$ are the adjacency matrices of edge-disjoint matchings whose union is a regular graph. 

\subsubsection{Proof}

The proof of \autoref{theorem:det-bbp} is a consequence of our multiplicative weights framework combined with \autoref{proposition:res-interpolation}. 
We state a more general version of our main result (\autoref{t:spectrum-full}) of this section.

\begin{theorem}[Deterministic Full Spectrum for Multiple Matrices] {\label{t:spectrum-full-extended}}
Suppose we are in the setting of \autoref{t:spectrum-full} with $r=1$, but instead of having one $A_0$ matrix, we have $A_0^{(1)},\ldots, A_0^{(k)}$. 
Then there is a polynomial time deterministic algorithm to compute a vector $x$ such that for all $j\in [k]$,
\[
\spec(A_0^{(j)} + A(x)) \subseteq \spec(A_0^{(j)} \otimes 1 + \XF) + [-\eps, \eps]
\quad \textrm{and} \quad
\spec(A_0^{(j)} \otimes 1 + \XF) \subseteq \spec(A_0^{(j)} + A(x)) + [-\eps, \eps],
\]
where $\eps \asymp \log^{\frac 34} d \cdot \sqrt{\sigma \cdot \nu} + \sqrt{\log{(kd)}} \cdot \sigma_*$.
\end{theorem}
\begin{proof}
We apply the same proof as in \autoref{t:spectrum-full}, but with one set of potential functions to control the spectrum of $A_0^{(j)} + A(x)$ for each $j\in [k]$. 
In particular, let
\[
\Phi^{(j)}_z(t,x) = \tot\big(\big|M_z(A_0^{(j)} \otimes 1 + \sqrt{1-t}\XF + A(x) \otimes 1)\big|^{2p}\big)^{\frac{1}{2p}}.
\]
By construction, for every $j \in [k]$, $\Phi^{(j)}_z$ satisfies the potential update bounds in \autoref{proposition:res-interpolation}. 
For each $j\in [k]$, we construct a net $\mathcal{N}^{(j)} = \{-b^{(j)}+t\sigma_* \mid t\in \{1,2,\ldots,\lceil 2b^{(j)}/\sigma_*\rceil\}\}$, where $b^{(j)} = \bignorm{A_0^{(j)}} + \max(2\sigma, \sigma_*\sqrt{n})$. 
We then apply \autoref{proposition:mw} with the set of potentials $\Phi_z^{(j)}(t,x) - \Phi_z^{(j)}(0,0)$ and $-\Phi_z^{(j)}(t,x) + \Phi_z^{(j)}(0,0)$ for each $z = \lambda + \eps i$ with $\lambda \in \mathcal{N}^{(j)}$ and $j\in [k]$. 
Applying the multiplicative weight update algorithm in \autoref{proposition:mw} with the potential update bound in \autoref{proposition:res-interpolation} then gives us a vector $x$ satisfying $|\Phi^{(j)}_z(1,x) - \Phi^{(j)}_z(0,0)|\leq \eps$ for all $j$ and all $z$ with real part in our net. 
Finally, applying \autoref{lemma:spec-cover} and \autoref{lemma:res-to-spec} completes the proof.
\end{proof}

Now \autoref{theorem:det-bbp} follows from two results in \cite{BCS+24}. 
The first result characterizes the maximum eigenvalue of the free model in the isotropic setting using Lehner's formula in \autoref{lemma:lehner}. 
The second result relates the eigenvalue correlation with the derivative of the function $\lambda_{\max}(\theta P + X)$ with respect to $\theta$.

\begin{theorem}[{\cite[Theorem 2.7]{BCS+24}}]\label{theorem:leh-formula}
Let $\Xf = A_0 \otimes 1 + \sum_{i=1}^nA_i\otimes s_i$ for self-adjoint matrices $A_0, A_1, \ldots, A_n$. 
Suppose $\sum_{i=1}^n A_i^2 = I$ and $A_0$ has rank at most $r$. Then
\[
|\lambda_{\max}(\Xf) - B(\lambda_{\max}(A_0))| \leq 2\sigma_* \cdot \sqrt{r}.
\]
\end{theorem}

\begin{lemma}[{\cite[Lemmas 6.4, 6.5]{BCS+24}}] \label{lemma:eig-cor}
Let $X$ be any (deterministic) matrix and $P$ be a projection matrix. Suppose, for some $t > 0$, 
\[
|\lambda_{\max}(X+(\theta+t)P)-B(\theta + t)| \leq \eps\quad\text{and}\quad  |\lambda_{\max}(X+(\theta-t)P)-B(\theta -t)|\leq \eps.
\]
Then, for any top eigenvector $v_{\max}$ of $X+\theta P$, 
\[
\Big|\bigip{v_{\max}}{Pv_{\max}} - \Big(1-\frac{1}{\theta^2}\Big)_+\Big| 
\leq \frac{\eps}{t} + t
\]
\end{lemma}

\begin{proofof}{\autoref{theorem:det-bbp}}
Let $\eps_0 \asymp \log^{3/4}{d} \cdot \nu^{1/2}+\sigma_*\sqrt{\log{(kd)}}$ and $\eps = \eps_0 + \sigma_*\sqrt{r}$. 
Apply \autoref{t:spectrum-full-extended} with the $A_0$ matrices $\{\theta_j P_j$, $(\theta_j+\sqrt{\eps})P_j,(\theta_j-\sqrt{\eps})P_j\}_{j=1}^k$ with error parameter $\eps_0$. 
Note that $\sigma = 1$ in this case as $\sum_{i=1}^nA_i^2= I$. 
Let $x$ be the output vector. 
Then \autoref{theorem:leh-formula} implies that $|\lambda_{\max}(\theta_jP_j+A(x)) - B(\theta_j)| \leq \eps$ for all $j$. 
Since this bound also holds for $\theta_j \pm \sqrt{\eps}$, 
\autoref{lemma:eig-cor} implies that $|\inner{v_{\max}^{(j)}}{P_jv_{\max}^{(j)}} - (1-\frac{1}{\theta_j^2})_+|\leq O(\sqrt{\eps})$.
\end{proofof}

\section{Computation on Free Semicircular Matrices} \label{s:computation}

We derive several formulas for computation on the space $\M_d(\C)\otimes \A$ in this section. 

In \autoref{section:computation}, we provide formulas for computing the moments of semicircular matrices and their resolvents. 
This implies polynomial time algorithms for computing all potential functions used in the paper. 

In \autoref{section:ibp}, we prove a non-commutative ``integration by parts'' formula for polynomials on $\M_d(\C)\otimes \A$. 
This will serve as a crucial tool for bounding the higher-order terms in our derandomization of the universality results of~\cite{BvH24} in the next section. 
We also show that the computational formulas in \autoref{section:computation} can be derived conveniently from this integration by parts formula.

In \autoref{section:ibp-proof}, we present the proof of this integration by parts formula.
Finally, in \autoref{section:rational-ibp}, we extend it to the setting of rational functions, which will be required in the resolvent analysis in the proof of \autoref{t:resolvent-uni}.

\subsection{Computation of Moments and Resolvents}\label{section:computation}

In this subsection, we provide formulas for computing moments and resolvents of random variables on $\M_d(\C)\otimes \A$. 

\begin{lemma}[Moments Computation] \label{lemma:moment-formula}
Let $\Xf = A_0\otimes 1 + \sum_{i=1}^n A_i\otimes s_i$ be a general semicircular matrix. 
For any integer $p\geq 2$, 
\[
\varphi[\Xf^p] 
= \varphi[\Xf^{p-1}] \cdot A_0+\sum_{i=1}^n\sum_{k=0}^{p-2}\varphi[\Xf^{k}] \cdot A_i \cdot \varphi[\Xf^{p-2-k}] \cdot A_i.
\]
Let $B\in \M_d(\C)$ be any matrix. 
For integers $p,q \geq 1$,
\begin{align*}
\varphi[\Xf^p (B \otimes 1) \Xf^q ] 
=  \varphi[\Xf^{p} (B \otimes 1) \Xf^{q-1}] \cdot A_0 
&+ \sum_{i=1}^n\sum_{k=0}^{p-1}\varphi[\Xf^{k}] \cdot A_i \cdot \varphi[\Xf^{p-1-k}(B \otimes 1) \Xf^{q-1}] \cdot A_i \\
&+\sum_{i=1}^n\sum_{k=0}^{q-2}\varphi[\Xf^{p}(B \otimes 1) \Xf^{k}] \cdot A_i \cdot \varphi[\Xf^{q-2-k}] \cdot A_i \cdot \1_{\{q\geq 2\}}.
\end{align*} 
\end{lemma}

These formulas can be derived from \autoref{lemma:sc-joint-moments}. 
For example, the first formula was used in some proofs in Van Handel's survey (see \cite[Lemma 4.4]{vH25}) but was not stated explicitly. 
In \autoref{section:ibp}, we will show that these formulas can be derived in a simple and systematic manner through the semicircular integration-by-parts formula.

Observe that the formulas lead to a polynomial time algorithm for computing moments.

\begin{lemma}[Efficient Computation for Moments] \label{lemma:computation}
Let $B_1,B_2$ be arbitrary $d\times d$ matrices. For any $p,q\in \N$, there is a polynomial time algorithm to compute the quantity $\tot((B_1 \otimes 1) \Xf^p (B_2 \otimes 1) \Xf^q)$.
\end{lemma}
\begin{proof}
Note that $\tot((B_1 \otimes 1) \Xf^p (B_2 \otimes 1) \Xf^q) 
= \tr(B_1 \cdot \varphi[(\Xf)^p (B_2 \otimes 1) (\Xf)^q])$, 
thus it suffices to evaluate $\varphi[\Xf^p (B_2 \otimes 1) \Xf^q]$. 
The formulas in \autoref{lemma:moment-formula} provide a natural recurrence for a dynamic programming algorithm to compute this quantity. 
Specifically, there are two types of subproblems: 
one of the form $\varphi[\Xf^k]$, 
and one of the form $\varphi[\Xf^{k} (B_2 \otimes 1) \Xf^\ell]$. 
Denote the size of each subproblem as $k+\ell$. 
Then we see from the recursive formulas that each subproblem only relies on solutions to subproblems of a smaller size. 
The base case for the first kind of subproblem is given by $\varphi[\Xf^0] = I$ and $\varphi[\Xf^1] = A_0$. 
The second type of subproblems eventually reduces to the first type when $k$ or $\ell$ reaches $0$. 
The total number of subproblems is at most $O(pq)$. 
This leads to a polynomial time dynamic programming algorithm.
\end{proof}

From \autoref{lemma:computation}, we can compute all moments of $\Xf$ in polynomial time by simply taking $B_1,B_2=I$. 
The more general formula with arbitrary $B_1,B_2$ is used for computation with the Hessian matrix of the potential function in \autoref{section:discrepancy}.

\subsubsection{From Moments to Resolvents}

Now, we show how the moments of $\Xf$ can be used to compute the moments of its resolvents. 
This is done via a re-centering trick that allows us to approximate the resolvent moment by a convergent power series. 

Let $X \in \M_d(\C) \otimes \A$ be a self-adjoint semicircular matrix.
For $z = \lambda + \eps i \not\in \spec(X)$, we would like to compute
\[
\tot\big(|z 1 - X|^{-2p}\big) = \tot\big(\big(\eps^2 1 + (\lambda 1 - X)^2\big)^{-p}\big),
\]
where we use the shorthand $1$ for $I \otimes 1$.
Note that $Y = (\lambda 1 - X)^2$ is an element in $\M_d(\C) \otimes \A$ with nonnegative spectrum and $\|Y\| \leq 2(\lambda^2 + \|X\|^2)$.
Therefore, it suffices to compute the moments of the resolvent of the form
\[
\tot\big(\big(\eps^2 1 + Y\big)^{-p}\big),
\]
where $Y$ has nonnegative spectrum and bounded norm.
The following lemma shows how to compute this using a convergent power series.

\begin{lemma}[Power Series for Resolvent] \label{lemma:res-computation}
Let $Y$ be an element in $\M_d(\C) \otimes \A$ with nonnegative spectrum with $\norm{Y} \leq u$. 
For any $\eps,\delta > 0$ and $p\in \N$, 
there is a deterministic algorithm to compute $\tot ((\eps^2 1 + Y)^{-p})$ up to an additive error $\delta$ in time $O(kT(k))$, where $k \in \poly(u,1/\eps^2,p,\log{(1/\delta)})$ and $T(k)$ is the time required to compute the $k$-th moment of $Y$.
\end{lemma}
\begin{proof}
We re-center the power series at the point $u1-Y$ such that
\begin{eqnarray*}
\tot((\eps^2 1 + Y)^{-p}) &=& \tot((\eps^2 1 + u1 - u1 + Y)^{-p})
\\
&=& (\eps^2+u)^{-p} \cdot \tot\Big( \Big(1-\frac{1}{\eps^2+u}(u 1-Y) \Big)^{-p} \Big).
\end{eqnarray*}
Since $Y$ has nonnegative spectrum and $u \geq \|Y\|$, 
it holds that $\norm{u1-Y} \leq u$
and thus $\frac{\norm{u1-Y}}{\eps^2+u} < 1$.
This implies that the $p$-th moment of the resolvent can be expressed as a converging power series using the Taylor expansion for the function $(1-x)^{-p}$ such that
\[
\tot((\eps^21 + Y)^{-p}) 
= \frac{1}{(\eps^2+u)^p} \cdot \sum_{k=0}^\infty {\binom{p+k-1}{k}}\frac{\tot((u1-Y)^k)}{(\eps^2+u)^k}.
\]
Since $\norm{u1-Y} \leq u$, the $k$-th term in the series is bounded by 
\[
\frac{1}{(\eps^2 + u)^p}{\binom{p+k-1}{k}}\frac{u^k}{(\eps^2 + u)^k} 
= \frac{1}{(\eps^2 + u)^p} {\binom{p+k-1}{p-1}}\Big(1-\frac{\eps^2}{\eps^2 + u} \Big)^k 
\leq \Big(\frac{k+p}{\eps^2 + u}\Big)^{p} \cdot \exp\Big(-\frac{k\eps^2}{\eps^2 + u}\Big),
\]
where we used $1 - y \leq e^{-y}$ in the last inequality.
Thus, the power series is dominated by a geometrically decreasing sequence. 
In particular, if $k$ satisfies $\frac{k}{\log{k}}\geq Cp\cdot \frac{\eps^2+u}{\eps^2}$ for a large enough constant $C$, 
then the rightmost expression is upper bounded by $\exp\big(-\frac{k\eps^2}{2(\eps^2+u)}\big)$. 
Truncating the power series at such a $k$ would yield error at most
\begin{eqnarray*}
\sum_{l \geq k}  \exp\Big(-\frac{l\eps^2}{2(\eps^2 + u)}\Big) 
& \leq & 
\Big(1 - \exp\Big(-\frac{\eps^2}{2(\eps^2 + u)} \Big)\Big)^{-1}
\exp\Big(-\frac{k\eps^2}{2(\eps^2 + u)}\Big)
\leq \frac{4(\eps^2 + u)}{\eps^2} 
\exp\Big(-\frac{k\eps^2}{2(\eps^2 + u)}\Big),
\end{eqnarray*}
where we used $1-e^{-y} \geq \frac{y}{2}$ for $y \in (0,1)$ in the last inequality. 
The truncation error is bounded by $\delta$ if $k$ also satisfies
\[
k \geq  \frac{2(u + \eps^2) }{\eps^2} \ln\Big(\frac{4(\eps^2+u)}{\delta \eps^2}\Big). 
\]
Finally, for each of the $j \leq k$ terms in the power series, we can compute 
\[\tot((u1-Y)^j) = \sum_{l=0}^j u^{j-l}(-1)^{l}{\binom{j}{l}}\tot(Y^l)\]
in at most $T(k)$ time.
\end{proof}

The power series leads to a polynomial time algorithm to compute the moments of the resolvent, which is required for the algorithm of \autoref{t:spectrum} in \autoref{section:resolvent}.
We also observe that the moments of resolvent with $z > \lambda_{\max} + \eps \in \R$ can be computed efficiently,
which will be used in the barrier method for proving the universality of operator norm in \autoref{section:uni}.

\begin{corollary}[Efficient Computation for Resolvents]
Let $X \in \M_d(\C) \otimes \A$ be a self-adjoint semicircular matrix with constant part $A_0 \otimes 1$. 
Let $z = \lambda +\eps i$ with $\eps > 0$. 
There is a deterministic algorithm to compute $\tot (|z1-X|^{-2p})$ up to an additive error $\delta$ in time $\poly(\|A_0\|, \sigma(X), d, \lambda, p, 1/\eps^2, \log(1/\delta))$. 

Furthermore, the same conclusion holds when $z = \lambda_{\max}(X) + \eps \in \R$.
\end{corollary}
\begin{proof}
We directly apply \autoref{lemma:res-computation} with $z = \lambda + \eps i$ and $Y = (\lambda 1 - X)^2$. 
Note that the moments of $Y$ can be computed from \autoref{lemma:computation} in polynomial time via binomial expansion,
where we use the upper bound $\norm{Y} \leq u = 2\lambda^2 + 2 (\|A_0\| + 2\sigma(X))^2$ provided by Pisier's bound in \autoref{lemma:pisier}.

For the furthermore part,
we write $\lambda 1 - X = \frac{\eps}{2} 1 + (\lambda-\frac{\eps}{2})1-X$. 
By construction, $W:=(\lambda-\frac{\eps}{2})1-X$ has nonnegative spectrum, with norm at most $2\|A_0\| + 4\sigma(X) + O(\eps)$. 
Then, the conclusion follows from \autoref{lemma:res-computation}, using \autoref{lemma:computation} to compute the moments of $W$.  
\end{proof}

\subsection{Integration by Parts}\label{section:ibp}

In this subsection, we prove a generalization of the Gaussian integration-by-parts formula in \autoref{lemma:gaussian-int} for free semicircular matrices. 

Before stating the formula, we first review its Gaussian analog with basic examples. 
Let $f$ be integrable with respect to Gaussian measure and $x$ be a centered Gaussian random variable with mean $0$ and variance $\sigma^2$. 
By iteratively applying the Gaussian integration by parts formula in \autoref{lemma:gaussian-int}, one can verify that
\begin{equation}
\E\big[x^mf(x)\big] = \sum_{\ell=0}^{m}\sigma^{m+\ell} {\binom{m}{\ell}}(m-\ell-1)!! \cdot \E\big[f^{(\ell)}(x)\big]\cdot \1_{\{\text{$m-\ell$ is even}\}}\label{eq:gibp}
\end{equation}
where we interpret $(-1)!! = 1$. 
Note that $ \E[x^{m-\ell}] = \sigma^{m-\ell}(m-\ell-1)!! \cdot \1_{\{m-\ell \text{ is even}\}}$. 
Thus, we can express \eqref{eq:gibp} as
\begin{equation}
\E\big[x^mf(x)\big]  
= \sum_{\ell=0}^{m} {\binom{m}{\ell}} \cdot \E\big[x^{m-\ell}\big] \cdot \E\big[x^2\big]^{\ell}\cdot \E\big[f^{(\ell)}(x)\big] 
= \sum_{\ell=0}^{m} {\binom{m}{\ell}} \cdot \E\Big[y^{m-\ell}\prod_{j=1}^{\ell}y_j^2\cdot f^{(\ell)}(x)\Big].  \label{eq:gibp2}
\end{equation}
where $y,y_1,\ldots, y_{\ell}$ are independent copies of $x$ and $\E[x^2]=\sigma^2$. 
While it may seem counter-intuitive to express \eqref{eq:gibp} as the form in \eqref{eq:gibp2}, this form actually lends itself to a natural generalization in the non-commutative setting. 
 We now state the non-commutative analog of \eqref{eq:gibp2} for semicircular matrices.

\begin{proposition}[Semicircular Integration by Parts Formula] 
\label{proposition:free-ibp}
Let $X_1, \ldots, X_n$ be freely independent, centered, and self-adjoint semicircular matrices. 
Let $\vec{X} = (X_1, \ldots, X_n)$. Let $F_1, \ldots, F_m \in \M_d(\C)\langle x_1,\ldots,x_n\rangle$ be non-commutative polynomials with matrix coefficients. Given any subset $S\subseteq [m]$ with $|S|=\ell$, define a new set of random variables: $X_{i,1},X_{i,2}, \ldots, X_{i,\ell}, X_{i,\overline{S}}$, which have the same distribution as $X_i$ but are freely independent from each other and from $X_1, \ldots, X_n$. 
For $S = \{j(1) < j(2)< \cdots < j(\ell)\}$, define $X_{i,j|S}$ for $j\in [m]$ as
\[
X_{i,j|S} = \begin{cases}
            X_{i,r}& \text{if $j = j(r)$ for some $r \in S$}, \\
            X_{i,\overline{S}} &\text{otherwise}.
            \end{cases}
\]
Then
\[
\varphi\big[F_1(\vec{X}) \cdot X_i \cdots F_m(\vec{X}) \cdot X_i\big] 
= \sum_{\ell=0}^m \sum_{S \in {{[m]}\choose \ell}} \varphi 
\Big[ \partial^\ell_{X_{i}} \Big(F_1(\vec{X}) \cdot X_{i,1|S}\cdots F_m(\vec{X}) \cdot X_{i,m|S}\Big)(X_{i,1}, \ldots, X_{i,\ell}) \Big].
\]
Note that $\varphi$ can be replaced with $\tot$ in the above equation by taking trace of both sides.
\end{proposition}

The proof will be provided in the next section. 
As a basic comparison between \autoref{proposition:free-ibp} and \eqref{eq:gibp2}, we note that the variable $X_{i,\overline{S}}$ is analogous to $y$ and $X_{i,1\ldots} X_{i,\ell}$ are analogous to $y_1,\ldots, y_\ell$ in \eqref{eq:gibp2}. 
In the non-commutative setting, the ${m\choose \ell}$ factor is replaced with a sum over ${m\choose \ell}$ ways of picking the positions of $X_{i,\overline{S}},X_{i,1},\ldots, X_{i,\ell}$, which is specified by the variables $X_{i,1|S},\ldots, X_{i,m|S}$. For example, if $m=6$, $S = \{2,3,6\}$, then $(X_{i,1|S}, X_{i,2|S}, \ldots, X_{i,6|S}) = (X_{i,\overline{S}},X_{i,1},X_{i,2}, X_{i,\overline{S}},X_{i,\overline{S}},X_{i,3})$.

Now we give some examples of how to apply \autoref{proposition:free-ibp}. 
These identities will be useful throughout our paper. 
First, we show what \autoref{proposition:free-ibp} looks like in the $m=1$ setting.

\begin{corollary}[First-Order Integration by Parts]\label{corollary:free-ibp1}
Let $X_1,X_2,\ldots, X_n$ be centered freely independent semicircular matrices.
Let $F\in \M_d(\C)\langle x_1,\ldots, x_n\rangle$ be a non-commutative polynomial with matrix coefficients. 
Then, for each $i\in [n]$,
\[
\varphi[ F(X_1,\ldots, X_n) \cdot X_i] 
= \varphi[\partial_{X_i}F(X_1,\ldots, X_n)(X_i')\cdot X_i'],
\]
where $X_i'$ has the same distribution as $X_i$ but is freely independent from $X_1,\ldots, X_n$. 
\end{corollary}
\begin{proof}
Applying \autoref{proposition:free-ibp} with $m=1$ and $F_1 = F$, it holds that 
\[
\varphi[F(X_1,\ldots, X_n) \cdot X_i] 
= \varphi[F(X_1, \ldots, X_n) \cdot X_i'] 
+ \varphi[\partial_{X_i} (F(X_1, \ldots, X_n) \cdot X_i')(X_i')],
\]
where the first term corresponds to the $\ell=0$ case and the second term corresponds to the $\ell=1$ case. 
But since $X_i'$ is centered and freely independent from $X_1,\ldots, X_n$, the first term vanishes. 
\end{proof}

Note that this is analogous to the Gaussian identity $E[x_i \cdot f(x_1,\ldots, x_n)] = \E[\partial_{x_i}f(x_1,\ldots, x_n)] \cdot \E[x_i^2]$. 
The key distinction is that we cannot factor out something like $\varphi[X_i^2]$ in the non-commutative setting, and this is the reason that we need to specify a separate free copy of $X_i$. 
But note that the order of $X_i'$ is two, which is analogous to the $E[x_i^2]$ term in the commutative setting.

Now, we will concretely apply \autoref{corollary:free-ibp1} to prove the computational formulas in the previous section.

\begin{proofof}{\autoref{lemma:moment-formula}}
For the first formula, 
define $X_0 := A_0 \otimes 1$, $X_i = A_i\otimes s_i$ for $i \in [n]$ and $F(X_1,\ldots, X_n) = (X_0+\sum_{i=1}^n X_i)^p = (\Xf)^{p}$. 
We use $X_i'$ to denote an independent copy of $X_i$ that is freely independent from $X_1,\ldots, X_n$. 
Then,
\begin{eqnarray*}
\varphi\big[\Xf^{p}\big] 
&=& \varphi\big[\Xf^{p-1} \cdot X_0\big] + \sum_{i=1}^n \varphi\big[\Xf^{p-1} \cdot X_i\big]
\\
&=& \varphi\big[\Xf^{p-1} \cdot X_0\big] + \sum_{i=1}^n\sum_{k=0}^{p-2}\varphi\big[\Xf^{k} \cdot X'_i \cdot \Xf^{p-2-k} \cdot X_i'\big]
\\
&=& \varphi\big[\Xf^{p-1}\big] \cdot A_0 + \sum_{i=1}^n\sum_{k=0}^{p-2} \varphi\big[\Xf^{k}\big] \cdot \varphi\big[X_i \cdot \big(\varphi\big[\Xf^{p-2-k}\big] \otimes 1\big) \cdot X_i\big]
\\
&=& \varphi[\Xf^{p-1}] \cdot A_0+\sum_{i=1}^n\sum_{k=0}^{p-2}\varphi[\Xf^{k}] \cdot A_i \cdot \varphi[\Xf^{p-2-k}] \cdot A_i.
\end{eqnarray*}
where the second line follows from \autoref{corollary:free-ibp1}, 
the third line is by \autoref{lemma:free-moments},
and the last line is by the identity $\varphi[X_i (M \otimes 1) X_i] = A_i M A_i$ since $X_i = A_i\otimes s_i$.

For the second formula, let $F(X_1,\ldots, X_n) = \Xf^p \cdot (B \otimes 1) \cdot \Xf^q$ where $X_i = A_i\otimes s_i$ and $\Xf = A_0 + \sum_{i=1}^nX_i$. 
Then
\[
\varphi\big[\Xf^p \cdot (B \otimes 1) \cdot \Xf^q\big] 
= \varphi\big[\Xf^p \cdot (B \otimes 1) \cdot \Xf^{q-1} \cdot (A_0 \otimes 1) \big] 
+ \sum_{i=1}^n\varphi\big[\Xf^p \cdot (B \otimes 1) \cdot \Xf^{q-1} \cdot X_i\big].
\]
Applying \autoref{corollary:free-ibp1} and the product rule on each summand gives
\begin{eqnarray*}
&&\varphi\big[\Xf^p \cdot (B \otimes 1) \cdot \Xf^{q-1} \cdot X_i\big] 
\\
&= &\sum_{k=0}^{p-1} \varphi\big[\Xf^{k} \cdot X_i' \cdot \Xf^{p-1-k} \cdot (B \otimes 1) \cdot \Xf^{q-1} \cdot X_i'\big] 
+ \sum_{k=0}^{q-2}\varphi\big[\Xf^{p} \cdot (B\otimes 1) \cdot \Xf^{k} \cdot X_i' \cdot \Xf^{q-2-k} \cdot X_i'\big] \cdot \1_{\{q\geq 2\}}
\\
&= & \sum_{k=0}^{p-1} \varphi\big[\Xf^{k}\big] \cdot A_i \cdot \varphi\big[\Xf^{p-1-k} \cdot (B \otimes 1) \cdot \Xf^{q-1}\big] \cdot A_i 
+ \sum_{k=0}^{q-2}\varphi\big[\Xf^{p} \cdot (B \otimes 1) \cdot \Xf^{k}\big] \cdot A_i \cdot \varphi[\Xf^{q-2-k}] \cdot A_i \cdot \1_{\{q\geq 2\}},
\end{eqnarray*}
where the second line follows by applying \autoref{lemma:free-moments} as for the first formula.
\end{proofof}

\subsection{Proof of Semicircular Integration-by-Parts Formula}\label{section:ibp-proof}

We prove \autoref{proposition:free-ibp} in this subsection.

\subsubsection{Intuition}

To build intuition for the proof, we first give a simple combinatorial proof of \eqref{eq:gibp2} in the special case where $f(x) = x^k$ is a monomial and $\sigma^2=1$. 
Recall from Wick's formula in \eqref{e:Wick} that $\E[x^k] = |P_2[k]|$ is exactly the number of pair-partitions of $k$ elements for positive integer $k$. 
We write the following recursive relation on pairing partitions:
\begin{equation}
    |P_2[m+k]| = \sum_{\ell=0}^{m}{m\choose \ell} \cdot |P_2[m-\ell]| \cdot k(k-1)\cdots (k-\ell+1)\cdot |P_2[k-\ell]| \cdot \1_{\{\ell \leq k\}}. \label{eq:gibp-comb}
\end{equation}
This follows by first splitting $[m+k]$ into a partition $T_1\cup T_2$, where $T_1$ contains the first $m$ elements and $T_2$ contains the last $k$ elements. 
Then, we condition on all the ways to split $T_1$ into two groups, $S$ and $\overline{S}$, where elements in $S$ must pair with elements in $T_2$ and elements in $\overline{S}$ must pair within $T_1$. 
In the above expression, $\ell$ is the size of $S$, ${m\choose\ell}$ is the number of ways to choose $S$, and the rest of the terms count the number of pairings given a choice of $S$. 
By noting that $k(k-1)\dots (k-\ell+1) \cdot |P_2[k-\ell]| = \E[f^{(\ell)}(x)]$ for  $f(x) = x^k$, we see that \eqref{eq:gibp2} is equivalent to \eqref{eq:gibp-comb}. 
By linearity, this combinatorial proof of \eqref{eq:gibp2} extends to all polynomials. 

While the notation in \autoref{proposition:free-ibp} is more complicated, the proof remains essentially the same. 
In particular, we use the fact that like how Gaussian moments count pair partitions, semicircular moments count non-crossing pair partitions. 
Consider when each $F_j$ in \autoref{proposition:free-ibp} is a monomial in $X_1,\ldots, X_n$. 
Then we can consider the entire expression on the LHS as a monomial, which is nothing but a word over the alphabet $X_1,\ldots, X_n$. 
We split the indices where $X_i$ occur into two groups: 
$T_1$ is the set of indices corresponding to the $X_i$'s sandwiched between the $F_j$'s, while $T_2$ is the set of indices within $F_1,\ldots, F_m$ where $X_i$ occurs. 
Then, we can similarly condition on the subset $S$ of $T_1$ that is paired with $T_2$. 
For the remainder of this section, we first devise the notation to make this argument formal, then we extend from monomials to arbitrary polynomials by linearity.

\subsubsection{Moments of Monomials} 

Let $\mathcal{T}$ be an arbitrary index set.
Consider a set of random variables $\{X_i \mid i\in \mathcal{T}\}$ where each $X_i\in \M_d(\C)\otimes \A$. 
We formally define the notion of monomials over these variables.

\begin{definition}[Monomials]
A monomial of degree $k$ over the variables $\{X_{i} \mid i\in \mathcal{T}\}$ is a polynomial of the form $(B_1 \otimes 1) X_{i_1}\cdots (B_{k} \otimes 1) \cdot X_{i_k} \cdot (B_{k+1} \otimes 1)$, where each $B_j\in \M_d(\C)$.
For ease of notation, we use the shorthand $B_1 X_{i_1} \cdots B_k X_{i_k} B_{k+1}$ to denote this monomial.
 
Note that the degree is $k$ because the $X_{i_j}$'s are variables and $B_j$'s are coefficients. 
Identify each monomial with a word $W$ over the alphabet $\mathcal{C}:=\{X_i:i\in \mathcal{T}\}\cup \M_d(\C)$.
\end{definition}

From this definition, it is clear that the set of all monomials form a basis for the vector space of polynomials over the variables $\{X_i \mid i\in \mathcal{T}\}$ with $\M_d(\C)$-valued coefficients. 
Moreover, we see that the expressions on the left and right hand sides of \autoref{proposition:free-ibp} are multi-linear in the polynomials $F_1,\ldots, F_m$.
Thus, by expanding out, it suffices to prove \autoref{proposition:free-ibp} in the case where each $F_j$ is a monomial. 
Henceforth, we assume that each $F_j$ is a monomial.

Now, we define the notion of indexing and applying the functionals $\varphi_\pi$ in \autoref{d:varphi-pi} on the monomials. 
We remark that the $\varphi_\pi$ functionals, used for the free Wick formula in \autoref{lemma:op-wick}, are crucial for the ``conditioning step'' later in the proof.

\begin{definition}[Linear Functional on Words]
Consider a monomial word
\[
W = B_1X_{i_1} \cdots B_{k}X_{i_k}B_{k+1}.
\]
For each $j\in [k]$, we let $W[j] = X_{i_j}$. 
For each partition $\pi \in NC_2[k]$, we let
\[
\varphi_\pi(W) = B_1 \cdot \varphi_\pi(X_1, B_2X_2,\ldots, B_{k}X_k) \cdot B_{k+1},
\]
where again we use $B_iX_i$ to denote $(B_i \otimes 1)X_i$.
We say that $\pi\sim W$ if $W[j] = W[j']$ for all $(j,j')\in \pi$.
\end{definition}

Let $W_1,\ldots, W_m$ be words corresponding to the monomials $F_1,\ldots, F_m$, with degrees $k_1,\ldots,k_m$ respectively. 
Let $W_C:= W_1X_iW_2X_i\cdots W_mX_i$ be the ``combined'' monomial with degree $k:=k_1+\cdots+ k_m+m$.  
Since the variables $X_1,\ldots, X_n$ are centered and freely independent, it follows from \autoref{lemma:sc-joint-moments} that
\begin{equation} \label{eq:ibp1}
\varphi[W_C] = \sum_{\substack{\pi\in NC_2[k]\\\pi\sim W_C}} \varphi_\pi(W_C).
\end{equation}
For the conditioning step, let $T = \{j \mid W_C[j]=X_i\}$ be the set of all indices in $W_C$ where $X_i$ occurs. 
Let $T_1 = \{k_1+1,k_1+k_2+2,\ldots,k_1+k_2+\cdots+ k_m+m\}$ be the set of indices of $W_C$ corresponding to the positions of the $m$ $X_i$'s that are in between $W_1,\ldots,W_m$. 
Let $T_2 = T\backslash T_1$. 
Since any $\pi \sim W_C$ must induce a pairing of indices in $T$, we condition on $S\subseteq T_1$ being the set of indices inside $T_1$ that pairs with indices in $T_2$. 
To precisely specify such a pairing, we select an $\ell\in [m]$ and subsets $S= \{j_1 < j_2<\cdots < j_\ell\}\subseteq  T_1$ and $S'= \{j'_1 < j'_2<\cdots < j'_\ell\} \subseteq T_2$.
Then, we pick a permutation $\sigma \in \mathcal{P}_\ell$ and pair $j_1$ with $j'_{\sigma(1)}$, pair $j_2$ with $j'_{\sigma(2)}$, and so on. 
To enforce that this is \textit{exactly} the pairing between indices in $T_1$ and indices in $T_2$, we define the partition of the indices $T_1\cup S'$ given by $\xi_{S,S',\sigma} :=  (\{j_1,j'_{\sigma(1)}\},\ldots,\{j_{\ell},j'_{\sigma(\ell)}\}, T_1\backslash S)$. 
Then, we see that $\pi$ respects the pairing we have selected if and only if $\pi|_{T_1\cup S'}$ (i.e., when $\pi$ is restricted to the indices in $T_1 \cup S'$) is still a pairing, and a \textit{refinement} of $\xi_{S,S',\sigma}$. We denote this event by $\pi_{|T_1\cup S'} \leq \xi_{S,S',\sigma}$. 
Thus, by conditioning over all possible choices of $S,S',\xi$, we have
\begin{equation}
    \sum_{\substack{\pi\in NC_2[k]\\\pi\sim W_C}} \varphi_\pi(W_C) = \sum_{\ell=0}^m\sum_{S\in {T_1\choose \ell}}\sum_{S'\in {T_2\choose \ell}} \sum_{\sigma \in \mathcal{P}_\ell} \sum_{\substack{\pi\in NC_2[k]\\ \pi\sim W_C\\ \pi|_{T_1\cup S'}\leq \xi_{S,S',\sigma}}}\varphi_\pi(W_C). \label{eq:ibp2}
\end{equation}

\subsubsection{Character Substitutions} 

Now, we will show how the conditioning formula in \eqref{eq:ibp2} can be interpreted in terms of derivatives. 
The main operation to introduce is the character substitution operation, which is a linear map that takes one monomial to another by replacing one variable with another.

\begin{definition}[Substitution]    
Let $W$ be a monomial of degree $k$. 
Given characters $H_1, H_2, \ldots, H_\ell \in \mathcal{C}$ and a subset of indices $S = \{1\leq j(1)<j(2)<\ldots<j(\ell) \leq k\}$, 
we define $ W[H_1,H_2, \ldots, H_\ell;S]$ as the word obtained by starting from $W$ and replacing the characters in $S$ with $H_1,\ldots, H_\ell$. 
Formally, if $W = B_1X_{i_1}\cdots B_{k}X_{i_k}B_{k+1}$, then $W[H_1,H_2, \ldots, H_\ell;S] = B_1Y_1\cdots B_{k}Y_kB_{k+1}$ where
\[
    Y_j=\begin{cases}
        X_{i_j} & j\notin S\\
        H_r & j=j(r)\in S.
    \end{cases}
\]
We also use $H^{(\ell)}$ to denote $H,H,\ldots H$ repeated $\ell$ times, so $W[H^{(\ell)},S]$ denotes substituting the same character, $H$, into the positions in $S$.
\end{definition}

For example, if $W = X_1BX_2X_1^2BX_3$, then $W[H^{(2)};\{1,3\}] = HBX_2HX_1BX_3$. We can also iteratively apply substitutions. For example, $W[H_1^{(2)};\{1,3\}][H_2;\{4\}] = H_1BX_2H_1H_2BX_3$.

We show two properties of substitutions. 
The first observation is that in a formula like the right hand side of \autoref{eq:ibp1}, conditioning on $\pi\leq \xi$ is equivalent to substituting the copies of $X_i$ in each part of $\xi$ with a distinct freely independent copy of $X_i$. 

\begin{claim}[Substitutions Respecting Partitions] \label{claim:free-sub}
Suppose $\{X_i \mid i\in \mathcal{T}\}$ are freely independent and centered semicircular matrices. 
Let $W$ be a monomial of degree $k$ and $i\in \mathcal{T}$. 
Let $S  \subseteq [k]$ be a subset of indices such that $W[j] = X_i$ for all $j\in S$. Let $\xi = (V_1,V_2, \ldots, V_q)$ be a partition of $S$. 
Then 
\[
\sum_{\substack{\pi\in NC_2[k],\\\pi\sim W\\\pi|_S\leq \xi}}\varphi_\pi(W) 
=\varphi\big(W[X^{(|V_1|)}_{i,1};V_1][X^{(|V_2|)}_{i,2};V_2] \cdots [X^{(|V_q|)}_{i,q};V_q]\big),
\]
where $X_{i,1}, \ldots, X_{i,q}$ are freely independent copies of $X_i$ (and are freely independent from $X_1, \ldots, X_n$). 
\end{claim}

\begin{proof}
Let $W' = W[X^{(|V_1|)}_{i,1};V_1][X^{(|V_2|)}_{i,2};V_2] \cdots [X^{(|V_q|)}_{i,q};V_q]$. 
Since $X_{i,1},\ldots, X_{i,q}$ all have the same distribution as $X_i$, 
$W[j]$ has the same distribution as $W'[j]$ for all $j \in [k]$. 
Thus, for any $\pi$ that satisfies both $\pi\sim W$ and $\pi\sim W'$,
\[
\varphi_\pi(W) = \varphi_\pi(W').
\]
Thus it suffices to show that $\pi\sim W$ and $\pi|_S \leq \xi$ if and only if $\pi\sim W'$. 
First, we note that if $\pi \sim W'$, then $\pi\sim W$ is also true, as $W'[u] = W'[v] \Rightarrow W[u] = W[v]$ for all $u,v\in [k]$ by construction. 
Moreover, for each $r\in [q]$, $W'$ has a copy of $X_i$ at each index in $V_r$ that is free from the random variables at all other indices outside $V_r$. 
Thus, $\pi \sim W'$ if and only if $\pi\sim W$ and for each $r\in [q]$, $\pi$ pairs all indices in $V_r$ with other indices in $V_r$. 
This is equivalent to saying that $\pi_S \leq\xi$. 
\end{proof}

The next observation is that the derivative of a monomial can be expressed in terms of substitutions via the chain rule.  

\begin{claim}[Derivative of Monomial] \label{claim:mono-derivative}
Let $F(X_1, \ldots, X_n)$ be a monomial of degree $k$ and $W$ be its corresponding word representation. 
Let $T\subseteq [k]$ be the indices of $W$ where $X_i$ occurs, 
i.e., $T=\{j: W[j] = X_i\}$. Then for all $\ell\geq 0$ we have
\[
\partial_{X_i}^{\ell}F(X_1, \ldots, X_n)(H_1, \ldots, H_\ell) 
= \sum_{S \in  {T\choose \ell}}\sum_{\sigma\in \mathcal{P}_\ell}W[H_{\sigma(1)}, \ldots, H_{\sigma(\ell)}; S]\cdot \1_{\{\ell\leq |T|\}}.
\]
\end{claim}
\begin{proof}
By the product rule, $\partial_{X_i}F(\vec{X})(H) = \sum_{j\in T}W[H; \{j\}]$, which is all the ways to select an index from $T$ and substitute it with $H_1$. 
By iteratively applying this $\ell$ times, we obtain a sum over all possible ways to select a set of $S$ indices from $T$ and substitute the terms $H_1,\ldots,H_\ell$ in $S$ in all possible orders. 
Finally, we note that if $\ell > |T|$, then the derivative is simply zero.
\end{proof}

\subsubsection{Completing the proof} 

Combining these two properties of substitutions will essentially complete the proof. 
By applying \autoref{claim:free-sub} to the right hand side of \eqref{eq:ibp2},
\begin{eqnarray} \label{eq:ibp3}
\sum_{\substack{\pi\in NC_2[k]\\\pi\sim W_C}} \varphi_\pi(W_C) 
&=& \sum_{\ell=0}^m\sum_{S\in {T_1\choose \ell}}\sum_{S'\in {T_2\choose \ell}} \sum_{\sigma \in \mathcal{P}_\ell} \sum_{\substack{\pi\in NC_2[k]\\ \pi|_{T_1\cup S'}\leq \xi_{S,S',\sigma}}}\varphi_\pi(W_C) 
\\
&=& \sum_{\ell=0}^m\sum_{\substack{S\in {T_1\choose \ell}\\S=\{j_1<\ldots<j_\ell\}}}\sum_{\substack{S'\in {T_2\choose \ell}\\S'=\{j_1'<\ldots< j_\ell'\}}} \sum_{\sigma \in \mathcal{P}_\ell} \varphi\big(W_C[X_{i,1}^{(2)};\{j_1, j'_{\sigma(1)}\}]\cdots[X_{i,\ell}^{(2)};\{j_\ell,j'_{\sigma(\ell)}\}][X_{i,\overline{S}}^{(m-\ell)};T_1\backslash S]\big) \nonumber
\end{eqnarray}
where $X_{i,1},\ldots, X_{i,\ell},X_{i,\overline{S}}$ are freely independent copies of $X_i$ introduced through the application of \autoref{claim:free-sub}. 
Moreover, we see that variables $X_{i,1|S},\ldots, X_{i,m|S}$ (defined in \autoref{proposition:free-ibp}) specify the position of these variables in the indices of $T_1$. The set $S'$ and the permutation $\sigma$ specify the positions of the variables $X_{i,1},\ldots, X_{i,\ell}$ in $T_2$. 
Thus, we can write
\begin{align}
W_C[X_{i,1}^{(2)};\{j_1, j'_{\sigma(1)}\}]\cdots &[X_{i,\ell}^{(2)};\{j_\ell,j'_{\sigma(\ell)}\}][X_{i,\overline{S}}^{(m-\ell)};T_1\backslash S] \nonumber 
\\
&= \Big(W_1 X_{i,1|S}W_2X_{i,2|S}\cdots W_{m}X_{i,m|S}\Big)[X_{i,\sigma(1)},\ldots, X_{i,\sigma(\ell)};S'].  \label{eq:ibp4}
\end{align}
Now, we define the auxiliary function $F(Y_1,\ldots, Y_m,X_1,\ldots, X_n) := F_1(\vec{X}) \cdot Y_1\cdots F_m(\vec{X}) \cdot Y_m$. 
Then, by definition, $W_1 X_{i,1|S}W_2X_{i,2|S}\cdots W_{m}X_{i,m|S}$ is the word representation of the monomial $F(X_{i,1|S},\ldots, X_{i,m|S},X_1,\ldots, X_n)$, and $T_2$ is exactly the set of indices where $X_i$ occurs in this monomial. 
Thus, we can further write
\begin{align}
\sum_{\substack{S'\in {T_2\choose \ell}\\S'=\{j_1' < \cdots < j_\ell'\}}} \sum_{\sigma \in \mathcal{P}_\ell} W_C&[X_{i,1}^{(2)};\{j_1, j'_{\sigma(1)}\}] \cdots [X_{i,\ell}^{(2)};\{j_\ell,j'_{\sigma(\ell)}\}][X_{i,\overline{S}}^{(m-\ell)};T_1\backslash S] \nonumber
\\
&=\sum_{\substack{S'\in {T_2\choose \ell}\}}} \sum_{\sigma \in \mathcal{P}_\ell} \Big(W_1 X_{i,1|S}W_2X_{i,2|S}\cdots W_{m}X_{i,m|S}\Big)[X_{i,\sigma(1)},\ldots, X_{i,\sigma(\ell)};S'] \nonumber
\\
&= \partial_{X_i}^\ell F(X_{i,1|S},\ldots, X_{i,m|S},X_{1},\ldots, X_{n})(X_{i,1},\ldots, X_{i,\ell}), \label{eq:ibp5}
    \end{align}
where the first identity is from \eqref{eq:ibp4} and the second identity is by applying \autoref{claim:mono-derivative}. 
Combining \eqref{eq:ibp5} with \eqref{eq:ibp3}, we have
\begin{align*}
\sum_{\substack{\pi\in NC_2[k]\\\pi\sim W_C}} \varphi_\pi(W_C) 
&= \sum_{\ell=0}^m\sum_{\substack{S\in {T_1\choose \ell}\\S=\{j_1<\cdots< j_\ell\}}}\sum_{\substack{S'\in {T_2\choose \ell}\\S'=\{j_1' < \cdots < j_\ell'\}}} \sum_{\sigma \in \mathcal{P}_\ell} \varphi\big(W_C[X_{i,1}^{(2)};\{j_1, j'_{\sigma(1)}\}]\cdots[X_{i,\ell}^{(2)};\{j_\ell,j'_{\sigma(\ell)}\}][X_{i,\overline{S}}^{(m-\ell)};T_1\backslash S]\big)
\\
&= \sum_{\ell=0}^m\sum_{\substack{S\in {T_1\choose \ell}}} \varphi\big[\partial_{X_i}^\ell F(X_{i,1|S},\ldots, X_{i,m|S},X_{1},\ldots, X_n)(X_{i,1},\ldots, X_{i,\ell})\big]. 
\end{align*}
The final conclusion then follows by noting that the left hand side of this equation is exactly 
\[\varphi[F_1(\vec{X})X_i\cdots F_m(\vec{X})X_i],\]
while each term in the summation on the right hand side is exactly 
\[\varphi[\partial^\ell_{X_i}\big(F_1(\vec{X})X_{i,1|S}\cdots F_m(\vec{X})X_{i,m|S}\big) (X_{i,1},\ldots, X_{i,\ell})].\]

\subsection{Extension to Rational Functions}\label{section:rational-ibp}

In \autoref{section:uni}, we also need to extend our integration by parts formula from polynomials to some rational functions. 
In particular, we will be concerned with polynomials of $\Xf$ and the resolvent $(z1-\Xf)^{-1}$. 

\begin{proposition}[Semicircular Integration by Parts Formula for Resolvents] \label{proposition:lbp-rational}
Let $F\in \M_d(\C)\langle x_1,\ldots, x_n\rangle$ be a self-adjoint polynomial with matrix coefficients. 
Let $\vec{X}=(X_1,X_2, \ldots, X_n)$ be a set of centered, self-adjoint, semicircular matrices. 
Let $\mathcal{I}$ be a compact interval containing $\spec(F(\vec{X}))$. 
Let $M_z(\vec{X}) = (z1 - F(\vec{X}))^{-1}$. 
Finally, let $Q_1,Q_2,\ldots, Q_m \in \M_d(\C) \langle x_1,\ldots x_n, r\rangle$. 
Then, for any $z\in \C\backslash \mathcal{I}$ and $i\in [n]$,
\begin{align*}
\tot\big(&Q_1\big(\vec{X},M_z(\vec{X})\big)X_i \cdots Q_m\big(\vec{X},M_z(\vec{X})\big)X_i\big)\\
&= \tot \bigg(\sum_{\ell=0}^{m}\sum_{S \in {[m]\choose \ell}}\partial^\ell_{X_i} \Big(Q_1(\vec{X},M_z(\vec{X}))X_{i,1|S}\cdots Q_m(\vec{X},M_z(\vec{X})) X_{i,m|S})\Big)(X_{i,1},\ldots, X_{i,\ell}) \bigg).
\end{align*}
Where the $X_{i,j|S}$'s are as in \autoref{proposition:free-ibp}, and free from $X_i$'s.
\end{proposition}
\begin{proof}
The function $z\mapsto (z1-F(\vec{X}))^{-1}$ is complex analytic, i.e., it is $\C$-differentiable. 
As the sum, the product, and the composition of analytic functions is analytic,
it can be readily verified that both the left and right hand side are analytic functions in $z$. 
By the coincidence principle, if $f$ and $g$ are analytic on a domain $U$, and $f=g$ on $S\subseteq U$, which contains a limit point, then $f=g$ on $U$. 
In particular, it suffices to check that the left hand side and the right hand side agree on all $z\in \C$ with $|z|\geq 2\bignorm{F(\vec{X})}$. 
   
In this regime, we can express $M_z(\vec{X})$ as a convergent (in operator norm) power series such that 
\[
M_z(\vec{X})=\sum_{k\geq0}z^{-{k+1}}F(\vec{X})^k.
\]
Moreover, this convergence continues to hold on an open ball $\mathcal{U}\subseteq (\M_d(\C)\otimes \A)^n$ containing $X_1,\ldots, X_n$. 
This means that $Q_j(\vec{X},M_z(\vec{X}))$ can be expressed as a convergent power series on $\mathcal{U}$ for each $j$. 
Formally, there exist polynomials $P_j^{N} \in \M_d(\C)\langle x_1,\ldots, x_n, x_1^*,\ldots, x_n^*, \rangle$ for $N\in\N$ such that $\sum_{N=0}^\infty P_j^N(\vec{X}')=  Q_j(\vec{X}' ,M_z(\vec{X}'))$ for all $\vec{X}'\in\mathcal{U}$. 
This means, in particular, that we can freely commute this infinite sum with the partial derivatives $\partial_{X_i}$ (see for example \cite[Theorem 3.6.1]{Car71}). 
The conclusion then follows by applying infinite series expansion to each $Q_j$, expanding out via multi-linearity of the product, and applying \autoref{proposition:free-ibp} term-by-term.
\end{proof}

To give better intuition about the statement, we derive a concrete formula which will be useful in the derandomization of \cite{BvH24}.

\begin{corollary}[Semicircular Integration by Parts Formula for Resolvents] \label{cor:res-ibp}
Let $X_1,\ldots, X_n$ be centered and self-adjoint semicircular matrices. 
Let $\Xf = A_0 \otimes 1 +\sum_{i=1}^nX_i$. 
If $\lambda > \lambda_{\max}(\Xf)$, then
\[
\tot\big(M_\lambda(\Xf)^{p} \cdot X_i\big) 
= \tr\Big(\sum_{k=1}^{p} \varphi\big[M_\lambda(\Xf)^k] \cdot \varphi[ X_i \cdot (\varphi[M_\lambda(\Xf)^{p-k+1}] \otimes 1) \cdot X_i\big]\Big)
\]
\end{corollary}
\begin{proof}
We apply \autoref{proposition:lbp-rational} with $m=1$ and $F(X_1,\ldots, X_n) = A_0 + X_1+\cdots + X_n$. In particular, by taking the derivative of the function $(\lambda 1 - (A_0 \otimes 1 + \sum_iX_i))^{-p}$ with respect to $X_i$,
\[
        \tot(M_\lambda(\Xf)^{p}X_i) = \tot(M_{\lambda}(\Xf)^{p} X'_i) + \sum_{k=1}^p\tot(M_\lambda(\Xf)^k X_i' M_\lambda(\Xf)^{p-k+1} X_i'),
\]
where $X_i'$ is a freely independent copy of $X_i$. Note that $\tot(M_{\lambda}(\Xf)^{p} X'_i)$ corresponds to the $\ell=0$ term in \autoref{proposition:lbp-rational}, which vanishes as $X'_i$ is free from $\Xf$.
The final conclusion then follows by applying \autoref{lemma:free-moments}.
\end{proof}

\section{Deterministic Universality}\label{section:uni}

In this section, we prove \autoref{t:moment-uni-intro} and \autoref{t:resolvent-uni} and show applications in deterministic constructions of expander graphs.

\subsection{Technical Statements} \label{ss:statements-uni}

We start by describing the basic settings.
Given a fixed $d \times d$ Hermitian matrix $A_0$ and random $d \times d$ matrices $Z_1, \ldots, Z_n$ with $\E[Z_i]=0$ for all $i \in [n]$,
the goal is to derive a matrix concentration inequality for the general random matrix model:
\[ 
Z = A_0 + \sum_{i=1}^n Z_i.
\]
The universality phenomenon proved in~\cite{BvH24} says that, under mild conditions, the spectrum of the random matrix $Z$ nearly coincides with that of the Gaussian random matrix
\[
G = A_0 + \sum_{i=1}^{n'} g_i A_i,
\]
where $g_i$'s are i.i.d.~standard Gaussians, $A_i$'s are $d \times d$ Hermitian matrices such that $Z$ and $G$ have the same covariance $\cov(Z) = \cov(G)$. 
Combining with the ``intrinsic freeness'' results in~\cite{BBvH23}, Brailovskaya and van Handel~\cite{BvH24} managed to capture the spectrum of $Z$ by the following free model corresponding to $G$:
\begin{equation} \label{eq:general-free}
    \Xf = A_0 \otimes 1 + \sum_{i=1}^{n'} A_i \otimes s_i,
\end{equation}
where $s_1, \ldots, s_{n'}$ are freely independent semicircular elements.
The following is their formal statement.

\begin{theorem}[Norm Universality {\cite[Theorem 2.16]{BvH24}}] \label{theorem:bvh24}
Let $\rho(Z) := \|\max_i \|Z_i\| \|_{\infty}$, where $\|Y\|_{\infty}$ denotes the essential supremum of the random variable $|Y|$ which is a uniform upper bound on $|Y|$.  Then
\[
\E [\norm{Z}] \leq \|\Xf\| + C \Big( \nu(Z)^{\frac12} \cdot \sigma(Z)^{\frac12} \cdot \log^{\frac34} d + \rho(Z)^{\frac13} \cdot \sigma(Z)^{\frac23} \cdot \log^{\frac23} d + \rho(Z) \cdot \log d \Big).
\]
\end{theorem}

We remark that the main result in~\cite{BvH24} is a much stronger form with a concentration result on the full spectrum.

In this section, we design polynomial time deterministic algorithm to find an outcome of $Z$ that nearly satisfies the bound in \autoref{theorem:bvh24}. 
For computational purpose, we make an additional assumption that the support of each random matrix $Z_i$ is discrete and
\begin{equation} \label{e:support-assumption}
    | \supp(Z_i) | \leq \poly(d), \quad \forall i \in [n],
\end{equation}
so that we can enumerate all possibilities of a random matrix $Z_i$ in polynomial time.
As we will show in \autoref{ss:uni-application},
this assumption is satisfied in combinatorial applications of \autoref{theorem:bvh24} such as deterministic constructions of expander graphs.

The following is the full version of \autoref{t:moment-uni-intro} about the Schatten $p$-norm of the output matrix.

\begin{theorem}[Deterministic Moment Universality] \label{t:moment-uni}
Assuming \eqref{e:support-assumption},
for any integer $p \geq 1$, there is a polynomial time deterministic algorithm runs in $\poly(n,d,p)$ to find a matrix $Z'\in \supp(Z)$ such that 
\[
\norm{Z'}_{2p} \leq \norm{\Xf}_{2p} + O\Big(1+\log{\frac{n \cdot \rho(Z)^2}{\sigma(Z)^2}} \Big) \cdot \big( p^{\frac34} \cdot \sigma(Z)^{\frac12} \cdot \nu(Z)^{\frac12} + p^{\frac23} \cdot \sigma(Z)^{\frac23} \cdot \rho(Z)^{\frac13} + p \cdot \rho(Z) \big).
\]
\end{theorem}

One may use \autoref{t:moment-uni} for the operator norm by setting $p \asymp \log d$, but this would lead to a leading constant $C$ such that $\norm{Z'} \leq C \cdot \norm{\Xf} + O(\cdots)$, which is not strong enough for some applications such as constructing near-Ramanujan graphs (see \autoref{ss:uni-application}).

The following is the full version of \autoref{t:resolvent-uni} about the operator norm of the output matrix, with leading constant being one.

\begin{theorem}[Deterministic Norm Universality] \label{t:norm-uni}
Assuming \eqref{e:support-assumption},
there is a polynomial time deterministic algorithm to find a matrix $Z'\in \supp(Z)$ such that 
    \begin{align*}
        \norm{Z'} \leq \norm{\Xf} + O\Big(1+\log{\frac{n \cdot \rho(Z)^2}{\sigma(Z)^2}} \Big) \cdot \big( \sigma(Z)^{\frac12} \cdot \nu(Z)^{\frac12} \cdot \log^{\frac34}d + \sigma(Z)^{\frac23} \cdot \rho(Z)^{\frac13} \cdot \log^{\frac23} d  + \rho(Z) \cdot \log d\big).
    \end{align*}
\end{theorem}

Notice that our derandomization schemes in \autoref{t:moment-uni} and \autoref{t:norm-uni} lose an additional factor of $\log (n \cdot \rho(Z)^2 / \sigma(Z)^2)$ comparing with the bound in \autoref{theorem:bvh24}.
This loss is negligible in many applications, 
as we will see in \autoref{ss:uni-application}. 
However, it would be interesting to see if we can get rid of this logarithmic factor, which is incurred in our randomized swap algorithm.

{\bf Notations:}
As the random model $Z$ is fixed throughout the section, 
we denote $\rho := \rho(Z)$ and $\sigma := \sigma(Z)$ for simplicity. We always assume, implicitly, that $\sigma > \rho$, as otherwise, the bounds in the theorems offer no improvements over classical results.
Also, for ease of notations, we write
\[
\Xf = A_0 \otimes 1 + X_1+X_2+ \cdots +X_n,
\]
where $\{X_i \in M_d(\C)\otimes \A\}_{i \in [n]}$ are freely independent semicircular matrices, and each $X_i$ has the same covariance as $Z_i$, i.e., $\cov(X_i) = \cov(Z_i)$ (see \autoref{d:cov-profile}). 
Moreover, we have $\cov(\Xf) = \cov(Z)$, as $\cov(\Xf) = \sum_i\cov(X_i)$ and $\cov(Z) = \sum_i \cov(Z_i)$.

\begin{remark*}[Explicit Representation]
Given $Z_1, \ldots, Z_n$, an explicit way to construct $X_1, \ldots, X_{n}$ is as follows.
For each $i$, let $\lambda_{i,1}, \ldots, \lambda_{i,d^2}$ and $v_{i,1}, \ldots, v_{i,d^2}$ be the eigenvalues and eigenvectors of $\cov(Z_i)$. 
Let $V_{i,1}, \ldots, V_{i,d^2}$ be $d\times d$ matrices so that $v_{i,j} = \vecc(V_{i,j})$. 
Then, we can explicitly write
\[
X_i = \sum_{j=1}^{d^2} \sqrt{\lambda_{i,j}} \cdot V_{i,j} \otimes s_{i,j},
\]
where $\{s_{i,j}:i\in[n],\,j\in[d^2]\}$ is a family of standard free semi-circular operators.
We note that this representation is not unique. Moreover, if $Z_i$ is self-adjoint, then $\cov(Z_i)$ can be interpreted as a linear operator on the space of self-adjoint matrices, which means its eigenvectors, $V_{i,1},\ldots V_{i,d^2}$ can also be chosen to be self-adjoint.
\end{remark*}

{\bf Organization:}
Since the proofs are quite technically involved,
we first present some interesting applications of our main results to constructing expander graphs in \autoref{ss:uni-application}. 
Then, we provide a technical overview of the proofs in \autoref{ss:outline-uni}.
The proof of \autoref{t:moment-uni} will be presented in \autoref{ss:moment-uni}, and the proof of \autoref{t:norm-uni} will be given in \autoref{ss:norm-uni}.

\subsection{Deterministic Expander Constructions} \label{ss:uni-application}

In this subsection, we present our results for explicit constructions of near-Ramanujan graphs in the moderately dense to dense regime. 
While many probabilistic constructions of expanders yield (nearly) optimal spectral gap \cite{Fri08,Bor20,CGT+25,HMY25}, a fundamental question in theoretical computer science is whether these constructions can be made explicit, or polynomial time deterministic. 
Explicit construction of expander graphs is an important and well-studied topic, with applications in algorithm design and complexity theory. 

Previously, Marcus, Spielman and Srivastava~\cite{MSS18} showed the existence of bipartite Ramanujan graphs for all degrees and sizes, and their construction was made polynomial time and deterministic by Cohen \cite{Coh16}. 
However, in the non-bipartite setting, much less is known about deterministic constructions compared to probabilistic constructions. 
In particular, all previous efficient deterministic constructions of near-Ramanujan non-bipartite graphs are restricted to the constant degree regime \cite{MOP20, OW20, JMO+22}, and not much is known when the degree is at least polylogarithmically large.

Using the derandomization results in this paper, we obtain deterministic constructions of (not necessarily bipartite) near-Ramanujan graphs in this regime for several random models, including the random edge-signing model in \autoref{section:signing}, the random permutation model in \autoref{section:perm}, and two random lift models in \autoref{section:lifts}.

The following corollary of \autoref{t:norm-uni} is used in this subsection for ease of applications.
The proof follows from \autoref{t:norm-uni} by applying Pisier's result in \autoref{lemma:pisier} and triangle inequality to bound the norm of the free model.

\begin{corollary}[Deterministic Norm Universality] \label{theorem:norm-uni-simple}
Let $Z = A_0+\sum_{i=1}^n Z_i$ be a random $d\times d$ matrix, where $Z_1,\ldots, Z_n$ are independent, centered, self-adjoint random matrices, where $\norm{Z_i}\leq \rho$ with probability one for all $i \in [n]$. 
Assuming~\eqref{e:support-assumption}, there is a polynomial time deterministic algorithm to find a matrix $Z'\in \supp(Z)$ such that
\[
\norm{Z'} \leq \norm{A_0} + 2\sigma(Z) + O\Big(1+\log{\frac{n \cdot \rho^2}{\sigma^2(Z)}} \Big)\cdot \big(\sigma(Z)^\frac12 \cdot \nu(Z)^\frac12 \cdot \log^{\frac34}d  + \sigma(Z)^\frac23 \cdot \rho^{\frac13} \cdot \log^{\frac23}d +\rho \cdot \log{d}\big).
\]
\end{corollary}

\subsubsection{Random Signings}\label{section:signing}

The first model that we will derandomize is the random signing of regular graphs. 
Given a $k$-regular graph $G$ over $d$ vertices, 
the goal is to find a signing of its edges so that its signed adjacency matrix has small operator norm.
 
This is an important model for expander construction due to its connection to 2-lifts, which will be elaborated more in \autoref{section:lifts}. 
The random signing model is a canonical example of the ``sparse Wigner model'' \cite{vH17}.
It is well known~\cite{BvH15} that if $k$ is at least $\Omega(\polylog{d})$, then a random signing of its adjacency matrix has norm at most $2\sqrt{k} \cdot (1+o_d(1))$. 
To our knowledge, there is no known polynomial time algorithm to deterministically construct such a signing in this regime,
and we present such a deterministic algorithm.

\begin{theorem}[Deterministic Edge Signing]\label{theorem:2-lift}
Let $G= ([d], E)$ be a $k$-regular graph over $d$ vertices with adjacency matrix $A$. 
If $k \gtrsim \log^{4}d$, then there is a polynomial time deterministic algorithm to find a signing $x\in \{-1,1\}^{|E|}$, so that the signed adjacency matrix $A(x)$ satisfies
\[
\norm{A(x)} \leq 2\sqrt{k} \cdot \Big(1+O\Big(\frac{\log^{\frac23}d}{k^{\frac16}}+\frac{\log^{\frac34}d}{k^{\frac14}}+\frac{\log{d}}{k^{\frac12}}\Big)\Big).
\]
\end{theorem}
\begin{proof}
Let $Z$ be the adjacency matrix of an independent random signing of the $dk/2$ edges of $G$. 
It can be readily verified that $\sigma^2(Z) = k$, $\nu^2(Z) = 2$ and $\rho(Z) = 1$. 
If we directly apply \autoref{theorem:norm-uni-simple} to the sum of $n = dk/2$ independent random signed adjacency matrices of all edges, the $\log \frac{n \rho^2}{\sigma^2} \asymp \log d$ dependence will give us an extra logarithmic factor. 
    
To remove the $\log d$ factor, we decompose $G$ into the union of at most $k+1$ edge-disjoint matchings $M_1,\ldots, M_{k+1}$ using Vizing's theorem on edge coloring. 
Consider the following alternate random matrix model, where we pick a \textit{pairwise independent} signing for the edges within each matching $M_i$, while the signings of the edges in different matchings $M_1,\ldots, M_{k+1}$ are mutually independent from each other. 
    
Let $\Tilde{Z}_i$ be the signed adjacency matrix of $M_i$ and $\Tilde{Z} = \sum_{i=1}^{k+1}\Tilde{Z}_i$. 
Since each $M_i$ is a matching, we have $\|\Tilde{Z}_i\| \leq 1$.
Meanwhile, as we pick pairwise independent signing for each $M_i$, the support size of each $\Tilde{Z}_i$ is in $\poly(d)$. 
Furthermore, since $\Tilde{Z}_i$'s are independent, the entries of $\Tilde{Z}$ are pairwise independent, and thus $\Tilde{Z}$ has the same covariance as $Z$, which implies that $\sigma^2(\Tilde{Z}) =k$ and $\nu^2(\Tilde{Z}) = 2$.     
Applying \autoref{theorem:norm-uni-simple} on $\Tilde{Z}_1, \ldots, \Tilde{Z}_{k+1}$, we obtain a signing $x$ satisfying
\[
\norm{A(x)} 
\leq 2\sqrt{k} + O\big( k^\frac14 \cdot \log^{\frac34}d + k^\frac13 \cdot \log^{\frac23}{d} +\log{d}\big) 
\leq 2\sqrt{k}\Big(1+O\Big(\frac{\log^{\frac23}d}{k^{\frac16}}+\frac{\log^{\frac34}d}{k^{\frac14}}+\frac{\log{d}}{k^{\frac12}}\Big)\Big). \qedhere
\]
\end{proof}

We note that the random signing model can also be framed as an instance of the matrix Spencer problem, 
where our input matrices are of the form $\{\chi_u\chi_v^\top + \chi_v\chi_u^\top:(u,v)\in E\}$. 
However, in this case, we need to directly find a full coloring with error around $2\sigma$ instead of having error depending on $n$ (which would be far too large).
This example gives an illustrative comparison between \autoref{theorem:norm-uni-simple} and \autoref{t:partial-coloring-full}:  
Given input matrices $A_1,\ldots, A_n$ with $\norm{A_i}\leq 1$ for $i \in [n]$, 
the former gives a signing $A(x)$ with very sharp bounds in terms of $\sigma$, but requires $\nu$ to be very small. 
If instead, we only have the much weaker assumption that $\max_i\norm{A_i}_F$ is small instead of $\nu$, then \autoref{t:partial-coloring-full} can still be applied recursively to give a full coloring with discrepancy $O(\sqrt{n})$.

\subsubsection{Random Permutations}\label{section:perm}

The next model that we will derandomize is the construction of regular expanders through the union of random perfect matchings. 
This is one of the most common and natural models for generating uniform random regular graphs. 
The classic result of Friedman~\cite{Fri08} showed that for any constant $k$ and $d$ large enough, random $k$-regular graphs sampled from this model have spectral radius at most $2\sqrt{k-1}+o_d(1)$ with high probability. 
In \cite[Theorem 3.8]{BvH15}, it was shown that the near-Ramanujan property in this model continues to hold when $k\gtrsim \log^4{d}$, which was the first time such a result was proven for any $k=\omega(\log{d}/\log\log{d})$. 
Our derandomization of their concentration inequality gives the following explicit construction result.

\begin{theorem}[Deterministic Permutation Model] \label{theorem:perm}
For any $k\gtrsim \log^4{d}$, there is a polynomial time deterministic algorithm to construct a $2k$-regular graph over $d$ vertices (possibly with parallel edges and self loops), whose adjacency matrix $A$ satisfies
\[
\Bignorm{A - \frac{2k}{d} \cdot \chi_d \chi_d^\top} 
\leq 2\sqrt{2k} \cdot \Big(1+ O\Big(\frac{\log^{\frac23}d}{k^{\frac16}} \Big)\Big),
\]
where $\chi_d$ denotes the $d$-dimensional all-one vector.
\end{theorem}
  
{\bf Proof}: While we used pairwise independent bits for the random edge signing model to ensure polynomial support size, we do not know how to obtain polynomial support size with pairwise independent permutations.
Fortunately, an approximate version of pairwise independent random permutation model is known in the literature to have polynomial support size, which we will need to use for the following applications.

\begin{definition}[Approximate $q$-wise Uniform Permutations]
A distribution $\mathcal{D}$ over $\mathcal{P}_m$ is called $(\delta,q)$-wise uniform if for every sequence $(i_1,\ldots, i_q)\in [m]^q$, the distribution of $(\sigma(i_1),\ldots,\sigma(i_q))$ has total variation distance at most $\delta$ from the uniform distribution on $[m]^q$. If $q=2$, then we say the permutation is approximately pairwise uniform.
\end{definition}
    
\begin{theorem}[Approximate $q$-wise Uniform Permutations~\cite{Kas07,KNR09}] \label{theorem:pwu}
For any $m,q,\delta>0$, there exists a $(\delta,q)$-wise uniform distribution over $\mathcal{P}_m$ whose support has size $\poly(m^q, \frac{1}{\delta})$ and can be computed in time $\poly(m^q, \frac{1}{\delta})$.
\end{theorem}

In the following claim, we show that replacing uniform permutations in a random matrix model with approximate pairwise uniform permutations does not alter the parameters $\sigma^2$ and $\nu^2$ by much.

\begin{claim}[Covariance Matrix of Pairwise Uniform Permutation] \label{corollary:pwu}
Let $\Pi$ be a uniform random $d\times d$ permutation matrix, and $\Tilde{\Pi}$ be a random $d\times d$ permutation matrix of a $(\delta,2)$-wise uniform permutation. Then 
\[\norm{\cov(\Tilde{\Pi})} \lesssim \frac{1}{d} + \delta d^2,
\quad\quad 
\norm{\E[\Pi^2] - \E[\Tilde{\Pi}^2]} \lesssim \delta d^2,
\quad\quad \norm{\E[\Tilde{\Pi}] - \frac{1}{d}\chi_d\chi_d^\top}\lesssim \delta d.
\]
\end{claim}
\begin{proof}
We begin by showing the last inequality that $\|\E[\Tilde{\Pi}] - \frac{1}{d}\chi_d\chi_d^\top\|\leq \delta d$. 
By the definition of $(\delta,2)$-wise independence, $|\E[\Tilde{\Pi}(i,j)]-1/d|\leq \delta$ for each $i,j$. 
Our desired bound then follows by using the fact that $\norm{M} \leq L$ for any $d \times d$ matrix $M$ when every row/column of $M$ has $\ell_1$-norm at most $L$. 

To bound the second order terms, we begin by noting that $\|\cov(\Pi)\| \lesssim 1/d$. 
This can be seen in \cite[Lemma 3.5]{BvH24}, or derived by direct computation via the row $\ell_1$-norm bound. 
By the definition of a $(\delta,2)$-wise uniform permutation, for each $i,j,k,l\in [d]$, 
\[
|\E[\Pi(i,j) \cdot \Pi(k,l)] - \E[\Tilde{\Pi}(i,j) \cdot \Tilde{\Pi}(k,l)]|\leq \delta.
\]
This implies that the entry-wise deviation of $\cov(\Tilde{\Pi})$ from $\cov(\Pi)$ is at most $O(\delta)$. 
Similarly, the entry-wise deviation of $\E[\Pi^2]$ from $\E[\Tilde{\Pi}^2]$ is at most $O(d \cdot \delta)$, as each entry of $\E[\Pi^2]$ is the sum of $d$ second-order terms. 
Therefore, by the row $\ell_1$-norm bound, we have $\|\cov(\Pi) - \cov(\Tilde{\Pi})\| \lesssim \delta d^2$ and $\|\E[\Pi^2] - \E[\Tilde{\Pi}^2]\| \lesssim \delta d^2$, proving the first two inequalities. 
\end{proof}

Now, we are ready to derandomize the random permutation model for constructing regular graphs. 
In \cite[Theorem~3.8]{BvH24}, it was shown that if $\Pi_1,\ldots, \Pi_k$ are independent random permutation matrices for $k\gtrsim \log^4d$, then
\[
    \E\Bignorm{\sum_{i=1}^k(\Pi_i +\Pi_i^\top) - \frac{2k}{d} \chi_d \chi_d ^\top} \leq 2\sqrt{2k} \cdot \Big(1+O\Big(\frac{\log^{\frac23}d}{k^{\frac16}} \Big) \Big).
\]
In other words, the $2k$-regular graph formed by the union of $k$ random $d\times d$ permutation matrices $\Pi_1,\ldots, \Pi_k$ is near Ramanujan with high probability. 

We derandomize this result using \autoref{theorem:norm-uni-simple} and approximate pairwise uniform permutations as follows.

\begin{proofof}{\autoref{theorem:perm}}
Let $Z = \sum_{i=1}^k Z_i$, 
where $Z_i = \Pi_i+\Pi_i^\top - \frac{2}d \chi_d \chi_d^\top$ and $\Pi_i$ is the permutation matrix of an independently sampled uniform random permutation over $[d]$. 
From the proof of \cite[Theorem~3.8]{BvH24} or by direct computation, 
$\sigma^2(Z) \leq 2k \cdot (1+\frac{1}{d-1})$. 
Now, let $\Tilde{\Pi}_1,\ldots, \Tilde{\Pi}_k$ be $(\delta,2)$-wise uniform permutation matrices that are independent from each other. 
Let $\Tilde{Z}_i = \Tilde{\Pi}_i + \Tilde{\Pi}_i^\top-\E[\Tilde{\Pi}_i] - \E[\Tilde{\Pi}^{\top}_i]$ (which satisfies the assumption \eqref{e:support-assumption}), and define
\[
\Tilde{Z} := \sum_{i=1}^k \big( \Tilde{\Pi}_i + \Tilde{\Pi}_i^\top \big) 
           - \frac{2k}{d}\chi_d\chi_d^\top =  A_0+\sum_{i=1}^k\Tilde{Z}_i
\]
where $A_0 = \E[\Tilde{Z}]= \sum_i (\E[\Tilde{\Pi_i}] + \E[\Tilde{\Pi}_i^{\top}]) -\frac{2k}{d}\chi_d\chi_d^\top $. 
By \autoref{corollary:pwu}, 
$\bignorm{\sum_i \big( \E[Z_i^2] - \E[\Tilde{Z_i}^2] \big) }\lesssim kd^2\delta$, 
thus 
\[ \sigma^2(\Tilde{Z}) = \biggnorm{\sum_{i=1}^k \E[\Tilde{Z}_i^2]}  \leq \biggnorm{\sum_{i=1}^k \E[Z_i^2] } + \biggnorm{\sum_{i=1}^k \big( \E[Z_i^2] - \E[\Tilde{Z_i}^2] \big) } \lesssim k + kd^2 \delta. \] 
Similarly, it follows from \autoref{corollary:pwu} that $\nu(\Tilde{Z})^2 = \bignorm{\cov(\Tilde{Z})} \lesssim \frac{k}{d} + kd^2\delta$
and $\norm{A_0}\leq kd\delta$.
    
Applying \autoref{theorem:norm-uni-simple} to the random matrix model $\Tilde{Z}$ with $n=k$ and $\rho = O(1)$, we can efficiently find a matrix $Z'\in \supp(\Tilde{Z})$ with $Z' = A- \frac{2k}{d} \chi_d \chi_d^\top$ such that
\[
\norm{Z'} \leq kd\delta + 2\sqrt{2k} + O\Big((k+kd^2 \delta)^{\frac14} \cdot \Big(\frac{k}{d}  + k d^2 \delta\Big)^{\frac14}\log^{\frac34}d + (k+kd^2 \delta)^{\frac13} \cdot \log^{\frac23}d + \log d \Big).
\]
By taking $\delta$ to be in the order of $1/\poly(k,d)$, we ensure that only the error term $k^{\frac13} \cdot \log^{\frac23}d$ dominates and the norm bound follows.
The runtime is polynomial in $k$ and $d$ by \autoref{theorem:pwu}.
\end{proofof}

\subsubsection{Random Lifts}\label{section:lifts}

A more general model of constructing expanders is by starting with a well-expanding base graph and taking a large lift of the base graph.

\begin{definition}[Lift of Graphs]
    Let $G = ([d_0],E)$ be a graph. A graph $H = ([d_0]\times [m], E')$ is an $m$-lift of $G$ if its edge set is comprised of a set of matchings where for each $(u,v)\in E$, $H$ has a perfect matching between $\{u\}\times [m]$ and $\{v\}\times [m]$.
\end{definition}

Generally, we can take lifts of graphs with multi-edges or self loops. In this view, the permutation model is simply a $d$-lift of a graph with a single vertex and $k$ self-loops. 
However, we restrict our attention to simple base graphs for sake of simplicity.

It can be readily verified that if $A_G$ and $A_H$ are the adjacency matrices of $G$ and $H$ respectively and $v$ is an eigenvector of $A_G$, 
then $v\otimes \frac{1}{\sqrt{m}}\chi_m$ is also an eigenvector of $A_H$ with the same eigenvalue.
 By removing these eigenvectors from $A_H$, the remaining matrix $A_H - A_G\otimes \frac{1}{m}\chi_m\chi_m^\top$ has the set of new eigenvalues introduced by the lift. 
A lift of a $k$ regular graph is called a ``Ramanujan lift'' if 
\[
\Bignorm{A_H - A_G\otimes \frac{1}{m}\chi_m\chi_m^\top}\leq 2\sqrt{k-1}.
\] 
In \autoref{section:signing}, we studied random edge signings of a regular graph. A key reason why this model is important is that the eigenvalues of the signed adjacency matrix of the base graph are exactly the new eigenvalues introduced by a corresponding 2-lift~\cite{BL06}. 
Thus, \autoref{theorem:2-lift} implies that for any $k$-regular graph $G$ with degree $\Omega(\log^4{d})$, 
we can deterministically compute a near-Ramanujan 2-lift of $G$. 

{\bf Previous Work}:
Bordenave~\cite{Bor20} showed that for any fixed $k$-regular base graph $G$, a random $m$-lift of $G$ is a $2\sqrt{k-1}+O(\frac{\log\log{m}}{\log{m}})$ near-Ramanujan lift with high probability, as long as $k\lesssim \frac{\log{m}}{\log\log{m}}$. 
Mohanty, O'Donnell, and Paredes~\cite{MOP20} gave a derandomization of Bordenave's result by replacing a single and arbitrarily large lift with an arbitrarily long sequence of smaller lifts. 
In particular, they start by applying Bordenave's method on a small enough lift size that can be derandomized by ``brute-force'' using $(\delta, q)$-wise uniform permutations. 
Then, they take iterated 2-lifts of this ``seed graph'', 
which are derandomized using $(\delta,q)$-wise independent bits, 
for $q$ on the order of $\sqrt{k}\log{d}$.  
This was a significant improvement of a similar approach by Bilu and Linial~\cite{BL06}. 
We note that these constructions in~\cite{MOP20} have runtime on the order of $d^{O(\sqrt{k})}$ and so they are only in polynomial time for constant $k$.

Brailovskaya and van Handel~\cite{BvH24} showed that the sharp matrix concentration inequalities can be applied in the complementary regime of $k\geq \polylog{m}$ to construct near-Ramanujan lifts. 
The proof of Bordenave does not work in this regime because it requires conditioning on large neighborhoods of each vertex having at most one cycle, which no longer holds with high probability in the dense setting. 

{\bf Our Work}:
By derandomizing the concentration inequalities in~\cite{BvH24}, we obtain explicit constructions of near-Ramanujan lifts when $k\gtrsim \log^6{d_0} \cdot \log^4d$ in polynomial time. 
In fact, we will derandomize two more general models, both of which will capture this result as a special case. 
The first model we derandomize is of lifting graphs with arbitrary degrees. 
The second model we derandomize is applying group-based lifts to regular graphs.
We will present the precise results in the next two subsubsections.

\subsubsection*{General Near-Ramanujan Lifts}

The notion of Ramanujan graphs can be generalized to irregular graphs,
by comparing the spectrum of a graph with that of its universal cover tree $\mathbbm{T}(G)$. 

\begin{definition}[Non-Regular Ramanujan Graphs]
Let $G$ be a (possibly irregular) graph and $A_{\mathbb{T}(G)}$ be the infinite dimensional adjacency operator of its universal cover tree. 
Then $G$ is called Ramanujan if its non-trivial eigenvalues are bounded in absolute value by $\bignorm{A_{\mathbb{T}(G)}}$.
\end{definition}

If $G$ is $k$-regular, then $\mathbb{T}(G)$ is the infinite $k$-regular tree, and it is well known that $\norm{A_{\mathbb{T}(G)}} = 2\sqrt{k-1}$. 
If $G$ is an irregular graph with maximum degree $k_{\max}$, then $\sqrt{k_{\max}}\leq \norm{A_{\mathbb{T}(G)}}\leq 2\sqrt{k_{\max}-1}$, where the upper bound is due to $\mathbb{T}(G)$ being a subgraph of the $k_{\max}$-regular infinite tree, and the lower bound is due to the star graph over $k_{\max}+1$ vertices being a subgraph of $\mathbb{T}(G)$. 
Since the work of Friedman~\cite{Fri08}, the construction of irregular near-Ramanujan graphs has been a long standing question.

Bordenave and Collins~\cite{BC19} resolved this question by showing that if $H$ is a random $m$-lift of a base graph $G$, 
then the set of new eigenvalues of $H$ are bounded by $\norm{A_{\mathbb{T}(G)}} + o_m(1)$ with high probability, 
as long as the size of the base graph is at most $O(\log m/\log\log{m})$. 
O'Donnell and Wu~\cite{OW20} provided a derandomization of this result,
but their method is an extension of those in \cite{MOP20}
and thus does not apply in the non-constant degree regime.

Brailovskaya and van Handel~\cite[Theorem 3.13]{BvH24} showed that the near-Ramanujan lift property continues to hold in the complementary $m\gtrsim \log^4(md_0)$ regime, and we derandomize their result.

\begin{theorem}[Deterministic General Near-Ramanujan Lifts] \label{theorem:lift}
Let $G = ([d_0],E)$ be a simple graph with maximum degree $k_{\max}$. 
Let $d=m \cdot d_0$ and suppose $k_{\max}\gtrsim \log^4d \cdot \log^6({|E|}/k_{\max})$. 
There is a polynomial time deterministic algorithm to compute $H$, an $m$-lift of $G$ over $d$ vertices, such that 
\[
\norm{A_H - \Big(A_{G}\otimes \frac{1}{m}\chi_m\chi_m^\top\Big)}\leq \norm{A_{\mathbb{T}(G)}} \cdot \Big(1+O\Big(\frac{\log (|E|/k_{\max})\cdot \log^{\frac23}d}{k_{\max}^{\frac16}}\Big) \Big).
\] 
In particular, when $G$ is $k$-regular, then
\[
\norm{A_H - \Big(A_{G}\otimes \frac{1}{m}\chi_m\chi_m^\top\Big) }\leq 2\sqrt{k} \cdot \Big(1+O\Big(\frac{\log d_0\cdot \log^{\frac23}d}{k^{\frac16}}\Big)\Big).
\]
\end{theorem}

{\bf Proof:}
The proof of~\cite[Theorem 3.13]{BvH24} was based on the formulas of Lehner~\cite{Leh99} to compare the maximum eigenvalue of $\Xf$ with that of $A_{\mathbb{T}(G)}$.

\begin{lemma}[{\cite[Lemma 9.3 and 9.7]{BvH24}}] \label{lemma:free-comparison}
Let $G = (V,E)$ be a simple graph with maximum degree $k_{\max}$.
Let $\{\Pi_e \mid e\in E\}$ be a set of independent uniform random permutation matrices. 
Consider the random matrix model 
\[
Z = \sum_{e=uv\in E} \bigg( \chi_u\chi_v^\top \otimes \Big(\Pi_{e}-\frac1m \chi_m\chi_m^\top \Big) + \chi_v\chi_u^\top \otimes \Big(\Pi_e^\top - \frac1m \chi_m\chi_m^\top \Big) \bigg),
\]
which is the adjacency matrix of a uniform random $m$ lift of $G$ restricted to the space orthogonal to the eigenspace of $A_G$. 
Let $\Xf$ be the free semicircular model corresponding to the matrix $Z$. Then 
\[
\norm{\Xf}\leq \norm{A_{\mathbb{T}(G)}} + O\big(k_{\max}^{\frac14}\big).
\]
\end{lemma}

\autoref{lemma:free-comparison} shows that for graphs whose maximum degree is moderately large, the spectral radius of the universal cover is well-approximated by that of the semicircular matrix $\Xf$. 
To derandomize this result, we need to show that the norm of the free model $\Xf$ does not change much if we replace the uniform random permutations with approximate pairwise uniform random permutations.

\begin{lemma}[Random Lifts by Pairwise Uniform Random Permutations] \label{lemma:fc-pairwise}
Let $Z$ be the random lift model as in \autoref{lemma:free-comparison}.
Let $\Tilde{Z}$ be the same model except each uniform permutation matrix $\Pi_e$ is replaced by a $(\delta,2)$-uniform permutation matrix $\Tilde{\Pi}_e$. 
Let $\Xf$ and $\tXf$ be the free models of $Z$ and $\Tilde{Z}$ respectively. 
Then 
\[
\bignorm{\tXf} \leq \norm{\Xf} + O\big(\sqrt{\delta} \cdot d^{\frac32}+\delta m |E|\big).
\]
\end{lemma}
\begin{proof}
From Lehner's formula in \autoref{lemma:lehner}, for all $Y\succ 0$,
\begin{eqnarray}
\bignorm{\tXf} & \leq & \max_{\eps \in \{\pm 1\}} \norm{ \eps \E\big[\Tilde{Z}\big]+Y^{-1} + \E\big[(\Tilde{Z}-\E[\Tilde{Z}]) \cdot Y \cdot (\Tilde{Z}-\E[\Tilde{Z}])\big]} \nonumber \\
    & \leq & \norm{Y^{-1}+\E\big[(\Tilde{Z}-\E[\Tilde{Z}]) \cdot Y \cdot (\Tilde{Z}-\E[\Tilde{Z}])\big]} + O(\delta |E|m), \label{eq:perturbation1}
\end{eqnarray}
where the second inequality follows by applying \autoref{corollary:pwu} to bound $\|\E[\Tilde{Z}]\| \leq 2|E|m\delta$. To prove the lemma, 
our plan is to find a matrix $Y$ so that $\norm{Y^{-1} + \E[ZYZ]}$ is close to $\norm{\Xf}$ and $Y$ itself is bounded. 
To find such a $Y$, we first note that since $\E[Z]=A_0=0$, 
the odd moments of $\Xf$ vanish, which follows by \autoref{lemma:sc-joint-moments} and the fact that there is no non-crossing pairing partition for odd moments. Therefore, the spectral distribution of $\Xf$ is symmetric. Thus, $\norm{\Xf} = \lambda_{\max}(\Xf)$. 
The key lemma we will use is the following relationship between matrix-valued transforms of $\Xf$. 

\begin{lemma}[Remark 2.3 in \cite{Leh99}]\label{lemma:gr-transform}
Suppose $\lambda > \lambda_{\max}({\Xf})$ and $Y = \varphi[(\lambda 1-\Xf)^{-1}] $. Then
\[
Y^{-1} + \E[ZYZ] = \lambda I.
\]
\end{lemma}

We set $\lambda = \lambda_{\max}(\Xf)+\eps$ for some $\eps$ to be fixed later and set $Y = \varphi[(\lambda 1-\Xf)^{-1}]$. 
By \autoref{lemma:gr-transform}, we have $Y^{-1} + \E[ZYZ] = (\lambda_{\max}(\Xf) + \eps)I$. 
Also, by our choice of $\lambda$, we have $\tr(Y) = \tot((\lambda 1 - \Xf)^{-1}) \leq \frac{1}{\eps}$. 
To carry out the perturbation analysis, note that
\begin{eqnarray}
& & \norm{Y^{-1}+ \E\big[(\Tilde{Z}-\E[\Tilde{Z}]) \cdot Y \cdot (\Tilde{Z}-\E[\Tilde{Z}])\big] - (\lambda_{\max}(\Xf)+\eps)I} \nonumber \\
&= & \norm{Y^{-1}+ \E\big[(\Tilde{Z}-\E[\Tilde{Z}]) \cdot Y \cdot (\Tilde{Z}-\E[\Tilde{Z}])\big]  - (Y^{-1}+\E[ZYZ])} \nonumber 
\\ 
&= & \norm{\E\big[(\Tilde{Z}-\E[\Tilde{Z}]) \cdot Y \cdot (\Tilde{Z}-\E[\Tilde{Z}])\big]  -\E[ZYZ]}. \label{eq:perturbation2}
\end{eqnarray}
To bound this error, we let $u_1,\ldots, u_d$ be the columns of $Z$ and $v_1,\ldots, v_d$ be the columns of $\Tilde{Z}-\E[\Tilde{Z}]$. 
Then the absolute value of the $(i,j)$-th entry of the matrix $\E\big[(\Tilde{Z}-\E[\Tilde{Z}]) \cdot Y \cdot (\Tilde{Z}-\E[\Tilde{Z}])\big]  -\E[ZYZ]$ is exactly
\begin{equation} \label{eq:perturbation3}
\big|\E\big[u_i^\top Yu_j - v_i^\top Yv_j\big]\big| 
= \big|\bigip{\E[u_ju_i^\top -v_jv_i^\top]}{Y}\big| 
\leq \norm{\E[u_ju_i^\top -v_jv_i^\top]} \cdot \Tr(Y).
\end{equation}
To bound $\norm{\E[u_ju_i^\top -v_jv_i^\top]}$, we will show that each of its entries is small in absolute value, and apply the row $\ell_1$-bound. In particular, we see that
\begin{eqnarray*}
\E[u_i(k) \cdot u_j(\ell) - v_i(k) \cdot v_j(\ell)] 
&=& \E[Z(i,k) \cdot Z(j,\ell)] - \E\big[(\Tilde{Z}(i,k) - \E[\Tilde{Z}(i,k)]) \cdot (\Tilde{Z}(j,\ell) - \E[\Tilde{Z}(j,\ell)])\big]
\\
&=& \cov(Z)(i,k,j,\ell) - \cov(\Tilde{Z})(i,k,j,\ell).
\end{eqnarray*}
Since $G$ is a simple graph, each possible edge of its lifted graph comes from exactly one corresponding entry of $\Pi_e$ for some $e\in E$. 
Thus, by $(\delta,2)$-wise uniformity, the entries of $\cov(Z)$ and $\cov(\Tilde{Z})$ differ by at most $O(\delta)$, which means $|\E[u_i(k) \cdot u_j(\ell) - v_i(k) \cdot v_j(\ell)]| \lesssim \delta$ for all $i,j,k,\ell$. 
Together with the fact that $\Tr(Y) = d \cdot \tr(Y) \leq d/\eps$, it follows from \eqref{eq:perturbation3} and the row $\ell_1$-bound that
$|\E[u_i^\top Y u_j - v_i^\top Y v_j]| \leq d^2\delta/\eps$. 
Therefore, by another row $\ell_1$-bound,
\begin{eqnarray*}
& & \norm{\E\big[(\Tilde{Z}-\E[\Tilde{Z}]) \cdot Y \cdot (\Tilde{Z}-\E[\Tilde{Z}])\big]  -\E[ZYZ]} \leq \frac{d^3\delta}{\eps}\\
&\implies& \quad
\norm{ Y^{-1}+ \E[(\Tilde{Z}-\E[\Tilde{Z}])Y(\Tilde{Z}-\E[\Tilde{Z}])]} \leq \lambda_{\max}(\Xf) + \eps + \frac{d^3\delta}{\eps},
\end{eqnarray*}
where the implication is by \eqref{eq:perturbation2}.
Taking $\eps$ to be $\sqrt{\delta } \cdot d^{\frac32}$ then completes the proof.
\end{proof}

Now, we obtain explicit constructions of general near-Ramanujan lifts in the moderately large degree setting by applying \autoref{t:norm-uni}.

\begin{proofof}{\autoref{theorem:lift}}
Let $Z$ be the projected random lift adjacency matrix as in \autoref{lemma:free-comparison} and $\Xf$ be its corresponding free model. 
Let $\Tilde{Z}$ be the $(\delta,2)$-wise approximation to $Z$ as in \autoref{lemma:fc-pairwise} for $\delta\leq 1/\poly(d)$, thus satisfying the assumption~\eqref{e:support-assumption}. 
Then $\Tilde{Z} = A_H - (A_G\otimes \frac1m \chi_m\chi_m^\top)$. 
To deterministically construct the lift $H$, 
it suffices to deterministically sample some $Z'\in \supp(\Tilde{Z})$ with bounded operator norm. 
To give our construction, 
we first bound $\sigma(\Tilde{Z})$ and $\nu(\Tilde{Z})$. 
Let $\Pi$ be a (fixed) permutation matrix. 
Note that 
\begin{eqnarray*}
&& \Big(\chi_u\chi_v^\top \otimes \Big(\Pi-\frac1m \chi_m\chi_m^\top\Big) + \chi_v\chi_u^\top \otimes \Big(\Pi^\top-\frac1m \chi_m\chi_m^\top \Big)\Big)^2 
\\
&=& \chi_u\chi_u^\top\otimes  \Big(\Pi\Pi^\top - \frac1m\chi_m\chi_m^\top \Big) +  \chi_v\chi_v^\top\otimes  \Big(\Pi^\top\Pi - \frac1m\chi_m\chi_m^\top \Big) 
\\
&=& \big(\chi_u\chi_u^\top + \chi_v\chi_v^\top\big) \otimes \Big(I_m-\frac{1}{m}\chi_m\chi_m^\top\Big),
\end{eqnarray*}
where we used the fact that $\Pi \chi_m = \Pi^\top \chi_m= \chi_m$ for the first equality. 
It then follows that
\[
\E[Z^2] 
= \sum_{uv \in E}(\chi_u\chi_u^\top + \chi_v\chi_v^\top) \otimes \Big(I_m-\frac{1}{m}\chi_m\chi_m^\top\Big) 
= D\otimes \Big(I_m-\frac{1}{m}\chi_m\chi_m^\top\Big),
\]
where $D$ is the diagonal degree matrix of $G$. 
Therefore, $\sigma^2(Z) = k_{\max}$.
By applying \autoref{corollary:pwu} as in the proof of \autoref{theorem:perm}, 
we have $\sigma^2(\Tilde{Z}) \leq k_{\max}+O(\delta |E|m^2) \leq 2k_{\max}$ for small enough $\delta$. 
By \autoref{corollary:pwu}, we also have that $\|\cov(\Tilde{\Pi}_e)\| \lesssim \frac{1}{m} + m^2\delta \lesssim \frac{1}{m}$ for each $e\in E$ for small enough $\delta$.
Moreover, since $G$ is a simple graph, the matrices $\chi_u\chi_v^\top \otimes \Pi_{u,v} + \chi_v\chi_u^\top \otimes \Pi_{u,v}^\top$ are independent, entry-wise disjoint, and have norm at most~$2$. 
Thus, $\nu^2(\Tilde{Z}) \lesssim \frac{1}{m}$ and $\rho \lesssim 1$. 
Note that $n\rho(Z)^2/\sigma(Z)^2 = |E|/k_{\max}$.

By applying \autoref{t:norm-uni} to the random matrix model $\Tilde{Z}$, 
we obtain a polynomial time deterministic algorithm to find $Z'\in \supp(\Tilde{Z})$ such that
\[
\norm{Z'} \leq \bignorm{\tXf}+ \log \Big( \frac{|E|}{k_{\max}} \Big) \cdot O\Big(\frac{\log^{\frac34}d}{m^{\frac14}}\cdot k_{\max}^{\frac14} + k_{\max}^{\frac13} \cdot \log^{\frac23}{d} + \log{d}\Big).
\]
Finally, by applying \autoref{lemma:fc-pairwise} and then \autoref{lemma:free-comparison}, we have
\[
\norm{\tXf} \leq \norm{\Xf} + O(\sqrt{\delta}d^{3/2} +m|E|\delta) \leq \norm{A_{\mathbb{T}(G)}} + O\big(k_{\max}^{\frac14} + \sqrt{\delta}d^{3/2} + m |E| \delta \big). 
\]
Thus, by taking $\delta \asymp 1/\poly(d,m,|E|)$, and using the fact that $\norm{A_{\mathbb{T}(G)}} \asymp \sqrt{k_{\max}}$, we conclude that
\[
\norm{Z'} \leq \norm{A_{\mathbb{T}(G)}} \cdot \Big(1+  \log \Big( \frac{|E|}{k_{\max}} \Big) \cdot O\Big(\frac{\log^{\frac34}d}{k_{\max}^{\frac14} \cdot m^{\frac14}}+ \frac{\log^{\frac23}{d}}{k_{\max}^{\frac16}} + \frac{\log{d}}{k_{\max}^{\frac12}}\Big)+ O\Big(\frac{1}{k_{\max}^{\frac14}}\Big)\Big).
\]
Then, if we assume $k_{\max} \gtrsim \log^6d_0 \cdot \log^4{d}$, then the term with $k_{\max}^{\frac16}$ dominates.
The runtime is polynomial by the choice of $\delta$ and \autoref{theorem:pwu}.
\end{proofof}

\subsubsection*{Group-Based Lifts} 

Another model of lifts that have recently garnered attention is the group-based lift, which can be viewed as a generalization of Cayley graphs. 

\begin{definition}[Group-Based Lifts] 
Let $G = ([d_0],E)$ be a graph. 
Let $\Gamma$ be a finite group of size $m$. 
Let $H = ([d_0]\times \Gamma, E')$ be an $m$-lift of $G$. 
We say that $H$ is a $\Gamma$-lift if for each $e = uv\in E$, 
there is a group element $h_e\in \Gamma$ such that the matching between $\{u\}\times \Gamma$ and $\{v\}\times \Gamma$ in $H$ is given by the edge set $\{(u,\alpha), (v,h_e\alpha): \alpha\in \Gamma\}$.
\end{definition}

In \cite{ACKM19}, it was shown that if $G$ is a $k$ regular $\lambda$-spectral expander over $d_0$ vertices with $k\lesssim \sqrt{d_0/\log{d_0}}$ and $\Gamma$ is the abelian group $\Z/m\Z$, 
then a uniform random group lift of size $m$ has spectral radius at most $O(\lambda)$.
This result was crucially used in the construction of quantum and classical LDPC codes in \cite{PK22}. 
Jeronmino, Mitall, Paredes, O'Donnell, and Tulsiani~\cite{JMO+22} gave explicit constructions of group-based lifts of $k$-regular graphs for constant $k$, 
and gave explicit constructions of quantum and classical LDPC codes as a consequence. 

In particular, in the regime where $m \leq \exp({O(\frac{\eps^2\log{d_0}}{k})})$, their algorithm constructs deterministic Abelian lifts of the base graph $G$ whose spectral radius is bounded by $2\sqrt{k-1}+\eps$. 
Their methods are extensions of those in \cite{MOP20}, and in the regime where $k$ is poly-logarithmic in $md_0$, are no longer of polynomial time nor providing optimal bounds. 
The results in \cite{BvH24} naturally capture the complementary dense regime for random group-lifts. 
Our algorithms give the following deterministic construction of near-Ramanujan group lifts in this regime. 
We note that our bounds do not require $\Gamma$ to be Abelian (or have any specific structure).

\begin{theorem}[Deterministic Group-Based Lifts]\label{theorem:group}
Let $G = ([d_0],E)$ be a $k$-regular simple graph and $\Gamma$ be a group of size $m$. 
Let $d = m \cdot d_0$. 
If $k\gtrsim \log^{4}d\cdot \log^6{d_0}$, then there is a polynomial time deterministic algorithm to compute a $\Gamma$-lift of $G$ over $d:= m \cdot d_0$ vertices such that
\[
\Bignorm{A_H - \Big(A_G\otimes \frac1m \chi_m \chi_m^\top\Big)} 
\leq 2\sqrt{k} \cdot \Big(1+O\Big(\frac{\log{d_0}\cdot \log^{\frac23}d}{k^{\frac16}} \Big)\Big).
\]
\end{theorem}
\begin{proof}
Consider the model $Z = \sum_{e\in E}Z_e$, where for $e=uv$, 
\[
Z_e = \chi_u\chi_v^\top \otimes \Big(\Pi_e - \frac1m \chi_m \chi_m^\top\Big) 
+ \chi_v\chi_u^\top \otimes \Big(\Pi_e^\top - \frac1m \chi_m \chi_m^\top\Big),
\]
where $\Pi_e$ is constructed as follows: 
Let $\{h_e\}_{e\in E}$ be an i.i.d.~uniform random sample from $\Gamma$, 
and let $\Pi_e$ be the matrix of the permutation on $\Gamma$ induced by left multiplication by $h_e$. 
Note that $\E[\Pi_e] = \frac1m\chi_m\chi_m^\top$, as each element is still equally likely to match to any other element by the group structure. 
As in the proof of \autoref{theorem:lift}, we have $\E[Z^2] = kI_{d_0} \otimes( I_m - \frac1m \chi_m\chi_m^\top)$, since this calculation did not depend on the distribution of each $\Pi_e$ except that its mean is $\frac1m \chi_m\chi_m^\top$. 
This implies that $\sigma(Z)^2 = k$ and $\rho \leq 2$.
 
To bound the quantity $\nu(Z)$, we first note that since the $Z_e$'s are entry-wise disjoint, we have $\nu(Z) = \max_{e\in E}\nu(Z_e)$. 
To bound each $\nu(Z_e)$, let $\alpha,\beta,\gamma,\zeta \in \Gamma$ be group elements, and $h$ be a uniformly random selected element in $\Gamma$. If both $h\alpha = \beta$ and $h\gamma = \zeta$, then we must have $\alpha\beta^{-1}\zeta=\gamma$ by rearranging and thus
\[
\Pr[h\alpha = \beta\cap h\gamma=\zeta] = \Pr[\alpha=h^{-1}\beta] \cdot \1_{\{\alpha\beta^{-1}\zeta=\gamma\}} 
= \frac{1}{m}\cdot \1_{\{\alpha\beta^{-1}\zeta=\gamma\}}. 
\]
Thus, for each fixed pair $\alpha,\beta$, there are exactly $m$ pairs $\gamma,\zeta$ for which $\Pr[h\alpha = \beta\cap h\gamma=\zeta] = \frac1m$, while $\Pr[h\alpha = \beta\cap h\gamma=\zeta] = 0$ for all other $\gamma,\zeta$. 
This means that for all $\alpha,\beta \in\Gamma$,
\[
\sum_{\gamma,\zeta \in\Gamma} \Big|  \Pr[h\alpha = \beta\cap h\gamma=\zeta] - \frac{1}{m^2}  \Big| 
\leq m\Big(\frac{1}{m} - \frac{1}{m^2}\Big) + m(m-1)\cdot\frac{1}{m^2} \lesssim 1.
\]
Thus, we have $\|\cov(\Pi_e)\| \lesssim 1$ for all $e$. 
Since the base graph is simple, the matrices $\chi_u\chi_v\otimes \Pi_{u,v}$ have disjoint support for all $uv\in E$. 
Thus $\nu(Z) \lesssim 1$. 

Since each $\Pi_e$ only has support size $m$, we can directly apply \autoref{theorem:norm-uni-simple} to find an outcome $Z'\in \supp (Z)$ satisfying
\[
\norm{Z'} \leq 2\sqrt{k} + O\big(\log d_0 \cdot\big( k^{\frac14} \cdot \log^{\frac34}d   + k^{\frac13} \cdot \log^{\frac23}d \big)\big).
\]
The conclusion then follows by noting that the second term dominates when $k\gtrsim \log^{4}d\cdot \log^{6}d_0$.
\end{proof}

By applying \autoref{theorem:group} to the group $\Z/m\Z$, 
we can also construct regular $m$-lifts of expansion $2\sqrt{k} \cdot (1+O((\log d_0\log^{\frac23}{d})/k^{\frac16}))$ for any $m$ in our regime. 
In fact, this is a simpler construction than that using \autoref{theorem:lift}, since we do not need to apply approximate pairwise uniform permutations. 
We also note that the $\log d_0$ factor can be removed using a similar method as in \autoref{theorem:2-lift}, where we decompose the base graph into $k+1$ edge-disjoint matchings and sample pairwise independent group elements for each matching. 
We omit the details on these improvements for the sake of brevity.

\subsection{Technical Outline} \label{ss:outline-uni}

To derandomize \autoref{theorem:bvh24}, 
consider the following natural iterative algorithm. 
Start with the free model $\Xf = A_0 \otimes 1 + \sum_{i=1}^n X_i$.
In each iteration $t$, we select an index $i_t$ to replace the semicircular matrix $X_{i_t}$ with a deterministic $Z'_{i_t} \otimes 1$ where $Z'_{i_t} \in \supp(Z_{i_t})$.
Through the whole process, we maintain a partially derandomized semicircular matrix
\[
    X_{f,t} = A_0 \otimes 1 + \sum_{i \not\in \mathcal{I}_t} Z'_i \otimes 1 + \sum_{j \in \mathcal{I}_t} X_j,
\]
where $\mathcal{I}_t \subseteq [n]$ is the set of indices that have not been derandomized at iteration $t$ yet.
After $n$ iterations, we end up with a deterministic $X_{f,n} = (A_0 + \sum_{i=1}^n Z'_i) \otimes 1$, and $Z' = A_0 + \sum_{i=1}^n Z'_i = \varphi[X_{f,n}]$ is our final solution.

The key in the derandomization process is to select an appropriate $i_t$ and $Z'_{i_t}$ in each iteration so that the deviation of $X_{f,t}$ from the free model $\Xf$ is upper bounded.
In order to control the deviation, we once again use a potential function: $\Phi: \M_d\otimes \A \rightarrow \C$, which controls the target spectral statistics.
In each iteration, we would like to bound the quantity
\[
\frac{1}{n-t} \cdot \E_{Z_i} \bigg[ \sum_{i\in \mathcal{I}_t} \Phi(X_{f,t} + Z_i \otimes 1 - X_i) \bigg],
\]
which is the expected potential value when we select an index $i_t \in \mathcal{I}_t$ uniformly at random and then select a deterministic $Z'_{i_t}$ following the distribution of $Z_i$.
Since the support of $Z_i$ is polynomial in $d$, we can always efficiently select $i_t$ and $Z'_{i_t}$ such that
\[
\Phi(X_{f,t+1}) = \Phi(X_{f,t} + Z'_{i_t} \otimes 1 - X_{i_t}) \leq \frac{1}{n-t} \cdot \E_{Z_i} \bigg[ \sum_{i\in \mathcal{I}_t} \Phi(X_{f,t} + Z_i \otimes 1 - X_i) \bigg].
\]

Therefore, the main task is to control the expected potential under a random perturbation $Z_i \otimes 1 - X_i$ for $i \in \mathcal{I}_t$.
A natural idea would be using a similar analysis as in \autoref{section:discrepancy} and \autoref{section:resolvent} to bound the change of potential.
However, there are two differences in the iterative swapping algorithm that require new ideas.

For the first one, bounding the perturbations in \autoref{section:discrepancy} and \autoref{section:resolvent} can be naturally reinterpreted as taking the difference between a finite dimensional random update $\sqrt{\eta}A(y)\otimes 1$ and an update with $\sqrt{\eta}\bar{X}_{\rm free}'$, a free copy of $\XF = \Xf - A_0 \otimes 1$. However, in the present setting, we are forced to directly work with the update $Z_i \otimes 1 - X_i$, where the $X_i$ term is not free from our current solution $X_{f,t}$.
As long as this $X_i$ is present in our update, we cannot directly apply the crucial tool \autoref{lemma:free-moments} to analyze the second order error term.
To overcome this issue, we will use the ``semicircular integration-by-parts'' formula in \autoref{proposition:free-ibp} to expand the perturbation error terms into a series of terms where we replace the occurrences of $X_i$ with free copies of it, so as to control the error terms.

For the second one, the perturbation sizes in \autoref{section:discrepancy} and \autoref{section:resolvent} are controlled by the step length $\eta$, which could be made arbitrarily small if necessary (but with the trade-off of increasing the running time). 
Because of this, it suffices to estimate the potential function with a second-order approximation and bound all the higher order error terms with a naive worst case bound via \autoref{corollary:general-holder}.
However, the perturbation $Z_i \otimes 1 - X_i$ in this section may not be small enough in the worst case, and so using a worst case bound to control the higher order error terms would not work anymore.
To overcome this, we follow a similar idea as in~\cite[Proposition~5.1]{BvH24} to control the higher order error terms in {\rm expectation} via an infinite dimensional trace inequality in \autoref{proposition:trace-holder}, but this leads to even more involved calculations than those in \autoref{section:discrepancy} and \autoref{section:resolvent}.

\begin{proposition}[Infinite Dimensional Trace Inequality] \label{proposition:trace-holder}
Let $Z_1, \ldots, Z_n$ be $d$-dimensional centered self-adjoint random matrices such that $\max_i \|Z_i\| \leq \rho$ with probability one and $\|\sum_{i=1}^n \E[Z_i^2]\| = \sigma^2$.
Let $\A$ be the $C^*$-algebra generated by all semicircular elements used, and $X_1, \ldots, X_n \in \M_d(\C) \otimes \A$ be centered semicircular matrices where each $X_i$ has the same covariance profile as $Z_i$ (i.e., satisfying \eqref{eq:cov-profile}).    
    
Let $m>2$, and $Y_1, \ldots, Y_m \in \M_d(\C)\otimes \A$, be a family of (not necessarily free) operators.
Suppose there exist integers $k_1, \ldots, k_m \geq 0$, and positive, self adjoint operator $Y\in \M_d(\C)\otimes \A$, such that  $\sum_{i=1}^m k_i = p$ and $|Y_j| \preccurlyeq Y^{k_j}$ for each $j \in [m]$.

For each $i\in[n]$ and $j\in[m]$, let $Z_{i,j}$ be a
self-adjoint operator. For each $j\in[m]$, suppose that either
$Z_{i,j}=Z_i\otimes1$ for every $i\in[n]$, or
$Z_{i,j}$ has the same distribution as $X_i$ and is free from
$Y_1,\ldots,Y_m$ for every $i\in[n]$.
Then, it holds that
\[
\bigg| \sum_{i=1}^n \E \big[ \tr\otimes\tau(Z_{i,1}Y_1\cdots Z_{i,m}Y_m) \big] \bigg| 
\leq \sigma^2 \cdot (2\rho)^{m-2} \cdot \tr\otimes\tau(Y^{p}),
\]
where the expectation is taken over those $Z_{i,j}$'s that are equal to $Z_i\otimes 1$.
\end{proposition}

While this bound is very similar to that in~\cite[Proposition~5.1]{BvH24}, the way in which it is applied in our argument is quite different. The form in the LHS of our expression arises from the Taylor expansion of the potential update formula in the random-swap algorithm. The $Z_i$ and $X_i$ terms arise from the update $Z_i-X_i$ corresponding to applying the swap to index $i$. Crucially, we use the simple observation that if $\norm{Z_i}\leq \rho$ then $\norm{X_i}\leq 2\rho$ by \autoref{lemma:semicircular-norm}. If we replaced $X_i$ in the above bound with finite dimensional $G_i$, where $G_i$ is a \textit{Gaussian} matrix with the same covariance profile as $Z_i$, then the bound above does not hold anymore since the spectrum of $G_i$ is not bounded almost surely.
We will provide a full proof of \autoref{proposition:trace-holder} in \autoref{a:trace-inequality} for completeness.

\subsubsection{Deterministic Moment Universality}

To elaborate, we take $\Phi(X) = \tot(X^{2p})$ as an example to demonstrate the technical issues in analyzing the moments.
The crucial quantity that we need to control is
\[
\E\big[ \tot(X_{f,t} + Z_i \otimes 1 - X_i)^{2p}\big], \qquad \forall i \in \mathcal{I}_t.
\]
After expanding $(X_{f,t} + Z_i \otimes 1 - X_i)^{2p}$, we obtain $3^{2p}$ terms where each is of the form
\[
\E\big[\tot (F(X_{f,t}, X_i, Z_i \otimes 1))\big],
\]
where $F(X_{f,t}, X_i, Z_i \otimes 1)$ is a multivariate non-commutative monomial in $X_{f,t}$, $X_i$, and $Z_i \otimes 1$.
Brailovskaya and van Handel~\cite{BvH24} provided a trace inequality (\cite[Proposition~5.1]{BvH24}, which plays a key role in their work) to control a similar quantity
\[
\bigg| \sum_{i=1}^n \E\big[ \tr(A_{i1} B_1 A_{i2} B_2 \cdots A_{ik} B_k)\big] \bigg|,
\]
where $A_{ij}$'s are a family of (possibly dependent) random matrices with identical distribution for each fixed $i$, and $B_1, \ldots, B_k$ is another family of (possibly dependent) random matrices that are independent from $A_{ij}$'s.
This is essentially what we need, but with two issues.
\begin{itemize}
    \item The first one is that their inequality is for finite dimensional random matrices. We need to use a version for infinite dimensional operators, i.e., \autoref{proposition:trace-holder}.

    \item The second one is a more serious issue. 
    In our setting, the monomial $F(X_{f,t}, X_i, Z_i \otimes 1)$ has two types of semicircular matrices that are not freely independent. 
    In particular, $X_{f,t}$ contains an $X_i$ that is to be removed, thus it is not freely independent from $X_i$. 
    This issue is handled via the ``semicircular integration by parts'' formula in \autoref{proposition:free-ibp}, which replaces these ``dependent'' $X_i$'s with freely independent copies of $X_i$, but requires expanding out into more higher-order terms.\footnote{We note that in the interpolation argument of \cite{BvH24}, applying the main trace inequality, proposition 5.1, also required a similar decoupling procedure, which was achieved using cumulant expansion on the $Z_i$'s instead.}.
\end{itemize}

\subsubsection{Deterministic Norm Universality}

Using the Schatten $2p$-norm bound in \autoref{t:moment-uni}, we can efficiently find matrices $Z'_1, \ldots, Z'_n$ satisfying the following bound on the operator norm by taking $p = \Theta(\log d)$:
\[
\biggnorm{A_0 + \sum_{i=1}^n Z_i'} 
\leq 
O\bigg(\biggnorm{A_0 \otimes 1 + \sum_{i=1}^n X_i}\bigg) 
+ O\Big(1+\log{\frac{n\rho^2}{\sigma^2}} \Big) \cdot \big( \sigma^{\frac12} \nu^{\frac12} \log^{\frac34} d + \sigma^{\frac23}\rho^{\frac13} \log^{\frac23} d+ \rho \log d \big).
\]
However, the dependence on $\norm{A_0 \otimes 1 + \sum_{i=1}^n X_i}$ is a constant factor away from the optimal, which is not strong enough for some applications such as constructing near-Ramanujan graphs.

To capture the spectral edge behavior, our idea is to use the barrier method developed by Batson, Spielman, Srivastava~\cite{BSS12}, where we shift a barrier $\lambda_t \in \R$ by $\delta_t > 0$ in each iteration such that $\lambda_t > \lambda_{\max}(X_{f,t})$ through the whole iterative swapping process.
Then, in the end, $\lambda_n$ serves as an upper bound on $\lambda_{\max}(X_{f,n})$ (note that it suffices to control $\lambda_{\max}$ in order to control the operator norm, see \autoref{ss:norm-uni}).
As discussed in the introduction, the barrier method only works for the free model but not for the Gaussian model as its maximum eigenvalue is not bounded.

The key in this algorithm is to select a good swap so that the shift of barrier $\delta_t$ is small in each iteration.
However, it is difficult to control the spectral edge $\lambda_{\max}$ directly as it is not a smooth quantity.
Thus, we control the alternative potential function, the $2p$-th moment of the resolvent, i.e., $\tot((\lambda_t (I \otimes 1) - X_{f,t})^{-2p})$ for some $p = \Theta(\log d)$.
We show that if the potential value is bounded by $(2\eps)^{-2p}$, then the barrier $\lambda_t \geq \lambda_{\max}(X_{f,t}) + \eps$ (see \autoref{lemma:barrier-invariance}).

Therefore, we reduce the problem to selecting an appropriate swap together with an appropriate shifting $\delta_t$ so that the potential value remains bounded.
It turns out that, if we choose a random swap $Z_i \otimes 1 - X_i$, then the key in controlling the shifting amount is in controlling the expected change of potential (see \autoref{lemma:potential-increase}) after the swap in the new barrier position $\lambda = \lambda + \delta_t$, i.e.,
\[
\E_{i, Z_i}\big[\tot\big(((\lambda + \delta_t)1 - (X_{f,t} + Z_i \otimes 1  - X_i))^{-2p}\big)\big] - \tot\big((\lambda 1 - X_{f,t})^{-2p}\big).
\]
The analysis of the expected change of potential in \autoref{lemma:potential-increase} is the most technical part, where we will have similar issues as in the deterministic moments universality analysis but with more complicated moments of resolvents.
The proof will follow a similar framework as in the analysis of potential change for moments in \autoref{ss:moment-uni} and combine with some analytical ideas in \autoref{section:resolvent}.

\subsection{Deterministic Moment Universality} \label{ss:moment-uni}

We start this section by formally describing the Iterative Swapping Algorithm for \autoref{t:moment-uni}.

    \begin{framed}{\bf{Iterative Swapping Algorithm for Moments}}
        \begin{itemize}
            \item \textbf{Initialize:} $\mathcal{I}_0 \gets [n]$ and $X_{f,0} \gets A_0 \otimes 1 + \sum_{i=1}^n X_i$;
            \item \textbf{For} $t=0$ to $n-1$ \textbf{do}
            \begin{enumerate}
                \item Find $i
                ^*\in \mathcal{I}_{t}$ and $Z'\in \supp(Z_{i^*})$ which minimizes
                \[
                    \|X_{f,t} + Z' \otimes 1 -X_{i^*}\|_{2p}.
                \]
                \item Update $X_{f,t+1}\gets X_{f,t} + Z' \otimes 1 - X_{i^*}$ and $\mathcal{I}_{t+1} \gets \mathcal{I}_t \backslash \{i^*\}$.
            \end{enumerate}
            \item \textbf{Return} $\varphi[X_{f,n}]$.
        \end{itemize}
    \end{framed}
    
    The algorithm is closely related to the method of conditional expectation.     
    Suppose we apply the conditional expectation method as follows: let $Z^{(1)}+ \cdots + Z^{(t)}$ be the deterministic matrices we have selected so far, and $\mathcal{I}_t$ is the set of indices where $Z_i$ remains random, then we take the conditional expectation
    \[
        \E\bigg[ \tr \bigg(A_0 + Z^{(1)}+ \cdots + Z^{(t)} + \sum_{i\in \mathcal{I}_t}Z_i \bigg)^{2p} \bigg]
    \]
    as the potential function. 
    We would like to choose an index $i \in \mathcal{I}_t$ and a matrix $Z^{(t+1)} \in \supp(Z_i)$ to ensure that the potential value is not increasing. 
    The issue, however, is that there may not be an efficient way to compute this potential function.

To overcome this issue, our idea is to use the alternative potential function based on the semicircular matrices:
\[
\biggnorm{\big(A_0 + Z^{(1)} + \cdots + Z^{(t)}\big) \otimes 1 + \sum_{i\in \mathcal{I}_t}X_i}_{2p} 
= \tot \bigg[\Big( \big(A_0 + Z^{(1)} + \cdots + Z^{(t)}\big) \otimes 1 + \sum_{i\in \mathcal{I}_t}X_i \Big)^{2p}\bigg]^{\frac{1}{2p}}.
\]
This potential function can be computed efficiently (see \autoref{lemma:moment-formula}) via the non-crossing combinatorial structure from free probability. 
The initial potential value $\| \Xf \|_{2p}$ is exactly our target, and we just need to control the accumulative errors through the iterations.

Similar to the analysis of the algorithms in \autoref{section:discrepancy} and \autoref{section:resolvent}, the key to the analysis of the Iterative Swapping Algorithm is to bound the increase of the potential function in one iteration. 

\begin{proposition}[Expected Potential Increase] \label{proposition:random-swap-moment}
Let $A_0 \in \M_d(\C)$ be a deterministic Hermitian matrix and let $Z_1, \ldots, Z_n \in \M_d(\C)$ be independent, zero-mean random Hermitian matrices. 
Denote $Z:= A_0 + \sum_{i=1}^n Z_i$, and $\sigma := \sigma(Z)$, $\nu := \nu(Z)$, and $\rho := \rho(Z)$. 
Let $X_f = A_0 \otimes 1 + \sum_{i=1}^n X_i$, where $X_1, \ldots, X_n$ are freely independent semicircular matrices such that each $X_i$ has the same covariance as $Z_i$.
Suppose $p \in \N$, and $\norm{X_f}_{2p} \geq \max\{ p^{\frac34} \sigma^{\frac12} \nu^{\frac12}, p^{\frac23} \sigma^{\frac23} \rho^{\frac13}, 64 p\rho\}$, then it holds that
\[
\frac{1}{n}\sum_{i=1}^n \E\big[\norm{X_f + Z_i \otimes 1 - X_i}_{2p} \big] - \norm{X_f}_{2p} \lesssim \frac{1}{n} \Big(p^{\frac34} \sigma^{\frac12} \nu^{\frac12} + p^{\frac23} \sigma^{\frac23} \rho^{\frac13} \Big),
\]
where the expectation is taken over $Z_1,\ldots, Z_n$.
\end{proposition}

Assuming \autoref{proposition:random-swap-moment}, we first prove \autoref{t:moment-uni}, the main theorem in this subsection by analyzing the Iterative Swapping Algorithm. 
The proof of \autoref{proposition:random-swap-moment} will be presented afterwards in \autoref{ss:potential-increase}.

\begin{proofof}{\autoref{t:moment-uni}}
We are going to apply \autoref{proposition:random-swap-moment} to each iteration $t$ of the algorithm.
Let 
\[
\sigma_t^2 
= \sigma\bigg(\sum_{i \in \mathcal{I}_t} Z_i\bigg)^2 
= \biggnorm{\sum_{i\in \mathcal{I}_t}\E[Z_i^2]}
\quad \textrm{and} \quad
\nu^2_t = \nu\bigg(\sum_{i \in \mathcal{I}_t} Z_i\bigg)^2
\] be the remaining total variance and the remaining covariance at time $t$.
Note that $\sigma_t \leq \sigma$ and $\nu_t \leq \nu$.

Let $\eps := \max\{ p^{\frac34} \sigma^{\frac12} \nu^{\frac12}, p^{\frac23} \sigma^{\frac23} \rho^{\frac13}, 64 p\rho\}$.
To meet the requirement in \autoref{proposition:random-swap-moment}, we need to ensure that $\norm{X_{f,t}}_{2p} \geq \eps$, which may not hold for every iteration $t$.
We consider two scenarios.
\begin{itemize}
\item Scenario 1: $\norm{X_{f,t}}_{2p} \geq \eps$ for all $t = 0,1,\ldots, n-1$.
\item Scenario 2: There exists some $t_0 \in \{0, \ldots, n-1\}$ such that $\norm{X_{f,t}}_{2p} < \eps$.
\end{itemize}
In Scenario 1, we can simply apply \autoref{proposition:random-swap-moment} to each iteration $t = 0, 1, \ldots, n-1$ and obtain that
\[
\E_{i\sim \mathcal{I}_t}\E_{Z_i}\norm{X_{f,t} + Z_i\otimes 1 -X_i}_{2p} - \norm{X_{f,t}}_{2p} 
\lesssim \frac{1}{n-t} \Big(p^{\frac34}\sigma_t^{\frac12} \nu_t^{\frac12} + p^{\frac23}\sigma_t^{\frac23} \rho^{\frac13}\Big).
\]
Therefore, it follows that 
\[
\norm{X_{f,n}}_{2p} - \norm{X_{f,0}}_{2p}
\leq \sum_{t=0}^{n-1} \frac{1}{n-t} \Big(p^{\frac34}\sigma_t^{\frac12} \nu_t^{\frac12} + p^{\frac23}\sigma_t^{\frac23} \rho^{\frac13}\Big).
\]
Since $\|Z_i\| \leq \rho$ for all $i \in [n]$, it is easy to see that $\sigma_t = \sqrt{\|\sum_{i \in \mathcal{I}_t} \E[Z_i^2]\|} \leq \min\{\sigma, \sqrt{n-t} \cdot \rho\}$. 
Together with the fact that $\nu_t \leq \nu$ and $X_{f,0} = \Xf$, we have
\begin{equation} \label{eq:2p-norm-uni}
\norm{X_{f,n}}_{2p} - \norm{\Xf}_{2p}~\lesssim~p^{\frac34} \nu^{\frac12} \cdot \underbrace{\sum_{t=0}^{n-1}  \min\Big\{ \frac{\sigma^{\frac12}}{n-t}, \frac{\rho^{\frac12}}{(n-t)^{\frac34}} \Big\}}_{(*)} 
+~p^{\frac23} \rho^{\frac13} \cdot \underbrace{\sum_{t=0}^{n-1} \min \Big\{ \frac{\sigma^{\frac23}}{n-t}, \frac{\rho^{\frac23}}{(n-t)^{\frac23}} \Big\}}_{(**)}.
\end{equation}
Note that both the summations in $(*)$ and $(**)$ have a similar structure.
We will use the following technical estimation to bound both of them.

\begin{claim}\label{claim:int-identity}
Let $\delta \in (0,1)$. It holds that
\[
\int_1^n \min\Big\{ \frac{\sigma^\delta}{t}, t^{\frac{\delta}{2}-1}\rho^\delta \Big\} dt \leq  \sigma^\delta \Big( \frac{2}{\delta} + \log{\frac{n\rho^2}{\sigma^2}} \Big).
\]
\end{claim}

\begin{proof}
We observe that the two functions $t^{-1} \sigma^{\delta}$ and $t^{\frac{\delta}{2}-1} \rho^{\delta}$ only intersect once at $t = \frac{\sigma^2}{\rho^2}$ over the whole positive real line $t > 0$.
Thus, it follows that $t^{\frac{\delta}{2}-1} \rho^{\delta} \leq t^{-1}\sigma^{\delta}$ for $t \in (0, \frac{\sigma^2}{\rho^2}]$ and $t^{\frac{\delta}{2}-1} \rho^{\delta} \geq t^{-1}\sigma^{\delta}$ for $t \geq \frac{\sigma^2}{\rho^2}$.
Therefore, 
\begin{eqnarray*}
\int_{1}^n\min \Big\{ \frac{\sigma^\delta}{t}, t^{\frac{\delta}{2}-1} \rho^\delta \Big\} dt 
&=& \int_{1}^{\frac{\sigma^2}{\rho^2}}t^{\frac{\delta}{2}-1}\rho^\delta dt 
+ \int_{\frac{\sigma^2}{\rho^2}}^n\frac{\sigma^\delta}{t}dt
\\
&=& \frac{2}{\delta} \Big( \Big(\frac{\sigma^2}{\rho^2}\Big)^{\frac{\delta}{2}}-1\Big)\rho^\delta + \Big(\log{n} - \log{\frac{\sigma^2}{\rho^2}} \Big)\sigma^\delta 
\leq \sigma^\delta \Big( \frac{2}{\delta} + \log{\frac{n\rho^2}{\sigma^2}}\Big).
\end{eqnarray*}
\end{proof}

We apply \autoref{claim:int-identity} with $\delta = \frac12$ and $\delta = \frac23$ to obtain an upper bound on $(*)$ and $(**)$ of \eqref{eq:2p-norm-uni} respectively:
\[
(*) \lesssim \sigma^{\frac12} \Big( 1 + \log \frac{n\rho^2}{\sigma^2} \Big)
\quad \textrm{and} \quad
(**) \lesssim \sigma^{\frac23} \Big( 1 + \log \frac{n\rho^2}{\sigma^2} \Big).
\]
Plugging back to \eqref{eq:2p-norm-uni}, it follows that
\[
\norm{X_{f,n}}_{2p} - \norm{\Xf}_{2p} \lesssim 
\Big(1+\log{\frac{n\rho^2}{\sigma^2}} \Big) \big( p^{\frac34}\sigma^{\frac12} \nu^{\frac12} + p^{\frac23}\sigma^{\frac23}\rho^{\frac13} \big).
\]    
In Scenario 2, let $t_0 \in \{0, \ldots, n-1\}$ be the last iteration with $\norm{X_{f,t_0}} < \eps$. 
Since $\norm{X_i} \leq 2\rho$ by \autoref{lemma:semicircular-norm}, 
it follows that $\norm{Z' \otimes 1 - X_i} \leq 3\rho$ for all $i \in [n]$ and $Z'\in \supp(Z_i)$. 
This implies that
\[ 
\eps \leq \norm{X_{f,t_0 + 1}}_{2p} \leq \eps + 3\rho.
\] 
Applying \autoref{proposition:random-swap-moment} to each $t \geq t_0 + 1$ iteration,
\[
\norm{X_{f,t+1}}_{2p} - \norm{X_{f,t}}_{2p} 
\lesssim \frac{1}{n-t} \Big( p^{\frac34} \sigma_t^\frac12 \nu^\frac12 + p^{\frac23}\sigma_t^{\frac{2}{3}} \rho^{\frac13} \Big).
\]
Summing over all $t \geq t_0 + 1$,
\begin{eqnarray*}
\norm{X_{f,n}}_{2p} & \leq & \norm{X_{f,t_0+1}}_{2p} + O\bigg( \sum_{t=t_0+1}^{n-1} \frac{1}{n-t} \Big(p^{\frac34}\sigma_t^{\frac12} \nu_t^{\frac12} + p^{\frac23}\sigma_t^{\frac23} \rho^{\frac13} \Big) \bigg) 
\\
& \leq & \eps + 3\rho + O\Big(1+\log{\frac{n\rho^2}{\sigma^2}} \Big) \Big( p^{\frac34}\sigma^{\frac12} \nu^{\frac12} + p^{\frac23}\sigma^{\frac23}\rho^{\frac13} \Big) 
\\
& \lesssim & \Big(1+\log{\frac{n\rho^2}{\sigma^2}} \Big) \cdot \Big( p^{\frac34}\sigma^{\frac12} \nu^{\frac12} + p^{\frac23}\sigma^{\frac23}\rho^{\frac13} + p \rho \Big).
\end{eqnarray*}
where the second inequality follows from the same argument as in scenario 1, 
and the last inequality follows from the definition of $\eps$.
\end{proofof}

\subsubsection{Analysis of the Potential Increase} \label{ss:potential-increase}

In this subsection, we prove \autoref{proposition:random-swap-moment} by analyzing the change of the potential value in each iteration of the Iterative Swapping Algorithm for Moments.

We start with an easy special case. When $p=1$, covariance matching implies that, for every $i\in[n]$,
\[
\E\big[\|X_f+Z_i\otimes1-X_i\|_2^2\big]=\|X_f\|_2^2.
\]
Hence, by Jensen's inequality,
\[
\frac1n\sum_{i=1}^n
\E\big[\|X_f+Z_i\otimes1-X_i\|_2\big]
\leq \|X_f\|_2.
\]
The conclusion of \autoref{proposition:random-swap-moment} therefore holds for $p=1$. 
Assume henceforth that $p\geq2$.

We apply Jensen's inequality twice to the expectation of the new potential value to obtain that
\begin{eqnarray}
\frac1n \sum_{i=1}^n \E\Big[ \norm{X_f + Z_i \otimes 1 - X_i }_{2p} \Big] 
& \leq & \frac{1}{n}\sum_{i=1}^n \E\big[\tr\otimes\tau (X_f+Z_i \otimes 1 - X_i )^{2p}\big]^{\frac{1}{2p}} \nonumber 
\\
& \leq & \Big( \frac{1}{n}\sum_{i=1}^n\E\big[\tr\otimes\tau(X_f+Z_i \otimes 1 - X_i)^{2p}\big] \Big)^{\frac{1}{2p}}. \label{eq:expected-potential-moments}
\end{eqnarray}
To simplify the notations, we denote
\[
C_i := Z_i \otimes 1 - X_i, \qquad \forall i \in [n].
\]
When we expand $\tot\big[(X_f + C_i)^{2p}\big]$, 
there are $\binom{2p}{r}$ terms containing $r$ copies of $C_i$'s, which can be written as
\[
\sum_{\substack{k_1+\cdots+k_{r+1}=2p-r \\ k_1, \ldots, k_{r+1} \geq 0}} \tr\otimes\tau(X_f^{k_1}C_i\cdots X_f^{k_r}C_i X_f^{k_{r+1}}).
\]
In particular, when $r=1$ and $r=2$, the expressions can be simplified to
\[
2p \cdot \tot \big(X_f^{2p-1} C_i \big) 
\qquad \text{and} \qquad 
\sum_{k=0}^{2p-2} p \cdot \tot \big(X_f^{k} C_i X_f^{2p-2-k} C_i\big).
\]
Therefore, the inner term of \eqref{eq:expected-potential-moments} can be written as
\begin{align*} 
\frac1n \sum_{i=1}^n \E \big[ \tr\otimes\tau(X_f+C_i)^{2p} \big] 
=\tr\otimes\tau\big(X_f^{2p}\big) & 
+ \underbrace{\frac{2p}{n} \sum_{i=1}^n \E \big[ \tot\big(X^{2p-1}_f C_i \big) \big]}_{R_1} 
+ \underbrace{\frac{p}{n} \sum_{i=1}^n \sum_{k=0}^{2p-2} \E \big[\tot\big( X^k_f C_i X^{2p-2-k}_f C_i\big) \big]}_{R_2} 
\\
& + \underbrace{\frac1n \sum_{i=1}^n \sum_{r \geq 3} \sum_{\substack{k_1+\cdots+k_{r+1}=2p-r \\ k_1, \ldots, k_{r+1} \geq 0}} \E \big[\tr\otimes\tau\big(X_f^{k_1}C_i\cdots X_f^{k_r}C_i X_f^{k_{r+1}}\big)\big]}_{R_{\geq 3}},
\end{align*}
which can also be treated as a consequence of Taylor expansion (\autoref{theorem:taylor-general}) of $\tot\big[(X_f + C_i)^{2p}\big]$.

In the following, we handle the first-order term $R_1$, the second-order term $R_2$, and the higher-order terms $R_{\geq 3}$ separately. 
A common theme in the analysis is that we start with using the ``semicircular integration-by-parts'' formula in \autoref{proposition:free-ibp} to decouple the ``dependence'' of the $C_i$'s and $X_f$ terms, which both contain $X_i$, into random variables that are freely independent from each other. 
Then we apply \autoref{proposition:trace-holder} to bound the \textit{expectation} of the higher order terms with respect to the choice of the random swaps.

\subsubsection*{First-Order Terms}

We start with handling the easiest first-order term $R_1$.
We observe that
\[
\E \big[\tot(X^{2p-1}_f C_i )\big] 
= \E \big[\tot\big(X_f^{2p-1} (Z_i \otimes 1)\big)\big] - \tot\big(X_f^{2p-1} X_i\big) 
= - \tot\big(X_f^{2p-1} X_i\big),
\]
where the first term vanishes as $\E[Z_i] = 0$, but the second term stays as $X_f$ and $X_i$ are not freely independent.

Now, we apply the first-order ``semicircular integration-by-parts'' formula \autoref{corollary:free-ibp1} to obtain that
\[
\tot(X_f^{2p-1} X_i) = \sum_{k=0}^{2p-2} \tot(X_f^k X'_i X_f^{2p-2-k} X'_i),
\]
where $X'_i$ is a freely independent copy of $X_i$.
Therefore, the term $R_1$ is exactly
\begin{equation} \label{eq:R1-moment-uni}
R_1 = - \frac{2p}{n} \sum_{i=1}^n \sum_{k=0}^{2p-2} \tot(X_f^k X'_i X_f^{2p-2-k} X'_i).
\end{equation}

\subsubsection*{Second-Order Terms}

Then, we consider the second-order term $R_2$. 
We expand the two $C_i$'s, and obtain that
\begin{eqnarray*}
\E \big[\tot \big( X^k_f C_i X^{2p-2-k}_f C_i \big)\big] 
& = & \E \big[\tot \big(X_f^k (Z_i \otimes 1) X_f^{2p-2-k} (Z_i \otimes 1) \big)\big] 
+ \tot \big(X_f^k X_i X_f^{2p-2-k} X_i \big) 
\\
&  & \quad  -~\E \big[\tot \big(X_f^k X_i X_f^{2p-2-k} (Z_i \otimes 1) \big) \big]- \E\big[\tot \big(X_f^k (Z_i \otimes 1) X_f^{2p-2-k} X_i \big)\big] 
\\
& = & \E\big[ \tot \big(X_f^k (Z_i \otimes 1) X_f^{2p-2-k} (Z_i \otimes 1) \big)\big] 
+ \tot (X_f^k X_i X_f^{2p-2-k} X_i ),
\end{eqnarray*}
where the second equality follows as $\E[Z_i] = 0$ and the cross terms vanish.
So, there are two parts in $R_2$:
\[
R_2 
= \frac{p}{n} \sum_{i=1}^n \sum_{k=0}^{2p-2} \E \big[\tot \big(X_f^k (Z_i \otimes 1) X_f^{2p-2-k} (Z_i \otimes 1) \big)\big] 
+ \underbrace{\frac{p}{n} \sum_{i=1}^n \sum_{k=0}^{2p-2} \tr\otimes\tau \big(X_f^{k}X_iX_f^{2p-2-k}X_i \big)}_{(*)}.
\]
We focus on the second part $(*)$.
Apply the ``semicircular integration-by-parts'' formula \autoref{proposition:free-ibp} to $(*)$, and note that the summands on the RHS expression is only non-zero for $\ell=0$ or $\ell=2$.
Indeed, the terms corresponding to $\ell=1$ in \autoref{proposition:free-ibp} vanish because $X_{i,\overline S}$ appears exactly once and is centered and free from the remaining factors. 
For $\ell=2$, expand the second derivative. Terms in which the two copies $X_i'$ and $X_i''$ alternate cyclically vanish by freeness and the semicircular moment formula. 
Every remaining term is cyclically equivalent, after possibly interchanging the names of $X_i'$ and $X_i''$, to a term in the second sum above. Each such term occurs four times, according to the two choices of the explicit occurrence of each free copy.
Therefore,
\begin{equation} \label{eq:free-2nd-order}
\begin{aligned}
(*) = \frac{p}{n} \sum_{i=1}^n \sum_{k=0}^{2p-2} \tot & \big(X_f^{k}X_iX_f^{2p-2-k}X_i \big) 
= \frac{p}{n} \sum_{i=1}^n \sum_{k=0}^{2p-2} \tot \big(X_f^{k}X_i'X_f^{2p-2-k}X_i' \big) 
\\
&\quad  + \frac{4p}{n} \sum_{i=1}^n \tr\otimes\tau \bigg(\sum_{k+l+m=2p-4}X_f^{k}X_i'X_f^{l}X_i'X_f^mX_i''X_f^{2p-4-k-l-m}X_i'' \bigg), 
\end{aligned}
\end{equation}
where both $X_i'$ and $X_i''$ are freely independent copies of $X_i$ that are also freely independent from $X_f$. 

Now, we apply the infinite dimensional trace inequality in \autoref{proposition:trace-holder} to the second part of \eqref{eq:free-2nd-order}, with $m=4$ parts and $X_f^k, X_f^l, X_f^m$ and $X_f^{2p-4-k-l-m}$ being the $Y_i$'s and $X'_i$ and $X''_i$ being the $Z_{i,j}$'s. 
We obtain that
\[
\bigg| \sum_{i=1}^n \tot\big(X_f^{k}X_i'X_f^{l}X_i'X_f^mX_i''X_f^{2p-4-k-l-m}X_i''\big) \bigg| 
\lesssim \sigma^2 \rho^2 \cdot \tot\big(X_f^{2p-4}\big),
\]
where we used $\|X_i\| \leq 2\rho$ by \autoref{lemma:semicircular-norm} and $\sigma\big(\sum_{i=1}^n X_i\big) = \sigma$ as $X_i$ and $Z_i$ have the same covariance.
By Jensen's inequality, 
\[
\tot\big(X_f^{2p-4}\big) 
\leq \tot\big(X_f^{2p}\big)^{1-\frac{2}{p}} 
= \frac{\tot\big(X_f^{2p}\big)}{\|X_f\|_{2p}^4}.
\]
Putting these back to \eqref{eq:free-2nd-order}, it follows that
\[
(*) \leq \frac{p}{n} \sum_{i=1}^n \sum_{k=0}^{2p-2} \tot \big(X_f^{k}X_i'X_f^{2p-2-k}X_i' \big)  + O\Big(\frac{p^4 \sigma^2 \rho^2}{n \|X_f\|_{2p}^4}\Big) \cdot \tot\big(X_f^{2p}\big).
\]
Putting $(*)$ back to $R_2$, we have
\begin{equation} \label{eq:R2-moment-uni}
\begin{aligned} 
R_2 \leq \frac{p}{n} & \sum_{i=1}^n \sum_{k=0}^{2p-2} \bigg( \E \big[\tot \big(X_f^k (Z_i \otimes 1) X_f^{2p-2-k} (Z_i \otimes 1) \big)\big] 
+ \tot \big(X_f^{k}X_i'X_f^{2p-2-k}X_i' \big) \bigg)   \\
& + O\Big(\frac{p^4 \sigma^2 \rho^2}{n \|X_f\|_{2p}^4} \Big) \cdot \tot(X_f^{2p}).
\end{aligned}
\end{equation}

\subsubsection*{Higher-Order Terms}

Finally, we will use the following lemma to deal with the higher-order terms in $R_3$.
        
\begin{lemma}[Higher Order Error Bound] \label{lemma:mu-taylor-bound}
In the setting of \autoref{proposition:random-swap-moment},
\[
\Bigg| \sum_{i=1}^n\sum_{r\geq 3}\sum_{\substack{k_1+\cdots+k_{r+1}=2p-r \\ k_1, \ldots, k_{r+1} \geq 0}} \E \big[\tot \big(X_f^{k_1}C_i\cdots X_f^{k_r}C_i X_f^{k_{r+1}} \big) \big] \Bigg| 
\lesssim \frac{p^3\sigma^2\rho}{\norm{X_f}_{2p}^{3}} \cdot \tot\big(X_f^{2p}\big).
\]
\end{lemma}

A direct consequence of \autoref{lemma:mu-taylor-bound} is that
\begin{equation} \label{eq:R3-moment-uni}
R_{\geq 3} \lesssim \frac{p^3\sigma^2\rho}{n \norm{X_f}_{2p}^{3}} \cdot \tot(X_f^{2p}).
\end{equation}

We postpone the proof of \autoref{lemma:mu-taylor-bound} to the end of this subsection, where we again need to use the ``semicircular integration-by-parts'' formula in \autoref{proposition:free-ibp} and the infinite dimensional trace inequality in \autoref{proposition:trace-holder}.

\subsubsection*{Combining All Terms Together}

Combining the upper bounds on $R_1$, $R_2$, and $R_{\geq 3}$ in \eqref{eq:R1-moment-uni}, \eqref{eq:R2-moment-uni}, and \eqref{eq:R3-moment-uni} respectively, 
\begin{eqnarray*} 
& & \frac1n \sum_{i=1}^n \E \big[ \tr\otimes\tau(X_f+C_i)^{2p} \big] 
\\
& \leq & \tr\otimes\tau(X_f^{2p}) + \underbrace{\frac{p}{n} \sum_{i=1}^n \sum_{k=0}^{2p-2} \Big( \E \big[ \tot(X_f^{k} (Z_i \otimes 1) X_f^{2p-2-k} (Z_i \otimes 1)) \big] - \tot(X_f^k X'_i X_f^{2p-2-k} X'_i) \Big)}_{(**)} 
\\
& & +~O\Big(\frac{p^4 \sigma^2 \rho^2}{n \|X_f\|_{2p}^4} \Big) \cdot \tot(X_f^{2p}) + O\Big(\frac{p^3\sigma^2\rho}{n \norm{X_f}_{2p}^{3}}\Big) \cdot \tr\otimes\tau(X_f^{2p}).
\end{eqnarray*}
It follows from \autoref{lemma:free-moments} that
\[
\tot \big( X_f^{k}X_i'X_f^{2p-2-k}X_i' \big)
= \tr ( \varphi\big[X_f^k\big] \cdot \varphi\big[X'_i \cdot \big(\varphi[X_f^{2p-2-k}]\otimes 1\big) \cdot X'_i\big] \big) 
= \E \big[\tr\big(\varphi[X_f^k]  \cdot Z_i \cdot \varphi[X_f^{2p-2-k}] \cdot Z_i\big)\big],
\]
where the last equality holds as $X'_i$ and $Z_i$ have the same covariance, i.e., satisfying \eqref{eq:cov-profile}.
        
Now, let $Y = Z_{\boldsymbol{i}}$, where $\boldsymbol{i}$ is an index selected uniformly at random from $[n]$. 
Then, we apply the crossing bound on the second-order difference to obtain that
\begin{eqnarray*}
(**) 
& = & \frac{p}{n}\sum_{i=1}^n\sum_{k=0}^{2p-2}\Big( \E \big[ \tot \big(X_f^k \cdot (Z_i \otimes 1) \cdot X_f^{2p-2-k} \cdot (Z_i \otimes 1)\big) \big] 
- \E \big[ \tr\big(\varphi[X_f^k] \cdot Z_i \cdot \varphi[X_f^{2p-2-k}] \cdot Z_i \big) \big] \Big)
\\
&=& p\sum_{k=0}^{2p-2} \E\big[\tot \big(X_f^k \cdot (Y \otimes 1) \cdot X_f^{2p-2-k} \cdot (Y\otimes 1)\big) \big] - \tr\big(\varphi[X_f^k] \cdot Y \cdot \varphi[X_f^{2p-2-k}] \cdot Y\big) \big]
\\
&\lesssim& \frac{p^4\sigma^2\nu^2}{n} \cdot \tot \big(X_f^{2p-4}\big) 
\\
&\leq& \frac{p^4\sigma^2\nu^2}{n\norm{X_f}_{2p}^4} \cdot \tot\big(X_f^{2p}\big),
\end{eqnarray*}
where the first inequality follows by applying \autoref{proposition:moment-crossing-bound2} with $\sigma(X_f) = \sigma$, $\nu(X_f) = \nu$, 
\[
\sigma(Y) = \biggnorm{\frac{1}{n}\sum_i\E[Z_i^2]}^{\frac12} = \frac{\sigma}{\sqrt{n}}
\quad \textrm{and} \quad 
\nu(Y) = \biggnorm{\sum_{i=1}^n\frac{1}{n}\E\big[\vecc(Z_i)\vecc(Z_i)^{*}\big]}^{\frac12} = \frac{\nu}{\sqrt{n}},
\] 
and the second inequality follows from Jensen's inequality. 
    
Finally, we put together these bounds and obtain that
\[
\frac{1}{n}\sum_{i=1}^n\E \big[ \tr\otimes\tau(X_f+C_i)^{2p} \big] 
\leq \tr\otimes\tau(X_f^{2p}) \cdot \bigg(1+O\bigg(\frac{p^4\sigma^2\nu^2}{n\norm{X_f}_{2p}^4} + \frac{p^4\sigma^2\rho^2}{n \norm{X_f}_{2p}^4} + \frac{p^3\sigma^2\rho}{n \norm{X_f}_{2p}^3}\bigg) \bigg).
\]
Using the assumption from \autoref{proposition:random-swap-moment} that
$\|X_f\|_{2p} \gtrsim \max\{p^{\frac34} \sigma^{\frac12} \nu^{\frac12}, p^{\frac23} \sigma^{\frac23} \rho^{\frac13}, p \rho\}$, we have
\[
\frac{p^4\sigma^2\nu^2}{\norm{X_f}_{2p}^4} \lesssim \frac{p^{\frac74} \sigma^{\frac12}\nu^{\frac12}}{\norm{X_f}_{2p}} \qquad \text{and} \qquad \frac{p^4\sigma^2\rho^2}{\norm{X_f}_{2p}^4} \lesssim \frac{p^3\sigma^2\rho}{\norm{X_f}_{2p}^3} \lesssim \frac{p^{\frac53} \sigma^{\frac23}\rho^{\frac13}}{\norm{X_f}_{2p}}.
\]
Therefore, it holds that
\[
\frac{1}{n}\sum_{i=1}^n\E \big[ \tr\otimes\tau(X_f+C_i)^{2p} \big] 
\leq \tr\otimes\tau(X_f^{2p}) \cdot \bigg(1+O\bigg(\frac{p^{\frac74} \sigma^{\frac12}\nu^{\frac12}}{n \norm{X_f}_{2p}} + \frac{p^{\frac53} \sigma^{\frac23}\rho^{\frac13}}{n \norm{X_f}_{2p}} \bigg)\bigg).
\]
Returning to the original expression in \eqref{eq:expected-potential-moments}, we conclude that
\begin{eqnarray*}
\bigg( \frac{1}{n}\sum_{i=1}^n\E \big[\tot(X_f+C_i)^{2p} \big] \bigg)^{\frac{1}{2p}} 
&\leq& \bigg(\tot\big(X_f^{2p}\big) \cdot \bigg(1+O\bigg(\frac{p^{\frac74} \sigma^{\frac12}\nu^{\frac12}}{n \norm{X_f}_{2p}} + \frac{p^{\frac53} \sigma^{\frac23} \rho^{\frac13} }{n \norm{X_f}_{2p}} \bigg) \bigg) \bigg)^{\frac{1}{2p}}
\\
&\leq& \norm{X_f}_{2p} \bigg(1+O\bigg(\frac{p^{\frac34} \sigma^{\frac12}\nu^{\frac12}}{n \norm{X_f}_{2p}} + \frac{p^{\frac23} \sigma^{\frac23} \rho^{\frac13}}{n\norm{X_f}_{2p}} \bigg) \bigg)
\\
&=& \norm{X_f}_{2p} + O\bigg(\frac{p^{\frac34} \sigma^\frac12 \nu^\frac12 + p^{\frac23}\sigma^{\frac23} \rho^{\frac13}}{n}\bigg),
\end{eqnarray*}
where the second inequality follows from Bernoulli's inequality that $(1+x)^r \leq 1+rx$ for $r \in [0,1]$ and $x \geq -1$.
This finishes the proof of \autoref{proposition:random-swap-moment}. \qed

\subsubsection*{Proof of \autoref{lemma:mu-taylor-bound}}

It remains to prove \autoref{lemma:mu-taylor-bound} to finish \autoref{ss:moment-uni}.

\begin{proofof}{\autoref{lemma:mu-taylor-bound}}
For each term of order $r$ in $C_i = Z_i \otimes 1 - X_i$, we expand it out into terms depending on $Z_i \otimes 1$ and $X_i$. For each $S\subseteq [r]$, and $j\in [r]$, we define 
\[
C_{i,j|S} : =\begin{cases}
        X_{i}& j\in S, \\
        Z_i \otimes 1 &j\notin S.
        \end{cases}
\]
Let $\mathcal{E}$ be the expression inside the absolute value in the LHS of the expression. Then we have
\begin{eqnarray*}
\mathcal{E} &=& 
\sum_{i=1}^n\sum_{r\geq 3}\sum_{\substack{k_1+\cdots+k_{r+1}=2p-r \\ k_1, \ldots, k_{r+1} \geq 0}} \E \big[ \tot \big(X_f^{k_1}C_i\cdots X_f^{k_r}C_i X_f^{k_{r+1}}\big) \big] 
\\
& = & \sum_{r\geq 3}\sum_{\substack{k_1+\cdots+k_{r+1} = 2p-r \\ k_1, \ldots, k_{r+1} \geq 0}}\sum_{S\subseteq [r]}(-1)^{|S|}\sum_{i=1}^n\E \big[ \tr\otimes\tau(X_f^{k_1}C_{i,1|S}\cdots X_f^{k_r}C_{i,r|S} X_f^{k_{r+1}}) \big].
\end{eqnarray*}
We deal with each inner sum $\sum_{i=1}^n\E[\tr\otimes\tau(X_f^{k_1}C_{i,1|S}\cdots X_f^{k_r}C_{i,r|S} X_f^{k_{r+1}})]$ individually.
We claim that
\begin{equation} \label{eq:mixed-moment-bound}
\bigg| \sum_{i=1}^n\E \big[\tot \big(X_f^{k_1}C_{i,1|S}\cdots X_f^{k_r}C_{i,r|S} X_f^{k_{r+1}} \big)\big] \bigg| 
\leq \sum_{\ell = 0}^{m\wedge (2p-r)}\sigma^2(2\rho)^{r+\ell-2}{m\choose \ell}{2p-r\choose\ell}{2\ell \choose \ell} \cdot \tot(|X_f|^{2p-r-\ell}).
\end{equation}
To prove the claim, we apply the ``semicircular integration by parts'' formula in \autoref{proposition:free-ibp} to replace the occurrence of $X_i$ in \eqref{eq:mixed-moment-bound} with freely independent copies of it, 
so that we can bound the resulting terms in the expansion by \autoref{proposition:trace-holder}. Fix an $S$ with $|S|=m$, then the term $X_f^{k_1}C_{i,1|S}\cdots X_f^{k_r}C_{i,r|S}X_f^{k_{r+1}}$ is a monomial of degree $2p-r$ in $X_f$, of degree $m$ in $X_i$, and of degree $r-m$ in $Z_i \otimes 1$. 
By the cyclic property of trace, we rewrite it without loss of generality as
\[ 
X_f^{k_1}C_{i,1|S}\cdots X_f^{k_r}C_{i,r|S} \to \prod_{j=1}^m F_j(X_f, Z_i \otimes 1) X_i,
\]
where each $F_j(X,Z)$ is a monomial of degree $p_j$ in $X$ and degree $q_j$ in $Z$, with $\sum_{j=1}^m p_j = 2p-r$ and $\sum_{j=1}^m q_j = r-m$. 
In order to make it easier to apply \autoref{proposition:free-ibp}, we define the composite function:
\[
F_{Y_1,\ldots, Y_m}(X_1, \ldots, X_n,Z_i \otimes 1) = \prod_{j=1}^m F_j(X_f,Z_i \otimes 1)Y_j.
\]
Note that $F_{Y_1, \ldots, Y_m}$ is a monomial of degree $2p-r$ in $X_f$ and degree $r-m$ in $Z_i \otimes 1$. Then, \autoref{proposition:free-ibp} states that
\begin{align*}
\sum_{i=1}^n\tr\otimes\tau & \bigg( \prod_{j=1}^m F_j(X_f,Z_i \otimes 1)X_i \bigg) \\
=~ & \sum_{i=1}^n\sum_{\ell=0}^{(2p-r) \wedge m} \sum_{S\in {[m]\choose \ell}}\tr\otimes\tau(\partial_{X_i}^\ell F_{X_{i,1|S},\ldots,X_{i,m|S}}(X_1,\ldots, X_n,Z_i \otimes 1)(X_{i,1},\ldots, X_{i,\ell})),
\end{align*}
where $X_{i,j|S}$'s are defined as in \autoref{proposition:free-ibp}, and $X_{i,1}, \ldots, X_{i,\ell}$ are freely independent copies of $X_i$ for each $i \in [n]$.
Observe that, if $\ell > 2p-r$, then the $\ell$-th derivative is $0$. Thus, it suffices to sum over $\ell$ up to $(2p-r) \wedge m$.
Note that $X_f = A_0 \otimes 1 + \sum_{i=1}^n X_i$ and $\partial_{X_i}(A_0 \otimes 1 + \sum_{i=1}^n X_i)(H) = H$. 
Thus, for $\ell \leq 2p-r$, the $\ell$-th partial derivative,
\begin{equation} \label{eq:derivative-poly}
\partial_{X_i}^\ell F_{X_{i,1|S}, \ldots, X_{i,m|S}}(X_1,\ldots, X_n,Z_i \otimes 1)(X_{i,1},\ldots, X_{i,\ell}),
\end{equation}  
is a polynomial of degree $2p$ with variables $X_f, Z_i \otimes 1,\{X_{i,j}\}_{j=1}^\ell, X_{i,\overline{S}}$ for all $i \in [n]$. 
In particular, each of the monomials in the polynomial in \eqref{eq:derivative-poly} has degree $2p-r-\ell$ in $X_f$, degree $r-m$ in $Z_i \otimes 1$, degree $2$ in each $X_{i,j}$ for $j\in [\ell]$, and degree $m-\ell$ in the variable $X_{i,\overline{S}}$, where $X_{i,j}$'s and $X_{i,\overline{S}}$'s are freely independent from $X_f$.

We apply the infinite dimensional trace inequality in \autoref{proposition:trace-holder} to each of the monomials in \eqref{eq:derivative-poly}, which are of the form $X_f^{k_1} Z_{i,1} \cdots X_f^{k_{r + \ell}}Z_{i,{r+\ell}} $, where $k_1+ \cdots + k_{r+\ell} = 2p-r-\ell$ and each $Z_{i,j}$ is either $Z_i \otimes 1$, $X_{i,\overline{S}}$, or $X_{i,j}$ for some $j\in [\ell]$.
Thus, by \autoref{proposition:trace-holder}, it holds that
\begin{equation} \label{eq:monomial-bound}
\bigg| \sum_{i=1}^n \E \big[ \tot (X_f^{k_1} Z_{i,1} \cdots X_f^{k_{r + \ell}}Z_{i,{r+\ell}}) \big] \bigg| \leq 
\sigma^2 \cdot (2\rho)^{r + \ell -2} \cdot \tot(|X_f|^{2p-r-\ell}),
\end{equation}
where we used $\|Z_i\| \leq \rho$ and $\norm{X_i}\leq 2\rho$ by \autoref{lemma:semicircular-norm}.

Next, we give an upper bound on the number of monomial terms in the polynomial \eqref{eq:derivative-poly}. 
A trivial upper bound is $\binom{2p-r}{\ell} \cdot \ell!$, which corresponds to the number of ways to pick $\ell$ indices containing $X_f$ and then substituting in the variables $X_{i,\sigma(1)}, \ldots, X_{i,\sigma(\ell)}$ for some permutation $\sigma\in \mathcal{P}_\ell$.
However, \autoref{lemma:sc-joint-moments} implies that at most ${2p-r \choose \ell}{2\ell\choose\ell}$ of those terms can have non-zero contribution: 
This is because a given substitution can have non-zero contribution only if the permutation of $X_{i,1}, \ldots, X_{i,\ell}$ in those selected indices does not create crossings when combined with the indices of the same variables that occur in $X_{i,1|S}, \ldots, X_{i,m|S}$. Thus, the number of valid permutations is at most the number of non-crossing partitions over $2\ell$ elements, with $|NC_{2}[2\ell]|\leq {2\ell\choose\ell}$. 

Summing \eqref{eq:monomial-bound} over all monomials in \eqref{eq:derivative-poly}, it follows that
\begin{eqnarray*}
& & \bigg| \sum_{i=1}^n\E \big[ \tot \big(\partial_{X_i}^\ell F(X_{i,1|S},\ldots,X_{i,m|S})(X_{i,1},\ldots,X_{i,\ell})\big) \big] \bigg| 
\\
& \leq & {m\choose\ell}{2p-r \choose \ell}{2\ell\choose\ell}\cdot \sigma^2 \cdot (2\rho)^{r+\ell-2} \cdot \tr\otimes\tau(|X_f|^{2p-r-\ell}),
\end{eqnarray*}        
which establishes the claim in \eqref{eq:mixed-moment-bound}.
        
Now, we put \eqref{eq:mixed-moment-bound} back to the LHS.
Observe that there are $\binom{2p}{r}$ ways to select $k_1, \ldots, k_{r+1}$, and thus
\begin{eqnarray*}
LHS &\leq& \sum_{r\geq 3}\sum_{\substack{k_1+\cdots+k_{r+1} = 2p-r \\ k_1, \ldots, k_{r+1} \geq 0}}\sum_{S\subseteq [r]} \bigg|\sum_{i=1}^n\E \big[\tot \big(X_f^{k_1}C_{i,1|S}\cdots X_f^{k_r}C_{i,r|S} X_f^{k_{r+1}}\big) \big] \bigg| 
\\
&\leq& \sum_{r\geq 3}{2p\choose r}\sum_{m=0}^r{r\choose m}\sum_{\ell=0}^{m \wedge (2p-r)}\sigma^2 \cdot (2\rho)^{r+\ell-2} \cdot {m\choose \ell}{2p-r\choose \ell}{2\ell\choose \ell} \cdot \tot(|X_f|^{2p-r-\ell}).
\end{eqnarray*}
Using the identity that ${2p\choose r}{2p-r\choose \ell}={2p\choose r+\ell}{r+\ell\choose r}$ and applying a change of variable $k = r+\ell$ (note that $\ell \leq m \leq r$, thus $\ell \leq \lfloor k/2 \rfloor$), we have
\begin{eqnarray*}
LHS & \leq & 
\sum_{r\geq 3}\sum_{m=0}^r\sum_{\ell=0}^{m \wedge (2p-r)} \sigma^2 \cdot (2\rho)^{r+\ell-2} \cdot {2p\choose r+\ell}{r+\ell\choose \ell}{r\choose m}{m\choose \ell}{2\ell\choose \ell} \cdot \tot\big(|X_f|^{2p-r-\ell}\big)\\
&= & \sigma^2\sum_{k\geq 3}(2\rho)^{k-2}{2p\choose k}\sum_{\ell=0}^{\lfloor k/2\rfloor}\sum_{m=\ell}^{k-\ell}{k\choose \ell}{k-\ell \choose m}{m\choose\ell}{2\ell\choose \ell} \cdot \tot\big(|X_f|^{2p-k}\big).
\end{eqnarray*}
By Jensen's inequality that $\tot\big(|X_f|^{2p-k}\big) \leq \tot\big(X_f^{2p}\big)^{1-\frac{k}{2p}}$, and the observation that ${2\ell \choose \ell}\leq {k\choose \ell}\leq 2^k$ (as $\ell \leq k/2$), it follows that
\begin{eqnarray*}
LHS &\leq &
\frac{4\sigma^2 \cdot \tot\big(X_f^{2p}\big)}{\norm{X_f}_{2p}^{2}}\sum_{k\geq 3}\Big(\frac{4\rho}{\norm{X_f}_{2p}}\Big)^{k-2}{2p\choose k}\sum_{\ell=0}^{\lfloor k/2\rfloor}\sum_{m=\ell}^{k-\ell}{k\choose \ell}{k-\ell \choose m}{m\choose\ell}
\\
&\leq& \frac{64\sigma^2\tr\otimes\tau(X_f^{2p})}{\norm{X_f}_{2p}^{2}}\sum_{k\geq 3}\Big(\frac{16\rho}{\norm{X_f}_{2p}}\Big)^{k-2}{2p\choose k},
\end{eqnarray*}
where the last line follows by the bound
\[
\sum_{\ell=0}^{\lfloor k/2\rfloor}\sum_{m=\ell}^{k-\ell}{k\choose \ell}{k-\ell \choose m}{m\choose\ell} = \sum_{\ell=0}^{\lfloor k/2\rfloor}\sum_{m=\ell}^{k-\ell} \frac{k!}{\ell! \ell! (k-\ell-m)! (m - \ell)!} \leq 4^k,
\]
where the LHS is the number of some particular way of partitioning $k$ elements into 4 groups and the RHS is the total number of such partitions.
       
Finally, our assumption $\norm{X_f}_{2p} \gtrsim p\rho$ in \autoref{proposition:random-swap-moment} implies that the sequence $\Big(\frac{16\rho}{\norm{X_f}_{2p}}\Big)^{k-2}{2p\choose k}$ is geometrically decreasing with constant rate. 
Therefore, the upper bound is dominated by the $k=3$ term, and we conclude that
\[
LHS \lesssim \frac{p^3 \sigma^2 \rho}{\|X_f\|_{2p}^3} \cdot \tot\big(X_f^{2p}\big).
\]
\end{proofof}

\subsection{Deterministic Operator Norm Universality} \label{ss:norm-uni}

The goal in this subsection is to prove \autoref{t:norm-uni}.

Note that it suffices to control $\lambda_{\max}$ in order to give a bound on the operator norm, as we can simply set
\[
    \tilde{Z} = \begin{pmatrix} A_0  & \\ & - A_0 \end{pmatrix}  + \sum_{i=1}^n \begin{pmatrix} Z_i & \\ & - Z_i \end{pmatrix} \quad \text{and} \quad \tilde{X}_{\rm free} = \begin{pmatrix} A_0 \otimes 1 & \\ & - A_0 \otimes 1 \end{pmatrix} + \sum_{i=1}^n \begin{pmatrix} X_i & \\ & - X_i    \end{pmatrix},
\]
which guarantees that $\lambda_{\max}(\Tilde{Z}) = \norm{Z}$ and $\lambda_{\max}(\tilde{X}_{f,t}) = \|X_{f,t}\|$.
Furthermore, as $\sigma(Z) = \sigma\Big(\begin{smallmatrix} Z & \\ & - Z\end{smallmatrix}\Big)$, $\sqrt{2} \cdot \nu(Z) = \nu\Big(\begin{smallmatrix} Z & \\ & - Z\end{smallmatrix}\Big)$, and $\rho(Z) = \rho\Big(\begin{smallmatrix} Z & \\ & - Z\end{smallmatrix}\Big)$, it suffices to work with $\lambda_{\max}$ to prove \autoref{t:norm-uni}.

We start by presenting a barrier method based iterative swapping algorithm for operator norm. Note that we use a similar notation for resolvent as in \autoref{section:resolvent}:
\[
    M_{\lambda}(X) := (\lambda 1 - X)^{-1},
\]
but we will only consider $\lambda \in \R$ satisfying $\lambda > \lambda_{\max}(X)$.

\begin{framed}{\bf{Barrier Method Based Iterative Swapping for Operator Norm}}
\begin{itemize}
\item \textbf{Initialize:} $\mathcal{I}_0 = [n]$; 
$X_{f,0} = A_0 \otimes 1 + \sum_{i=1}^n X_i$, 
and $\eps = C\max\big\{ p^{\frac34} \sigma^{\frac12} \nu^{\frac12}, p^{\frac23}\sigma^{\frac23}\rho^{\frac13}, p\rho\big\}$ for some constant $C$ and $p = \Theta(\log d)$; 
$\lambda_0 = \lambda_{\max}(X_{f,0}) + 2\eps$. 
To compute $\lambda_{\max}(X_{f,0})$ in polynomial time, we can solve the formula in \autoref{lemma:lehner} using standard convex optimization techniques.
            
\item \textbf{For} $t=0$ to $n-1$ \textbf{do}
\begin{enumerate}
\item Let $\sigma^2_t := \norm{\sum_{i \in \mathcal{I}_t}\E[Z_i^2]}$, 
$\nu_t := \nu\big(\sum_{i \in \mathcal{I}_t} Z_i\big)$, and for some constant $C'$, let
\[
\delta_t := \frac{C'}{n-t} \cdot (p^{\frac34} \sigma_t^{\frac12} \nu_t^{\frac12} + p^{\frac23} \sigma_t^{\frac23} \rho^{\frac13});
\]
\item Update $\lambda_{t+1} \gets \lambda_t+\delta_t$;
\item Find $i^*\in \mathcal{I}_{t}$ and $Z'\in \supp(Z_{i^*})$ that minimizes
\[
\tot\big[M_{\lambda_{t+1}}\big(X_{f,t} + Z' \otimes 1 -X_{i^*}\big)^{2p}\big];
\]
\item Update $X_{f,t+1}\gets X_{f,t} + Z'\otimes 1 - X_{i^*}$ and $\mathcal{I}_{t+1} \gets \mathcal{I}_t \backslash \{i^*\}$
\end{enumerate}

\item Return $\varphi[X_{f,n}]$.
\end{itemize}
\end{framed}

Suppose $\lambda_n > \lambda_{\max}(X_{f,n})$ in the end, then $\lambda_0 + \sum_{t=0}^{n-1} \delta_t$ is an upper bound on the final solution $\lambda_{\max}(\varphi[X_{f,n}])$.
In order to achieve $\lambda_n > \lambda_{\max}(X_{f,n})$ after the last iteration, we maintain two invariants throughout the whole process:
\begin{enumerate}
\item The barrier $\lambda_t$ is well above the spectral edge of $X_{f,t}$, such that $\lambda_t \geq \lambda_{\max}(X_{f,t}) + \eps$;
\item The potential value is bounded, such that $\tot(M_{\lambda_t}(X_{f,t})^{2p}) \leq (2\eps)^{-2p}$.
\end{enumerate}
Observe that the two invariants are satisfied at the beginning of the algorithm, due to our choice of $\lambda_0 = \lambda_{\max}(X_{f,0}) + 2\eps$. 
Thus, to maintain the second invariant, it suffices to ensure that the potential value never increases.

To see the connection between the two invariants and our goal, 
note that the first invariant is actually stronger than what we need. 
However, to ensure the second invariant can be satisfied by the update, we need $\lambda_t$ to be bounded away from $\spec(X_{f,t})$.
In turn, the following lemma shows that the second invariant also helps to maintain the first invariant.

\begin{lemma}[Potential Controls Largest Eigenvalue] \label{lemma:barrier-invariance}
Let $X,Y\in \M_d(\C) \otimes \A$ be self-adjoint operators. 
Suppose for some $\eps \ge \sqrt{\sigma(X) \nu(X)}$ it holds that $\lambda \geq \lambda_{\max}(X)+\eps$ and $\norm{Y}\leq \frac{\eps}{2}$.
If $\delta>0$ satisfies that
\[
\tot(M_{\lambda +\delta}(X+Y)^{2p}) \leq (2\eps)^{-2p},
\]
where $p \asymp \log d$,
then it holds that $\lambda + \delta \geq \lambda_{\max}(X+Y) + \eps$.
\end{lemma}
\begin{proof}
By the assumptions that $\lambda \geq \lambda_{\max}(X) + \eps$ and $\|Y\| \leq \frac{\eps}{2}$, 
it holds that $\lambda +\delta - \lambda_{\max}(X+Y) \geq \lambda-\lambda_{\max}(X)-\|Y\| \geq \frac{\eps}{2}$. 
Thus, by applying the ultracontractivity bound in \autoref{lemma:bar-ultra} with $r = \frac{500}{\eps} \cdot \sigma_*(X+Y)$, and $q = \frac{p}{2}$, it holds that
\begin{eqnarray*}
\norm{M_{\lambda+\delta}(X+Y)} 
&\leq& \Big(d \Big(\frac{10^3p\sigma_*(X + Y)}{\eps} + 1\Big) \Big)^{\frac{3}{2p}}\Big(\|M_{\lambda+\delta}(X+Y)\|_{2p}+\frac{1}{4\eps} \Big)
\\
&\leq& \Big(d \Big(\frac{10^3p\sigma_*(X)}{\eps} + 500p + 1 \Big) \Big)^{\frac{3}{2p}}\cdot \Big( \frac{1}{2\eps} + \frac{1}{4\eps}\Big) \\
& \leq & (d 2 \times 10^3 p + 1)^{\frac{3}{2p}} \cdot \frac{3}{4\eps}
\end{eqnarray*}
where the second inequality follows by the assumption on the potential bound and the fact that $\sigma_*(X+Y)\leq \sigma_*(X) + \sigma_*(Y) \leq \sigma_*(X) + \eps/2$ due to \autoref{l:sigma*},
and the last inequality holds by the assumption that $\eps \ge \sqrt{\sigma(X) \nu(X)}$ and the fact that $\sigma_*(X) \leq \min\{\sigma(X), \nu(X)\}$ (see, e.g., Section 2.1 in~\cite{BBvH23}).

By taking $p = C \log d$ with large enough $C$ such that $(d (1 + 2 \times 10^3 \cdot p))^{\frac{3}{2p}} \le \frac{4}{3}$, we have
\[ \norm{M_{\lambda+\delta}(X+Y)} \leq \frac{1}{\eps}, \]
which implies that $\lambda +\delta \geq \lambda_{\max}(X+Y) + \eps$.
\end{proof}

Therefore, the key in maintaining the second invariant is to decide how much we need to shift the barrier so that the potential value does not increase.
The following lemma provides an upper bound on the potential change when we shift the barrier by $\delta$ and make an update $Y$.

\begin{lemma}[Potential Update] \label{lemma:barrier-shift}
Let $X,Y\in \M_d(\C)\otimes \A$ be self adjoint operators and $\lambda > \lambda_{\max}(X)$. 
Then,
\begin{align*}
\tr\otimes\tau\big(M_{\lambda+\delta}(X+&Y)^{2p}\big) - \tot\big(M_\lambda(X)^{2p}\big) 
\\
\leq~ &\tot\big(M_{\lambda+\delta}(X+Y)^{2p}\big) - \tot\big(M_{\lambda+\delta}(X)^{2p}\big) - 2p\delta \cdot \tot\big(M_{\lambda+\delta}(X)^{2p+1}\big).
\end{align*}
\end{lemma}
\begin{proof}
We rewrite the change of the potential as
\begin{align*}
\tot\big(&M_{\lambda+\delta}(X+Y)^{2p}\big) - \tot\big(M_\lambda(X)^{2p}\big) 
\\
& = \tot\big(M_{\lambda+\delta}(X+Y)^{2p}\big) - \tot\big(M_{\lambda+\delta}(X)^{2p}\big) + \tot\big(M_{\lambda+\delta}(X)^{2p}\big) - \tot\big(M_\lambda(X)^{2p}\big).
\end{align*}
Then, note that for all $x\in \spec(X)$,
\begin{align*}
\frac{1}{(\lambda-x)^{2p}} - \frac{1}{(\lambda-x+\delta)^{2p}} 
= \frac{(\lambda-x+\delta)^{2p} - (\lambda-x)^{2p}}{(\lambda-x)^{2p}(\lambda-x+\delta)^{2p}} 
= \frac{\big(1+\frac{\delta}{\lambda-x}\big)^{2p}-1}{(\lambda-x+\delta)^{2p}} 
& \geq \frac{2p\delta}{(\lambda-x)(\lambda-x+\delta)^{2p}} 
\\
&\geq \frac{2p\delta}{(\lambda-x+\delta)^{2p+1}},
\end{align*}
where we used the fact that $\lambda - x > 0$ and the Bernoulli's inequality $(1+y)^{r} \geq 1+ry$ for all $y \geq -1$ and $r \geq 1$ for the first inequality, and $\delta > 0$ for the last inequality.
Then, the lemma follows as
\[
\tot(M_{\lambda+\delta}(X)^{2p}) - \tot(M_\lambda(X)^{2p}) 
\leq -2p\delta \cdot \tot(M_{\lambda+\delta}(X)^{2p+1}). \qedhere
\]
\end{proof}

The following is the main lemma that controls the expected change of potential after a random swap,
which can be derandomized to find a desired swap by simply selecting the best one. 
The proof of the lemma will be postponed to the end of this section.

\begin{lemma}[Expected Potential Change] \label{lemma:potential-increase}
Let $X_f = A_0 \otimes 1 + \sum_{i=1}^nX_i$, where $A_0\in \M_d(\C)$ and $X_1,\ldots, X_n \in \M_d(\C) \otimes \A$ are centered, self-adjoint semicircular matrices with the same mean and covariance as the random matrices $Z_1,\ldots, Z_n \in \M_d(\C)$. 
Suppose $p\geq 2$, and $\lambda - \lambda_{\max}(X_f) > \eps > 4p\rho$. Then,
\[
\frac{1}{n}\sum_{i=1}^n\E_{Z_i} \big[ \tot\big(M_\lambda(X_f + Z_i \otimes 1 -X_i)^{2p}\big) \big] 
- \tot\big(M_\lambda(X_f)^{2p}\big) 
\lesssim \Big( \frac{p^4\sigma^2\nu^2}{\eps^3n}+\frac{p^3\sigma^2\rho}{\eps^2n} \Big) \cdot \tot\big(M_\lambda(X_f)^{2p+1}\big).
\]
\end{lemma}

Assuming \autoref{lemma:potential-increase}, we prove the main theorem in this section via analyzing the Barrier Method Based Iterative Swapping Algorithm.

\begin{proofof}{\autoref{t:norm-uni}}
As shown in the beginning of this subsection, 
it suffices to work with $\lambda_{\max}$ instead of the operator norm, 
and the goal is to prove that the two invariants that
\[
(a)\quad\lambda_t \geq \lambda_{\max}(X_{f,t})+\eps\quad\quad \text{and} \quad\quad 
(b)\quad\tot\big(M_{\lambda_t}(X_{f,t})^{2p}\big) \leq (2\eps)^{-2p}
\]
are maintained at each iteration $t$.
Clearly both invariants are satisfied when $t=0$ by the choice of $\lambda_0$. 
    
Now, assume by induction that both these invariants hold at iteration $t$. 
Since the invariant $(a)$ holds, \autoref{lemma:barrier-shift} implies that, for any $i \in \mathcal{I}_t$ and any $Z'_i \in \supp(Z_i)$,
\begin{equation} \label{eq:resolvent-bound}
\begin{aligned}
& \tot\big(M_{\lambda_{t+1}}(X_{f,t} + Z'_i \otimes 1 - X_i)^{2p}\big) - \tot\big(M_{\lambda_t}(X_{f,t})^{2p}\big)
\\
\leq~ & \tot\big(M_{\lambda_{t+1}}(X_{f,t}+ Z'_i \otimes 1 - X_i)^{2p}\big) - \tot\big(M_{\lambda_{t+1}}(X_{f,t})^{2p}\big) - 2p\delta_t \cdot \tot\big(M_{\lambda_{t+1}}(X_{f,t})^{2p+1}\big).
\end{aligned}
\end{equation}
By the inductive hypothesis $\lambda_{t+1} \geq \lambda_t \geq \lambda_{\max}(X_{f,t}) + \eps$ and $\eps \geq C\rho\cdot p$ due to the parameter choice,
we can apply \autoref{lemma:potential-increase} with $\lambda = \lambda_{t+1}$ to bound
\begin{eqnarray*}
& & \bigg(\frac{1}{n-t}\sum_{i\in \mathcal{I}_t}\E \big[\tot\big(M_{\lambda_{t+1}}(X_{f,t}+Z_{i} \otimes 1 -X_i)^{2p}\big)\big] \bigg) - \tot\big(M_{\lambda_{t+1}}(X_{f,t})^{2p}\big) 
\\
& \lesssim & \bigg(\frac{p^4\sigma_t^2\nu_t^2}{\eps^3 (n-t)} + \frac{p^3\sigma_t^2\rho}{\eps^2(n-t)}\bigg) \cdot \tot\big(M_{\lambda_{t+1}}(X_{f,t})^{2p+1}\big) 
\\
& \lesssim & \bigg(\frac{p^{\frac74} \sigma_t^{\frac12} \nu_t^{\frac12} + p^{\frac53} \sigma_t^{\frac23} \rho^{\frac13}}{n-t}\bigg) \cdot \tot\big(M_{\lambda_{t+1}}(X_{f,t})^{2p+1}\big),
\end{eqnarray*}
where we used $\eps \geq \max\{ p^{\frac34}\sigma^{\frac12}\nu^{\frac12},p^{\frac23}\sigma^{\frac23}\rho^{\frac13} \}$ in the last inequality. 
    
Thus, by taking $\delta_t = \frac{C'}{n-t} \big(p^{\frac34} \sigma_t^\frac12 \nu^\frac12 + p^{\frac23} \sigma_t^{\frac23} \rho^{\frac13} \big)$ for some large enough constant $C'$ in \eqref{eq:resolvent-bound}, there is always an $i^* \in \mathcal{I}_t$ and $Z'_{i^*} \in \supp(Z_{i^*})$ such that
\[
\tot(M_{\lambda_{t+1}}(X_{f,t} + Z'_{i^*} \otimes 1 - X_{i^*})^{2p}) 
\leq \tot(M_{\lambda_t}(X_{f,t})^{2p}) \leq (2\eps)^{-2p},
\]
where the last inequality follows by the induction hypothesis. 
    
Thus, invariant $(b)$ is satisfied at iteration $t+1$. 
Then, by applying \autoref{lemma:barrier-invariance}, we can guarantee that invariant $(a)$ is also satisfied at iteration $t+1$. 
Note that, to apply \autoref{lemma:barrier-invariance}, we need to choose $p \asymp \log d$.
The two invariant conditions ensure that
\begin{eqnarray*}
\lambda_{\max}(X_{f,n}) - \lambda_{\max}(X_{f,0})\leq 2 \eps + \sum_{t=0}^{n-1}\delta_t 
& \lesssim & \eps + \sum_{t=0}^{n-1}\frac{p^{\frac34} \sigma_t^{\frac12} \nu_t^{\frac12} + p^{\frac23} \sigma_t^{\frac23} \rho^{\frac13}}{n-t} 
\\
& \lesssim & \Big(1+\log{\frac{n\rho^2}{\sigma^2}} \Big) \cdot \big( p^{\frac34}\sigma^{\frac12} \nu^{\frac12} + p^{\frac23} \sigma^{\frac23} \rho^{\frac13} + p \rho \big),
\end{eqnarray*}
where the last inequality follows by the choice of $\eps$ and a similar argument as in the proof of \autoref{t:moment-uni} (e.g., the log term comes from \autoref{claim:int-identity}).
\end{proofof}

\subsubsection{Bounding Potential Change}

In this subsection, we prove \autoref{lemma:potential-increase} by analyzing the change of the potential value in one iteration in the Barrier Method Based Iterative Swapping Algorithm for Operator Norm.

We start by expanding $\tot\big(M_{\lambda}(X_f + Z_i \otimes 1 - X_i)^{2p}\big)$ with the Taylor series.
For ease of notation, we write $Y = Z'_i \otimes 1 - X_i$ for some fixed $Z'_i \in \supp(Z_i)$. 
Then, \autoref{theorem:taylor-general} shows that
\begin{equation} \label{eq:taylor-res-moment}
\tot\big(M_\lambda(X_f+Y)^{2p}\big) 
= \tot\big(M_\lambda(X_f)^{2p}\big) 
+ \sum_{r=1}^\infty\frac{1}{r!}\tot\big(D^rM_\lambda(X_f)^{2p}(Y,\ldots, Y)\big),
\end{equation}
whenever the series converge.
Combining \autoref{lemma:res-higher-der} and \autoref{corollary:general-holder}, each term in the summation is
\begin{eqnarray*}
\bigg| \frac{1}{r!}\tot(D^rM_\lambda(X_f)^{2p}(Y,\ldots, Y)) \bigg| 
& \leq & {2p+r-1\choose r} \cdot \norm{Y}^{r} \cdot \tot\big(M_\lambda(X_f)^{2p+r}\big) 
\\
& \leq & \Big( \frac{\norm{Y}}{\eps} \Big)^r \cdot {2p+r-1\choose r} \cdot \tot\big(M_\lambda(X_f)^{2p}\big) 
\\
& \leq & 2^{-r} {2p+r-1\choose r} \cdot \tot\big(M_\lambda(X_f)^{2p}\big),
\end{eqnarray*}
where the second inequality follows as $\|M_\lambda(X_f)\| \leq 1/\eps$ when $\lambda > \lambda_{\max}(X_f) + \eps$, 
and the last inequality follows as $\|Y\| \leq 3\rho \leq \frac{\eps}{2}$ by the assumption $\eps \geq 4p \rho$ and $\|X_i\| \leq 2\rho$ by \autoref{lemma:semicircular-norm}.

Thus, when $r \geq 2p$ the series in \eqref{eq:taylor-res-moment} is dominated by a geometric series with the common ratio strictly less than one, so it indeed converges.
Therefore, it holds that
\[
\frac{1}{n}\sum_{i=1}^n \E \big[ \tot\big( M_\lambda(X_f+Z_i \otimes 1-X_i)^{2p} \big) \big] 
= \tot\big(M_\lambda(X_f)^{2p}\big) + R_1 + R_2 + R_{\geq 3}, \quad \text{where}
\]
\begin{eqnarray*}
&& R_1 := \frac1n \sum_{i=1}^n \E \big[\tot\big(D (M_{\lambda}(X_f)^{2p}) (Z_i \otimes 1 - X_i)\big) \big], \\
&& R_2 :=  \frac{1}{2n} \sum_{i=1}^n \E \big[ \tot\big(D^2 (M_{\lambda}(X_f)^{2p}) (Z_i \otimes 1 - X_i, Z_i \otimes 1 - X_i)\big) \big] 
\\
&& R_{\geq 3} := \frac{1}{n}\sum_{i=1}^n\ \E\bigg[\sum_{r\geq 3}\frac{1}{r!} \tot\big(D^rM_\lambda(X_f)^{2p}(Z_i \otimes 1-X_i, \ldots, Z_i \otimes 1 - X_i)\big)\bigg].
\end{eqnarray*}

We bound first-order term $R_1$, the second-order term $R_2$, and the higher-order term $R_{\geq 3}$ separately. 
The analysis follows a similar framework as in the proof of \autoref{proposition:random-swap-moment}
    
\subsubsection*{First-Order Term}

We start with the first-order term $R_1$.
It follows from the derivative formula for resolvent \autoref{lemma:res-higher-der} that
\[
\E \big[ \tot\big(D (M_{\lambda}(X_f)^{2p}) (Z_i \otimes 1 - X_i)\big)\big]  
= 2p \cdot \E \big[ \tot\big(M_{\lambda}(X_f)^{2p+1} (Z_i \otimes 1 - X_i) \big)\big] 
= -2p \cdot \tot\big(M_{\lambda}(X_f)^{2p+1} X_i\big),
\]
where the last equality follows as $\E[Z_i]=0$.

Then, we apply the ``semicircular integration-by-parts'' formula for resolvent in \autoref{cor:res-ibp} to obtain that
\[
\tot\big(M_{\lambda}(X_f)^{2p+1} X_i\big) 
= \sum_{k=1}^{2p+1} \tot\big(M_{\lambda}(X_f)^{k} \cdot X'_i \cdot M_{\lambda}(X_f)^{2p+2-k} \cdot X'_i \big),
\]
where $X'_i$ is a freely independent copy of $X_i$.
Therefore, the term $R_1$ is exactly
\begin{equation} \label{eq:R1-res-uni}
R_1 = - \frac{2p}{n} \sum_{i=1}^n \sum_{k=1}^{2p+1} \tot\big(M_{\lambda}(X_f)^{k} \cdot X'_i \cdot M_{\lambda}(X_f)^{2p+2-k} \cdot X'_i\big).
\end{equation}

\subsubsection*{Second-Order Term}

For the second-order term $R_2$, again it follows from \autoref{lemma:res-higher-der} that
\begin{eqnarray*}
R_2 & = & 
\frac{p}{n} \sum_{i=1}^n \sum_{k=1}^{2p+1}  \E \big[ \tot\big(M_\lambda(X_f)^{k} \cdot (Z_i \otimes 1 - X_i) \cdot M_\lambda(X_f)^{2p-k+2} \cdot (Z_i \otimes 1 - X_i)\big) \big] 
\\
& = & \frac{p}{n} \sum_{i=1}^n \sum_{k=1}^{2p+1} \E \big[ \tot\big(M_\lambda(X_f)^{k} \cdot (Z_i \otimes 1) \cdot M_\lambda(X_f)^{2p-k+2} \cdot (Z_i\otimes 1)\big) \big] 
\\
&&~ + \underbrace{\frac{p}{n} \sum_{i=1}^n \sum_{k=1}^{2p+1} \tot\big(M_\lambda(X_f)^{k} \cdot X_i \cdot M_\lambda(X_f)^{2p-k+2} \cdot X_i \big)}_{(*)},
\end{eqnarray*}
where we used $\E[Z_i] = 0$ to get rid of those terms containing only one $Z_i$ in the last equality.

Applying the ``semicircular integration-by-parts'' formula for resolvents in \autoref{proposition:lbp-rational} to $(*)$ gives
\begin{eqnarray*}
(*) & = & \frac{p}{n} \sum_{i=1}^n \sum_{k=1}^{2p+1}\tot\big( M_\lambda(X_f)^{k} \cdot X_i \cdot M_\lambda(X_f)^{2p+2-k} \cdot X_i \big) 
\\
& = & \frac{p}{n} \sum_{i=1}^n \bigg(\sum_{k=1}^{2p+1} \tot\big(M_\lambda(X_f)^{k} \cdot X_i' \cdot M_\lambda(X_f)^{2p+2-k} \cdot X_i'\big) 
\\
& & \quad + \sum_{\substack{k_1+ \cdots +k_4=2p+4\\k_1, k_2,k_3,k_4 \geq 1}} \tot\big(M_\lambda(X_f)^{k_1} \cdot X_i' \cdot M_\lambda(X_f)^{k_2} \cdot X_i' \cdot M_\lambda(X_f)^{k_3} \cdot X_i'' \cdot M_\lambda(X_f)^{k_4} \cdot X_i''\big)\bigg) ,
\end{eqnarray*}
where $X_i'$ and $X_i''$ are both freely independent copies of $X_i$. 
Note that the $\ell = 1$ term in \autoref{proposition:lbp-rational} vanishes as there is always a freely independent copy of $X_i$ that cannot be paired up.

Since $X'_i$ and $X''_i$ are freely independent from $X_f$, 
after summing over all $i \in [n]$ the above $4$-th order term can be controlled by the infinite dimensional trace inequality \autoref{proposition:trace-holder}, which gives
\[
\sum_{i=1}^n \tot\big(M_\lambda(X_f)^{k_1} \cdot X_i' \cdot M_\lambda(X_f)^{k_2} \cdot X_i' \cdot M_\lambda(X_f)^{k_3} \cdot X_i'' \cdot M_\lambda(X_f)^{k_4} \cdot X_i''\big) \lesssim \sigma^2\rho^2 \cdot \tot\big(M_\lambda(X_f)^{2p+4}\big).
\]
Therefore, $R_2$ can be bounded by
\begin{equation} \label{eq:R2-res-uni}
\begin{aligned} 
R_2 &~ = ~\frac{p}{n} \sum_{i=1}^n \sum_{k=1}^{2p+1} \E \big[ \tot\big(M_\lambda(X_f)^{k} \cdot (Z_i \otimes 1) \cdot M_\lambda(X_f)^{2p-k+2} \cdot (Z_i \otimes 1)\big) \big]  
\\
& \quad +  \frac{p}{n} \sum_{i=1}^n \sum_{k=1}^{2p+1} \tot\big(M_\lambda(X_f)^{k} \cdot X'_i \cdot M_\lambda(X_f)^{2p-k+2} \cdot X'_i\big)
\\
& \quad + O\Big(\frac{p^4\sigma^2\rho^2}{n}\Big) \cdot \tot\big(M_\lambda(X_f)^{2p+4}\big).
\end{aligned}
\end{equation}

\subsubsection*{Higher-Order Term}

We use the following lemma to deal with the higher-order term $R_{\geq 3}$.

\begin{lemma}[Higher-Order Error Bound] \label{lemma:rhobound}
Let $X_f = A_0 \otimes 1 + \sum_{i=1}^nX_i$. If $\lambda -\lambda_{\max}(X_f)\geq \eps  \gtrsim p\cdot\rho$, then
\[
\bigg| \sum_{i=1}^n \E\bigg[ \sum_{r \geq 3} \frac{1}{r!} \tot\Big(D^r \big(M_\lambda(X_f)^{2p}\big) \big( Z_i \otimes 1 -X_i,\ldots, Z_i \otimes 1 -X_i\big)\Big) \bigg] \bigg| 
\lesssim \frac{p^3\sigma^2 \rho}{\eps^2} \cdot \tot\big( M_{\lambda}(X_f)^{2p+1}\big).
\]
\end{lemma}

\autoref{lemma:rhobound} is an analog of \autoref{lemma:mu-taylor-bound} in the deterministic moment universality theorem, and we postpone the proof to the end of this section. 
We will proceed with the following direct consequence of the lemma that
\begin{equation} \label{eq:R3-res-uni}
R_{\geq 3} \lesssim \frac{p^3\sigma^2\rho}{\eps^2n} \cdot \tot\big(M_\lambda(X_f)^{2p+1}\big).
\end{equation}

\subsubsection*{Combining All Terms Together}

Combining \eqref{eq:R1-res-uni} and \eqref{eq:R2-res-uni}, it follows that
\begin{eqnarray*}
& & R_1 + R_2 
\\
& = & \frac{p}{n} \sum_{i=1}^n \sum_{k=1}^{2p+1} \Big( \E\big[ \tot\big(M_\lambda(X_f)^{k} \cdot Z_i \otimes 1 \cdot M_\lambda(X_f)^{2p-k+2} \cdot Z_i \otimes 1\big) \big] - \tot\big(M_\lambda(X_f)^{k} \cdot X_i' \cdot M_\lambda(X_f)^{2p+2-k} \cdot X_i'\big) \Big) 
\\
& & \qquad +~O\Big( \frac{p^4\sigma^2\rho^2}{n} \Big) \cdot \tot\big(M_\lambda(X_f)^{2p+4}\big).
\end{eqnarray*}
By our assumption in \autoref{lemma:potential-increase}, $\|M_{\lambda}(X_f)\| = 1/(\lambda - \lambda_{\max}(X_f)) \leq 1/\eps$, which implies that
\[
\frac{p^4\sigma^2\rho^2}{n} \cdot \tot(M_\lambda(X_f)^{2p+4}) \leq \frac{p^4\sigma^2\rho^2}{\eps^3 n} \cdot \tot(M_\lambda(X_f)^{2p+1}).
\]
Then, since $X'_i$ is freely independent from $X_f$, it follows from \autoref{lemma:free-moments} that
\begin{eqnarray*}
\tot\big(M_\lambda(X_f)^{k} \cdot X_i' \cdot M_\lambda(X_f)^{2p+2-k} \cdot X_i' \big) 
& = & 
\tr\big(\varphi\big[M_\lambda(X_f)^{k}\big] \cdot \varphi\big[X'_i \cdot \big(\varphi\big[M_\lambda(X_f)^{2p-k+2}\big]\otimes 1\big) \cdot X'_i\big]\big) 
\\
& = & \E\big[ \tr\big(\varphi\big[M_\lambda(X_f)^{k}\big] \cdot Z_i \cdot \varphi\big[M_\lambda(X_f)^{2p-k+2}\big] \cdot Z_i\big)\big],
\end{eqnarray*}
where the last equality follows as $Z_i$ and $X_i'$ have the same covariance. 
Therefore,
\begin{eqnarray*}
R_1 + R_2 & = & 
\frac{p}{n} \sum_{i=1}^n \sum_{k=1}^{2p+1} \Big( \E\big[ \tot\big(M_\lambda(X_f)^{k} \cdot Z_i \otimes 1 \cdot M_\lambda(X_f)^{2p-k+2} \cdot Z_i \otimes 1 \big) \big] 
\\
& & \underbrace{\quad \qquad \qquad - \E \big[ \tr\big(\varphi\big[M_\lambda(X_f)^{k}\big] \cdot Z_i \cdot \varphi\big[M_\lambda(X_f)^{2p-k+2}\big] \cdot Z_i\big) \big] \Big)}_{(**)} 
\\
& & +~O\Big( \frac{p^4\sigma^2\rho^2}{\eps^3 n} \Big) \cdot \tot\big(M_\lambda(X_f)^{2p+1}\big).
\end{eqnarray*}
To bound $(**)$, we define $Y = Z_{\boldsymbol{i}}$, where $\boldsymbol{i} \in [n]$ is a uniform random index, as in the proof of \autoref{proposition:random-swap-moment}. Then, it holds that
\begin{eqnarray*}
(**) & = & 
p \sum_{k=1}^{2p+1} \Big( \E \big[\tot\big(M_\lambda(X_f)^{k} \cdot Y\otimes 1 \cdot M_\lambda(X_f)^{2p-k+2} \cdot Y\otimes 1\big)\big] 
- \E\big[\tr\big(\varphi[M_\lambda(X_f)^{k}] \cdot Y \cdot \varphi[M_\lambda(X_f)^{2p-k+2}] \cdot Y\big)\big] \Big) 
\\
& \lesssim & p \sum_{k=1}^{2p+1}
\frac{p^2 \cdot \sigma^2 \cdot \nu^2}{n}
\cdot \tot\big(M_\lambda(X_f)^{2p+4}\big) 
\\
& \lesssim & \frac{p^4\sigma^2\nu^2}{\eps^3 n} \cdot \tot\big(M_\lambda(X_f)^{2p+1}\big), 
\end{eqnarray*}
where the first inequality follows by applying \autoref{cor:res-crossing-bound} with $\lambda > \lambda_{\max}(X_f)$. The second inequality follows as $\sigma(X_f) = \sigma$, $\nu(X_f) = \nu$, $\sigma(Y)=\sigma/\sqrt{n}$, and $\nu(Y) = \nu/\sqrt{n}$, as in the proof of \autoref{proposition:random-swap-moment}. Finally, the last inequality follows from $\|M_{\lambda}(X_f)\| = 1/(\lambda - \lambda_{\max}(X_f)) \leq 1/\eps$.

Combining together with the bound on $R_{\geq 3}$ in \eqref{eq:R3-res-uni}, we conclude that
\begin{eqnarray*}
& & \frac{1}{n}\sum_{i=1}^n\E \big[ \tot\big( M_\lambda(X_f+Z_i \otimes 1-X_i)^{2p} \big) \big] - \tot\big(M_\lambda(X_f)^{2p}\big) = R_1 + R_2 + R_{\geq 3}\\
& \lesssim & \Big(\frac{p^4\sigma^2\nu^2}{\eps^3n} + \frac{p^4 \sigma^2 \rho^2}{\eps^3 n} + \frac{p^3\sigma^2\rho}{\eps^2n}\Big) \cdot \tot\big(M_\lambda(X_f)^{2p+1}\big) 
\lesssim \Big(\frac{p^4\sigma^2\nu^2}{\eps^3n} + \frac{p^3\sigma^2\rho}{\eps^2n}\Big)\cdot \tot\big(M_\lambda(X_f)^{2p+1}\big),
\end{eqnarray*}
where the last inequality follows by the assumption $\eps \gtrsim p \rho$. This finishes the proof of \autoref{lemma:potential-increase}.
\qed

\subsubsection*{Proof of \autoref{lemma:rhobound}}

It remains to prove \autoref{lemma:rhobound} to finish \autoref{ss:norm-uni}.

\begin{proofof}{\autoref{lemma:rhobound}}
Note that the multilinear map $D^r(M_{\lambda}(X_f)^{2p})$ is symmetric, so we can rewrite
\begin{equation} \label{eq:trace-derivative}
\begin{aligned}
\tot\big(D^r (M_\lambda(X_f)^{2p}\big)(Z_i \otimes 1&-X_i, \ldots, Z_i\otimes 1-X_i)) 
\\
& = \sum_{m=0}^r{r\choose m} (-1)^m \cdot \tot \big(D^{r} (M_\lambda(X_f)^{2p}\big) \big((Z_i \otimes 1)^{(r-m)},X_i^{(m)}\big),
\end{aligned}
\end{equation}
where we recall that $X^{(m)}$ denotes repeating $X$ for $m$ times in the input.
    
By \autoref{lemma:res-higher-der}, $D^{r}\big(M_\lambda(X_f)^{2p}\big)\big((Z_i \otimes 1)^{(r-m)},X_i^{(m)}\big)$ is a sum over $r!{2p+r-1\choose r}$ monomials, and each of the monomials is of degree $2p+r$ in $M_\lambda(X_f)$, of degree $r-m$ in $Z_i \otimes 1$, and of degree $m$ in $X_i$. 
The monomials can be written in the following form
\[
\prod_{j=1}^m F_j(M_\lambda(X_f), Z_i \otimes 1)X_i,
\]
where each $F_j(M_{\lambda}(X_f), Z_i \otimes 1)$ is a monomial of degree $p_j$ in $M_{\lambda}(X_f)$ and of degree $q_j$ in $Z_i \otimes 1$, where $\sum_{j=1}^m p_j = 2p+r$ and $\sum_{j=1}^m q_j = r-m$.

Then, we claim that, when summing over all $i \in [n]$, it holds that
\begin{equation} \label{eq:mixed-res-moment-bound}
\begin{aligned} 
\bigg| \sum_{i=1}^n \E \Big[ \tot\Big(& \prod_{j=1}^m F_j\big(M_\lambda(X_f), Z_i \otimes 1\big) X_i \Big) \Big] \bigg| 
\\
& \leq \sigma^2\sum_{\ell=0}^{m}(2\rho)^{r+\ell-2}{m\choose\ell}{2p+r+\ell-1\choose\ell}{2\ell\choose \ell} \cdot \tot\big(M_\lambda(X_f)^{2p+r+\ell}\big).
\end{aligned}
\end{equation}
The claim is just a counterpart of \eqref{eq:mixed-moment-bound} in the previous subsection. 
To prove the claim, first observe that we can apply the ``semicircular integration-by-parts'' formula for resolvent in \autoref{proposition:lbp-rational} to obtain that
\[
\tot\Big( \prod_{j=1}^m F_j\big(M_\lambda(X_f), Z_i \otimes 1\big) X_i \Big) 
= \tot \bigg( \sum_{\ell=0}^m \sum_{S \in {[m]\choose \ell}}\partial^\ell_{X_i}\Big(\prod_{j=1}^m F_j \big(M_\lambda(X_f),Z_i \otimes 1\big) X_{i,j|S}\Big)(X_{i,1},\ldots, X_{i,\ell}) \bigg),
\]
where the notation $X_{i,j|S}$ is defined in \autoref{proposition:free-ibp}, 
which indicates a freely independent copy of $X_i$.

Fix some $\ell \in [m]$ and $S \in \binom{[m]}{\ell}$. We deal with each $\partial^\ell_{X_i}\Big(\prod_{j=1}^m F_j \big(M_\lambda(X_f),Z_i \otimes 1\big) X_{i,j|S}\Big)(X_{i,1},\ldots, X_{i,\ell})$ separately.
Note that only the $M_{\lambda}(X_f)$ terms in the monomial $\prod_{j=1}^m F_j \big(M_\lambda(X_f),Z_i \otimes 1\big) X_{i,j|S}$ depend on $X_i$. 
By the product rule, the operation $\partial_{X_i} (\cdot) (H)$ over this monomial is replacing each of the $2p+r$ occurrences of $M_{\lambda}(X_f)$ by $M_{\lambda}(X_f) \cdot H \cdot M_{\lambda}(X_f)$ (which increases the number of $M_{\lambda}(X_f)$ terms by one) in a sequence, and then sum all of these $2p+r$ terms up.
If we repeat the operation for $\ell$ times, then similar to the higher derivatives formula for resolvent in \autoref{lemma:res-higher-der}, it holds that
\begin{equation} \label{eq:partial-l}
        \partial^\ell_{X_i}\Big(\prod_{j=1}^m F_j \big(M_\lambda(X_f),Z_i \otimes 1\big) X_{i,j|S}\Big)(X_{i,1},\ldots, X_{i,\ell}) = \sum_{\pi \in \mathcal P_{\ell}}\sum_{\alpha=1}^{{2p+r+\ell-1 \choose \ell}}W_{\alpha,\pi,i},
\end{equation}
where each $W_{\alpha,\pi,i}$ is a monomial of degree $2p+r+\ell$ in $M_\lambda(X_f)$, of degree $r-m$ in $Z_i \otimes 1$, of degree $2$ in each of the $X_{i,j}$ (one appearance from $X_{i,j|S}$'s and the other from the substitution of the partial derivative $\partial^{\ell}_{X_i}(\cdot)(X_{i,1}, \ldots, X_{i,\ell})$), and of degree $m-\ell$ in $X_{i,\overline{S}}$. 
Moreover, the positions of the $X_{i,1}, \ldots, X_{i,\ell}$ from the derivative substitutions are determined by the permutation $\pi$.

Then, we apply the infinite dimensional trace inequality \autoref{proposition:trace-holder} to each $W_{\alpha,\pi,i}$ so that
\[
\bigg| \sum_{i=1}^n \E \big[\tot(W_{\alpha, \pi,i}) \big] \bigg| \leq \sigma^2(2\rho)^{r+\ell-2} \cdot \tot\big(M_\lambda(X_f)^{2p+r+\ell}\big),
\]
where we used the free independence between $X_{i,j}, X_{i,\overline{S}}$ and $X_f$ and the total degree of $M_{\lambda}(X_f)$ in $W_{\alpha,\pi,i}$ is $2p+r+\ell$.

Now, we count how many $W_{\alpha, \pi,i}$ terms are there in \eqref{eq:partial-l}.
Similar to what happened in the proof of \autoref{lemma:mu-taylor-bound}, a naive bound would be $\ell! \cdot \binom{2p+r+\ell-1}{\ell}$.
But, whenever the $2\ell$ appearances of $X_{i,j}$ ($j \in [\ell]$) have some crossings, $\varphi[W_{\alpha, \pi, i}] = 0$.
As there are at most $\binom{2\ell}{\ell}$ non-crossing partitions over $2\ell$ elements, 
we have a better bound $\binom{2p+r+\ell-1}{\ell} \binom{2\ell}{\ell}$.
Thus, we have established the claim in \eqref{eq:mixed-res-moment-bound} to control the expectation of each of the $r! \cdot \binom{2p+r-1}{r}$ monomials in the polynomial $D^{r}\big(M_\lambda(X_f)^{2p}\big)\big((Z_i \otimes 1)^{(r-m)},X_i^{(m)}\big)$.
Therefore, 
\begin{eqnarray}
& & \bigg| \frac{1}{r!} \sum_{i=1}^n\E \Big[ \tot \Big(D^r \big(M_\lambda(X_f)^{2p}\big) \big( (Z_i \otimes 1)^{(r-m)},X_i^{(m)}\big) \Big) \Big] \bigg| \nonumber 
\\ 
& \leq &  {2p+r-1\choose r}\sigma^{2}\sum_{\ell=0}^m(2\rho)^{r+\ell-2}{m\choose \ell}{2p+r+\ell-1 \choose \ell}{2\ell \choose \ell} \cdot \tot\big(M_\lambda(X_f)^{2p+r+\ell}\big) \nonumber 
\\
& \leq & \frac{\sigma^2}{\eps}\sum_{\ell=0}^m\bigg(\frac{2\rho}{\eps}\bigg)^{r+\ell-2}{m\choose \ell}{2p+r+\ell-1 \choose r+\ell}{r+\ell\choose r}\cdot 4^{\ell} \cdot \tot\big(M_\lambda(X_f)^{2p+1}\big), \label{eq:rterm-bound}
\end{eqnarray}
where in the last line, we have used ${2\ell\choose \ell}\leq 4^\ell$, the identity that $ \binom{2p+r+\ell-1}{\ell}\binom{2p+r-1}{r} = \binom{2p+r+\ell-1}{r+\ell}\binom{r+\ell}{r}$, and the fact that $\|M_{\lambda}(X_f)\| \leq 1/\eps$ when $\lambda > \lambda_{\max}(X_f) + \eps$. 
    
We are ready to bound the higher-order error term in the statement of \autoref{lemma:rhobound}.
Since all $Z_i$'s have finite support, the expectation commutes with the infinite sum. Combining with \eqref{eq:trace-derivative}, the higher-order error is bounded by
\begin{eqnarray*}
R_{\geq 3}
& = & \frac{1}{n} \bigg| \sum_{i=1}^n \E \bigg[ \sum_{r\geq 3} \frac{1}{r!} \tot\Big(D^r \big(M_\lambda(X_f)^{2p}\big) \big((Z_i \otimes 1) -X_i, \ldots, (Z_i \otimes 1) - X_i \big)\Big) \bigg] \bigg| 
\\
&\leq & \frac{1}{n} \sum_{r\geq 3} \sum_{m=0}^r{r\choose m} \bigg| \frac{1}{r!} \sum_{i=1}^n \E \Big[ \tot \Big(D^{r} \big(M_\lambda(X_f)^{2p}\big) \big((Z_i \otimes 1)^{(r-m)},X_i^{(m)}\big)\Big) \Big] \bigg|.
\end{eqnarray*}
Plugging in \eqref{eq:rterm-bound} and doing a change of variable $k \gets r+\ell$ (note that $m \in [k-r,r]$ after the change of variable), we have
\begin{eqnarray*}
R_{\geq 3} & \leq & 
\frac{\sigma^2}{n \eps}\sum_{r\geq 3} \sum_{m=0}^{r} {r\choose m}\sum_{\ell=0}^m\bigg(\frac{2\rho}{\eps}\bigg)^{r+\ell-2}{m\choose \ell}{2p+r+\ell-1 \choose r+\ell}{r+\ell\choose r}\cdot 4^{\ell} \cdot \tot\big(M_\lambda(X_f)^{2p+1}\big)
\\
& = & \frac{\sigma^2}{n \eps}\tot\big(M_\lambda(X_f)^{2p+1}\big)\sum_{k\geq 3}{2p+k-1\choose k}\bigg(\frac{2\rho}{\eps}\bigg)^{k-2}\sum_{r\leq k}\sum_{k-r\leq m\leq r}{k\choose r}{r\choose m}{m\choose k-r}\cdot 4^{k-r}.
\end{eqnarray*}
Observe that 
\[
\sum_{r\leq k}\sum_{k-r\leq m\leq r}{k\choose r}{r\choose m}{m\choose k-r} \leq 4^{k}
\]
since the LHS is a particular way of partitioning $k$ elements into 4 groups.
Thus, it follows that
\[
R_{\geq 3}
\leq \frac{\sigma^2}{n\eps}\tot\big(M_\lambda(X_f)^{2p+1}\big)\sum_{k\geq 3}{2p+k-1\choose k}\bigg(\frac{2\rho}{\eps}\bigg)^{k-2}\cdot 16^{k}.
\]
Finally, note that ${2p+k-1\choose k} \leq (2p)^k$. 
Hence, for $\eps \gtrsim p \rho$ in the assumption of \autoref{lemma:rhobound}, 
the sequence ${2p+k-1\choose k}(\frac{2\rho}{\eps})^{k-2}\cdot 16^{k}$ decreases geometrically with a rate less than $1/2$.
Therefore, the sum is dominated by the $k=3$ term, which implies that
\[
R_{\geq 3} \lesssim \frac{p^3\sigma^2\rho}{n \eps^2} \cdot \tot\big(M_\lambda(X_f)^{2p+1}\big). \qedhere
\]
\end{proofof}

%%%%%%%%%%%%%%%%%%%%%%%%%%%%%%%%%%%%%%%%%%%%%%%%%%

\subsection{An Infinite Dimensional Trace Inequality} \label{a:trace-inequality}

Brailovskaya and van Handel~\cite[Proposition 5.1]{BvH24} proved a trace inequality to control a higher-order quantity of the following form
\begin{equation} \label{eq:bvh24-trace}
    \bigg| \sum_{i=1}^n \E[ \tr(A_{i1} B_1 A_{i2} B_2 \cdots A_{ik} B_k)] \bigg|,
\end{equation}
where $A_{ij}$'s are a family of (possibly dependent) random matrices with identical distribution for each fixed $i$, and $B_1, \ldots, B_k$ is another family of (possibly dependent) random matrices that are independent from $A_{ij}$'s.

As mentioned in \autoref{s:overview}, when analyzing the swapping algorithm for derandomizing the general random matrix model, we need to bound the expected change of the potential function under a random swap perturbation of the form $Z_i \otimes 1 - X_i$, where $Z_i$ is a random matrix and $X_i$ is an infinite dimensional semicircular matrix.
The expectation of higher-order error terms under the perturbation is in a similar form as in \eqref{eq:bvh24-trace}, but involves infinite dimensional semicircular matrix $X_i$.
Therefore, we need to derive an infinite dimensional variant of the trace inequality in~\cite[Proposition~5.1]{BvH24} (see \autoref{ss:outline-uni} for more details on the motivations).

\begin{proposition*}[Restatement of the Infinite Dimensional Trace Inequality in \autoref{proposition:trace-holder}]
Let $Z_1, \ldots, Z_n$ be $d$-dimensional centered self-adjoint random matrices such that $\max_i \|Z_i\| \leq \rho$ with probability one and $\|\sum_{i=1}^n \E[Z_i^2]\| = \sigma^2$.
Let $X_1, \ldots, X_n \in \M_d(\C) \otimes \A$ be centered semicircular matrices where each $X_i$ has the same covariance profile as $Z_i$ (i.e., satisfying \eqref{eq:cov-profile}).    
    
Let $m>2$, and $Y_1, \ldots, Y_m \in \M_d(\C)\otimes \A$, be a family of (not necessarily free) operators.
Suppose there exist integers $k_1, \ldots, k_m \geq 0$, and positive, self adjoint operator $Y\in \M_d(\C)\otimes \A$, such that  $\sum_{i=1}^m k_i = p$ and $|Y_j| \preccurlyeq Y^{k_j}$ for each $j \in [m]$.

For each $i\in[n]$ and $j\in[m]$, let $Z_{i,j}$ be a
self-adjoint operator. For each $j\in[m]$, suppose that either
$Z_{i,j}=Z_i\otimes1$ for every $i\in[n]$, or
$Z_{i,j}$ has the same distribution as $X_i$ and is free from
$Y_1,\ldots,Y_m$ for every $i\in[n]$.
Then, it holds that
\[
\bigg| \sum_{i=1}^n \E \big[ \tr\otimes\tau(Z_{i,1}Y_1\cdots Z_{i,m}Y_m) \big] \bigg| 
\leq \sigma^2 \cdot (2\rho)^{m-2} \cdot \tr\otimes\tau(Y^{p}),
\]
where the expectation is taken over those $Z_{i,j}$'s that are equal to $Z_i\otimes 1$.
\end{proposition*}

Similar to the proof of the generalized H\"older's inequality in \autoref{corollary:general-holder},
we will first prove a finite dimensional version (\autoref{lemma:finite-dim-trace-holder}) and then take the dimension to infinity to prove \autoref{proposition:trace-holder}.
Our proof for the finite dimensional version basically follows the same framework in the proof of Proposition 5.1 in~\cite{BvH24}, which will be presented in the following section.

\begin{lemma}[Variant of Proposition 5.1 in~\cite{BvH24}] \label{lemma:finite-dim-trace-holder}
Let $Y_1, \ldots, Y_m \in \M_d(\C)$, $m>2$, be Hermitian random matrices and $Z_{i,1}, \ldots, Z_{i,m}$ for $i \in [n]$ be another family of Hermitian random matrices that are independent from $Y_1, \ldots, Y_m$. Let $\eps \in (0,1)$. Define the matrix parameters:
\[
\sigma_\eps^2 = \max_{j\in [m]}\biggnorm{\sum_{i=1}^n \E\big[|Z_{i,j}|^{\frac{2}{1-\eps}}\big]}^{1-\eps} 
\quad \text{and} \quad 
\rho_\eps = \max_{\substack{i\in [n]},j\in [m]} \E \Big[\tr\Big(|Z_{i,j}|^{\frac{2(m-2)}{\eps}}\Big)\Big]^{\frac{\eps}{2(m-2)}}.
\]
Then, for all $m > 2$, and $p_1,p_2, \ldots, p_m\geq 1$ such that $\sum_{i=1}^m \frac{1}{p_i} = 1-\eps$, 
\[
\bigg| \sum_{i=1}^n \E\tr (Z_{i,1}Y_1 \cdots Z_{i,m}Y_m ) \bigg| 
\leq n^{\eps} \cdot \sigma_\eps^2 \cdot \rho_\eps^{m-2} \cdot \norm{Y_1}_{p_1} \cdots \norm{Y_{m}}_{p_m}.
\]
\end{lemma}
\begin{remark*}
This is almost the same as \cite[Proposition 5.1]{BvH24}, except that it has no assumption that $Z_{i,1},\ldots,Z_{i,m}$ are identically distributed. Instead, we just take a uniform bound on all of the relevant spectral statistics, so that it can be used to prove \autoref{proposition:trace-holder}.
\end{remark*}

Now we extend this finite dimensional bound to the infinite dimensional case.

\begin{proofof}{\autoref{proposition:trace-holder}}
First, assume that each $Y_j$ is a polynomial of semicircular elements, i.e. $Y_j\in M_d(\C)\langle s_1,\ldots s_n\rangle$. Let $\A_N$ be the algebra generated by an arbitrarily large set of independent $N\times N$ GOE matrices $\{G^N_i\}_{i=1}^{n'}$ that are also independent from $Z_1,\ldots, Z_n$. 
Let $\big(Y_j^{N}\big)_{j=1}^m \in \M_d(\C) \langle G_1^N,\ldots G_{n'}^N\rangle$ be $dN\times dN$ random matrices that converge in distribution to $Y_1, \ldots, Y_m$.
Let $\big(Z_{i,j}^N\big)_{i \in [n],j \in [m]} \in \M_d(\C) \otimes \A_N$ be $dN\times dN$ random matrices that converge in distribution to either $Z_i \otimes 1$ or $X_i$.
Specifically, let $S_F = \{j \mid Z_{i,j} \overset{d}{=} X_i\;\forall i \in [n] \}$. 
Then we have the following:
\begin{itemize}
\item If $j\in S_F$, then $Z_{i,j}^N \in \M_d(\C)\otimes \A_N$ is a Gaussian random matrix satisfying $Z^N_{i,j} \overset{d}{\rightarrow} X_i$ for all $i \in [n]$, and they are independent from $Y_1^N, \ldots, Y_m^N$;
\item If $j\notin S_F$,  then $Z^N_{i,j} = Z_i\otimes I_N$. 
\end{itemize}
We will also define the following matrix parameters: For each $\eps \in (0,1)$,
\begin{align*}
\sigma_{\eps,N}^2 = \max_{j\in [m]}&\biggnorm{\sum_{i=1}^n \E\big[|Z^N_{i,j}|^{\frac{2}{1-\eps}}\big]}^{1-\eps}, 
\hspace{2cm}\rho_{\eps,N} = \max_{i\in [n],j\in [m]}\E\tr\Big(|Z_{i,j}^N|^{\frac{2(m-2)}{\eps}}\Big)^{\frac{\eps}{2(m-2)}}, 
\\
&\sigma_{\eps,\infty}^2 = \max\bigg\{\biggnorm{\sum_{i=1}^n \E\big[|Z_{i}|^{\frac{2}{1-\eps}}\big]}^{1-\eps},\biggnorm{\sum_{i=1}^n\varphi\big[|X_{i}|^{\frac{2}{1-\eps}}\big]}^{1-\eps}\bigg\}, 
\\
&\rho_{\eps,\infty} = \max_{i\in [n]}\Big\{\max\Big\{\E\tr\big(|Z_{i}|^{\frac{2(m-2)}{\eps}}\big), \tr\otimes\tau\big(|X_i|^{\frac{2(m-2)}{\eps}}\big)\Big\}\Big\}^{\frac{\eps}{2(m-2)}}.
\end{align*}
Since Gaussian matrices have uniformly bounded moments, the weak convergence in \autoref{cor:cond-convergence-cont} and \autoref{lem:weak-converge-cont} imply that $\sigma_{\eps,N}\rightarrow \sigma_{\eps,\infty}$ and $\rho_{\eps,N}\rightarrow \rho_{\eps,\infty}$ as $N\rightarrow\infty$. 
In addition, since $\norm{Z_i}\leq \rho$, it holds that $\norm{X_i}\leq 2\rho$ by \autoref{lemma:semicircular-norm}, which implies that $\rho_{\eps,\infty}\leq 2\rho$ for all $\eps \in (0,1)$. 
Moreover, we have
\[
\bigg| \sum_{i=1}^n \E\tr\otimes\tau (Y_1Z_{i,1}\cdots Y_mZ_{i,m}) \bigg| = \lim_{N\rightarrow\infty} \bigg| \sum_{i=1}^n \E\tr(Y_1^NZ_{i,1}^N\cdots Y_m^NZ_{i,m}^N) \bigg|.
\]
Now, we apply \autoref{lemma:finite-dim-trace-holder} with $p_j = \frac{p}{k_j(1-\eps)}$ (note that we may assume $k_j > 0$ wlog as otherwise we can ignore $Y_j$) so that $\sum_j\frac{1}{p_j} = \sum_{j}\frac{k_j(1-\eps)}{p} = 1-\eps$, which gives
\begin{eqnarray*}
\bigg| \sum_{i=1}^n \E\tr\otimes\tau (Y_1Z_{i,1}\cdots Y_mZ_{i,m}) \bigg| 
& \leq & \lim_{N\rightarrow\infty} n^{\eps} \cdot \sigma_{\eps,N}^2 \cdot \rho_{\eps,N}^{m-2} \cdot \prod_{j=1}^m\E\tr\Big[\big|Y^N_j\big|^{\frac{p}{k_j(1-\eps)}}\Big]^{\frac{k_j(1-\eps)}{p}}
\\
& = & n^{\eps} \cdot \sigma_{\eps,\infty}^2 \cdot \rho_{\eps,\infty}^{m-2} \cdot \prod_{j=1}^m\tot\Big(\big|Y_j\big|^{\frac{p}{k_j(1-\eps)}}\Big)^{\frac{k_j(1-\eps)}{p}}
\end{eqnarray*} 
By taking polynomial limits, the above inequality immediately extends to all general $Y_j \in M_d(\C)\otimes \A$. Moreover, since the inequality holds for all $\eps\in (0,1)$, we can take the limit as $\eps \rightarrow0$. In particular, note that $\varphi[X_i^2] = \E[Z_i^2]$ as $X_i$ and $Z_i$ have the same covariance. Thus, we have $n^{\eps}\sigma_{\eps,\infty}^2\rightarrow \sigma^2$ as $\eps \rightarrow 0$, and we conclude that
\[
\bigg| \sum_{i=1}^n \E\tr\otimes\tau (Y_1Z_{i,1}\cdots Y_mZ_{i,m}) \bigg| 
\leq \sigma^2 \cdot (2\rho)^{m-2} \cdot \tot(Y^{p}),
\]
where we have used $|Y_j|\preceq Y^{k_j}$ for all $j$.
\end{proofof}

%%%%%%%%%%%%%%%%%%%%%%%%%%%%%%%%%%%%%%%%%%%%%%%%%%%%%%%%

\subsubsection*{Proof of \autoref{lemma:finite-dim-trace-holder}}

Our proof follows the same framework as in~\cite[Proposition~5.1]{BvH24}, 
which requires the following known facts.
The first one is a variant of the Riesz-Thorin interpolation theorem.
\begin{lemma}[See {\cite[Lemma 5.2]{BvH24}}]\label{lemma:log-convex}
Let $F$ be a multilinear functional that maps $d\times d$ random matrices with finite moments of all orders to $\C$. 
Then, the map
\[
\bigg(\frac{1}{p_1},\frac{1}{p_2},\ldots, \frac{1}{p_m}\bigg) \mapsto \log\sup_{\substack{Y_1, \ldots, Y_m\in L^+_\infty(\M_d(C))\\Y_1, \ldots, Y_m\neq 0}}\frac{|F(Y_1,Y_2,\ldots,Y_m)|}{\norm{Y_1}_{p_1}\norm{Y_2}_{p_2} \cdots \norm{Y_m}_{p_m}}
\]
is convex on $[0,1]^m$, where $\norm{Y}_{p} = \E[\tr(|Y|^p)]^{\frac{1}{p}}$.
\end{lemma}

The second one is the Lieb-Thirring inequality.
\begin{lemma}[See {\cite[Lemma 5.4]{BvH24}}] \label{lemma:leib-thirring}
Let $A,B$ be PSD matrices. Then, for all $p\geq 1$,
\[
\tr ((ABA)^{p}) \leq \tr(A^pB^pA^p).
\]
\end{lemma}

\begin{proofof}{\autoref{lemma:finite-dim-trace-holder}}
Let $\boldsymbol{i}$ be an index in $[n]$ selected uniformly at random. 
We define the new random matrices $(\bZ_1,\ldots,\bZ_m)$ so that $\bZ_j = Z_{\boldsymbol{i},j}$ for all $j\in [m]$. 
Then, it suffices to show that
\[
\E \tr (Y_1\bZ_1 \cdots Y_m\bZ_m ) 
\leq \frac{1}{n^{1-\eps}} \cdot \sigma_\eps^2 \cdot \rho_\eps^{m-2} \cdot \prod_{j=1}^m \norm{Y_j}_{p_j},
\]
where the expectation is taken over both $\boldsymbol{i}$ and the $Z_{i,j}$ matrices. 
Consider the linear functional
\[
F(Y_1,Y_2, \ldots, Y_m) := \E\tr(\bZ_1Y_1 \cdots \bZ_mY_m).
\]
It suffices to show that
\[
\sup_{Y_1,\ldots,Y_m\neq 0}\frac{|F(Y_1,\ldots,Y_m)|}{\norm{Y_1}_{p_1}\cdots \norm{Y_m}_{p_m}} 
\leq \frac{\sigma_\eps^2 \cdot \rho_\eps^{m-2}}{n^{1-\eps}}.
\]
By \autoref{lemma:log-convex}, the LHS is log-convex on the simplex $\big\{\frac{1}{p_1} + \cdots + \frac{1}{p_m} = 1-\eps\big\}$. 
Thus, it is maximized when there is some $j^*\in[m]$ such that $p_{j^*} = \frac{1}{1-\eps}$ and $p_{j} = \infty$ for $j\neq j^*$. 
Moreover, by cyclic invariance of trace, we can assume without loss that $j^* = 1$. 
Thus, it suffices to show that
\[
\sup_{\norm{Y_1}_{\frac{1}{1-\eps}} = 1,\;\norm{Y_2}_\infty= \ldots \norm{Y_m}_\infty=1} |F(Y_1, \ldots, Y_m)|
\leq \frac{\sigma_\eps^2 \cdot \rho_\eps^{m-2}}{n^{1-\eps}}.
\]
We let $Y_1, \ldots, Y_m$ be any set of random matrices satisfying $\norm{Y_1}_{\frac{1}{1-\eps}} = \norm{Y_2}_\infty = \cdots = \norm{Y_m}_\infty = 1$. Let $Y_1 = V|Y_1|$ be the polar decomposition of $Y_1$. 
Thus, we have
\begin{eqnarray*}
|F(Y_1, \ldots, Y_m)| 
&=& |\E\tr(\bZ_1Y_1\bZ_2Y_2 \cdots \bZ_mY_m)|
\\
&=& |\E\tr(\bZ_1V|Y_1|\bZ_2Y_2 \cdots \bZ_mY_m)|
\\
&\leq& \E[\tr(\bZ_1V|Y_1|V^*\bZ_1)]^{\frac12} \cdot \E[\tr(Y_m^*\bZ_m\cdots Y_2^*\bZ_2|Y_1|\bZ_2Y_2\cdots Y_m\bZ_m)]^{\frac12}.
\end{eqnarray*}
where the last inequality follows by applying Cauchy-Schwarz. 
Now we bound each of the above terms individually. 
Starting with the left term, we apply the polar decomposition $\bZ_1 = U|\bZ_1|$. 
Since $\bZ_1$ is self adjoint, $U$ is also self adjoint and commutes with $|\bZ_1|$. 
Thus,
\begin{eqnarray*}
\E[\tr(\bZ_1V|Y_1|V^*\bZ_1)]^{\frac12} 
&=& \E[\tr(|\bZ_1|UV|Y_1|V^*U|\bZ_1|)]^{\frac12}
\\
&\leq& \E\Big[\tr\big((|\bZ_1|UV|Y_1|V^*U|\bZ_1|)^\frac{1}{1-\eps}\big)\Big]^{\frac{1-\eps}{2}}
\\
&\leq& \E\big[\tr\big(|\bZ_1|^{\frac{1}{1-\eps}}UV|Y_1|^{\frac{1}{1-\eps}}V^*U|\bZ_1|^{\frac{1}{1-\eps}}\big)\big]^{\frac{1-\eps}{2}}
\\
&=& \E\big[\tr\big(|\bZ_1|^{\frac{2}{1-\eps}}V|Y_1|^{\frac{1}{1-\eps}}V^*\big)\big]^{\frac{1-\eps}{2}}
\\
&=& \big(\tr\big(\E\big[|\bZ_1|^{\frac{2}{1-\eps}}\big]\E\big[V|Y_1|^{\frac{1}{1-\eps}}V^*\big]\big)\big)^{\frac{1-\eps}{2}},
\end{eqnarray*}
where the second line is by Jensen's inequality and the fact that the matrix inside the $\E\tr$ operator is positive semidefinite, 
the third line is by Lieb-Thirring (\autoref{lemma:leib-thirring}), 
the fourth line is by the fact that $U$ commutes with $\bZ_1$ and $U^2 = I$, 
and the last line is by the fact that $\bZ_1$ is independent from $Y_1$. 
Note that the matrix $\E[V|Y_1|^\frac{1}{1-\eps}V^*]$ is positive semidefinite and has trace 1, because
\[
\tr\big(\E\big[V|Y_1|^{\frac{1}{1-\eps}}V^*\big]\big) 
= \E\tr\big(V|Y_1|^\frac{1}{1-\eps}V^*\big) 
= \E \tr\big(|Y_1|^{\frac{1}{1-\eps}}\big) = 1.
\]
This implies that
\[
\tr\big(\E\big[|\bZ_1|^{\frac{2}{1-\eps}}\big] \E\big[V|Y_1|^{\frac{1}{1-\eps}}V^*\big]\big)^{\frac{1-\eps}{2}} 
\leq \Bignorm{\E\big[|\bZ_1|^{\frac{2}{1-\eps}}\big]}^{\frac{1-\eps}{2}} 
= \biggnorm{\frac{1}{n}\sum_{i=1}^n\E\big[|Z_{i,1}|^{\frac{2}{1-\eps}}\big]}^{\frac{1-\eps}{2}} 
\leq \frac{\sigma_\eps}{\sqrt{n^{1-\eps}}}.
\]
Next, we proceed with bounding the term on the right side. 
Here, we will apply the matrix H\"older's inequality in \autoref{lemma:matrix-holder}. 
For each $j > 1$, we give the power $\infty$ to each occurrence of $Y_j$. 
For each $j > 2$, we give the power $\frac{2(m-2)}{\eps}$ to each occurrence of $\bZ_j$. 
Finally, we give the power $\frac{1}{1-\eps}$ to $\bZ_2|Y_1|\bZ_2$. 
In total, there are $2(m-2)$ terms of the form $\bZ_j$ for $j > 2$. Thus, we can verify that the sum of the inverse powers satisfies:
\[
2(m-2)\cdot\frac{\eps}{2(m-2)} + 1-\eps = 1.
\]
Applying the H\"older bound then gives
\[
\E\tr(Y_m^*\bZ_m\cdots Y_2^*\bZ_2|Y_1|\bZ_2Y_2\cdots Y_m\bZ_m)^{\frac12} 
\leq \bigg(\prod_{j>2}\big(\E\tr\big[|\bZ_j|^{\frac{2(m-2)}{\eps}}\big]^{\frac{\eps}{2(m-2)}}\big)^2\cdot \norm{Y_j}_\infty^2\E\tr\big[(\bZ_2|Y_1|\bZ_2)^{\frac{1}{1-\eps}}\big]^{1-\eps}\bigg)^{\frac12}.
\]
By our hypothesis, $\norm{Y_j}_\infty = 1$. 
Also, by definition, $\E[\tr(|\bZ_j|^{\frac{2(m-2)}{\eps}})]^{\frac{\eps}{2(m-2)}} \leq \rho_\eps$ for each $j \in [m]$. This means
\[
\bigg(\prod_{j>2}\big(\E\tr\big[|\bZ_j|^{\frac{2(m-2)}{\eps}}\big]^{\frac{\eps}{2(m-2)}}\big)^2\cdot \norm{Y_j}_\infty^2\bigg)^{\frac12} \leq  \rho_{\eps}^{m-2}.
\]
By applying Lieb-Thirring and polar decomposition as in when bounding the left term, 
we also have 
\[
\E\tr\big[(\bZ_2|Y_1|\bZ_2)^{\frac{1}{1-\eps}}\big]^{\frac{1-\eps}{2}} 
\leq \biggnorm{\frac{1}{n}\sum_{i=1}^n\E\big[|Z_{i,2}|^{\frac{2}{1-\eps}}\big]}^{\frac{1-\eps}{2}} 
\leq \frac{\sigma_\eps}{\sqrt{n^{1-\eps}}}.
\]
Putting everything together, we conclude that
\[
|F(Y_1, \ldots, Y_m)| \leq \frac{\sigma_\eps^2 \rho_\eps^{m-2}}{n^{1-\eps}}.
\]
\end{proofof}

\newpage

\bibliographystyle{alpha}
\bibliography{ref}

\end{document}